\documentclass[11pt,letterpaper]{article}
\usepackage{jheppub}
\usepackage{amsmath,amssymb,amsfonts,bm}
\usepackage{graphicx,wrapfig}
\usepackage{multirow}
\usepackage{verbatim}
\usepackage{appendix}
\usepackage{rotating}

\graphicspath{{./figures/}}

\numberwithin{equation}{section}
\renewcommand\afterTocRuleSpace{\bigskip}

\newcommand{\fs}{{\mathfrak s}}
\newcommand{\fc}{{\mathfrak c}}

\newcommand{\bea}{\begin{eqnarray}}
\newcommand{\eea}{\end{eqnarray}}
\newcommand{\be}{\begin{equation}}
\newcommand{\ee}{\end{equation}}
\newcommand{\bse}{\begin{subequations}}
\newcommand{\ese}{\end{subequations}}
\newcommand{\mb}{\mathbf}

\newtheorem{theorem}{Theorem}

\newcommand{\wt}{\widetilde}

\newcommand{\ol}{\overline}
\newcommand{\ds}{\displaystyle}

\newcommand{\eg}{\emph{e.g.}}
\newcommand{\ie}{\emph{i.e.}}
\newcommand{\cf}{\emph{cf.}}

\newcommand{\Z}{{\mathbb Z}}
\newcommand{\R}{{\mathbb R}}
\newcommand{\C}{{\mathbb C}}

\newcommand{\cp}{{\mathbb{CP}}}

\newcommand{\Tr}{{\rm Tr \,}}
\renewcommand{\Re}{{\rm Re}}
\renewcommand{\Im}{{\rm Im}}
\newcommand{\bs}{\backslash}
\newcommand{\pd}{\partial}

\newcommand{\CA}{\mathcal{A}}
\newcommand{\CB}{\mathcal{B}}
\newcommand{\CC}{\mathcal{C}}
\newcommand{\CD}{\mathcal{D}}

\newcommand{\CK}{\mathcal{K}}
\newcommand{\CL}{\mathcal{L}}

\newcommand{\CN}{\mathcal{N}}
\newcommand{\CO}{\mathcal{O}}
\newcommand{\CP}{\mathcal{P}}

\newcommand{\CR}{\mathcal{R}}
\newcommand{\CS}{\mathcal{S}}
\newcommand{\CT}{\mathcal{T}}

\newcommand{\CX}{\mathcal{X}}

\newcommand{\CZ}{\mathcal{Z}}

\title{RG Domain Walls and Hybrid Triangulations}

\author[1]{Tudor Dimofte}
\author[1,2]{Davide Gaiotto}
\author[3]{Roland van der Veen}

\affiliation[1]{Institute for Advanced Study, Einstein Dr., Princeton, NJ 08540, USA}
\affiliation[2]{Perimeter Institute for Theoretical Physics, %31 Caroline St. N., 
Waterloo, Ontario, Canada N2L 2Y5}
\affiliation[3]{Korteweg-de Vries Institute for Mathematics, University of Amsterdam, P.O. Box 94248, 1090 GE Amsterdam, The Netherlands}

\abstract{This paper studies the interplay between the $\CN=2$ gauge theories in three and four dimensions that have a geometric description in terms of twisted compactification of the six-dimensional (2,0) SCFT. Our main goal is to construct the three-dimensional domain walls associated to any three-dimensional cobordism. We find that we can build a variety of 3d theories that represent the local degrees of freedom at a given domain wall in various 4d duality frames, including both UV S-dual frames and IR Seiberg-Witten electric-magnetic dual frames. We pay special attention to Janus domain walls, defined by four-dimensional Lagrangians with position-dependent couplings. If the couplings on either side of the wall are weak in different UV duality frames, Janus domain walls reduce to S-duality walls, \ie\ domain walls that encode the properties of UV dualities. If the couplings on one side are weak in the IR and on the other weak in the UV, Janus domain walls reduce to RG walls, \ie\ domain walls that encode the properties of RG flows. We derive the 3d geometries associated to both types of domain wall, and test their properties in simple examples, both through basic field-theoretic considerations and via comparison with quantum Teichm\"uller theory. Our main mathematical tool is a parametrization and quantization of framed flat $SL(K)$ connections on these geometries based on ideal triangulations.
}

\begin{document}

\maketitle

%%%%%%%%%%%%%%%%%%%%%%%%%%%%%%%%%%%%%%%%%%%%%%%%%%%%%%%%%%%%%%%%%%%%%%%%%%

\section{Introduction}
\label{sec:intro}

Over the last few years, a wealth of new results and insights about supersymmetric gauge theories in three and four dimensions 
have followed from two main directions of research. Protected quantities such as indices and correlation functions in curved space were computed 
exactly, testing known dualities, and predicting new ones. Moreover, the twisted compactification of six-dimensional $(2,0)$ superconformal field theories
provided an organizing principle for very large classes of lower dimensional field theories and their BPS defects and operators.
The interplay between these two ideas is striking: protected calculations in a lower dimensional field theory often  
manifest the theory's six-dimensional origin. The prominent example is the twisted compactification of the six-dimensional theories on a punctured Riemann surface, which produces 
four dimensional theories in ``class ${\cal S}$'' \cite{Gaiotto-dualities, GMN}. The superconformal index and the sphere partition function 
of a class ${\cal S}$ theory take the form of two-dimensional correlation functions on the Riemann surface \cite{AGT, Rastelli-2dQFT}. 

The situation in three dimensions is somewhat analogous. It is possible to give an algebraic definition of a vast class of three-dimensional ${\cal N}=2$ SCFT's
built in such a way that protected quantities have a natural interpretation in terms of the geometry of an auxiliary three-manifold. 
These SCFTs have UV Lagrangian descriptions as abelian Chern-Simons-Matter theories deformed by superpotential terms, 
which may contain monopole operators \cite{DGG, DGG-Kdec}. 
The properties of such theories, which we will denote as ``class ${\cal R}$,'' loosely match what one could expect to 
obtain from the twisted compactification of the six-dimensional $A_{K-1}$ theories on three-manifolds with networks of line defects.%
\footnote{The class $\CR$ of abelian Chern-Simons-Matter theories was defined purely algebraically in \cite{DGG, DGG-index}, with no reference to three-dimensional geometries or Lie algebras. It was then shown (also in \cite{DGG-Kdec}) that for Lie algebras $\mathfrak g=A_{K-1}$ one can construct three-manifold theories that belong to this class.} %
The expectation was reinforced by the M-theory analysis of \cite{CCV}, which explains how the abelian CSM structure emerges in the IR.

However, in order to give a precise six-dimensional construction of a given class $\CR$ theory, several subtleties remained to be addressed. 
In this paper, we will attempt to clarify them. Our motivation is not just to dot all i's and cross all t's.  
The three-dimensional theories are expected to define natural boundary conditions and domain walls for the four-dimensional theories. 
In order to make these relations manifest, we find it necessary to first clarify the above-mentioned subtleties. 

The six-dimensional intuition predicts that to every cobordism between two punctured Riemann surfaces, 
one should be able to associate a natural domain wall between the corresponding class ${\cal S}$ theories.
Our main result is that the same domain wall can be given not one but several descriptions as distinct class ${\cal R}$ theories, 
coupled to either a full UV description of the four-dimensional theories, or to an effective IR Seiberg-Witten description of the same theories, or to a hybrid mixture. Different descriptions will be useful in different regions of parameter space, depending on which 
bulk degrees of freedom are weakly coupled.

This includes two special cases. The first, which we expected to find from the beginning, is the class of ``S-duality walls,'' which 
implement the equivalence between different S-dual descriptions of the same four-dimensional theory \cite{GW-Sduality, DGG-defects}. The second, which was unexpected, 
is a class of ``RG walls,'' which implement the relation between a UV and an IR description of the same theory. The discovery of RG walls inspired the related two-dimensional work of~\cite{Gaiotto-2dRG}.
We also sketch a general argument explaining why such domain walls for any ${\cal N}=2$ theory should admit class ${\cal R}$ descriptions, 
with no reference to six dimensions. This can be thought of as a field theory version of the arguments in \cite{CCV}. 

There are also some ancillary mathematical payoffs to our exploration.
Using S-duality walls, we find a precise 3d interpretation of the integral kernels that implement 
the Moore-Seiberg groupoid in Liouville theory, or, equivalently, a mapping-class-group action on quantized Fenchel-Nielsen coordinates of Teichm\"uller theory.
From RG domain walls, we find a 3d
interpretation of the eigenfunctions of geodesic length operators in quantum Teichm\"uller theory \cite{Kash-kernel, Kash-Teich, FockChekhov, Teschner-TeichMod}. For higher-rank Lie algebras $A_{K-1}$, these generalize to operations in Toda theory and higher Teichm\"uller theory \cite{FG-Teich}, respectively.

\subsection{Framed 3-manifolds and the six-dimensional dictionary}
\label{sec:6ddict}

\begin{wrapfigure}{r}{2in}
\centering
\vspace{-.1in}
\includegraphics[width=1.8in]{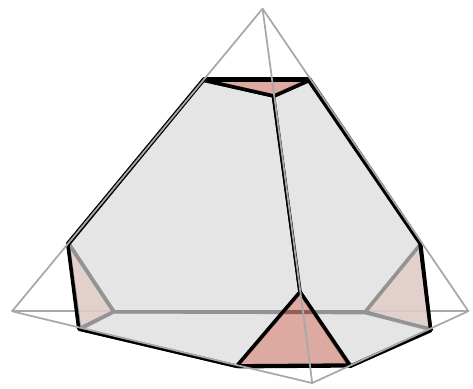}
\caption{Truncated tetrahedron.}
\label{fig:trunc}
\end{wrapfigure}

\begin{figure}[htb]
\centering
\includegraphics[width=5.5in]{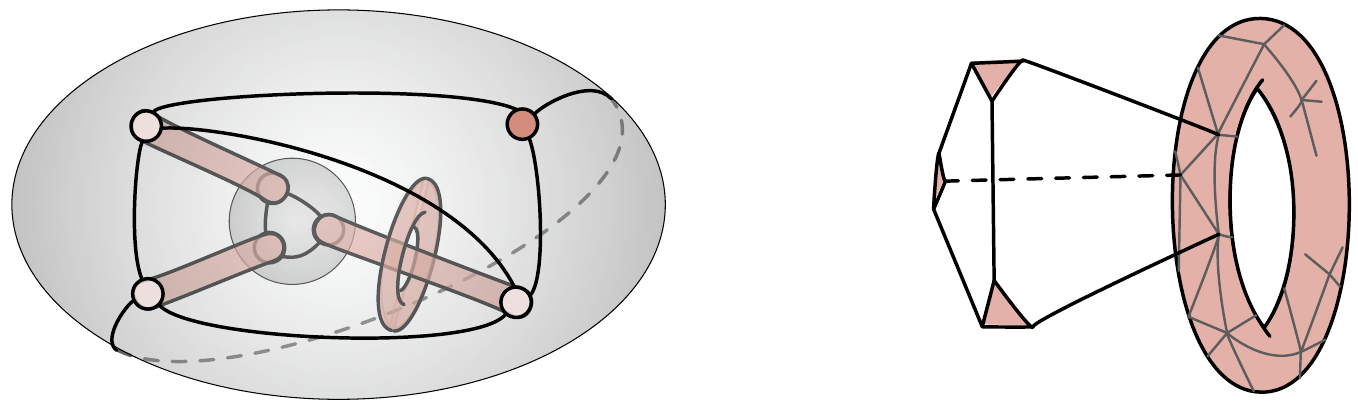}
\caption{Left: a 3-manifold that admits a decomposition into truncated tetrahedra. It has two big boundaries (a 4-punctured sphere and a 3-punctured sphere) with fixed 2d triangulation $\mb t$; as well as a small disc (filling in one puncture), three small annuli, and a small torus. Topologically, this manifold has just two connected boundary components, the torus and a surface of genus two. Right: schematic of how truncated tetrahedra are assembled around a small torus boundary.}
\label{fig:admM}
\end{figure}

The oriented three-manifolds $M$ that label theories in class $\CR$ are assembled by gluing together truncated tetrahedra (Figure \ref{fig:trunc}) \cite{DGG, DGG-Kdec}. Some (not necessarily all) of the big hexagonal faces of tetrahedra are glued together in pairs, while the small triangular faces at the truncated vertices are never glued. Therefore, $M$ acquires two kinds of boundary: a \emph{big} boundary tiled by unglued big hexagons, and a \emph{small} boundary tiled by small triangles. One additionally requires that the small boundary components have the topology of discs, annuli, or tori. Every hole in the big boundary is either filled in by a small disc, or connected to another hole by a small annulus. One such manifold is shown in Figure \ref{fig:admM}.

We say that a three-manifold $M$ whose boundary is separated into big and small pieces this way is \emph{framed}. We will only consider framed three-manifolds in this paper, and will assume that they admit finite decompositions into truncated tetrahedra.%
\footnote{We call our 3-manifolds ``framed'' because they are perfectly suited for the study of ``framed'' flat connections in the sense of Fock and Goncharov \cite{FG-Teich}. This use of ``framing'' should not be confused with a choice of trivialization of the tangent bundle $TM$; the latter will not play an important role here.}

For example, a truncated tetrahedron itself is a framed three-manifold, whose big boundary is a sphere with four holes, and whose small boundary consists topologically of four discs that fill in the holes. Another example is a knot complement, formed by gluing together tetrahedra so that the only remaining boundary is a small torus.

Notice that the big boundary of such a 3-manifold inherits a 2d ideal triangulation: a triangulation whose edges all begin and end at holes, or ``punctures.'' Then the 3d SCFT associated to $M$, for any Lie algebra $\mathfrak g = A_{K-1}$, depends on
\begin{itemize}
\item the 2d ideal triangulation $\mb t$ of the big boundary; and
\item a choice of \emph{polarization} $\Pi$ for all boundaries, compatible with this triangulation.
\end{itemize}
Thus we can denote the 3d theory as $T_K[M;\mb t;\Pi]$. Notably, the theory does \emph{not} depend on the 3d bulk triangulation of $M$, as long as that triangulation is compatible with the big-boundary triangulation $\mb t$. Indeed, different bulk triangulations lead to mirror-symmetric descriptions of $T_K[M;\mb t;\Pi]$ as a class-$\CR$ theory.

Geometrically, $\Pi$ is a polarization for an open subset of the symplectic moduli space of framed flat $PGL(K)$ connections on $\pd M$, which we will review later. For small tori, the polarization is a choice of A and B cycles. For small annuli, the polarization is canonical --- since there is a canonical non-contractible A-cycle on an annulus. For big boundaries $\CC$, however, there is typically an $Sp(2d,\Z)$ worth of choices, where $2d$ is the dimension of the moduli space of flat $PGL(K)$ connections on $\CC$.

One of the central themes of this paper is that \emph{the 3d $\CN=2$ theory $T_K[M,\mb t,\Pi]$ can provide a half-BPS boundary condition for 4d $\CN=2$ theories of class $\CS$ in many different ways}. One of these ways was described%
\footnote{In \cite{DGG} mainly the $\mathfrak g=A_1$ case was discussed. The generalization to $A_{K-1}$ is straightforward following \cite{DGG-Kdec}.} %
in \cite{DGG}, and is closely related to constructions of \cite{CCV}: every big boundary $\CC$ of $M$ can be coupled to a 4d Seiberg-Witten theory $SW_K[\CC]$. The Seiberg-Witten theory is the IR description of the non-abelian theory $T_K[\CC]$ obtained by twisted compactification of the 6d $A_{K-1}$ theory on $\CC$. $SW_K[\CC]$ has gauge group $U(1)^d$, and an $Sp(2d,\Z)$ electric-magnetic duality group.
The gauge fields of $SW_K[\CC]$ can be used to gauge abelian flavor symmetries of $T_K[M,\mb t,\Pi]$, after which the dependence on big-boundary polarization $\Pi$ is erased.%
\footnote{More precisely, different choices of polarization become related by $Sp(2d,\Z)$ electric-magnetic duality transformations of the combined 3d-4d system. We will review how $\mb t$ and $\Pi$ are ``erased'' in Section \ref{sec:bdycouple}.} %
Subsequently, electric BPS hypermultiplets of $SW_K[\CC]$ can be coupled to chiral operators of $T_K[M,\mb t,\Pi]$, after which the dependence on the big-boundary triangulation $\mb t$ is erased.

\begin{figure}[htb]
\centering
\includegraphics[width=5.8in]{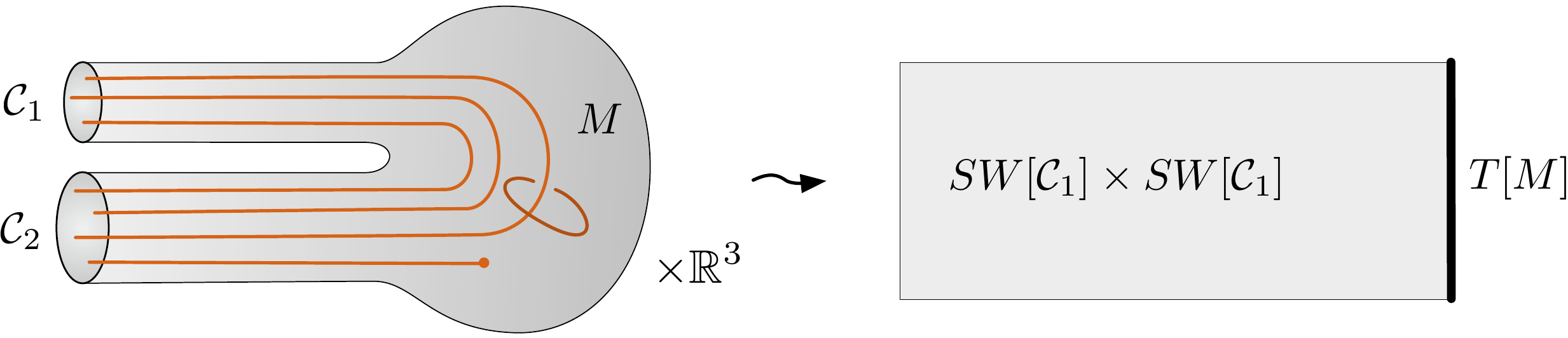}
\caption{Adjusting the metric on the manifold $M$ from Figure \ref{fig:admM} and replacing small boundaries with line defects so that the compactification of the 6d theory on $M\times \R^3$ produces Seiberg-Witten theories on a half-space coupled to $T_K[M]$ as a boundary condition.}
\label{fig:regSW}
\end{figure}

We can reproduce this 3d-4d coupling from six dimensions (Figure \ref{fig:regSW}). To do so, we first choose a metric on $M$ such that all big boundaries $\CC_i$ are pulled out to infinity, forming semi-infinite cylindrical regions that asympotitically take the form $\R_+\times \CC_i$. Then we must ``remove the regulator'' from the small boundaries of $M$. That is, we replace each small boundary with a codimension-two defect of the 6d theory.
Each small disc, filling in a hole on a big boundary $\CC_i$, becomes a semi-infinite regular line defect stretching out into the corresponding cylindrical region. Each small annulus is shrunk to a regular line defect connecting punctures in the cylindrical regions. And each small torus is shrunk to closed regular line defect, in such a way that the A-cycle of the torus becomes an infinitesimally small loop linking the defect.

Upon compactifying the 6d $A_{K-1}$ theory on $M\times \R^3$ with this metric, and flowing to the far infrared, one obtains a (non-interacting) product of Seiberg-Witten theories $\prod_i SW_K[\CC_i]$ on a half-space $\R_+\times \R^3$, all coupled to the theory $T_K[M,\mb t,\Pi]$ on the boundary $\R^3$. The coupling turns out to be precisely the one described above --- we will review how this arises in Sections \ref{sec:couplings}--\ref{sec:Janus}. The full 3d-4d system does not depend on $\mb t$ or on big-boundary polarization~$\Pi$.

Now, for this picture to make sense, two things must be true:
\begin{enumerate}
\item The regular line defects of the 6d $A_{K-1}$ theory must be able to end. (Otherwise the dictionary between small discs and semi-infinite defects breaks down.)
\item The abelian flavor symmetries of $T_K[M,\mb t,\Pi]$ associated to the A-cycles of small annuli and tori, a priori $U(1)^{K-1}$, must be enhanced%
\footnote{This precise statement holds for \emph{maximal} of \emph{full} defects of the 6d $A_{K-1}$ theory. Otherwise one expects enhancement to non-abelian subgroups of $SU(K)$. For $K=2$ the regular defects are unique, so there is no ambiguity.} %
to $SU(K)$. (Because closed or infinite regular defects of the 6d theory carry non-abelian $SU(K)$ flavor symmetry.) More precisely, these flavor symmetries of $T_K[M]$ must be enhanced after coupling to Seiberg-Witten theories as above.
\end{enumerate}
A large part of this paper will be devoted to understanding the second point, i.e. the non-abelian enhancement of flavor symmetry and its geometric origin. Constructions from our work \cite{DG-E7} will play a key role. We will mainly focus on the $A_1$ theory, obtaining enhancements from $U(1)$ to $SU(2)$. The case of general $K$ and maximal defects can begin to be analyzed using methods of~\cite{DGG-Kdec}.

The first point will not play a further role in this paper. Although we believe it is likely that codimension-two regular defects can end on 
a special quarter-BPS codimension-three defect, we will not attempt to define such an object here. We will only encounter small disks organized in a very specific configuration, which will be given an alternative six-dimensional interpretation in terms of irregular defects.
We will explain momentarily that a rank $r$ irregular defect in the six-dimensional $A_1$ theory joining two big boundary components 
can be naturally ``regularized'' by gluing the two boundary components along a $2r$-sided polygon with small disks 
at the vertices. 

\subsection{Non-abelian couplings}

Once we know that the flavor symmetries of $T_K[M,\mb t,\Pi]$ that are associated to small annulus and torus boundaries have non-abelian enhancements, we can attempt to interpret this 3d theory as a boundary condition for a more universal 4d bulk.

Let us set $K=2$ for concreteness. Then for every small torus boundary, $T_2[M,\mb t,\Pi]$ can provide a quarter-BPS boundary condition for 4d $\CN=4$ $SU(2)$ super-Yang-Mills theory $T_2[T^2]$. In particular, the $SU(2)$ flavor symmetry associated to the A-cycle of the small torus is identified with the gauge symmetry of $T_2[T^2]$. This coupling has a six-dimensional origin that is closely related to the original argument for the flavor symmetry of defects. After using the dictionary above to shrink a small torus to closed regular defect in $M\times \R^2$, we can choose a metric for $M$ that pulls the closed defect out to infinity, producing cylindrical regions of the form $T^2\times \R_+\times \R^3$. Compactification on $T^2$ leads to $\CN=4$ SYM on a half-space, coupled to $T_2[M,\mb t,\Pi]$.

More interesting are the small annuli. Recall that a small annulus connects two punctures (or holes) on the big boundary, which is made of surfaces $\CC_i$.
After we couple $T_2[M,\mb t,\Pi]$ to Seiberg-Witten theories $\prod_i SW_2[\CC_i]$, the $SU(2)$ flavor symmetry associated to an annulus should be identified with the two $SU(2)$ flavor symmetries of the Seiberg-Witten theories associated to the big boundaries $\CC_i$ that the annulus ends on. This flavor symmetry can then be gauged in the product of SW theories in the 4d bulk.

\begin{figure}[htb]
\centering
\includegraphics[width=3.5in]{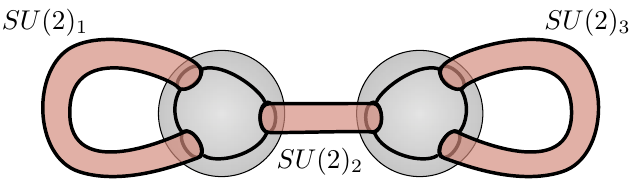}
\caption{Forming a genus-two boundary from two big three-punctured spheres connected by small annuli. This equivalent to a pants decomposition.}
\label{fig:pants}
\end{figure}

For example, consider a closed topological boundary component $\wt \CC$ of $M$ consisting of a collection of big three-punctured spheres connected by small annuli (Figure \ref{fig:pants}). The separation into big and small pieces is equivalent to a pair-of-pants decomposition of $\wt \CC$. Each big three-punctured sphere is associated to a free 4d half-hypermultiplet in the tri-fundamental representation of an $SU(2)^3$ flavor group. We can then identify and gauge the pairs of $SU(2)$ flavor groups associated to punctures that are connected by small annuli. What we get is precisely the non-abelian UV description of the 4d class-$\CS$ theory $T_2[\wt \CC]$, in a UV duality frame corresponding to the given pants decomposition of $\wt \CC$ \cite{Gaiotto-dualities}. Now $T_2[M,\mb t,\Pi]$ should provide a half-BPS boundary condition for $T_2[\wt \CC]$\,! This can also be motivated with a 6d construction.

The general picture now becomes the following. To the boundary of any framed 3-manifold $M$ we claim that we can associate a universal 4d $\CN=2$ theory $T_2[\pd M]$, for which $T_2[M]$ provides a half-BPS boundary condition. Each closed topological component of the boundary $\pd M$ contributes an independent factor to $T_2[\pd M]$. Small tori contribute $\CN=4$ SYM factors. The remaining boundaries $\wt \CC_i$ are decomposed into big punctured surfaces connected by small annuli, and thus contribute products of Seiberg-Witten theories in which pairs of $SU(2)$ flavor groups have been identified and gauged. Notice that this can be interpreted as the class-$\CS$ theory $T_2[\wt \CC_i]$ after a partial flow to the IR.
In coupling $T_2[M]$ to $T_2[\pd M]$, \emph{all} the flavor symmetries of $T_2[M]$ are gauged, and chiral operators of $T_2[M]$ couple via superpotentials to BPS hypermultiplets of $T_2[\pd M]$.

Starting from this general setup, it is easy to recover other boundary conditions by sending various 4d gauge couplings to zero (and masses to infinity). We can picture this operation as ``cutting'' the boundary $\pd M$ into pieces. For example, by cutting all small annuli (sending the couplings of their $SU(2)$'s to zero) we recover a boundary condition for a product of pure Seiberg-Witten theories. Of course, we can also leave some of the small annuli intact, leading to boundary conditions for products of pure Seiberg-Witten theories and non-abelian UV theories.

\subsection{Irregular defects}

We can even go further and cut the big boundaries of $M$ into pieces along closed polygonal edge-paths in the boundary triangulation. We require that the paths' vertices lie at small discs (not annuli). Cutting $\CC\to \CC_1\cup \CC_2$ along a polygon corresponds to taking a limit in the parameter space of $SW_2[\CC]$ so that BPS states associated to the edges of the polygon become infinitely massive while BPS states associated to all other edges (in a triangulation of $\CC$) remain light. Following the 2d dictionary of \cite{GMNII}, the new polygonal boundaries of $\CC_1$ and $\CC_2$ should now be interpreted as \emph{irregular} punctures in these surfaces.
The theory $SW_2[\CC]$ degenerates into a product $SW_2[\CC_1]\times SW_2[\CC_2]$, where each $SW_2[\CC_i]$ results from compactification of the 6d $A_1$ theory with irregular defects.
The 3d theory $T_2[M]$ provides a BPS boundary condition for the product.
The polygon can be thought of as a regularization of the irregular defect joining the irregular punctures on $\CC_1$ and $\CC_2$.

\begin{figure}[htb]
\centering
\includegraphics[width=4.5in]{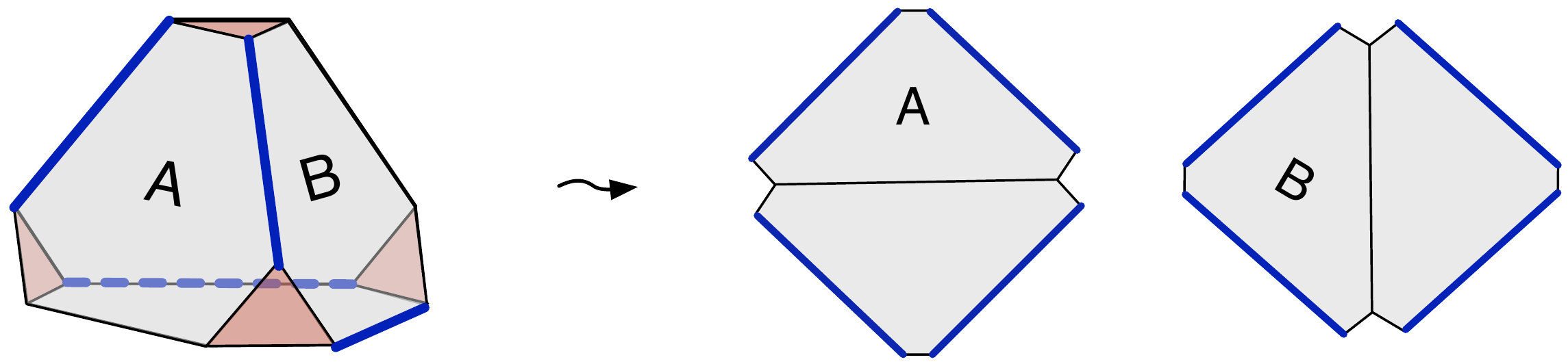}
\caption{Cutting the big boundary of a tetrahedron in half along a four-sided polygon (in blue).}
\label{fig:tetcut}
\end{figure}

The simplest illustration of this last scenario is for $M=\Delta$ a single tetrahedron (Figure \ref{fig:tetcut}). Since the big boundary of $\Delta$ is a four-punctured sphere, the theory $T_2[\pd \Delta]$ is the Seiberg-Witten description of 4d $\CN=2$ $N_f=4$ SQCD. The IR gauge group is $U(1)$. In a ``strong coupling'' region of the $u$-plane the theory has a finite collection of BPS hypermultiplets with electric, magnetic, and dyonic charge. Cutting $\pd \Delta$ in half along a 4-sided polygon corresponds to taking an Argyres-Douglas type limit, in which only two electric particles survive. They are the BPS states of two ``$A_1$'' Argyres-Douglas theories \cite{AD}. The tetrahedron theory $T_\Delta=T_2[\Delta,\mb t,\Pi]$ then provides a boundary condition for a product of the two Argyres-Douglas theories --- or equivalently a domain wall between one and the other. This domain wall was the basic object of study in \cite{CCV}.
We will discuss it further in section \ref{sec:JanusSW}. 

\subsection{Duality and RG domain walls}

\begin{figure}[htb]
\centering
\includegraphics[width=5.5in]{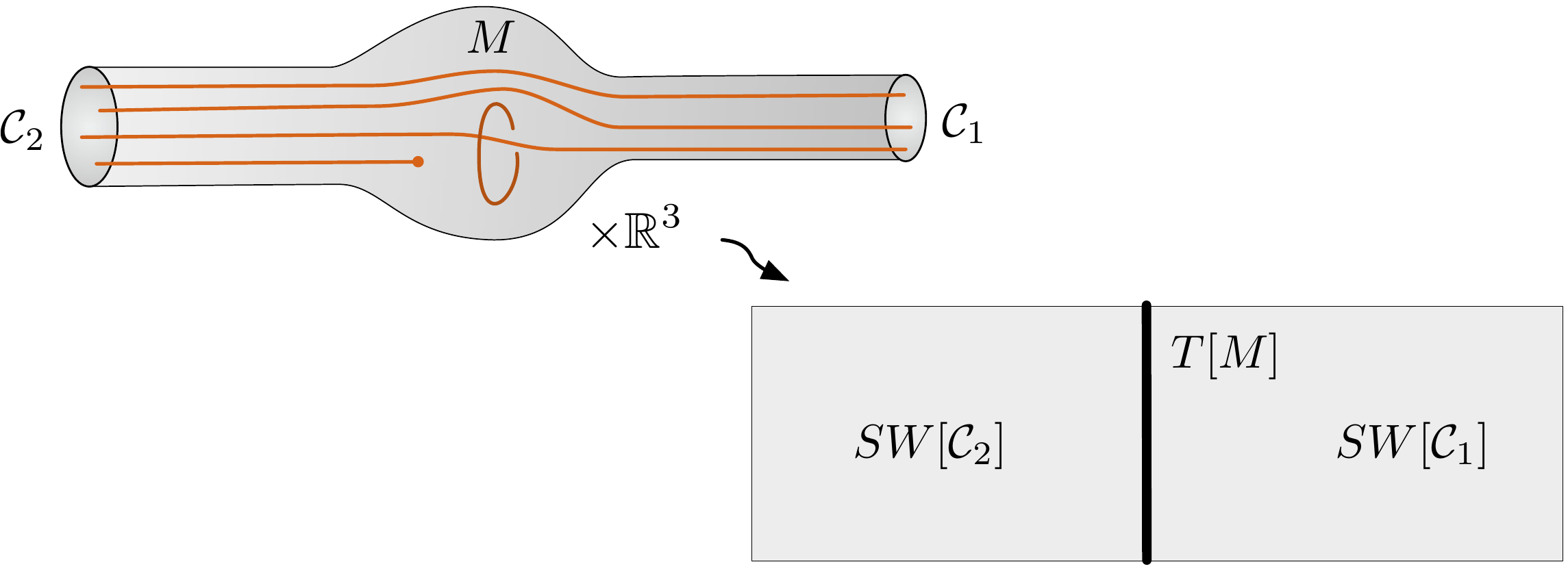}
\caption{Reinterpreting the boundary condition of Figure \ref{fig:regSW} as a domain wall.}
\label{fig:domSW}
\end{figure}

Notice that a BPS boundary condition for a product of theories $T_1\times T_2$ is equivalent to a BPS domain wall between $T_1$ and $T_2$. Graphically, this might be thought of as in Figure \ref{fig:domSW}. Therefore, when a 3-manifold $M$ either has topologically disjoint boundary components or a boundary that has been ``cut'' into pieces as above, the theory $T_K[M,\mb t,\Pi]$ should provide a BPS domain wall.
In this interpretation, two particular kinds of 3-manifolds that were anticipated in \cite{DG-Sdual} play a special role. (For simplicity, we'll continue to take $K=2$.)

Let $\CC$ be a punctured surface, and construct a ``trivial'' 3-manifold $M=\CC\times I$ that has two big boundaries $\CC,\,\ol\CC\simeq \CC$, and a collection of small annuli connecting the punctures of $\CC$ directly to the punctures of $\ol \CC$. We consider $\pd M$ to be ``cut'' along these annuli.
We have claimed that $T_2[M,\mb t,\Pi]$ provides a boundary condition for $SW_2[\CC]\times SW_2[\ol\CC]$, and thus a domain wall between the 4d Seiberg-Witten theory $SW_2[\CC]$ and itself. It is a totally reflecting boundary, or a totally transparent domain wall.

\begin{figure}[htb]
\centering
\includegraphics[width=5.8in]{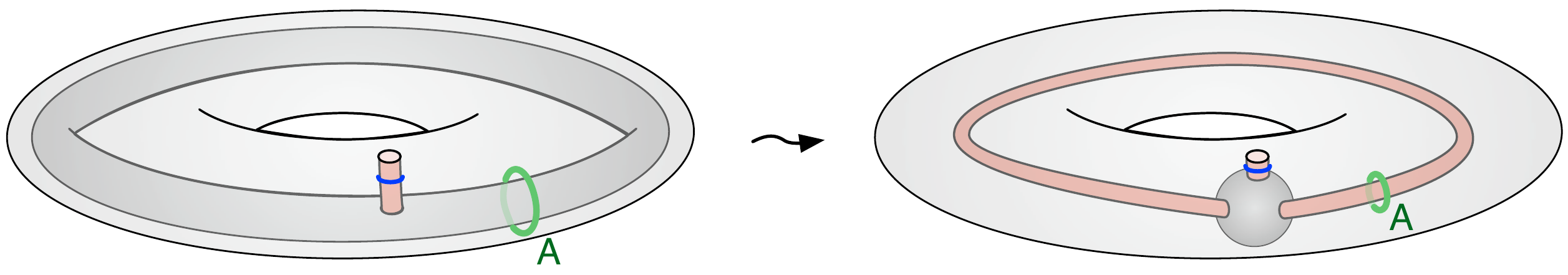}
\caption{Left: trivial cobordism $M=\CC\times I$ with $\CC$ a big punctured torus. Right: shrinking an A-cycle on the ``inner'' boundary $\ol \CC$ to create a new manifold $M_{\mb p}$; its inner boundary is now a big three-punctured sphere with one puncture connected to the outside and the other two connected to each other by small annuli.}
\label{fig:torusp}
\end{figure}

To obtain something more interesting, let us choose a pants decomposition $\mb p$ for $\ol \CC$ (\ie\ a maximal set of non-intersecting A-cycles), and form a new three-manifold $M_{\mb p}$ by shrinking the A-cycles, as in Figure \ref{fig:torusp}. Specifically, although $M_{\mb p}$ and $M$ are equivalent topologically, we have replaced the boundary $\ol \CC$ in $M$ by a network of big three-punctured spheres connected by small annuli. The 4d theory associated to this new boundary is the non-abelian class-$\CS$ theory $T_2[\CC]$. We now expect the 3d theory $T_2[M_{\mb p}]$ to couple to $SW_2[\CC]\times T_2[\CC]$, and thus provide a natural domain wall between IR and UV descriptions of the same 4d theory --- an RG domain wall!

We will construct RG domain walls in Section \ref{sec:Janus} purely in field theory by starting from Janus configurations where 4d parameters vary in space. In the limit that parameters vary very quickly, over a short interval, we will recover the 3d domain-wall theory $T_2[M_{\mb p}]$ coupled 4d UV and IR theories in the bulk.

The most basic property we would expect from a BPS RG domain wall 
is that UV bulk BPS line operators brought to the interface should match with their bulk IR description, brought to the interface from the opposite side. 
The correspondence between UV and IR descriptions of a BPS line defect was discussed in detail in \cite{GMNIII}, and thus we 
have in our hands an infinite set of constraints which the RG domain walls should obey. These constraints 
can be made more manageable by inserting the line defects inside a protected calculation, such as 
an ellipsoid partition function, or a sphere index. Alternatively, we can compare their vevs when the theory is compactified on a circle.   

The correspondence between UV and IR operators will work automatically for our candidate RG domain walls,
for a simple geometric reason. The UV line operators are associated to closed loops on the Riemann surface $\CC$ \cite{DMO}, 
and their vevs to the trace of the holonomy along the loops of a flat $PSL(2,\C)$ connection on $\CC$ \cite{GMNIII}. The 3d geometry is selected in such a way that the same holonomy can be computed along a corresponding loop 
on the big boundary region in terms of certain ``edge coordinates.'' The edge coordinates coincide with the vevs of IR line operators and the relation between holonomies and edge coordinates is known to encode the relation between UV and IR line defects \cite{GMNIII}. The role of the 3d geometry is to allow us to transport the UV line operators from the UV end of the geometry to the IR end of the geometry, where the triangulation and edge coordinates live.

\begin{figure}[htb]
\centering
\includegraphics[width=5.5in]{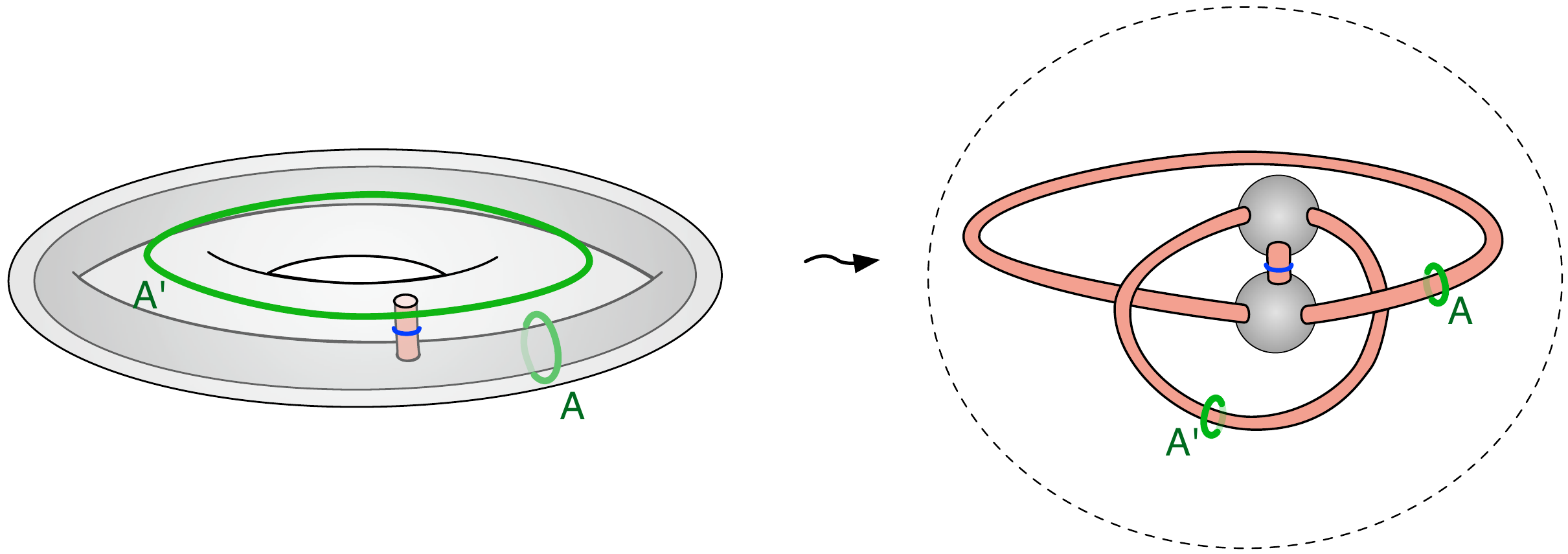}
\caption{Shrinking A- and A$'$-cycles on both the inner and outer boundaries of $M$ to create $M_{\mb p,\mb p'}$. The resulting $M_{\mb p,\mb p'}$ here can be thought of as the complement of a trivalent ``Hopf network'' in $S^3$. It is the 3-manifold that gives rise to the theory $T[SU(2)]$.}
\label{fig:toruspp}
\end{figure}

A closely related modification of the trivial cobordism $M=\CC\times I$ shrinks a set of A-cycles on both ends, according to two different pants decompositions of $\CC$: $\mb p$ and $\mb p'$. Call the resulting manifold $M_{\mb p,\mb p'}$. Then the 3d theory $T_2[M_{\mb p,\mb p'}]$ couples naturally to two copies of $T_2[\CC]$ in different weakly coupled UV descriptions, corresponding to $\mb p$ and $\mb p'$. We obtain a concrete Lagrangian formulation for the domain wall that implements the $\CN=2$ S-duality of \cite{Gaiotto-dualities}, in direct analogy with the $\CN=4$ S-duality domain walls of \cite{GW-Sduality}.

The simplest example of such an S-duality domain wall is for 4d $\CN=2^*$ theory. The 3d domain-wall theory is an $\CN=2$ deformation of the theory called $T[SU(2)]$ in \cite{GW-Sduality}. We will recover $T[SU(2)]$ as the theory corresponding to the 3-manifold in Figure \ref{fig:toruspp}. In fact, we will obtain several different mirror-symmetric descriptions of $T[SU(2)]$, including one whose $SU(2)\times SU(2)$ flavor symmetry is manifest.

\subsection{Organization}

Most of this paper will specialize to theories coming from the compactification of the $A_1$ six-dimensional (2,0) SCFT, \ie\ to $K=2$.

We begin in Section \ref{sec:couplings} by reviewing geometric properties of Seiberg-Witten theories of class $\CS$ and presenting their natural coupling to 3d theories of class $\CR$ at IR interfaces. In Section \ref{sec:Janus}, we discuss properties of Janus configurations for 4d $\CN=2$ theories and explain (in principle) how to extract from them 3d domain-wall theories and their bulk-boundary couplings. Generalizing the constructions of \cite{CCV}, we interpret Lagrangian definitions of the 3d theories in terms of triangulated 3-manifold geometries.

We apply our understanding of Janus configurations in Section \ref{sec:RG} to build the most fundamental example of an RG domain wall, for pure $\CN=2$ $SU(2)$ gauge theory. We interpret it in terms of a framed 3-manifold geometry, establishing the basic geometric properties of UV (non-abelian) bulk-boundary couplings. Then in Section \ref{sec:enhance} we generalize the geometry of UV bulk-boundary couplings to any 4d $(A_1)$ theory of class $\CS$ coupled to a 3d theory of class $\CR$. We focus in particular on how the (a priori) abelian flavor symmetries in 3d get enhanced so that they can be identified with non-abelian symmetries in a 4d bulk. 

Finally, in Sections \ref{sec:N=2*} and \ref{sec:N4} we demonstrate our various constructions in two detailed examples, describing RG walls and duality walls for 4d $SU(2)$ $\CN=2^*$ theory (coming from compactification on a punctured torus) and for 4d $SU(2)$ theory with $N_f=4$ flavors of matter (coming from compactification on a four-punctured sphere).

In Appendix \ref{app:coords} we include mathematical definitions of various moduli spaces of framed flat connections used in the paper, their coordinates (including a novel definition of complexified Fenchel-Nielsen twists), and their quantization. In Appendix \ref{app:triang} we explain how to build a useful 3d triangulation for any framed 3-manifold corresponding to an RG or duality wall.

%%%%%%%%%%%%%%%%%%%%%%%%%%%%%%%%%%%%%%%%%%%%%%%%%%%%%%%%%%%%%%%%%%%%%%%%%%%%%
%%%%%%%%%%%%%%%%%%%%%%%%%%%%%%%%%%%%%%%%%%%%%%%%%%%%%%%%%%%%%%%%%%%%%%%%%%%%%
%%%%%%%%%%%%%%%%%%%%%%%%%%%%%%%%%%%%%%%%%%%%%%%%%%%%%%%%%%%%%%%%%%%%%%%%%%%%%

\section{The simplest couplings}
\label{sec:couplings}

In this section, we seek to describe the natural couplings between a 3d theory of class $\CR$ and an abelian Seiberg-Witten theory of class $\CS$. These couplings are associated to the big boundary of a framed 3-manifold. The main point is that at fixed values of its moduli a 4d Seiberg-Witten theory $SW_2[\CC]$ comes with a family of natural WKB triangulations of the surface $\CC$, which encode much of the 4d physics. By matching a WKB triangulation of $\CC$ with the big-boundary triangulation $\mb t$ of a 3-manifold $M$, we can encode the physics of the coupling between $SW_2[\CC]$ and $T_2[M,\mb t]$ --- including a proper identification of gauge and flavor symmetries, and superpotential terms involving BPS hypermultiplets.
We always have $K=2$, studying theories coming from the $A_1$ $(2,0)$ theory in 6d.

\subsection{Big boundaries and Seiberg-Witten theory}
\label{sec:2d}

We begin by reviewing some basic properties of Seiberg-Witten theories in class $\CS$, and their relation to triangulations and polarizations of surfaces. These ideas were developed in \cite{GMNII}, following \cite{SW-I,SW-II, Witten-M}, \cite{Gaiotto-dualities, GMN}, and the mathematical constructions of \cite{FG-Teich}.

We would like to consider the six-dimensional $A_1$ $(2,0)$ theory in the background $\CC\times \R^4$, where $\CC$ is a real two-dimensional oriented surface.%
\footnote{We will use both Euclidean and Lorentzian signature at various points, and only distinguish between the two when necessary. In the case of Lorentzian signature, time is always one of the directions in $\R^4$.} %
We allow real codimension-two defects that fill all of $\R^4$, and show up as ``punctures'' or ``holes'' on $\CC$. An appropriate topological twist%
\footnote{This twist identifies the $SO(2)$ isometry group of $\CC$ with the first factor in the subgroup $SO(2)_R\times SO(3)_R$ of the $SO(5)_R$ R-symmetry group of the 6d theory.} %
along $\CC$ allows eight supercharges to be preserved, along with an $SU(2)_R\times U(1)_r$ R-symmetry. At low energies, we expect to find an effective 4d $\CN=2$ theory $T_2[\CC]$. It is a superconformal theory, and $U(1)_r$ remains non-anomalous, as long as all defects are regular, which we will assume in this discussion.

One may choose an arbitrary background metric for $\CC$, though after flowing to the IR the theory $T_2[\CC]$ only depends on its complex (or conformal) structure. If $\CC$ has genus $g$ and $h$ holes, the complex structure moduli space has complex dimension
\be d = d(g,h) := 3g-3+h\,. \ee
An additional $h$ parameters are used to prescribe boundary conditions for the complex structure at the holes.
In the limits where $\CC$ is stretched into pairs of pants, $T_2[\CC]$ acquires a weakly coupled Lagrangian description as an $SU(2)^d$ gauge theory, with $SU(2)^h$ flavor symmetry. The complex structure of each stretched internal leg in $\CC$ provides the gauge coupling for a dynamical $SU(2)$, while each global $SU(2)$ is associated with a bounday hole.

If we move onto the Coulomb branch and flow further to the infrared, $T_2[\CC]$ can be described as a Seiberg-Witten theory $SW_2[\CC]$, with gauge group broken to $U(1)^d$. Of course the $SU(2)^h$ flavor symmetry persists. The Seiberg-Witten curve $\Sigma$ is a double cover of $\CC$ that is branched at exactly $4g-4+2h=-2\chi(\CC)$ points. The curve depends on the $d$ marginal couplings from the UV and $h$ mass parameters, as well as $d$ new Coulomb moduli.

\begin{figure}[htb]
\centering
\includegraphics[width=3in]{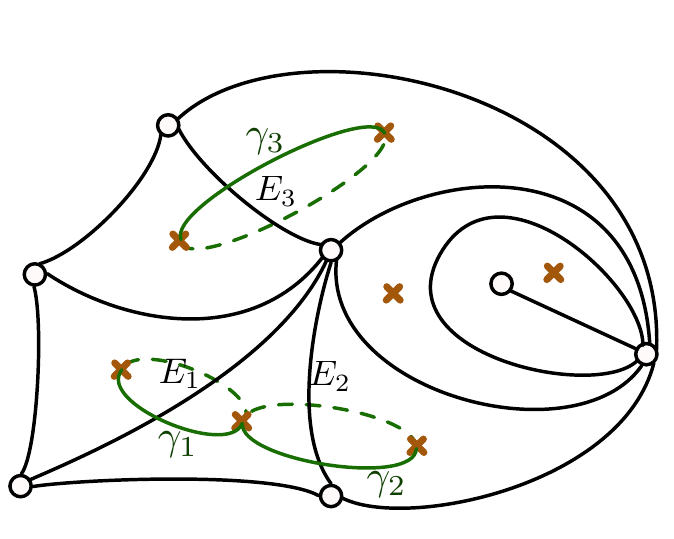}
\caption{A typical WKB triangulation of a part of $\CC$. The SW fibration $\Sigma\to \CC$ has a branch point in each triangle, and each edge $E_i$ corresponds to a constant-phase path $\gamma_i$ on $\Sigma$. Here (for example) $\langle \gamma_1,\gamma_2\rangle=+1$, while $\langle \gamma_1,\gamma_3\rangle=0$, etc.}
\label{fig:WKB}
\end{figure}

Many properties of $SW_2[\CC]$ can be understood by considering \emph{WKB triangulations} of $\CC$ \cite{GMNII}. These are ideal triangulations of $\CC$ --- meaning that all edges in the triangulation start and end at holes%
\footnote{For these triangulations to make sense, we must assume that $h>0$: there is always at least one hole.} %
--- that depend on a point $u$ on the Coulomb branch and a phase $\theta$. (More generally, the triangulation depends on mass parameters $m$ and marginal UV couplings $\tau$; we include these parameters in `$u$'.)  A WKB triangulation $\mb t_{u,\theta}$ at generic $(u,\theta)$ is obtained by looking at trajectories on $\CC$ such that the Seiberg-Witten form $\lambda_{\rm SW}$ has constant phase $\theta$; each triangle is a region of $\CC$ that contains a unique branch point of the cover $\Sigma\to \CC$ and whose boundaries the trajectories never cross (Figure \ref{fig:WKB}).
From a topological perspective, the dependence of the triangulation on $(u,\theta)$ is piecewise constant.

At fixed $u$, the charge lattice $\Gamma_\CC$ of $SW_2[\CC]$ consists roughly of elements in $H_1(\Sigma,\Z)$ that are odd under deck transformations.%
\footnote{We refer to \cite{GMNIII} for a detailed description of the charge lattice.} %
It includes a sub lattice $\Gamma_f \sim \Z^h$ of flavor charges, which map to 
small loops around the punctures of $\Sigma$. The lattice of gauge charges $\Gamma_g = \Gamma_\CC/\Gamma_f \sim \Z^{2d}$ is the odd part of $H_1(\ol \Sigma,\Z)$, where $\ol \Sigma$ denotes $\Sigma$ with its punctures removed.
The electric-magnetic skew-symmetric product $\langle \gamma,\gamma'\rangle$ is just the intersection product for 1-cycles in $\Sigma$; it is non-degenerate on $\Gamma_g$ and vanishes for any pure-flavor charge $\gamma \in \Gamma_f$. The central charge of a state with gauge/flavor charge $\gamma \in \Gamma_\CC$ can be written as the period of the SW differential 
\be Z_\gamma = \gamma \cdot (a,a_D,m) = \int_\gamma \lambda_{SW}
\,,\ee
where $m$ are the complex masses associated with the flavor symmetry. Abelian electric-magnetic duality acts by $Sp(2d,\Z)$ transformations on the gauge lattice $\Gamma_g$, and dual transformations on $(a,a_D)$.

Any WKB triangulation $\mb t_{u,\theta}$ of $\CC$ provides a basis for $\Gamma_\CC$.%
\footnote{More precisely, the WKB triangulation provides a basis for a sublattice of $\Gamma_\CC$ generated by the charges of BPS particles in the theory. Sometimes this is a proper sublattice of $\Gamma_\CC$.} %
To understand this, note that any edge $E$ of the triangulation separates two triangles, each of which contains a branch point. The unique open on $\CC$ that connects the branch points and crosses $E$ lifts to a cycle $\wt \gamma_E$ on $\Sigma$, and thus defines a charge $\gamma_E$. (With some care, the orientation of $\CC$ can be used to consistently assign an orientation to $\wt \gamma_E$, and thus to fix the sign of $\gamma_E$.) It is not hard to see that
\begin{enumerate}

\item The product $\langle \gamma_{E},\gamma_{E'}\rangle$ equals the (signed) number of oriented triangles shared by edges $E$ and $E'$. Thus $\langle \gamma_{E},\gamma_{E'}\rangle \in \{0,\pm 1,\pm 2\}$.

\item The sum $\gamma_v:= \sum_{\text{$E$ touches $v$}}\gamma_E$ of edge charges for the edges that touch a puncture $v$ equals twice the flavor charge associated to that puncture. Thus $\langle \gamma_v,*\rangle=0.$

\end{enumerate}

\begin{wrapfigure}{r}{2.2in}
\centering \vspace{-.8cm}
\includegraphics[width=2in]{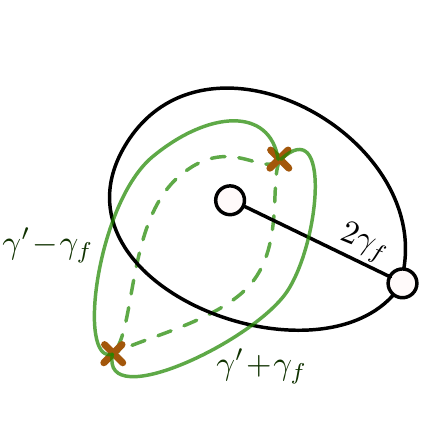}
\caption{A univalent edge in a degenerate triangle.}
\label{fig:lollipop}
\end{wrapfigure}

In addition, every edge $E$ of a WKB triangulation $\mb t_{u,\theta}$ that is not the univalent edge in a degenerate triangle as in Figure \ref{fig:lollipop} also corresponds to a BPS hypermultiplet $\Phi_E$ of $SW_2[\CC]$. The hypermultiplet has charge $\gamma_E$ and is stable in a region of the Coulomb branch that includes $u$. Geometrically, it arises because the topological cycle $\wt \gamma_E\subset \Sigma$ can be realized as a unique path of minimal length on $\Sigma$, on which $\lambda_{SW}$ has some constant phase.%
\footnote{In the M-theory construction of $SW_2[\CC]$, where an M5 brane wraps $\Sigma\times \R^4$, M2 branes can end on the cycles $\tilde \gamma_W$ without breaking SUSY, giving rise to BPS particles in $\R^4$.}

The univalent edge $E_f$ in a degenerate triangle is somewhat special. It carries pure flavor charge $2\gamma_f$ associated to the central puncture, in the normalization where a doublet for the $SU(2)_f$ flavor symmetry at the puncture has charges $\pm\gamma_f$. There is no BPS hypermultiplet of charge $2\gamma_f$ in the spectrum of $SW_2[\CC]$, because there is no minimal-length path $\wt \gamma_{E_f}\subset \Sigma$ that just winds around the puncture. Instead, there are \emph{two} hypermultiplets $(\Phi_f,\wt\Phi_f)$ with charges $\gamma'\pm\gamma_f$, in a doublet of $SU(2)_f$, where the charge $\gamma_{E'}=\gamma'-\gamma_f$ is naturally associated to the outer edge $E'$ that surrounds the degenerate triangle.
Notice that the full BPS spectrum of the theory will be $SU(2)_f$ invariant only for sufficiently small values of the 
$SU(2)_f$ mass parameter. For general values of the mass parameter, wall-crossing can break the $SU(2)_f$ multiplets. 
The presence of the doublet of hypers is thus a bit special. To understand it, and many other facts, we need to review the notion of flip.

\begin{figure}[htb]
\centering
\includegraphics[width=4.3in]{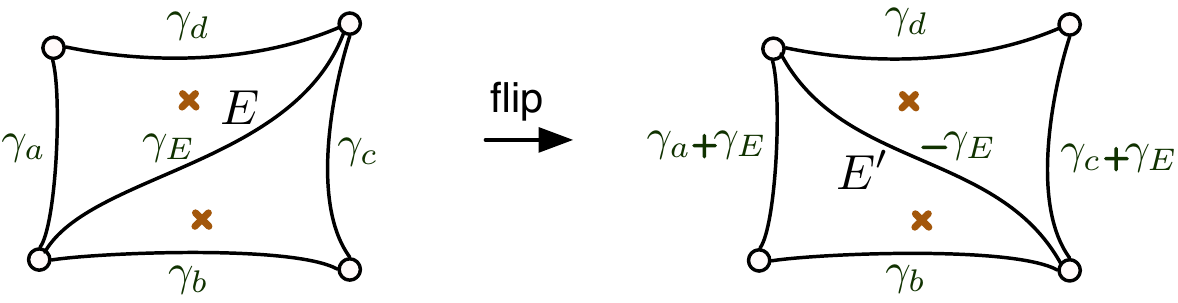}
\caption{A flip of the WKB triangulation. Here the nonzero intersection products are $\langle \gamma_a,\gamma_E\rangle=\langle \gamma_c, \gamma_E\rangle = -1$, $\langle \gamma_b,\gamma_E\rangle=\langle\gamma_d,\gamma_E\rangle=\langle \gamma_a,\gamma_d\rangle=\langle \gamma_c,\gamma_b\rangle=1$. Note how the transformed charges after the flip continue to have the right intersection products.}
\label{fig:flip2d}
\end{figure}

As the phase $\theta$ is varied, a WKB triangulation $\mb t_{u,\theta}$ may jump. In particular, at a critical value $\theta^*$ that allows some pair of branch points to be connected by a phase--$\theta^*$ trajectory, an edge of the WKB triangulation ``flips'' ($E\to E'$) as in Figure \ref{fig:flip2d}. The BPS hypermultiplets associated to the edges $E,E'$ on either side of the flip are a particle/anti-particle pair, with $\gamma_{E'}=-\gamma_E$; they both correspond to the cycle $\wt\gamma_{E}$ of phase $\theta^*$ on $\Sigma$, in two opposite orientations. The charges of the four other edges of the quadrilateral involved in the flip also change by multiples of $\gamma_E$.

There is another special phenomenon, the ``pop,'' which will happen at the critical values in $\theta$ aligned with the 
phase of the mass parameter at a puncture. It helps understand why 
the degenerate triangle is associated to a doublet of hypermultiplets. The pop switches the sign of the flavor charge $\gamma_f$ 
and thus exchanges the roles of the two hypers in the doublet as the BPS particle associated to the edge $E'$. 
As we vary $\theta$ by $\pi$, the pop will happen exactly once, and the triangulation at a pop always includes a degenerate 
triangle at the puncture. If this is our original degenerate triangle, we are done. If not, a flip must have happened at some point 
to the edge $E'$ of our original degenerate triangle. After that flip, there are two edges going into the puncture, associated to charges 
$\gamma_f \pm \gamma'$, and thus we have again the desired doublet of hypers in the BPS spectrum.

\subsubsection{Framed flat connections}
\label{sec:flat2d}

We need one more important idea to relate the combinatorics of theories of class $\CR$ to boundary SW theories of class $\CS$.

When a theory of class $\CS$ is compactified on a circle $S^1_\beta$ of finite radius, one obtains in the far infrared a 3d sigma model to a hyperk\"ahler moduli space $\CP_2(\CC)$ of real dimension $4d$. There are several nice ways to understand it. Very roughly $\CP_2(\CC)$ is obtained by treating the $d$-complex-dimensional Coulomb branch of $SW_2[\CC]$ as a real moduli space and using electric and magnetic Wilson lines around $S^1_\beta$ to complexify it again. A more accurate description is that $\CP_2(\CC)$ is the $SU(2)$ Hitchin moduli space associated to $\CC$. Indeed, upon compactification on $S^1_\beta$ the 6d $A_1$ theory becomes maximally supersymmetric 5d Yang-Mills theory. In the presence of the topological twist on $\CC$, its BPS equations are Hitchin equation for a real $SU(2)$ connection $A$ and a complex adjoint-valued one-form field $\varphi$.

The complex structure on $\CP_2(\CC)$ is parametrized by a twistor coordinate $\zeta\in \cp^1$. The most useful fact for us is that when $\zeta\neq 0,\infty$ the solution to the Hitchin equations can be repackaged as a flat complex connection
\be  \label{Acx}
  \CA = A + \frac{i\beta}{2}\big(\zeta^{-1}\varphi+\zeta\ol\varphi\big)\,.
\ee
We will usually take $\zeta=e^{i\theta}$ to be a pure phase, so that $\CA=A+i\,\Re(\zeta^{-1}\varphi)$.
Then $\CP_2(\CC)$ is \emph{essentially} isomorphic, as a complex symplectic manifold, to the moduli space of flat $PGL(2,\C)$ connections on $\CC$.
This is how we will think of it from now on. On smooth parts of the moduli space, the canonical holomorphic symplectic form is given by the Atiyah-Bott-Goldman formula \cite{AtiyahBott-YM, Goldman-symplectic}
\be \Omega = \int_\CC \Tr \delta\CA\wedge\delta \CA\,. \label{WP} \ee
The smooth parts of $\CP_2(\CC)$ have complex dimension $2d=6g-6+2h$.

The isomorphism between $\CP_2(\CC)$ and the moduli space of flat complex connections is subject to two important caveats. One will be addressed in Section \ref{sec:PGL}: we must sometimes consider lifts of flat connections to $SL(2,\C)$, and sometimes connections halfway between $PGL(2,\C)$ and $SL(2,\C)$, depending on the precise choice of UV gauge groups in $T_2[\CC]$. The other caveat is that the isomorphism only holds when the space of flat connections is smooth.%
\footnote{We refer again to \cite{GMNIII} for more detailed discussion.}

To be more precise, in the presence of punctures, the \emph{three} real mass parameters for $SW_2[\CC]$ on $\R^2\times S^1_\beta$ associated to the flavor symmetry at each puncture 
control the K\"ahler class and complex structure 
of the moduli space. When complex-structure mass parameters vanish, the standard moduli space of flat $PGL(2,\C)$ connections would develop a singularity; but the additional nonzero K\"ahler mass partially resolves the actual moduli space $\CP_2(\CC)$.
A simple shortcut to deal with this resolution is to supplement the complex flat connections by a choice of \emph{framing}.

Indeed, the eigenvalues of the holonomy around each puncture are fixed in terms of the complex-structure masses. But in addition we introduce on $\CC$ an associated bundle of flags in $\C^2$, and choose in the neighborhood of each puncture a flat section that is invariant under the $PGL(2,\C)$ holonomy. Since a flag in $\C^2$ is just a complex line, this amounts to choosing an eigenline. Then $\CP_2(\CC)$ is identified as the space of framed flat connections on $\CC$:
\be \CP_2(\CC) \simeq \{\text{framed $PGL(2,\C)$ connections on $\CC$}\}\,. \ee

\begin{figure}[htb]
\centering
\includegraphics[width=1.5in]{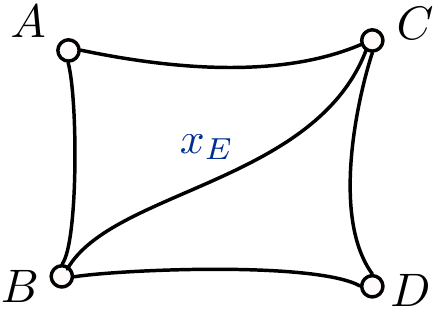}\hspace{.5in} \raisebox{.45in}{$\ds x_E = \frac{\langle A\wedge B\rangle\langle C\wedge D\rangle}{\langle A\wedge C\rangle\langle B\wedge D\rangle}\,.$}
\caption{Defining the edge coordinate $x_E$ as a cross-ratio of four framing flags $A,B,C,D$. The flags must be parallel-transported to a common point inside the quadrilateral in order to evaluate the cross-ratio.}
\label{fig:xrat}
\end{figure}

The choice of framing accomplishes the desired de-singularization of $\CP_2(\CC)$. If the squares of the two eigenvalues $\lambda^{\pm 2}$ at a given puncture are distinct (\ie\ $\lambda\neq \pm 1$), the framing just chooses one or the other, so the framed moduli space is a two-fold cover of the standard one. However, if the eigenvalues coincide, the choice of framing adds a $\cp^1$ to the moduli space, thereby blowing up a singularity. 
Given an ideal triangulation of $\CC$ (and a specific framing), one can define a set of coordinates for a Zariski-open patch of $\CP_2(\CC)$ (isomorphic to $(\C^*)^{2d}$) that are associated to the edges of a WKB triangulation \cite{FG-Teich}.%
\footnote{These are a natural generalization of Thurston's shear coordinates in 2d hyperbolic geometry, thereafter studied by Fock, Penner, and others.} %
Mathematically, the coordinate $x_E$ for edge $E$ is defined by taking the cross-ratio of four framing lines at the vertices of the quadrilateral containing $E$ (Figure \ref{fig:xrat}).%
\footnote{Conventions for cross-ratios are as in \cite{DGG-Kdec}. In particular, the cross-ratios used in this paper, which are natural from the 3d perspective, are \emph{minus} the positive coordinates of \cite{FG-Teich}.} %
Physically, $x_E$ is the expectation value of a half-BPS line operator of charge $\gamma_E$ that wraps $S^1_\beta$ in the Seiberg-Witten theory $SW_2[\CC]$. Thus, $\CP_2(\CC)$ is parametrized by vevs of IR line operators. The subalgebra of the $\CN=2$ supersymmetry algebra preserved by the line operators is characterized by the same phase $\zeta$ that determined the complex structure of $\CP_2(\CC)$.%
\footnote{In contrast to domain walls, BPS line defects typically preserve a full $SU(2)_R$ as well.}

The holomorphic edge coordinates $x_E$ obey Poisson brackets induced by \eqref{WP}, which are simply determined in terms of the electric-magnetic product. Namely,
\be \{x_E,x_{E'}\} = \langle \gamma_E,\gamma_{E'}\rangle x_Ex_{E'}\,. \ee
In particular, the product of the edge coordinates surrounding a given puncture is a central element. It is equal to the square of the $PGL(2)$ eigenvalue (as chosen by the framing) at that puncture:
\be \prod_{\text{$E$ ending at $v$}}(-x_E) = \lambda^2\,. \label{prodx}\ee
This follows easily by multiplying cross-ratios as in Figure \ref{fig:xrat}.

The line operators can be \emph{quantized} by adding angular momentum to the $\R^3\times S^1_\beta$ geometry \cite{Ramified, GMNIII}. One considers a fibered product $\R\times (\R^2\times_q S^1_\beta)$ such that a complex coordinate $z$ on $\R^2$ undergoes a rotation $z\to q z$ after a turn around $S^1_\beta$. An additional R-symmetry twist allows this background to preserve half the supersymmetry. BPS line operators are then constrained to live at the origin of $\R^2$ and any point on $\R$, while wrapping $S^1_\beta$. They satisfy relations of a quantum torus algebra
\be \hat x_E\hat x_{E'} = q^{\langle \gamma_E,\gamma_{E'}\rangle} \hat x_{E'}\hat x_E\,, \ee
where the ordering of $\hat x_E$'s is precisely the ordering of line operators along $\R$.

It is often convenient to use formal logarithmic variables $\hat X_E$ so that $\hat x_E = \exp \hat X_E$. Then
\be [\hat X_E,\hat X_{E'}] = \langle \gamma_E,\gamma_{E'}\rangle\hbar\,,\qquad q = e^\hbar\,;\ee
and now there is also a simple formula for the central elements (\ie\ pure flavor Wilson lines)
\be  \sum_{\text{$E$ ending at $v$}} \big(\hat X_E-i\pi -\tfrac\hbar 2\big) = 2\Lambda \,, \label{sumX} \ee
where $\lambda=\exp\Lambda$ is the holonomy eigenvalue at the puncture $v$.
The factors of $\hbar$ (and the less significant $i\pi$'s) in this formula arise from the R-symmetry twist in the physical geometry.

\subsubsection{SL(2) vs PGL(2)}
\label{sec:PGL}

We mentioned above that the precise moduli space of (framed) flat connections isomorphic to the space of vacua $\CP_2(\C)$ is sometimes neither an $SL(2)$ moduli space nor a $PGL(2)\simeq PSL(2) \simeq SL(2)/\Z_2$ moduli space, but rather a space halfway inbetween. The subtle choice descends from choices of UV gauge groups.

\begin{wrapfigure}{r}{2in}
\vspace{-.5cm}
\centering
\includegraphics[width=1.9in]{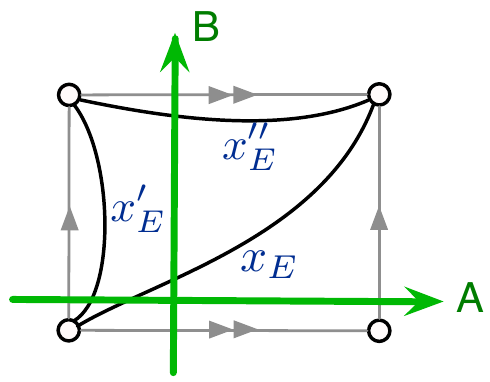}
\caption{Punctured torus, with triangulation and A,B-cycles.}
\label{fig:puncT}
\end{wrapfigure}

The simplest example to illustrate this is the ${\cal N}=2^*$ $SU(2)$ gauge theory, 
\ie\ the theory associated to a one-punctured torus. It is well known that S-duality exchanges the theory with $SU(2)$ gauge group and the theory with $PSU(2)\simeq SU(2)/\Z_2$ gauge group. 
S-duality simply exchanges the A- and B-cycles of the torus. The expectation values of Wilson loop operators for the $SU(2)$ gauge theory wrapping the circle $S^1_\beta$
map to traces of the holonomy of the flat connection on the A-cycle of the one-punctured torus in the corresponding representations of $SL(2)$. The expectation value of 't Hooft loop operators map to traces of the B-cycle holonomy. 

In the ${\cal N}=2^*$ $SU(2)$ gauge theory, the Wilson loop in a fundamental representation is allowed, but the fundamental 't Hooft loop is not. 
The converse is true for a ${\cal N}=2^*$ $PSU(2)$ gauge theory. On the other hand, the theories have the same set of BPS particles, which 
 have electric and magnetic charges (associated to IR edge-coordinates $x_E$) with the same quantization as Wilson and 'tHooft 
loop operators of integer spin.

We arrive at the following picture: 
\begin{itemize}
\item The moduli space of $PSL(2)$ connections on the one-punctured torus is ``too small'': it admits functions that correspond to vevs of Wilson and 'tHooft 
loop operators of integer spin only. The vevs of these operators can be expressed as Laurent polynomials in the $x_E$ variables for the three edges of a triangulated 
one-punctured torus. The intersection pairing of the $\gamma_E$ cycles is even: $\langle \gamma_E,\gamma_E'\rangle=\langle \gamma_E',\gamma_E''\rangle=\langle \gamma_E'',\gamma_E\rangle=2$.
\item The moduli space of $SL(2)$ flat connections on the one-punctured torus is ``too big'': it admits functions which correspond to vevs of both fundamental 
Wilson and 't Hooft loop operators. These vevs can be expressed in terms of the $x_E$ only if we allow certain square roots of $x_E$ monomials. 
The square roots create sign problems in defining things like the quantum torus algebra.
\item The moduli space of vacua for the ${\cal N}=2^*$ $SU(2)$ gauge theory is an intermediate quotient of the moduli space of $SL(2)$ flat connections,
where we keep track of the A-cycle holonomy in $SL(2)$, but only of the $PSL(2)$ image of the B-cycle holonomy. Some square roots of monomials of the $X_E$ 
have to be allowed, but just enough to keep signs under control, and the quantum torus algebra well defined. 
\end{itemize}

Simple generalizations of these statements hold for Riemann surfaces of higher genus $g$. The gauge group has a $\Z_2^{g}$ center that acts trivially on matter fields, and thus can be left ungauged or gauged in various patterns. 
The moduli space of $SL(2)$ flat connections can quotiented in various ways, so that 
one keeps track of holonomies in $SL(2)$ for a maximal set of cycles with even mutual intersection, and only in $PSL(2)$ for other cycles. 
The correspondence between four-dimensional choices of gauge group and two-dimensional choices of 
$SL(2)$ vs $PSL(2)$ can be established through the map between four-dimensional line defects and 
closed curves in the two-dimensional geometry \cite{GMNIII}.

%%%%%%%%%%%%%%%%%%%%%%%%%%%%%%%%%%%%%%%%%%%%%%%%%%%%%%%%%%%%%%%%%%%%%%%%%%%%%

\subsection{3-manifolds and class $\CR$}
\label{sec:3d}

The 3d $\CN=2$ theory in class $\CR$ associated to a triangulated 3-manifold $M$ (of the type described in the introduction) is constructed algorithmically in a way that mimics the construction of framed $PGL(2,\C)$ flat connections on $M$. We review here the main results of the construction, and the basic relations between the geometry of flat connections and 3d physics.

\subsubsection{Framed flat connections in 3d}

To begin, we recall that in three dimensions the choice of framing for a flat connection should be associated to the \emph{small boundary} of $M$ \cite{DGG, DGG-Kdec}. Since the small boundaries have abelian fundamental group, any standard flat connection on $M$ admits a choice of framing --- \ie\ a choice of invariant flag on the small boundary.
One then considers several algebraic moduli spaces:%
\footnote{A complete summary of these spaces and their coordinates is included in Appendix \ref{app:coords}.}%
\be 
\begin{array}{r@{\quad}l}
\bullet & \CP_2(\pd M) = \{\text{framed flat $PGL(2,\C)$ connections on $\pd M$}\} \\[.1cm]
\bullet & \wt\CL_2(M) = \{\text{framed flat $PGL(2,\C)$ connections on $M$}\} \\[.1cm]
\bullet & \CL_2(M) = \{\text{framed flat connections on $\pd M$ that extend to $M$}\} \;\subset\; \CP_2(\pd M)\,.
\end{array}
\ee
Notice that $\CL_2(M)$ is just the image of the natural projection $\wt\CL_2(M)\to \CP_2(\pd M)$; often $\wt \CL_2(M)$ and $\CL_2(M)$ are birationally isomorphic. The space $\CL_2(M)$ is ultimately the one we are most interested in. Mathematically, a certain Zariski-open subset of $\CL_2(M)$ is a Lagrangian submanifold inside $\CP_2(\pd M)$.%
\footnote{In fact, $\CL_2(M)\subset \CP_2(\pd M)$ is a subvariety with the rather special property that a canonical class in the second algebraic $K$-theory group of the function field of $\CP_2(\pd M)$ vanishes when restricted to $\CL_2(M)$. Thus it is a ``$K_2$ Lagrangian.'' This property is reviewed in \cite{GS-quant, DGG-Kdec} and follows immediately from the symplectic reduction of tetrahedron moduli spaces described below.} %
Physically, (this subset of) $\CL_2(M)$ ends up describing the space of supersymmetric vacua of $T_2[M]$ on $\R^2\times S^1$. We need a few more details to state this properly.

% One considers the moduli space $\wt\CL_2(M)$ of flat $PGL(2,\C)$ connections augmented by a choice of flat invariant flag along each of the small boundaries. Since the small boundaries have abelian fundamental group, the existence of such invariant flags (given any standard flat connection on $M$) is guaranteed. The space $\wt \CL_2(M)$ projects to a subvariety in the space $\CP_2(\pd M)$ of framed flat connections on the boundary --- it is the subvariety of connections on the boundary that extend to the bulk. A suitable open subset $\CL_2(M)$ of this subvariety is a complex Lagrangian submanifold of $\CP_2(\pd M)$.

Suppose that we fix a 2d ideal triangulation $\mb t$ for the big boundary of $M$, whose $i\text{-}$th component has (say) genus $g_i$ and $h_i$ holes. Suppose also that $M$ has $a$ small annular boundaries and $t$ small torus boundaries. Then $\CP_2(\pd M)$ generically has dimension $\sum_i 2d_i + 2a+2t$, with $d_i=3g_i-3+h_i$ as usual. Using the triangulation $\mb t$, the framing data on the boundary can be used to construct coordinates on an open patch of $\CP_2(\pd M)$, which we can call $\CP_2(\pd M,\mb t)$, that is isomorphic to a complex torus
\be \CP_2(\pd M,\mb t)\,\simeq\, \Big[\prod_i(\C^*)^{2d_i}\Big] \times (\C^*)^{2a}\times (\C^*)^{2t}\,. \label{PdM} \ee
The first factor simply corresponds to big-boundary moduli spaces, with coordinates $x_E$ as in Section \ref{sec:flat2d}. The last factor is completely decoupled, and is parametrized by A- and B-cycle eigenvalues for each small torus (which are canonically conjugate with respect to the complex symplectic form on $\CP_2(\pd M)$).

The middle factor in \eqref{PdM} is rather more interesting. It is parameterized by $a$ A-cycle eigenvalues along with $a$ canonically conjugate ``twist'' coordinates. These are analogous to Fenchel-Nielsen length and twist coordinates in 2d hyperbolic geometry \cite{Wolpert-deformation, Wolpert-symplectic}. Since the construction of these coordinates for framed flat connections has not been fully described before, we present it in detail in Appendix \ref{app:twist}.

\begin{figure}[htb]
\centering \vspace{-.3in}
\includegraphics[width=2.2in]{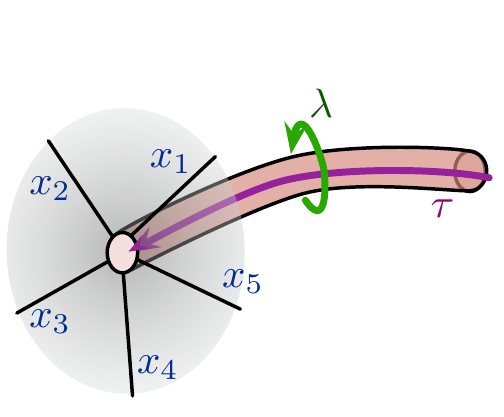}\hspace{.1in}\raisebox{.8in}{$\ds \prod_{i}(-x_i)=\lambda^2\,;\quad \begin{array}{c}\{\tau,\lambda\}=\tau\lambda\\[.2cm] \{\tau,x_i\}=\{\lambda,x_i\}=0\end{array}$}
\caption{Relation between annulus coordinates (length $\lambda$, twist $\tau$) and big-boundary eigenvalues.}
\label{fig:twistbdy}
\end{figure}

Note that the length coordinates for each small annulus (the A-cycle eigenvalues) provide the puncture eigenvalues at the two punctures on the big boundary where the annulus ends (Figure \ref{fig:twistbdy}). From the perspective of the big boundary, these are fixed. Moreover, the holonomy around any puncture on the big boundary that is filled in by a small disc is defined to be unipotent. If it were not so, it would be impossible to extend flat connections from $\pd M$ to $M$.

\begin{wrapfigure}{r}{1.5in}
\centering
\includegraphics[width=1.3in]{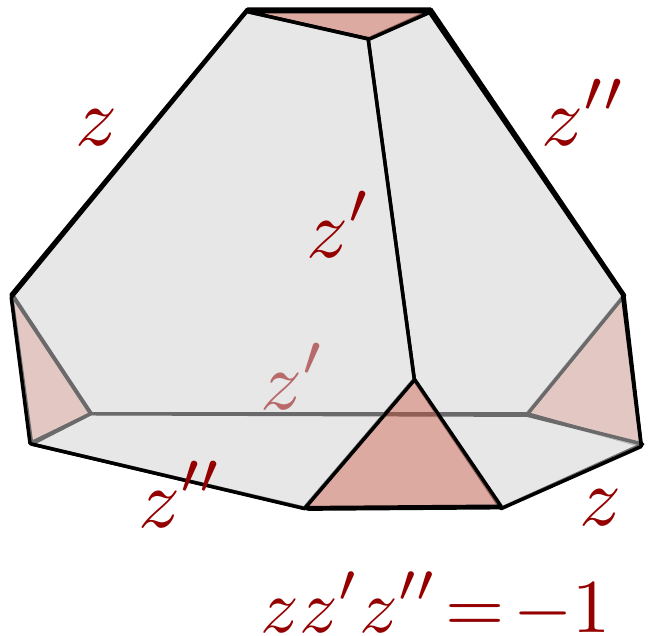}
\caption{$\CP_2(\pd \Delta)$}
\label{fig:tetz}
\end{wrapfigure}

A standard example of a pair $\CL_2(M)\subset \CP_2(\pd M)$ occurs when $M=S^3\bs \CK$ is a knot complement \cite{gukov-2003}. Then $\CP_2(\pd M) \simeq \C^*\times \C^*$ and $\CL_2(M)$ is cut out by a polynomial in two variables, the A-polynomial of the knot \cite{cooper-1994}. A much more fundamental example is for $M=\Delta$ a tetrahedron. The tetrahedron has a canonical boundary triangulation $\mb t$, leading to a phase space
\be \begin{array}{c} \CP_{\pd\Delta} := \CP_2(\pd\Delta,\mb t) = \{z,z',z''\in \C^*\,|\, zz'z''=-1\}\simeq \C^*\times \C^*\,; \\[.2cm]
\{z,z'\}=zz'\,,\; \{z',z''\}=z'z''\,,\;\{z'',z\}=z''z\,.
\end{array} \label{Ptet} \ee
The coordinates $z,z',z''$ label pairs of opposite edges of the tetrahedron. The fact that the product equals $-1$ enforces unipotent holonomy, \cf\ \eqref{prodx}. The Lagrangian submanifold is~\cite{Dimofte-QRS, DGG}
\be \CL_\Delta := \CL_2(\Delta,\mb t) = \{z''+z^{-1}-1=0\}\,. \label{Ltet}\ee
Notice that the pair $\CL_\Delta\subset \CP_{\pd \Delta}$ would not make sense without the addition of framing! Indeed, the standard space of flat connections on a tetrahedron is trivial. In this case, all information lies in the framing, and identifies $\CL_\Delta$ with the configuration space of four lines in $\C^2$.

In general, in each of the complex-torus patches $\CP_2(\pd M,\mb t)$ of a boundary phase space, labelled by a boundary triangulation $\mb t$, we can consider the Lagrangian submanifold $\CL_2(M,\mb t) :=  \CL_2(M)\cap \CP_2(\pd M,\mb t)$. A central property of the Lagrangian pairs $\CL_2(M,\mb t)\subset \CP_2(\pd M,\mb t)$ is that they can be glued together by \emph{symplectic reduction}.

Namely, if one glues together two manifolds $M_1,M_2$ along part of a common big boundary (where their boundary triangulations $\mb t_1$, $\mb t_2$ agree) to obtain a new manifold $M$, then \cite{Dimofte-QRS}
\be \CP_2(\pd M,\mb t) \simeq \big[\CP_2(\pd M_1,\mb t_1)\times \CP_2(\pd M_2,\mb t_2)\big]\big/\!\!\big/ (\C^*)^{\text{\# internal edges}}\,, \label{redP}\ee
taking a symplectic quotient with respect to a $\C^*\simeq GL(1,\C)$ action for every new internal edge created during the gluing. The products of big-boundary coordinates $x_E$ around these internal edges are the moment maps for the reduction, which must be set to $1$ (since the holonomy of a flat connection around an internal edge is trivial). Similarly, every new external edge $E'$ of $\pd M$ formed by gluing together edges $E_1,E_2$ gets a $\C^*$ coordinate $x_{E'}=x_{E_1}x_{E_2}$.%
\footnote{If new small annuli or tori are created during the gluing, the $\C^*$ coordinates associated to them in $\CP_2(\pd M,\mb t)$ are also Laurent monomials of the $\C^*$ coordinates in $\CP_2(\pd M_1,\mb t_1)\times \CP_2(\pd M_2,\mb t_2)$. See Appendix~\ref{app:coords}.} %
Most importantly, in parallel with the reduction of boundary phase spaces, the Lagrangian $\CL_2(M,\mb t)$ is obtained by pulling the product $\CL_2(M_1,\mb t_1)\times \CL_2(M_2,\mb t_2)$ through the quotient in \eqref{redP}.

As an application of the gluing formula, one can fully decompose $M$ into tetrahedra $\Delta_i$, so that
\be \CP_2(\pd M,\mb t) = \Big[\prod_i \CP_{\pd \Delta_i}\Big]\big/\!\!\big/ (\C^*)^{\text{\# internal edges in $M$}}\,, \label{PtetM} \ee
while $\CL_2(M,\mb t)$ is the reduction of a product of canonical Lagrangians \eqref{Ltet}. Generically (that is, for suitably refined triangulations) the resulting pair $\CL_2(M,\mb t)\subset \CP_2(\pd M,\mb t)$ is independent of how $M$ is decomposed.
The very special symplectic properties of edge-coordinates that underlie the reductions \eqref{redP}--\eqref{PtetM} were first studied in 3d hyperbolic geometry by Neumann and Zagier \cite{NZ} (and later Neumann \cite{Neumann-combinatorics}), following work of Thurston \cite{thurston-1980}.

\subsubsection{The 3d theories}

Now let us take a 3-manifold $M$ with big-boundary triangulation $\mb t$. Let $d = \sum_i d_i$ be the dimension of the phase spaces associated to big boundaries, and let $a,t$ denote the number of small annuli and tori. Let us also choose a polarization $\Pi$ for $\CP_2(\pd M,\mb t)\simeq (\C^*)^{2d}\times(\C^*)^{2a}\times (\C^*)^{2t}$, splitting the coordinates in each factor into canonically conjugate pairs of positions and momenta, or ``electric'' and ``magnetic'' coordinates. For big boundaries, this is equivalent to a splitting of the electric-magnetic charge lattice $\Z^{2d}$ from Section \ref{sec:2d}.

The 3d $\CN=2$ superconformal theory $T_2[M,\mb t,\Pi]:= T_2[M,\mb t,\Pi]$ constructed in \cite{DGG} turns out to have the following basic properties.
\begin{enumerate}

\item $T_2[M,\mb t,\Pi]$ has a UV Lagrangian description as an abelian Chern-Simons-matter theory; \ie, it is in class $\CR$.

\item $T_2[M,\mb t,\Pi]$ has a manifest $U(1)^{d+a+t}$ flavor symmetry as well as a $U(1)_R$ R-symmetry. Each $U(1)$ is associated to one of the electric coordinates in the polarization $\Pi$.

\item $T_2[M,\mb t,\Pi]$ has a chiral operator $\CO_E$ for every edge $E$ on the big boundary of $M$.

3a.\;\,If the edge $E$ is electric (meaning, \eg, that $x_E$ is a monomial of purely electric coordinates in the big-boundary phase space) then $\CO_E$ is an ordinary chiral operator, transforming with charge $+1$ under the $U(1)$ flavor symmetry corresponding to $E$.

3b.\;\,If $E$ has nontrivial magnetic charge, then $\CO_E$ exists in the presence of a magnetic-monopole background for an appropriate $U(1)$ flavor symmetry.%
\footnote{Background monopole operators like this and their anomalous dimensions were recently analyzed in \cite{PS-mon}.}

\end{enumerate}
Note that degenerate edges as in Figure \ref{fig:lollipop} are excluded from property (3).%
\footnote{From a combinatorial 3d perspective, such degenerate edges are excluded because, no matter how a 3-manifold is triangulated, it is impossible for a degenerate edge to contain only electric edges ($z$ and not $z',z''$) of the individual tetrahedra. Then, as discussed in \cite[Sec 4.1]{DGG}, it is impossible to define a corresponding operator~$\CO_E$.} %
We will eventually see that they come with doublets of chiral operators under enhanced $SU(2)$ flavor symmetries. 

A change of polarization $\Pi\to \Pi'$ is implemented by a symplectic transformation $g\in Sp(2N,\Z)$, with $N=d+a+t$. In the context of quantization (which we shall touch upon momentarily) this can be extended to an affine symplectic $ISp(2N,\Z)\simeq Sp(2N,\Z)\ltimes \Z^{2N}$ action that includes multiplicative ``shifts'' of $\C^*$ coordinates by $-q^{1/2}$. The affine symplectic group also acts on 3d $\CN=2$ SCFT's with a $U(1)_R$ symmetry \cite{Witten-sl2}, providing a natural way to change the polarization of $T_2[M]$:
\be T_2[M,\mb t,g\circ \Pi] = g \circ T_2[M,\mb t,\Pi]\,.\ee
Specifically, ``$T$-type'' elements of the symplectic group add background Chern-Simons couplings for flavor symmetries (redefining flavor currents); ``$S$-type'' elements gauge a flavor symmetry, replacing it with a new topological $U(1)_J$ global symmetry (then flowing to the IR); while affine shifts add flavor currents to the R-current.%
\footnote{See the appendix of \cite{DGG-Kdec} for a thorough review.}
This action implies an extension of the standard property (3a) above to the more general statement (3b).

The mathematical operation of symplectic reduction also translates nicely to 3d $\CN=2$ SCFT's \cite{DGG}. Basically, a quotient by a $\C^*$ action as in \eqref{redP} becomes the operation of adding an operator to the $\CN=2$ superpotential to break a corresponding $U(1)$ symmetry, then flowing to the IR. (For this to make sense, one must make sure to use a polarization for which the moment map of the $\C^*$ action is electric.) Therefore, gluing together two manifolds to form $M=M_1\cup M_2$ translates to taking a product of theories, changing polarization if needed, and adding superpotential operators corresponding to the new internal edges created in the gluing
\be T_2[M,\mb t,\Pi] = g\circ\big(T_2[M_1,\mb t_1,\Pi_1]\times T_2[M_2,\mb t_2,\Pi_2]\big) \quad+\;\; \Big\{ W = \sum_{E_I}\CO_{E_I}\Big\}\,.\ee

An immediate consequence of the gluing prescription is that any three-manifold theory can be glued together from elementary tetrahedron theories
\be T_\Delta := T_2[\Delta,\mb t,\Pi_z]: \left\{\begin{array}{l}
\text{free chiral multiplet $\phi$} \\ \text{with charges (+1,0) under $U(1)_{flavor}\times U(1)_R$ symmetry\,;} \\
\text{level -1/2 background Chern-Simons term for $(A_{flavor}-A_R)$.}
\end{array}\right.
\ee
The canonical tetrahedron polarization $\Pi_z$ is chosen so that $z$ is an electric coordinate and $z''$ is its magnetic conjugate. Then the operator $\CO_z=\phi$ is associated to the pair of opposite edges of the tetrahedron in Figure \ref{fig:tetz} labelled ``$z$''. By gluing $T_2[M,\mb t,\Pi]$ from a product of tetrahedron theories, one naturally finds an abelian class-$\CR$ Lagrangian in the UV.

\begin{figure}[htb]
\centering
\includegraphics[width=4.7in]{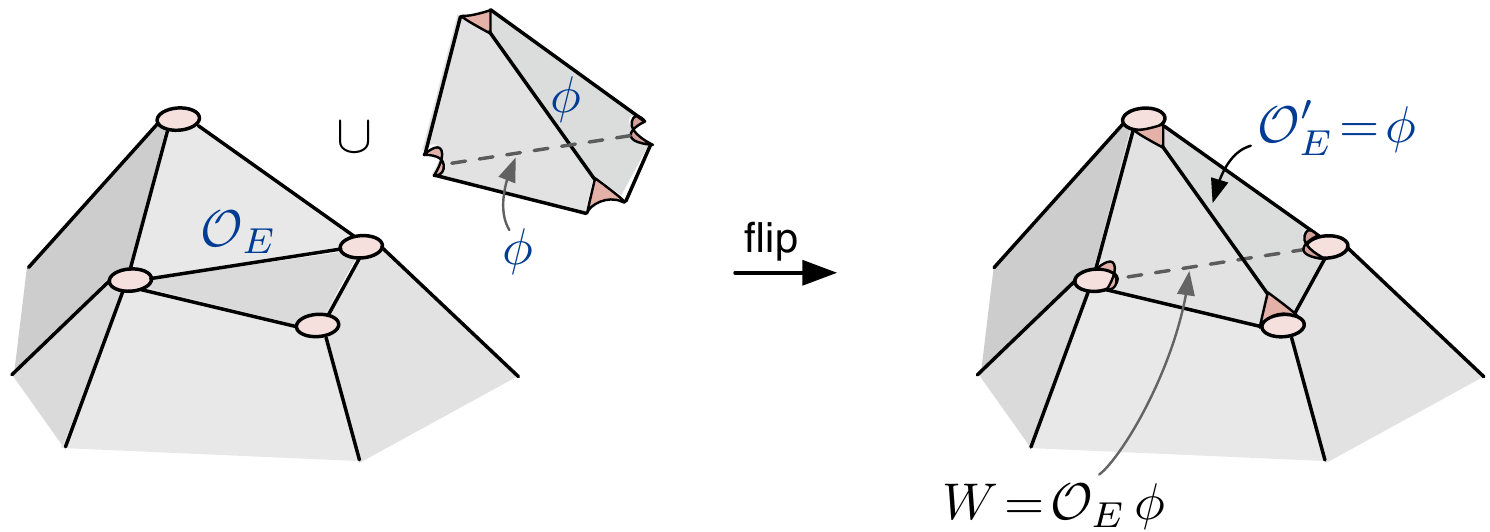}
\caption{Adding a tetrahedron (with operator $\phi$) to effect a flip on $\pd M$.}
\label{fig:flip3d}
\end{figure}

Another easy application of the gluing rule is to describe how a theory changes when an edge of the big-boundary triangulation flips. In 3d, a flip $F_E:\mb t\to \mb t'$ on an edge $E$ is implemented by gluing a tetrahedron onto the big boundary (Figure \ref{fig:flip3d}). The tetrahedron is glued along two adjacent faces, so that a new internal edge $E_I$ is created. Working in a polarization so that $E$ is electric (and using a polarization for the tetrahedron so that its electric edge is glued to $E$), we find that the theory $T_2[M,\mb t,\Pi]$ with associated operator $\CO_E$ transforms to
\be \text{flip}:\quad T_2[M,\mb t',\Pi] = T_2[M,\mb t,\Pi]\times T_\Delta \quad +\;\;\; \left\{W=\CO_E\,\phi\right\}\,.\qquad \ee
The internal-edge operator we have added is $\CO_{E_I}=\CO_E\,\phi$. The new theory has an operator $\CO_{E'}=\phi$ corresponding to the flipped edge $E'$, with charges exactly opposite those of $\CO_E$.

\begin{figure}[htb]
\centering
\includegraphics[width=3.5in]{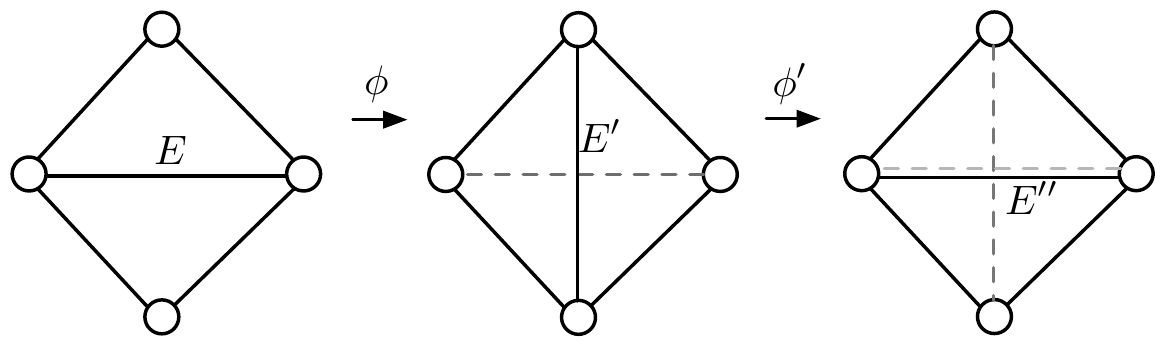}
\caption{Double-flip, a trivial operation.}
\label{fig:doubleflip}
\end{figure}

Note that flipping twice is a trivial operation once we flow to the IR. Indeed, given a superpotential
\be W = \CO_E\,\phi+\phi \phi'\,, \ee
where we have in mind that $\phi=\CO_{E'}$ (the first flip) and $\phi' = \CO_{E''}$ (the second flip), we can simply integrate out the massive fundamental field $\phi$. We find that $\CO_E = \phi'=\CO_{E''}$, reflecting the fact that the doubly-flipped edge $E''$ is identical to $E$ (Figure \ref{fig:doubleflip}).

\subsubsection{Line operators}
\label{sec:lineops}

To complete the circle of 3d ideas, let us recall how the Lagrangian $\CL_2(M,\mb t)$ arises in 3d theories.

If the theory $T_2[M,\mb t,\Pi]$ is compactified on a circle $S^1_\beta$, the twisted masses associated to every $U(1)$ flavor symmetry are complexified by Wilson lines, and can directly be identified with the position coordinates in $\Pi$. The canonically conjugate momentum coordinates in $\Pi$ are vevs of the complexified moment map operators for these $U(1)$'s. 
It is useful to couple the 3d theory supersymmetrically to a $\CN=2$ abelian four-dimensional gauge theory defined on half space. Then 
the coordinates on $\CP_2(\pd M,\mb t)$ are vevs of half-BPS flavor Wilson line operators and dual flavor `t-Hooft lines in the four-dimensional theory. 
They are not independent: we can bring the four-dimensional operators on the boundary, and there the vevs must lie on the Lagrangian $\CL_2(M,\mb t)$, due to the coupling to $T_2[M,\mb t,\Pi]$.

Indeed, upon compactification on a circle, the four-dimensional gauge theory reduces at low energy to a 2d $\CN=4$ sigma model whose target is the complex symplectic manifold 
$\CP_2(\pd M,\mb t)$. The boundary condition defined by coupling to $T_2[M,\mb t,\Pi]$ constrains the low energy fields to live on a Lagrangian submanifold. The boundary condition preserves 
$\CN=(2,2)$ supersymmetry in two dimension. The generating function of the Lagrangian $\CL_2(M,\mb t)$ coincides with the the effective twisted superpotential $\wt W$ 
for the two-dimensional degrees of freedom at the boundary, a function of complex masses $x$ and complex gauge scalars $\sigma$. Thus the equations for $\CL_2(M, \mb t)$ arise as
\cite{DG-Sdual, DGG}
\be \exp \bigg(x\frac{\pd \wt W}{\pd x}\bigg) = p\;\bigg|_{\pd \tilde W/\pd \sigma=0}\,,\ee
where $p$ are the momenta conjugate to $x$.

The line operators can be ``quantized'' by adding angular momentum, changing the geometry to $\R^2\times_q S^1_\beta$. Then the order in which line operators are brought to the boundary 
to act on the three-dimensional theory $T_2[M,\mb t,\Pi]$ matters, and conjugate pairs obey $\hat p\hat x=q\hat x\hat p$. Moreover, they satisfy a Ward identity that is a quantization of the $\CL_2(M,\mb t)$ equations. For example, for a tetrahedron the Ward identity is
\be \hat\CL_\Delta:\quad \hat z''+ \hat z^{-1}-1 \simeq 0\,.\ee
It should be viewed as the generator of a left ideal in the operator algebra.
Mathematically, the quantization of Lagrangians $\CL_2(M,\mb t)$ is uniquely defined by pulling products of $\hat \CL_\Delta$'s through symplectic reduction.

It is again convenient to work with logarithmic coordinates in order to properly keep track of $q$ (or $\hbar=\log q$) corrections. In this case, the affine extension of the symplectic action on polarizations or 3d theories is crucial --- affine shifts correspond to multiplying coordinates powers of $q$. This is easy to understand physically, since affine shifts change the R-current and there is an R-symmetry twist in the geometry $\R^2\times S^1_\beta$. Further details appear in \cite{DGG} and the appendix of \cite{DGG-Kdec}.

\subsection{Half-BPS boundary conditions}
\label{sec:bdycouple}

The properties of the theory $T_2[M,\mb t,\Pi]$ suggest a natural way to couple it to a 4d Seiberg-Witten theory $SW_2[\CC]$ corresponding to the big boundary of $M$, as a 3d boundary condition.

A half-BPS boundary condition for a supersymmetric theory generally splits supermultiplets in half (according to the broken supersymmetry), giving one half Neumann boundary conditions (N b.c.) and the other half Dirichlet (D b.c.) \cite{dWFO, GW-boundary}. More precisely, half the bosonic fields get N b.c. and the other half D b.c., while half the fermions are set to zero.

In the present case, we are interested in superconformal boundary conditions for a 4d $\CN=2$ Seiberg-Witten theory, which break SUSY to $\CN=2$ in 3d. The boundary conditions should preserve a $U(1)_R\subset SU(2)_R$ R-symmetry and break $U(1)_r$. The breaking pattern is thus characterized by parameters $\omega \in SU(2)_R/U(1)_R \simeq \cp^1$ and by a phase $\zeta \in U(1)_r$.

We put the theory on $\R^3\times \R_+$, with $x^3\geq 0$ parameterizing $\R_+$. The standard Neumann boundary condition 
for gauge fields essentially gives N b.c. to the components $A_\parallel$ of the gauge field parallel to the boundary, and D b.c. to the component $A_\perp$ perpendicular to the boundary. Correspondingly, the complex scalar $\Phi$ in the gauge multiplet gets split roughly  into real and imaginary parts, with half N b.c. and half D b.c., so that
\be \pd_3\, \Re\big( \zeta^{-1} \Phi\big)\big|_\pd = 0\,,\qquad\, \Re\big( \zeta^{-1} \tau \Phi\big)\big|_\pd =0\,, \label{bdyadj} \ee
where ``$|_\pd$'' denotes restriction to the boundary $x^3=0$\footnote{For standard Dirichlet b.c. for the gauge fields, the role of the two 
parts of $\Phi$ are exchanged}. 
Altogether, the free boundary values of the fields $A_\parallel$ and $\Re\big( \zeta^{-1} \varphi\big)$ compose a 3d $\CN=2$ gauge multiplet. In agreement with the boundary condition on adjoint scalars, the real 3d central charge of any state that transforms under bulk symmetries must be
\be Z_{\rm 3d} = \Re\big(\zeta^{-1}Z_{\rm 4d}\big)\,. \label{bdyZ} \ee

In addition, each hypermultiplet $\Phi$ is split into a pair of chiral/anti-chiral fields $(X,Y^\dagger)$ with well-defined $U(1)_R$ charges $\pm 1$ and identical flavor quantum numbers. More commonly, we write this as a pair of chirals $(X,Y)$ with $R=1$ and opposite flavor symmetry. Then the basic boundary conditions, chosen independently for each hyper, are to give N b.c. to (the bosonic fields in) $X$ and D b.c. to $Y$, or vice versa. We'll denote this as (\cf\ \cite{DG-E7})
\be \label{NDchiral}
 \CB_X:\begin{array}{c} \pd_3 X|_\pd = 0\\[.1cm]
  Y|_\pd = 0 \end{array} \qquad\text{or}\qquad \CB_Y: \begin{array}{c} \pd_3 Y|_\pd = 0\\[.1cm]
  X|_\pd = 0 \end{array}\,.
\ee
From a 3d perspective, the theory on $\R^4$ has a crucial $\CN=2$ superpotential
\be  W = \int_{\R_+} \pd_3 X\, Y = -\int_{\R_+} X\,\pd_3Y\,. \label{WXY} \ee
which shows how $\pd_3 X$ appears as the F-term in the chiral multiplet whose lowest component is $Y$ and $\pd_3 Y$ as the F-term for $X$.
This explains the complementary pairs of boundary conditions. 

The basic boundary conditions just described for a bulk $\CN=2$ theory $T_{\rm bulk}$ can be deformed in the presence of a 3d $\CN=2$ theory $T_{\rm bdy}$ on the boundary. First, any flavor symmetries of $T_{\rm bdy}$ may be either gauged in the bulk or identified with bulk flavor symmetries. That is, we can identify 3d background vector multiplets with boundary values of the pieces of 4d vector multiplets (dynamical or background) that have N b.c.

Thereafter, we may use boundary superpotentials to couple the boundary value of any bulk chiral $X$ or $Y$ that has N b.c. to a chiral operator of $T_{\rm bdy}$ with dual gauge/flavor charges:
\be W_{\rm bdy} = X|_\pd\cdot \CO \qquad\text{or}\qquad W_{\rm bdy} = Y|_\pd\cdot\CO'\,. \label{Wbdy} \ee
Due to the superpotential \eqref{WXY}, a boundary coupling $W_{\rm bdy} = X|_\pd\cdot \CO$ actually forces the D b.c. on $Y$ to be relaxed to $Y|_\pd = \CO$, and similarly for $X\leftrightarrow Y$. Altogether, the deformed b.c. for bulk hypermultiplets in the presence of $T_{\rm bdy}$ may be summarized as
\be \label{NDX}
 \CB_X[\CO]:\left\{\begin{array}{c} \pd_3 X|_\pd = F_\CO \\[.1cm]
  Y|_\pd = \CO \end{array}\right. \quad\text{or}\quad \CB_Y[\CO]: \left\{\begin{array}{c} \pd_3 Y|_\pd = F_{\CO'}\\[.1cm]
  X|_\pd = \CO' \end{array}\right..
\ee

Now, suppose that our bulk theory is $SW_2[\CC]$ and that our boundary theory is $T_2[M,\mb t,\Pi]$, where $\CC$ is a big boundary of $M$. The properties summarized in Sections \ref{sec:2d}--\ref{sec:3d} suggest a natural way to couple the two.

Let us choose any point $u$ on the Coulomb branch of $SW_2[\CC]$ (again we include masses and marginal UV couplings in `$u$') and an angle $\theta$ so that the WKB triangulation $\mb t_{u,\theta}$ of $\CC$ agrees with $\mb t$. We consider the bulk theory in an electric-magnetic duality frame that agrees with the polarization $\Pi$ on the big boundary. Then:
\begin{enumerate}

\item We break 4d $\CN=2$ supersymmetry in the bulk to a 3d $\CN=2$ subalgebra characterized by the phase $\zeta=e^{i\theta}$ (and $\omega\in \cp^1$) and impose the basic boundary conditions \eqref{bdyadj} for the (electric) bulk vector multiplets.

\item We use the surviving 3d $\CN=2$ $U(1)$ gauge multiplets at the boundary to gauge $d$ corresponding $U(1)$ flavor symmetries of $T_2[M,\mb t,\Pi]$.

\item The bulk theory has BPS hypers $\Phi_E=(X_E,Y_E)$ of charge $\gamma_E$ for every edge $E$ in the WKB triangulation. We impose N b.c. for the chiral halves $X_E$ (say) of these multiplets with charge $-\gamma_E$. Then we add a coupling $\CB_{X_E}[\CO_E]$ to the chiral operators $\CO_E$ with charge $+\gamma_E$ that must exist in $T_2[M,\mb t,\Pi]$.

\end{enumerate}

The third item in this list requires a little explanation. Obviously, there is no IR Lagrangian description of SW theory that can simultaneously accommodate 
all BPS hypermultiplets as elementary fields. At most, we can pick a maximal set of electric BPS particles in an adequate duality frame, maybe adjust the parameters so that these electric particles are much lighter than other dyonic particles, and include them in the effective Lagrangian. Then it makes sense to give them boundary conditions and couple them to the chiral operators for electric edges. On the other hand, heavy dyonic particles behave as line defects in the bulk SW theory. It makes sense to ask if such line defects  can end supersymmetrically on the boundary, where they look like monopole operators for a flavor symmetry of the 3d boundary theory. The coupling $\CB_{X_E}[\CO_E]$ for the corresponding edges should be interpreted as allowing such line defect to end on $\CO_E$. 

Away from loci where electric particles are light, we can adopt the line-defect point of view for all edge couplings. It smoothly reduces to the Lagrangian description 
of the coupling in the regions of parameter space with mutually local light particles. However, Argyres-Douglas regions of parameter space \cite{AD, APSW}, 
where mutually non-local particles are light, 
lack a complete Lagrangian description in the bulk; then we cannot hope to fully describe a boundary condition in an elementary way.

We should also note that the charges $\gamma_E$ do not exhaust the charges of all BPS particles. Rather, the charges of all BPS particles in the theory 
can be written as a linear combination of $\gamma_E$ with non-negative coefficients \cite{GMNIII}. In a certain sense (made precise by the BPS quivers program \cite{BPSquivers-complete, BPSquivers}, \cite{Kontsevich-PI})
all BPS particles can be thought as bound states of the ones with $\gamma_E$ charges. We should probably think that the bound states of charges $\sum_E n_E \gamma_E$
will couple to operators $\prod_E \CO_E^{n_E}$ at the boundary. This ansatz is compatible with the results of our next section, where we demonstrate that our couplings are covariant under changes of triangulation. 

Finally, in order to define couplings as above, it is not really necessary for $\mb t$ to agree with a WKB triangulation of $\CC$: any triangulation whose edges are associated to a positive basis of bulk BPS states will do. However, the WKB triangulation is especially natural, and leads to a nice interpretation of boundary conditions in terms of Janus configurations (Section \ref{sec:Janus}).

The couplings we have described may, in general, break the $SU(2)$ flavor symmetries of the bulk to $U(1)$'s. The $U(1)$ subgroups should be identified with $U(1)$ flavor symmetries of $T_2[M,\mb t,\Pi]$ coming from annuli of $M$ (that attach to holes on $\CC$). We will eventually argue in Section \ref{sec:enhance} that whenever a hole of $\CC$ is attached to an annulus (as opposed to being capped off by a small disc), full $SU(2)$ flavor symmetry of the coupled 3d-4d system is in fact restored.

\subsubsection{Dissolving polarization and triangulation}
\label{sec:dissolve}

Given any 3-manifold $M$ with big boundary $\CC$, or possibly a disjoint union of multiple big boundaries, the above rules define a boundary coupling between $SW_2[\CC]$ and $T_2[M,\mb t,\Pi]$. We could call the full 3d-4d theory $\CB_{SW}[M,\mb t,\Pi]$. Quite pleasantly, however, it turns out that in the IR the coupled system depends neither on a choice of big-boundary triangulation nor on a choice of polarization \cite{DGG}. Thus, it may unambiguously be called $\CB_{SW}[M]$.

Let us try to understand why this is true. First, looking only at gauge multiplets, consider what happens if we act with electric-magnetic duality on a combined 3d-4d system. The bulk electric-magnetic duality group $Sp(2d,\Z)$ simply acts by changing polarization on the boundary. This is precisely how the $Sp(2d,\Z)$ action on 3d CFT's was obtained in \cite{Witten-sl2}. Therefore, the two coupled systems

\be  SW_2[\CC]\text{---} T_2[M,\mb t,\Pi] \qquad\simeq \qquad g\circ SW_2[\CC] \text{---} T_2[M,\mb t,g\circ\Pi] \ee
are equivalent in the IR.
In one case, bulk electric $U(1)$'s gauge boundary flavor symmetries; while in the other case bulk magnetic $U(1)$'s gauge ``dual'' boundary flavor symmetries. 

\begin{figure}[htb]
\centering
\includegraphics[height=1.25in]{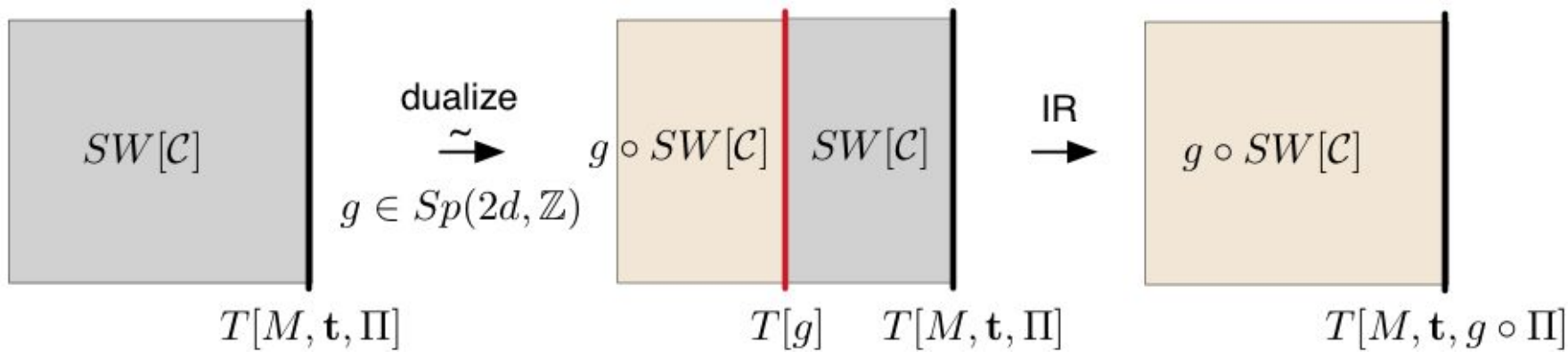}
\caption{Passing a duality wall through the 3d-4d system.}
\label{fig:wallPol}
\end{figure}

A nice way to picture the equivalence is by starting with $SW_2[\CC]$ coupled to $T_2[M,\mb t,\Pi]$ on a half-space, then dualizing the bulk theory on a slightly ``smaller'' half-space, as in Figure \ref{fig:wallPol}. This is equivalent to coupling the magnetic theory on the far left to an appropriate 3d abelian Chern-Simons theory on a BPS domain wall, then coupling the domain wall to a slice of the original electric theory on its right. For example, for a standard $g=S=\left(\begin{smallmatrix}0&-1\\1&0\end{smallmatrix}\right)$ action, the duality domain wall is ``theory'' of two background $U(1)$ flavor multiplets with a mixed Chern-Simons coupling $A'dA$. After flowing to the IR, the duality domain wall collides with the boundary condition, effecting a change of polarization on the boundary.

Keeping track of hypermultiplets and their couplings to chiral operators in the electric-magnetic duality requires a little extra care. For fixed triangulation $\mb t$, the duality action changes the set of edges that are electric, and thus the subset of couplings that can (and do) appear in a bulk-boundary superpotential. 

For example, suppose that an electric hyper $(X,Y)$ of $SW_2[\CC]$ couples to a chiral operator $\CO$ of $T_2[M,\mb t,\Pi]$, with $W_{\rm bdy} = X|_\pd\,\CO$. Let us dualize the system by passing an $S$ duality wall through the bulk and colliding it with the boundary, as in Figure \ref{fig:wallPol}. In addition to carrying an abelian Chern-Simons theory, the duality wall has the property that a magnetic line defect of $S\circ SW_2[\CC]$ on the left can end on a chiral operator $X|_{\rm wall}$ on the right --- put differently, that magnetic BPS particles passing through from the left become electric particles on the right. After colliding the duality wall with the boundary, we simply find that magnetic line defects in the bulk can end on a new chiral operator $\CO'$ in $T_2[M,\mb t,S\circ \Pi]$.

Similarly, when we collide an $S$-wall with the boundary, any magnetic line defect of $SW_2[\CC]$ that can end on both the boundary and the duality wall becomes trapped. It combines with the magnetic operator $\CO_m$ of $T_2[M,\mb t,\Pi]$ that it ends on to form a new electric operator $\CO_m'$ of $T_2[M,\mb t,S\circ\Pi]$. This operator $\CO_m'$ is coupled to an electric hyper $(X',Y')$ of $S\circ SW_2[\CC]$. Such mechanisms, involving line defects, demonstrate that bulk-boundary superpotential couplings transform covariantly, as desired, under a change of polarization.

To see that the system $\CB_{SW}[M,\mb t,\Pi]$ is independent of triangulation as well as polarization, we consider how the flip of an edge in $\mb t$ acts on our standard couplings \cite{DG-E7}. Suppose that the flipped edge $E$ is electric (we can adjust the polarization to make it so), and that before the flip half of a 4d hyper $(X,Y)$ is coupled to $\CO_E$ via the superpotential $W_{\rm bdy} = X\big|_\pd\,\CO_E$. After the flip, we introduce a new 3d chiral $\phi$ (from a tetrahedron theory) and the superpotential becomes
\be W_{\rm bdy} = Y\big|_{\pd}\,\phi + \phi\, \CO_E\,. \label{bdyflip}\ee
Note how the flip effectively switches N and D b.c. for the bulk hyper, so that the field $Y$, which carries (minus) the 4d charge of the flipped edge $-\gamma_{E'}=\gamma_E$ can couple to the new boundary operator $\CO_{E'}=\phi$. Concurrently, due to the bulk superpotential \eqref{WXY} we find modified D b.c. $X|_\pd=\phi$.
The coupling \eqref{bdyflip}, however, has made the 3d field $\phi$ massive. Flowing to the IR we may integrate it out, and simply get back to the original coupling $W_{\rm bdy} = X\big|_\pd\,\CO_E$. Therefore, $\CB_{SW}[M,\mb t,\Pi] = \CB_{SW}[M,\mb t',\Pi]$.

\begin{figure}[htb]
\centering
\includegraphics[height=1.25in]{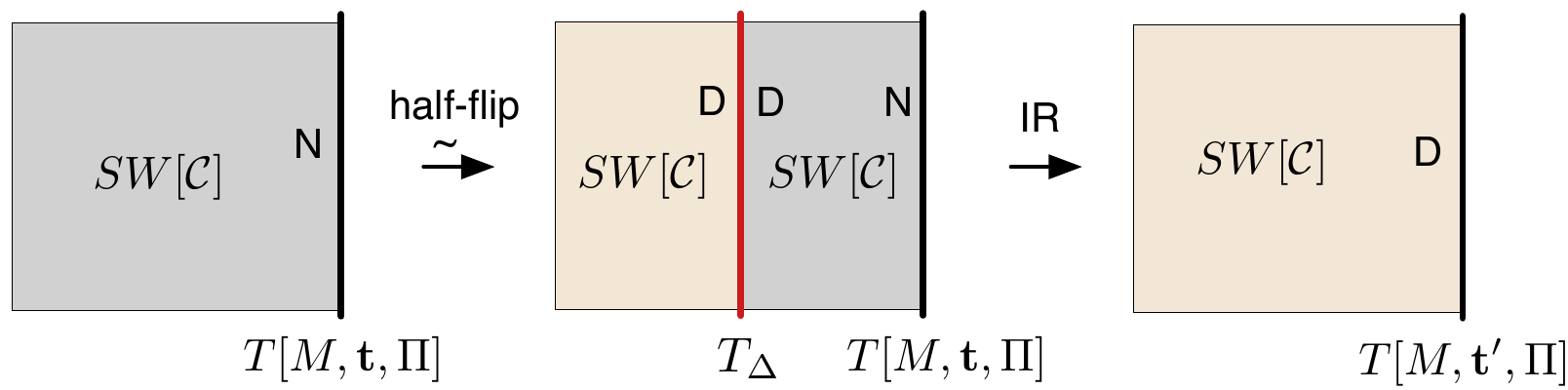}
\caption{Flipping the 3d-4d system.}
\label{fig:wallFlip}
\end{figure}

There is an interesting subtlety at play here, concerning the charges of magnetic chiral operators before and after the flip. 
For example, in the present $A_1$ case the flip of an edge of the triangulation modifies the charges carried by the nearby edges 
in a specific way. Physically, adding a chiral field charged under a flavor symmetry modifies the quantum numbers of monopole operators for that flavor symmetry because of the quantization of fermionic zero-modes. It would be interesting to 
study these charge shifts in a general theory, and to reproduce the expected
transformation of charges of fundamental BPS particles, encoded by a tropical cluster mutation.

We can picture the action of a flip using duality walls, in much the same way that we understood changes of polarization. Namely, we note that the tetrahedron theory $T_\Delta$ can trivially be inserted as a wall in any theory $SW_2[\CC]$, coupling to some hypermultiplet $(X,Y)$ via
\be W_{\rm wall}  = Y\big|_\pd\, \phi + \phi\, Y'\big|_\pd\,. \label{TDwall} \ee
Here $Y,\, Y'$ denote the half-hypers with Neumann b.c. on the two sides of the wall; their superpartners have Dirichlet b.c. $X|_\pd = \phi = X'|_\pd$. After integrating out $\phi$ in the IR, this wall becomes trivial. By inserting a $T_\Delta$ wall into a theory on a half-space (Figure \ref{fig:wallFlip}) and colliding it with the boundary, we effectively perform a boundary flip.

We have shown that the coupled system $\CB_{SW}[M]$ only depends on the topology of $M$ and its separated big/small boundary. Nevertheless, it should be clear that any Lagrangian realization of this theory does require a choice of $\mb t$ and $\Pi$. For example, if we ever want to make the 4d bulk theory infinitely weakly coupled, so that we just leave behind a dynamical 3d boundary theory, we must choose an electric-magnetic duality frame $\Pi$. These choices then appear as data for the 3d theory $T_2[M,\mb t, \Pi]$ that remains.

%%%%%%%%%%%%%%%%%%%%%%%%%%%%%%%%%%%%%%%%%%%%%%%%%%%%%%%%%%%%%%%%%%%%%%%%%%%%%
%%%%%%%%%%%%%%%%%%%%%%%%%%%%%%%%%%%%%%%%%%%%%%%%%%%%%%%%%%%%%%%%%%%%%%%%%%%%%
%%%%%%%%%%%%%%%%%%%%%%%%%%%%%%%%%%%%%%%%%%%%%%%%%%%%%%%%%%%%%%%%%%%%%%%%%%%%%

\section{Janus domain walls in 4d ${\cal N}=2$ gauge theories}
\label{sec:Janus}

In this section, we would like to review the idea of BPS Janus configurations for four-dimensional $\CN=2$ theories. Janus configurations provide the simplest examples of domain walls and boundary conditions associated to 3-manifolds. Moreover, in special limits, one can sometimes extract the field content of an effective 3d boundary theory from the data of a Janus configuration  --- thus confirming our more abstract, combinatorial constructions of 3-manifold theories. For example, in Section \ref{sec:JanusSW} we will use a Janus configuration to rediscover the basic boundary couplings in a 3d-4d system $\CB_{SW}[M]$. Later in Section \ref{sec:RG} we will introduce Janus configurations for RG walls.

\subsection{General concepts}
\label{sec:Janusgen}

A BPS  Janus configuration for a four-dimensional $\CN=2$ gauge theory is a modification of the Lagrangian (assuming one exists) that allows the gauge couplings $\tau$ and mass parameters $m$ to vary in an arbitrary way along 
a direction $x^3$, preserving three-dimensional ${\cal N}=2$ supersymmetry. %It is easy to write down explicitly the modified Lagrangian. 
The three-dimensional superalgebra has real central charges that are the real part of the four-dimensional central charges \eqref{bdyZ}. (We'll set $\zeta=1$ for simplicity.) In a supersymmetric vacuum, the 3d central charges must be constant.
Then the equations
\begin{equation} \label{d3Z}
\partial_3\, \mathrm{Re}\, Z^{\rm 4d}_{\gamma}\big(u(x^3), m(x^3), \tau(x^3)\big) =0
\end{equation} 
fix the allowed evolution of the Coulomb-branch moduli $u$ as a function of $x^3$. Indeed, the number of real constraints \eqref{d3Z} equals the real dimension of the Coulomb branch.

Janus configurations have a well-defined limit to a domain-wall configuration where the jumps in $m,\tau$ occurs suddenly. 
Furthermore, it is useful to observe that many protected quantities are homotopy invariant, unaffected by continuous deformations of the path $m(x^3),\tau(x^3)$. This includes sphere partition functions and indices, and (we expect) the infrared SCFT limits of the 3d theories themselves that become trapped on domain walls.

In \cite{CCV} a special case of Janus domain walls was considered, the ``R-flow'' where the variation of $m,\tau$ was arranged in such a way that 
the relative order of the phases of the $Z^{\rm 4d}_\gamma$ central charges would be constant, and the $Z^{\rm 4d}_\gamma$ would go to infinity at large positive or large negative $x^3$.
In this and later sections we will not impose such a constraint, but we will still take inspiration from the R-flow analysis. 

The notion of Janus domain wall is intimately related to the notion of S-duality wall. If we take a Janus configuration that flows from a weak coupling region 
to a region where an S-dual weak coupling description exists, it is natural to act with S-duality on a half-space, 
and have a domain wall interpolating between different weakly coupled descriptions of the theory. Up to D-term deformations, we will have a duality 
wall for each element of the S-duality groupoid of the theory, and duality walls will compose appropriately by collision. 

As the four-dimensional theories on the two sides are weakly coupled, we can sensibly talk about degrees of freedom living at the wall, and describe the duality wall as 
a specific 3d ${\cal N}=2$ SCFT coupled to the two dual descriptions of the same four-dimensional theory.
The canonical example is (mass deformed) $T[G]$, a 3d theory which appears on the S-duality wall between 
${\cal N}=2^*$ SYM with a gauge groups $G$ and its Langlands dual ${}^L G$. 
For $G=SU(2)$, $T[SU(2)]$ is simply ${\cal N}=4$ SQED with ${\cal N}_f=2$, a self-mirror theory 
with two $SU(2)$ flavor symmetries, acting on the Coulomb and Higgs branches respectively. The two bulk theories couple to these 
two $SU(2)$ flavor symmetries. 

The correspondence between Janus domain walls and S-duality walls is an exact UV statement. 
It is also useful to ask in general how a Janus configuration would look in the IR Seiberg-Witten description of the theory. There are a few things that can be said in complete generality (in particular, without specializing to theories of class $\CS$), and we would like to point out some interesting facts and open questions.

The abelian gauge field Lagrangian simply has a profile of the IR gauge coupling determined by $u(x^3)$ and $m(x^3),\tau(x^3)$. 
The massive BPS particles have a more interesting behavior: for generic $x^3$, they are not BPS anymore, 
as their mass $|Z_\gamma|$ is larger than the 3d central charge $\Re\,Z^{\rm 4d}_{\gamma}$. But at the special 
locations where $\Im\, Z^{\rm 4d}_{\gamma}=0$, a 4d BPS particle has a chance to be trapped, and behave as a 3d BPS particle. 
We expect that a 4d BPS hypermultiplet of charge $\gamma$, for example, will give rise to a 3d chiral multiplet of charge $\pm \gamma$, 
depending on the sign of $\partial_3 \mathrm{Im} Z_{\gamma}$. It would be interesting to verify this statement with a detailed calculation. 

We can use these observations to argue that the 3d ${\cal N}=2$ theories produced from Janus domain walls in 4d ${\cal N}=2$ gauge theories should 
always have a mirror description as abelian Chern-Simons-matter theories, \ie, theories of class $\CR$\,!
Indeed, the IR description of the Janus configuration can be easily converted to an abelian Chern-Simons-matter description.
The chiral matter arises from the trapped 4d BPS particles. We need to make each BPS particle locally electrically charged. 
This can be accomplished by picking an appropriate electric-magnetic duality frame in the neighborhood of the wall 
$\Im\, Z^{\rm 4d}_{\gamma}=0$, such that $\gamma$ is electric. Changes in the electric-duality frame for the four-dimensional gauge fields 
are implemented by abelian duality walls, which add appropriate Chern-Simons couplings  to the abelian gauge fields. 

Thus the matter content and the gauge Lagrangian of the 3d theories are easily derived in terms of the $x^3$ dependence of the $Z_\gamma$.
The latter may be hard to derive from $\tau(x^3)$ though, which was one reason for the restrictions imposed on R-flows in \cite{CCV}.
Furthermore, there is a final, crucial ingredient to be derived: the superpotential couplings. 
From the four-dimensional point of view, these should arise from non-local instanton effects that allow the chiral matter at different locations to interact. 
We do not know how to derive such contributions in general, and we feel it is a very interesting open problem. 
We owe much of our intuition on such instantonic processes to an ongoing project on Janus domain walls in 2d field theories \cite{GMW-graphs}.

It is easy to argue that such instanton effects must be present: they implement homotopy invariance for the Janus configuration. 
We can give two illuminating examples. First, consider a homotopy that interpolates between a Janus configuration where, 
for some $\gamma$,  $\Im\, Z^{\rm 4d}_{\gamma}$ is positive in a certain region, and a Janus configuration where $\Im\, Z^{\rm 4d}_{\gamma}$
becomes negative and then immediately positive again in the same region, so that we have two locations with $\Im\, Z^{\rm 4d}_{\gamma}=0$,
hosting two chiral particles $\phi_\pm$ of opposite charge. It is clear that in order to have homotopy invariance, 
the second configuration should include a superpotential coupling $\phi_+ \phi_-$, which makes the particles disappear in the IR. 
Such a superpotential may arise from an instanton process, where a 4d particle moves from one of the two locations.

A second example is a homotopy that crosses a basic wall of marginal stability for four-dimensional BPS particles,
interpolating between a configuration with two relevant BPS particles, of charges $\gamma_1$ and $\gamma_2$, giving rise to two 3d chirals, 
and a configuration with three relevant 4d BPS particles, of charges $\gamma_1$, $\gamma_2$ and $\gamma_1 + \gamma_2$, giving rise to three 3d chirals. 
As observed in \cite{CCV}, a natural way to insure homotopy invariance is to have a superpotential coupling of the $XYZ$ type between the three chirals, 
related by the basic 3d mirror symmetry to the configuration of two chirals. 

A full analysis of this setup would hopefully give the broadest possible generalization of our results. Leaving that for future work, we specialize now to 
four-dimensional theories in class ${\cal S}$.

\subsection{Janus configurations for class $\CS$}
\label{sec:JanusS}

We briefly review a few facts about the 6d interpretation of the IR description of Januses, along the lines of \cite{CCV}.

The Seiberg-Witten curve $\Sigma$ of a 4d theory $SW_K[\CC]$ of class $\CS$ is naturally expressed as a $K$-fold branched cover of the UV curve $\CC$, as discussed in Section \ref{sec:2d}.
In turn, a Janus configuration for $SW_K[\CC]$ amounts to varying the cover $\Sigma\overset{K}{\to} \CC$ in the $x^3$ direction, so as to sweep out a cover
\be \wt M\overset{K}\to M \ee
of a 3-manifold $M$ that topologically is equivalent to $\CC\times \R$.

As the 4d central charges are periods of the Seiberg-Witten differential $\lambda_{\rm SW}$, the condition \eqref{d3Z} implies that the real part of $\pd_3 \lambda_{\rm SW}$ is exact, \ie
\be
\partial_3 \Re\, \lambda_{\rm SW} =d \rho
\ee
for some real function $\rho$. In other words, the  1-form 
\be
v = \Re\, \lambda_{\rm SW} + \rho\, dx^3  \label{vSW}
\ee
is closed. Notice that $v$ can be viewed either as a single-valued 1-form on $\wt M$, or a multi-valued 1-form on $M$.

The 1-form $v$ parameterizes the Coulomb branch of the twisted 6d SCFT on $M$. Indeed, the twisting makes three of the scalar fields of the abelian 6d theory into a 1-form $v'$ in the same cohomology class as $v$. Given a metric $g$ on $M$, the BPS equations of the 6d theory imply that $v'$ is harmonic
\be d*_g v'= 0\,,\ee
\ie\ $v'$ is the harmonic representative of $[v]\in H^1(M,\R)$. The three components of $v'$ parameterize the deformations of the cover $\wt M$ in the cotangent bundle $T^*M$. In regions of $M$ where the cover becomes independent of $x^3$, so that $\pd_3v'=0$, we simply have $v=v'= \Re\,\lambda_{\rm SW}$, and the holomorphic Seiberg-Witten form $\lambda_{\rm SW}$ on $\Sigma$ can be reconstructed from its real part.

We can also understand the harmonic $v'$ in terms of an M-theory compactification. If we wrap $K$ M5 branes on the supersymmetric cycle $M\times \R^3$ in the 11d geometry $T^*M\times \R^5$, with any given metric $g$ on $M$, then at low energies the branes may separate in the fibers of the cotangent bundle, recombining (say) into a $K$-fold cover $\wt M$ of $M$. In  order to preserve supersymmetry, this cover, just like $M$, must be a special Lagrangian submanifold in $T^*M$. Deformations of special Lagrangians, however, are precisely parameterized by harmonic 1-forms --- in the present case, the 1-form $v'$.

If we compactify the entire setup on a circle in $\R^3$, we can even find a non-abelian origin for $v'$. We end up with D4 branes wrapping $M\times \R^2$, which support a twisted 5d super-Yang-Mills theory. The BPS equations are 3d analogues of Hitchin equations along $M$, just as in Section \ref{sec:flat2d}. They can be written compactly as
\be [D_i,D_j]=0\,,\qquad g^{ij}[D_i,D_j^\dagger]=0\,,\ee
where $D = d+\CA$ is a covariant derivative on $M$ formed from the complexified $SL(K)$ connection $\CA=A+i\,\Re\,\varphi$ as in \eqref{Acx}. In components, the equations read $F_A=[\varphi,\varphi]$ and $d_A\varphi=d_A*\varphi=0$, and thus imply that the adjoint-valued 1-form $\varphi$ is covariantly harmonic. If we are able to diagonalize $\varphi$ locally --- corresponding to a well-defined separation of the D4 branes in $T^*M$ --- then the eigenvalues of $\varphi$ become a standard (multivalued) harmonic 1-form $v'$ on $M$.

Now let us follow the analysis of \cite{CCV} further, recast in a field-theory language, to describe the BPS content of an effective theory $T_K[M]$.
First, recall that four-dimensional BPS particles are represented by (webs) of strands of 6d BPS strings.
In 2d compactifications, the BPS strings follow trajectories of constant phase of $\lambda_{\rm SW}$, ending at branch points where two sheets of the branched cover $\Sigma \to C$ meet. These trajectories lift to 1-cycles in $\Sigma$, \cf\ Figure~\ref{fig:WKB}.

In a 3d compactification, the BPS strings must follow trajectories on which the harmonic 1-form $v'$ restricts to the volume form, in a given background metric. These trajectories end on the branch-lines of the cover $\wt M\to M$, swept out by the $x^3$ evolution of branch points on $\CC$. They lift to minimum-volume, or ``calibrated'' 1-cycles on $\wt M$. In the limit where $\CC$ varies slowly, the minimum-volume condition reduces to the $\Im\,Z^{\rm 4d}_\gamma=0$ condition for a trapped BPS state. 
The wrapped BPS strings then give rise to 3d chiral multiplets.

An important payoff of the 6d description is that the instantons that give rise to superpotential terms 
can also be described in terms of BPS strings that trace out minimum-volume discs in $M$. These discs have boundary along branch lines and along the minimum-volume trajectories that gave rise to BPS states.
In \emph{principle}, by fully analyzing the allowed configurations of 6d BPS strings, one can recover a full abelian Chern-Simons-matter description of a domain-wall theory.

In practice, in order to be able to identify BPS cycles, one must judiciously choose the profile $m(x^3),\tau(x^3)$, or the background metric on $M$, or both. In particular, it is useful to arrange that branch lines be well separated throughout most of $M$, and only coming close together briefly. Then each region of $M$ where a pair of branch lines pass near each other supports a single BPS chiral --- and can be shown to map to one of the combinatorial tetrahedra%
\footnote{For $K>2$, these regions maps to octahedra in the $K$-decomposition of a tetrahedron, as described in \cite{DGG-Kdec}.} %
reviewed in Section \ref{sec:couplings}. Of course, there may be many ways to put the branch lines in such a ``nice'' configuration, just as there are many ways to triangulate a 3-manifold; the different Chern-Simons matter Lagrangians obtained should all be related by 3d mirror symmetry, flowing to the same IR SCFT.

\subsection{Seiberg-Witten walls and couplings}
\label{sec:JanusSW}

To be a little more specific, let us review the basic IR Janus configuration of \cite{CCV}, and show how it generalizes to a $T_\Delta$ duality wall in a 3d-4d system, with couplings as in Section \ref{sec:bdycouple}.

We can model a local region of $M$ where a pair of branch lines come close together by evolving the Seiberg-Witten curve
\be \Sigma:\quad w^2 = -z^2 + m\,,\qquad \lambda_{SW} = w\,dz\,. \ee
The curve is fibered over the $z$-plane, which represents an open neighborhood of $\CC$. This happens to be the Seiberg-Witten curve of the $A_1$ Argyres-Douglas theory \cite{AD}.

For fixed $m$, there is an obvious trajectory of constant phase that connects the branch points at $z=\pm \sqrt m$, and lifts to a 1-cycle $\gamma$ in $\Sigma$. Thus the 4d theory $SW_2[\CC]$ has a BPS hypermultiplet $\Phi=(X,Y)$ of central charge
\be Z^{\rm 4d}_\gamma = \frac1\pi\oint_\gamma \lambda_{\rm SW} = \frac2\pi\int_{-\sqrt m}^{\sqrt m}\lambda_{\rm SW} = m\,,\ee
and mass $|m|$. The hyper is charged under a $U(1)$ flavor symmetry (which could be gauged in the global Seiberg-Witten theory $SW_2[\CC]$).

We may form a Janus configuration by giving $m$ a profile
\be m(x^3) = m_0 + ix^3\,,  \label{Janusm} \ee
for fixed real $m_0$. This is an R-flow configuration. There is a trapped 3d BPS chiral $\phi\equiv X$ (say) at $x^3=0$, where $\Im\,Z^{\rm 4d}_\gamma = 0$. Its 3d central charge is $m_0$. Since $|m|\to\infty$ as $x^3\to \pm \infty$, the 4d bulk theory completely decouples on either side of the domain wall and we are left with an effective 3d theory $T_\Delta$. If we had set $m(x^3) = m_0 - ix^3$ instead, we would have obtained a 3d chiral $\phi'\equiv Y$ of opposite charge, and 3d central charge $-m_0$.

\begin{figure}[htb]
\centering
\includegraphics[width=5in]{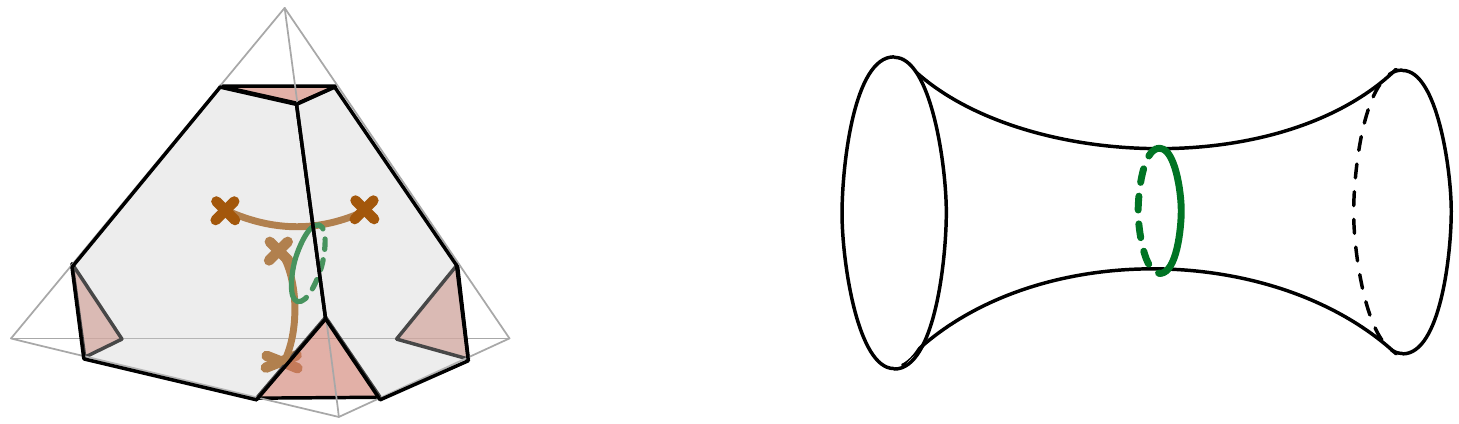}
\caption{Left: the branch lines in a double cover $\wt \Delta$ of the tetrahedron. Right: schematic 2d slice of the cover $\wt \Delta$ and the minimal-volume cycle giving rise to a 3d chiral $\phi$.}
\label{fig:tetlines}
\end{figure}

The Janus configuration \eqref{Janusm} leads to the local geometry $\wt M\to M$ represented by a tetrahedron, with four branch points on its boundary (one on each face) connected pairwise by branch lines, as in Figure \ref{fig:tetlines}. It is helpful to note that the WKB triangulation of $\CC$ in the neighborhood of our two branch points looks like a square (see Figure \ref{fig:flip2d} on page \pageref{fig:flip2d}). Indeed, this is triangulation of the UV curve of the $A_1$ Argyres-Douglas theory, which is just a disc with an irregular singularity/puncture on its boundary \cite{GMNII}. There are two choices for how the square is triangulated, one appropriate for $\Im\, m > 0$ and one for $\Im\, m < 0$. The 3d tetrahedron flips the triangulation, as in Figure \ref{fig:flip3d}.

\begin{figure}[htb]
\centering
\vspace{.2in}
\includegraphics[width=5.8in]{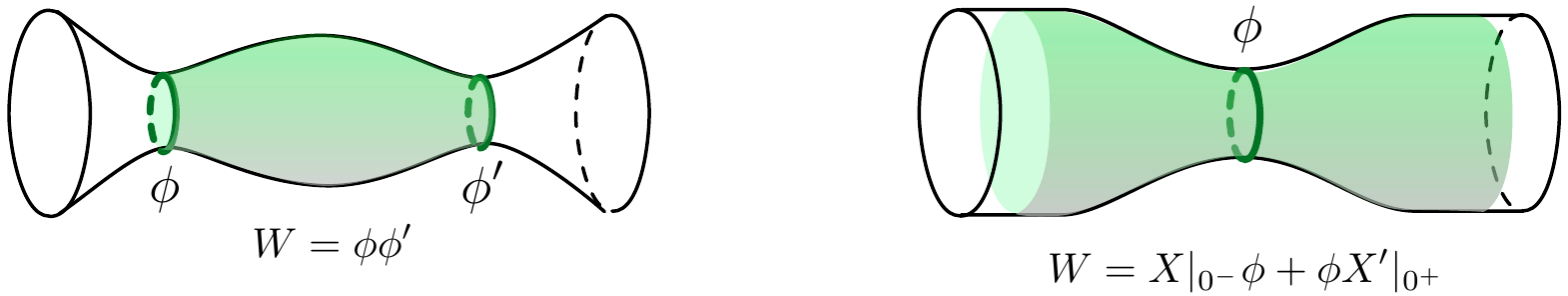}
\caption{Schematic 2d slices of the covers $\wt M$ for flows corresponding to $m(x^3) = m_0 +i\big[c-(x^3)^2\big]$ with $c>0$ (left) and $m(x^3) = m_0 + i \tanh(x^3)$ (right).}
\label{fig:Rflow}
\end{figure}

By introducing a more complicated profile for the Janus configuration
\be m(x^3) = m_0 +i\big[c-(x^3)^2\big]\,,\ee
which is no longer an R-flow, but still has $|m|\to \infty$ asymptotically, we find more interesting trapped 3d theories. Now for $c<0$ the theory has no trapped 3d particles (it is impossible to satisfy $\Im\, m =0$); while for $c>0$ there are two BPS chirals $\phi$ and $\phi'$ of opposite charge, because the locus $\Im\, m=0$ is crossed twice. In the latter case, there is also an instanton that mediates an interaction between these chirals coming from a 6d BPS string that wraps an annulus in $\wt M$ (a disc in $M$), as in Figure \ref{fig:Rflow}. The instanton generates the 3d superpotential
\be W = \phi \phi'\,, \ee
so that ultimately, in the IR, the 3d theory is again empty. This example demonstrates the basic homotopy invariance that was discussed in Section \ref{sec:Janusgen}.

Finally, suppose that we do \emph{not} send $|m|\to\infty$ asymptotically, so that we do not decouple the 4d bulk. We could have a profile
\be m(x^3) = m_0 + i \tanh(x^3)\,.\ee
We cross $\Im\,m=0$ once, so we get a 3d BPS chiral $\phi \simeq Y|_{x^3=0}$, but we still keep the 4d hypermultiplet in the asymptotic regions $x^3\to \pm \infty$. Let us call the 4d hyper $\Phi=(X,Y)$ at $x^3\ll0$ and $\Phi'=(X',Y')$ at $x^3\gg0$. Now in the 6d theory there are two BPS strings that wrap annuli with one boundary on the minimal-volume cycle $\gamma$, and another boundary in an asymptotic region. In a limit where the jump in the Janus configuration happens instantaneously, we find two copies of the 4d theory $SW_2[\CC]$ on half-spaces $x^3< 0$ and $x^3>0$, along with bulk-boundary superpotential couplings generated by the string-instantons:
\be  W_{\rm wall} = X\big|_{x^3=0^-}\phi + \phi\,X'\big|_{x^3=0^+}\,. \ee
This are precisely the kind of couplings we described combinatorially in Section \ref{sec:bdycouple}, and specifically correspond to the $T_\Delta$ wall discussed around \eqref{TDwall}.

%%%%%%%%%%%%%%%%%%%%%%%%%%%%%%%%%%%%%%%%%%%%%%%%%%%%%%%%%%%%%%%%%%%%%%%%%%%%%%%
%%%%%%%%%%%%%%%%%%%%%%%%%%%%%%%%%%%%%%%%%%%%%%%%%%%%%%%%%%%%%%%%%%%%%%%%%%%%%%%
%%%%%%%%%%%%%%%%%%%%%%%%%%%%%%%%%%%%%%%%%%%%%%%%%%%%%%%%%%%%%%%%%%%%%%%%%%%%%%%

\section{RG walls}
\label{sec:RG}

We now turn to RG domain walls and the framed 3-manifolds that give rise to them. The basic manifold $M_{\rm RG}$ defining an RG wall for pure $SU(2)$ Seiberg-Witten theory will turn out to play a crucial role in the construction of more general RG and S-duality walls, and the general analysis of enhanced flavor symmetry for theories of class $\CR$.

\begin{figure}[htb]
\centering
\includegraphics[width=3in]{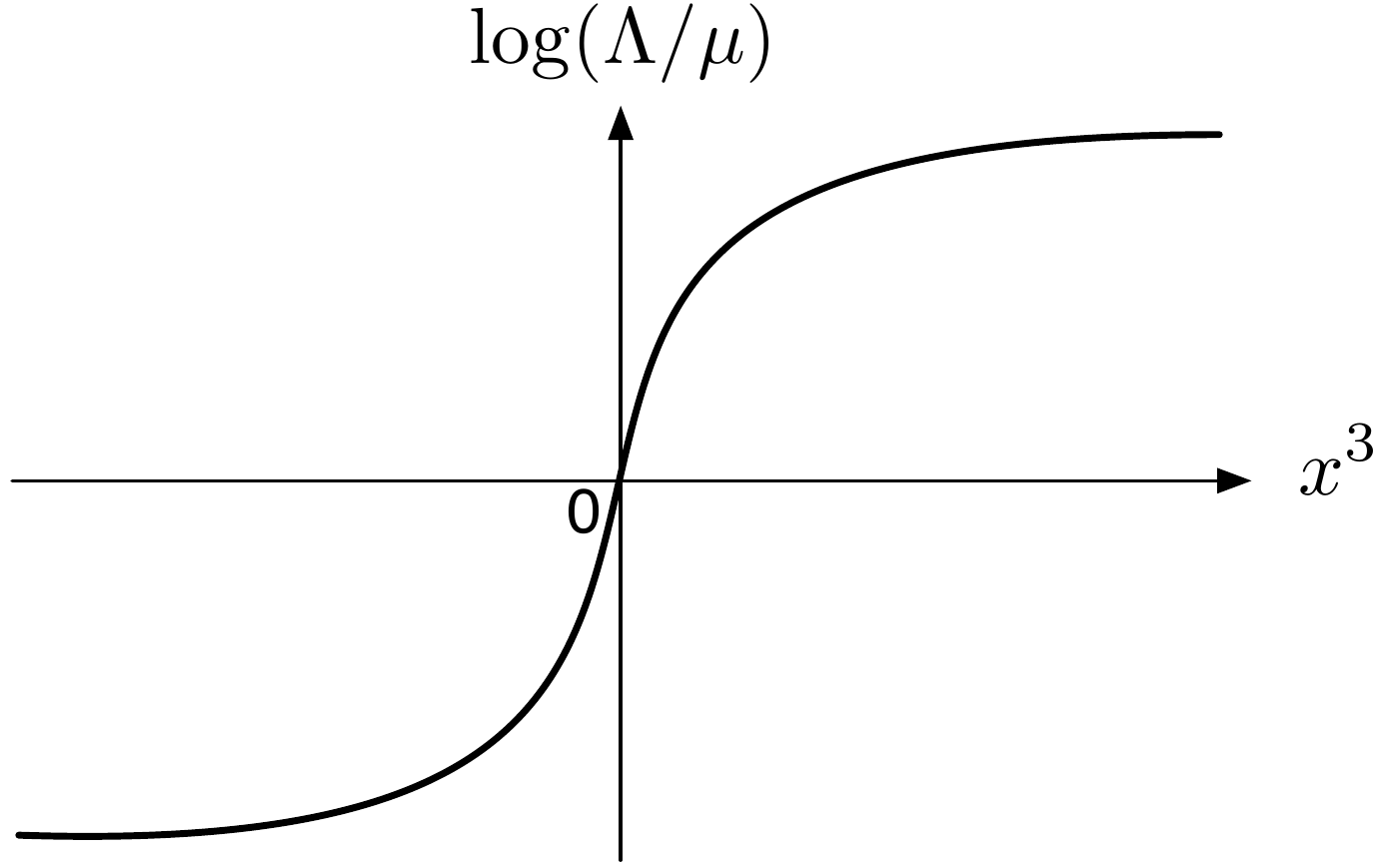}
\caption{Schematic Janus configuration corresponding to an effective RG flow.}
\label{fig:RGJanus}
\end{figure}

One way to define an RG domain wall is by using Janus configurations. For example, in a 4d $\CN=2$ theory with an asymptotically free gauge group, we can fix an energy scale $\mu$ at which to observe the theory and vary the strong coupling scale $\Lambda(x^3)$ from zero to infinity relative to $\mu$ (Figure \ref{fig:RGJanus}), while preserving the BPS condition \eqref{d3Z}. At $x^3\ll 0$, with $\Lambda \ll \mu$, the theory is effectively non-abelian; while at $x^3\gg 0$, with $\Lambda \gg \mu$, the theory will best be described as an abelian Seiberg-Witten theory far out on its Coulomb branch. We will quantify the latter claim in Section \ref{sec:attractor}.

More generally, one can consider any 4d $\CN=2$ theory and vary a combination of marginal UV couplings $\tau(x^3)$, strong-coupling scales $\Lambda(x^3)$, and masses $m(x^3)$, in such a way that at $x^3\ll 0$ the theory is near the origin of the Coulomb branch for a chosen set of gauge groups, while at $x^3\gg 0$ the theory is far out on the Coulomb branch. In this case, what we are calling an RG wall might also be termed an ``abelianization'' wall.

In the limit that the jump in $\Lambda$ (or other parameters) occurs very quickly, and effective gauge couplings become very weak at $x^3<0$ and $x^3>0$, one might hope to trap a well-defined 3d $\CN=2$ theory at $x^3=0$. 
The 3d theory should have both abelian and non-abelian flavor symmetries to allow it to couple to the 4d bulk. Moreover, it should also have electric and magnetic chiral operators that couple to bulk BPS hypermultiplets on the abelian side, as in Section \ref{sec:bdycouple}. On the non-abelian side, we similarly expect to find additional chiral operators in non-trivial flavor multiplets that couple to the bulk hypermultiplets in the UV Lagrangian.

In Section \ref{sec:flag} we will argue on general grounds that, in the infrared, the 3d RG-wall theory associated to the breaking $G\to \mathbb T$ of a gauge group to its maximal torus is roughly a $G/\mathbb T$ sigma-model. For example, for pure $SU(2)$ Seiberg-Witten theory in the bulk, we expect a 3d $\cp^1$ sigma-model. In this description, however, the abelian flavor symmetries of the 3d theory are not entirely manifest.

In Sections \ref{sec:MRG}--\ref{sec:RGline}, we will introduce a fundamental framed 3-manifold $M_0$, and show that the 3d theory $T_2[M_0]$ derived from it combinatorially has all the right properties to be the RG-wall theory for pure $SU(2)$ Seiberg-Witten theory in the bulk. Later, we will look at more complex examples of theories with flavors.

\subsection{Janus attractors}
\label{sec:attractor}

The best way to understand some of the expected properties of RG Janus configurations is to study a simple, concrete example. We focus here on pure $SU(2)$ $\CN=2$ gauge theory in four dimensions.

Recall that this is a 4d theory of class $\CS$, whose UV curve $\CC$ is a sphere with two irregular punctures of the mildest possible type \cite{GMN, GMNII}. One might also describe $\CC$ as an annulus. The square of the Seiberg-Witten differential is
\be \lambda_{\rm SW}^2 = \Big( \frac{\Lambda^2}{z}+2u + \Lambda^2 z\Big)\frac{dz^2}{z^2}\,, \ee
with strong-coupling scale $\Lambda$ and Coulomb modulus $u=\langle \Tr \Phi^2\rangle$. Here the $z$-plane $\CC$ has rank 1/2 irregular singularities at $z=0$ and $z=\infty$.

\begin{wrapfigure}[17]{r}{2.6in}
\centering
\includegraphics[width=2.5in]{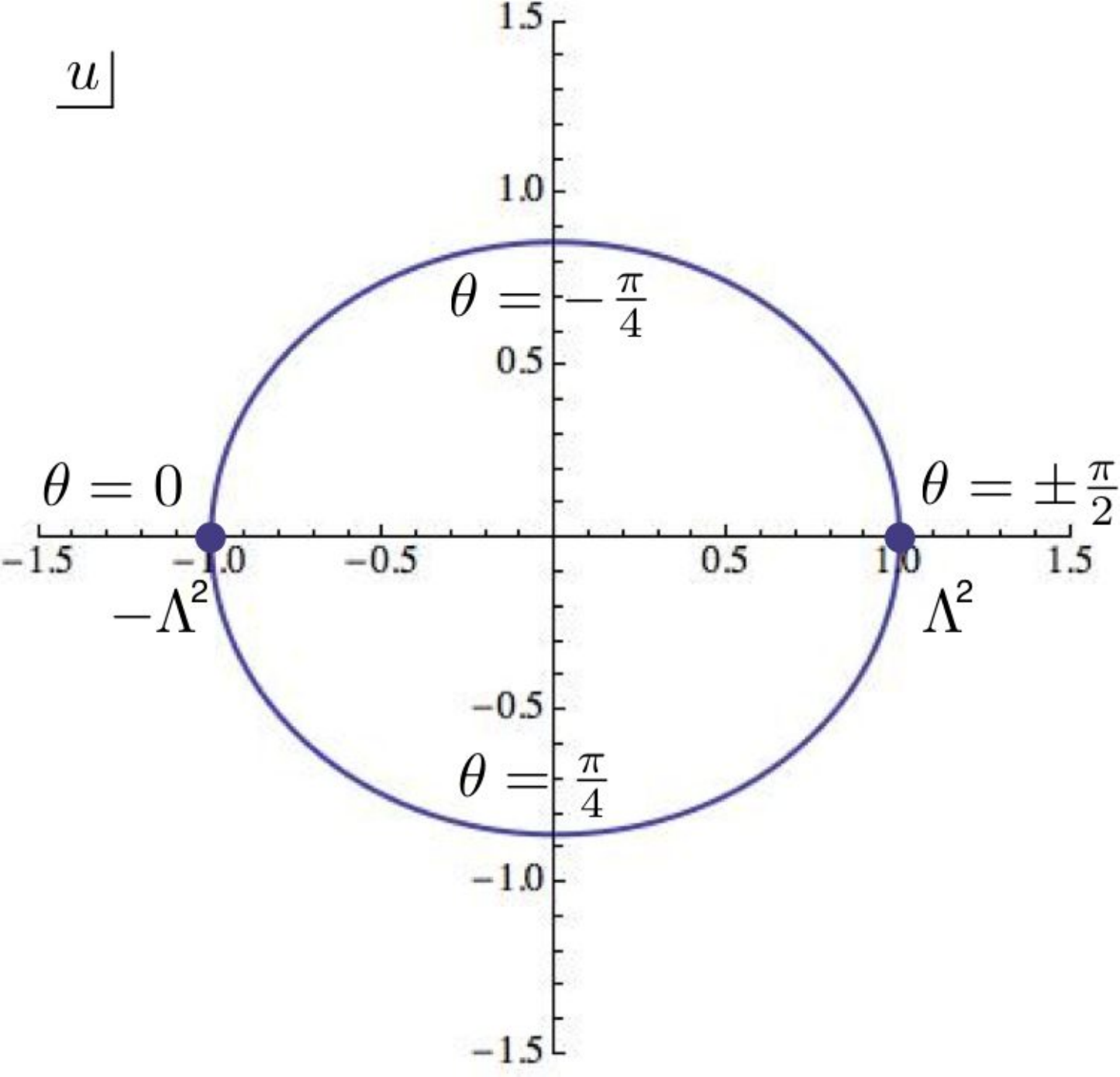}
\caption{The $u$-plane and line of marginal stability, for $\Lambda=1$.
}
\label{fig:uplane}
\end{wrapfigure}

The abelian charge lattice of $SW_2[\CC]$ consists only of gauge charges. We call the fundamental magnetic, dyonic, and electric charges $\gamma_m,\,\gamma_d,$ and $\gamma_e$, with
\be \gamma_m + \gamma_d = 2\gamma_e\,,\qquad \langle \gamma_m,\gamma_d\rangle = 2\,. \ee
BPS states only carry even electric charge, so the BPS lattice is generated by $\gamma_m,\gamma_d$ alone. In the $u$-plane, there are two stability chambers separated by (roughly) an ellipsoidal curve of marginal stability as in Figure \ref{fig:uplane}, defined by the condition
\be \arg a = \arg a_D \;=: \theta\,, \label{MStheta} \ee
where $a$ and $a_D$ are the electric and magnetic central charges. The point on the curve of marginal stability where \eqref{MStheta} holds for fixed $\theta$ lies approximately at polar angle $2\theta-\pi$.

\begin{figure}[htb]
\centering
\includegraphics[width=4.5in]{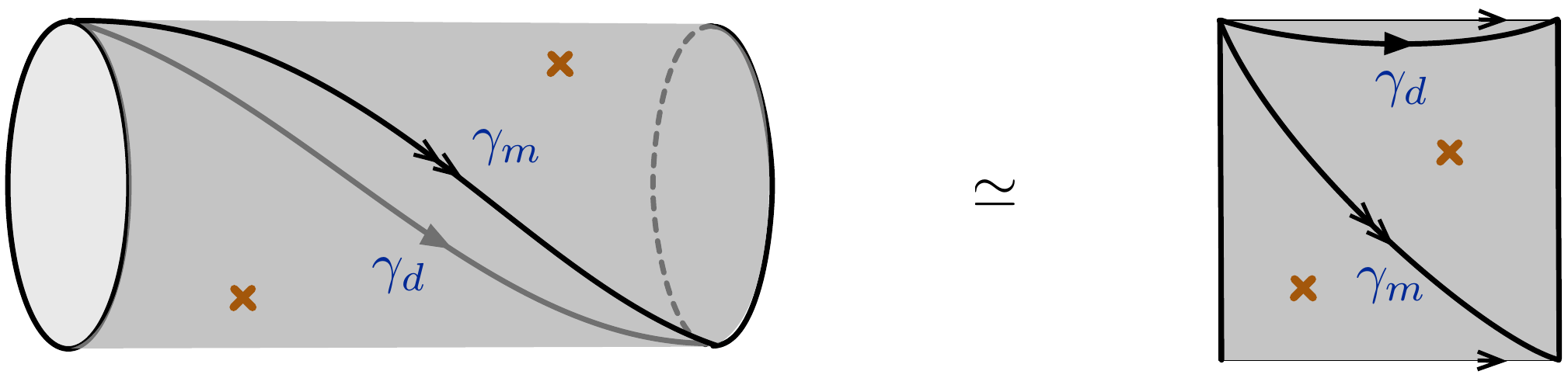}
\caption{WKB triangulation of the annulus $\CC$, with branch points at $z=-\tilde u\pm \sqrt{\tilde u^2-1}$, $\tilde u:=u/\Lambda^2$.}
\label{fig:annulus}
\end{figure}

All WKB triangulations $\mb t_{u,\theta}$ of the curve $\CC$ look identical topologically --- there really exists only one triangulation for the annulus, shown in Figure \ref{fig:annulus}. The two vertices of the triangulation lie on the $S^1$ boundaries of the annulus, in accordance with the fact that there are rank-1/2 irregular singularities there. There are two non-boundary edges in the triangulation, which separate the two branch points of $\lambda_{\rm SW}$.
 If $u$ is inside the curve of marginal stability, the two non-boundary edges are simply labelled by charges $\pm \gamma_m$ and $\pm \gamma_d$ (the precise assignment depends on $\theta$). These charges correspond to magnetic and dyonic cycles in the double cover $\Sigma\to \CC$. When $u$ is outside the curve of marginal stability, the edges of the WKB triangulation can also correspond to pairs $\pm\big((n+1)\gamma_m+n\gamma_d,n\gamma_m+(n-1)\gamma_d\big)$, $n\in \Z$, reflecting the fact that there are now an infinite number of BPS states. Notice that flips of the WKB triangulation, which do not affect its topological type, do change $n$.

We could distinguish ``different'' triangulations by holding the $S^1$ boundaries of $\CC$ fixed, and not allowing them to rotate with respect to each other. Then performing pairs of consecutive flips is equivalent to twisting the annulus by full turns.

Now let us imagine varying $\Lambda(x^3)$ from zero to infinity, while keeping $\Re(e^{-i\theta}a)$ and $\Re(e^{-i\theta}a_D)$ fixed, so as to build an RG Janus configuration. We allow ourselves an extra phase $\zeta=e^{i\theta}$ in selecting which 3d $\CN=2$ superalgebra to preserve. The central charges can be written as
\be a(u,\Lambda) = \int_{\gamma_e}\lambda_{\rm SW}= \Lambda\, f(u/\Lambda^2)\,,\qquad a_D(u,\Lambda) = \int_{\gamma_m}\lambda_{\rm SW}= \Lambda\, f_D(u/\Lambda^2) \ee
for some (locally) holomorphic functions $f$, $f_D$. It follows generically from this that no matter what values of $a,a_D$ we take at $x^3\ll 0$ (when $\Lambda \ll \mu$), after sending $\Lambda/\mu\to \infty$ we will find $|a|,|a_D|\sim\Lambda \gg \mu$. Thus an RG Janus configuration necessarily forces the theory far onto the Coulomb branch.

We can actually do much better, and identify precisely where on the Coulomb branch we land. In terms of the dimensionless variable $\tilde u:=u/\Lambda^2$, we want to find $\tilde u(x^3=\infty)$. Writing $\pd_3 a = \Lambda f'(\tilde u)\, \pd_3 \tilde u+f(\tilde u)\,\pd_3\Lambda$ and similarly for $a_D$, we easily obtain from the Janus condition $\pd_3 \Re\,\zeta^{-1}(a,a_D)=0$ that
\be \label{attractor}
\pd_3\begin{pmatrix}  \Re\,\tilde u \\  \Im\,\tilde u \end{pmatrix} = 
 \begin{pmatrix} -\Re(\zeta^{-1}f') & \Im (\zeta^{-1}f') \\
  -\Re(\zeta^{-1}f_D') & \,\Im (\zeta^{-1}f_D') \end{pmatrix}^{\!-1}
 \begin{pmatrix} \Re(\zeta^{-1} f) \\ \Re (\zeta^{-1}f_D) \end{pmatrix} \pd_3 \log\frac{\Lambda}{\mu}\,.
\ee
This flow equation has an obvious fixed point given by $ \Re(\zeta^{-1} f(\tilde u))= \Re(\zeta^{-1} f_D(\tilde u))=0$, which turns out to be the unique fixed point in the $\tilde u$ plane, and is attractive for increasing $\Lambda$.
Moreover, by comparison to \eqref{MStheta}, this fixed point must lie precisely on the curve of marginal stability, with phase $\zeta=e^{i\theta}$! Thus as $\Lambda\to \infty$, all flows are attracted to marginal stability.

\begin{figure}[htb]%{r}{2.5in}
\centering
\includegraphics[width=2.5in]{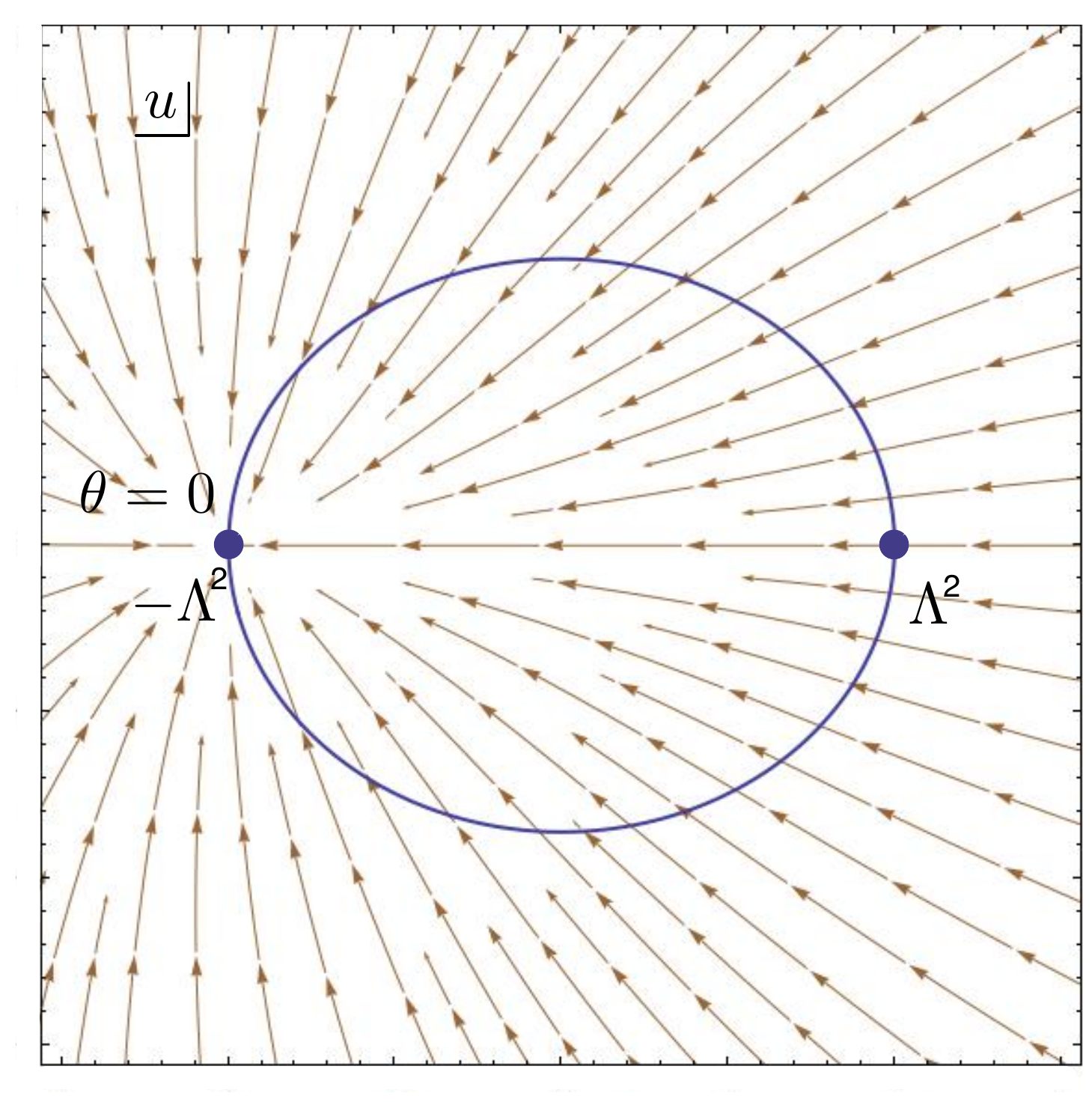}
\caption{Janus flows to the dyon point.}
\label{fig:flows}
\end{figure}

For example, when $\zeta=\pm 1$, the attractor point is at the dyon point $\tilde u=-1$ or $u = -\Lambda^2$. The exact flows to the dyon point are shown in Figure \ref{fig:flows}. Since $f$ is nonvanishing here, we necessarily have $|a| \sim \Lambda |f(-1)| \sim \Lambda$ as $\Lambda\to \infty$. Thus the theory $T_2[\CC]\to SW_2[\CC]$ is abelianized as promised.
The attractor moreover identifies a natural $Sp(2,\Z)$ abelian duality frame for $SW_2[\CC]$: it is the frame in which the light dyon becomes a fundamental ``electric'' hypermultiplet.

Some basic aspects of this RG flow attractor mechanism remain true for any domain wall that scales to infinity in a uniform way 
all the dimensional couplings of a generic $\CN=2$ theory (masses and asymptotically free gauge couplings). If we write all coupling as 
$\Lambda_i = c_i \Lambda(x^3)^{d_i}$ in terms of constant dimensionless ratios $c_i$, and the Coulomb branch parameters in terms of dimensionless parameters as
as $u_i = \tilde u_i \Lambda(x^3)^{D_i}$, we can write the periods as $Z_\gamma= \Lambda z_\gamma(\tilde u,c)$ 
and the BPS condition will force the flow to be attracted to the point $\Re (\zeta^{-1} z_\gamma)=0$. 

This is a rather interesting point, which plays a key role in the recent work by  Kontsevich \cite{Kontsevich-PI} on wall-crossing. 
It is the unique fixed point for the gradient flow of an interesting function on the Coulomb branch. The useful role in wall-crossing
follows from the observation that the central charge of all BPS particles is aligned at this point, and thus all BPS particles can be seen as marginal bound states of a 
basic set of stable objects.

In a sense, this property supports our strategy to build the couplings of domain wall theories to the IR Seiberg-Witten theory 
by specifying only the couplings to the basic set of stable objects, which is in correspondence to the edges of the WKB triangulation. 

\subsection{Trapped Goldstone bosons}
\label{sec:flag}

We can use a simple thought exercise to guess what 3d degrees of freedom might be trapped on an RG domain wall. In order to isolate a 3d theory, we want to send the dynamical 4d gauge couplings to zero on both sides of the wall --- for a gauge group $G$ on the UV side and its maximal torus $\mathbb T$ on the IR side. Suppose that we do this first on the UV side, $x^3\leq 0$. Then the $G$ gauge symmetry becomes a flavor symmetry. As we move onto the Coulomb branch on the IR side, this \emph{flavor} symmetry is broken to $\mathbb T$, and we might expect to find trapped Goldstone bosons at the wall itself, parametrizing a $G/\mathbb T$ moduli space.

We can make this a little more concrete by starting with nonabelian $G$ gauge theory on the whole 4d space, and replacing the region $x^3<0$ with a half-BPS Dirichlet boundary condition for the gauge fields. This should have the same effect as sending the UV gauge coupling to zero there.
As reviewed in Section \ref{sec:bdycouple}, the basic supersymmetric Dirichlet boundary condition pairs the three components $A_\parallel$ of the gauge field parallel to the boundary with the real part of Higgs field $\Phi$, setting%
\footnote{One could also deform the Dirichlet b.c. on the Higgs field to $\Re\,\Phi|_\pd = const$, which introduces mass terms for the $G/\mathbb T$ sigma models discussed below.}
\be A_\parallel\big|_\pd = 0\,,\qquad \Re\,  \Phi\big|_\pd = 0\,,\ee
while giving Neumann b.c. to $\Re\, \tau_{UV} \Phi$. (For clarity, we will just set $\zeta=1$ here.)

In order to find the effective degrees of freedom on the RG wall, we ask: \emph{how does the UV Dirichlet boundary condition look in the IR}? 
We can move onto the Coulomb branch by turning on the Cartan part of the Higgs field at infinity, compatibly with the 
with Neumann-Dirichlet boundary conditions at $x^3=0$. 
In the 4d bulk $(x^3>0)$, this is a standard Higgsing, and Goldstone bosons are eaten up by gauge fields to give massive W-bosons.
At the boundary, however, the gauge field $A_\parallel$ is frozen, so massless Goldstone bosons survive, parametrizing the desired $G/\mathbb T$.

We can see evidence of the boundary degrees of freedom if we look at the moduli space of supersymmetric vacua of the theory. In the UV, 
the boundary condition at infinity forces the boundary value of $\Re\, \tau_{UV} \Phi$ to lie in a specific conjugacy class, 
but different points inside that $G/\mathbb T$ manifold correspond to different vacua of the theory. Thus in the IR we need to have some boundary degrees of freedom which reproduce this space of vacua. The precise set of degrees of freedom depends on the choice of IR duality frame.

Consider the simplest example, $G=SU(2)$. The most obvious way to obtain a $\cp^1$ moduli space of vacua is to couple the IR theory to a 3d $\cp^1$ sigma-model, with $\CN=2$ supersymmetry. There is also a more subtle way. We need to remember the boundary conditions imposed by coupling the IR gauge fields to some matter: 
the vev of $\Re\,  a \big|_\pd$ acts as a 3d mass parameter, while $\Re\,  a_D \big|_\pd$ acts as an FI parameter
\be \Re\,  a_D \big|_\pd = \mu \,,\ee
where $\mu$ is the moment map for the 3d flavor symmetry we are gauging. Thus if the boundary theory consisted, say, of a doublet of chiral fields $q_\alpha$ 
of charge $-1$ under the gauge group, the moment map condition together with the $U(1)$ gauge symmetry would reproduce precisely a $\cp^1$ moduli space of vacua.

In order to test this simple candidate theory we can couple it to both a non-abelian $SU(2)$ gauge theory and the abelian $U(1)$ gauge theory in order to 
engineer a potential RG domain wall. This chiral doublet interface between the non-abelian theory and the abelian IR theory gives rather reasonable boundary conditions. For example, we have 
\be  \Re\,  \tau_{UV} \Phi_{\alpha \beta} \big|_\pd = \bar q_{(\alpha} q_{\beta )}\, \qquad \Re\,  a_D \big|_\pd = |q|^2 \,,\ee
which insures that $q^\alpha$ and $\bar q^\alpha$ are eigenvectors of $\Re\,  \tau_{UV} \Phi$ with eigenvalues $\Re\,  a_D$. 
As the eigenvalues of $\tau_{UV} \Phi$ are essentially the classical values of $a_D$, this condition agrees well with the condition that $\Re\,  a_D$
should remain constant from the UV to the IR side of a Janus. 

On the other hand, in order to allow the $q_\alpha$ vevs without breaking SUSY, we need the 3d mass matrix 
\be \Re\,  \Phi_{\alpha \beta} \big|_\pd - \Re \,  a \,\epsilon_{\alpha \beta} \big|_\pd \,,\ee
to have $q_\alpha$ as a zero eigenvector. Thus $\Re \,  a$ must coincide with an eigenvalue of 
$\Re\,  \Phi$ and we recover the condition that $\Re\,  a$
should remain constant from the UV to the IR side of a Janus.

If we move to other IR duality frames, which are more natural if the attractor RG fixed point is close to the monopole ($\zeta \sim i$) or dyon ($\zeta \sim 1$) points, 
we should do an appropriate $Sp(2,\Z)$ transformation on the 3d degrees of freedom. This will produce 
a 3d GLSM: a theory of two chiral multiplets $\phi_1,\phi_2$ with an axial $U(1)$ symmetry that has been gauged. The remaining flavor symmetry is $SU(2)$ (rotating the chirals as a doublet) times a topological $U(1)_J$. The $U(1)_J$ will be coupled to the gauge fields in the new duality frames. 
The abelian boundary conditions are covariant under $Sp(2,\Z)$ transformations, and thus the new setups will work as well to mimic the desired boundary conditions. 

In our later analysis based on the explicit RG manifold, we will recover these descriptions of the RG domain wall, and thus automatically test it further, by insuring the existence of appropriate couplings to the monopole and dyon particles and the correct behavior of line defects at the interface. This will provide rather robust evidence that the chiral doublet 
interface reflects faithfully the low energy properties of the RG domain wall.

\subsection{The basic RG manifold $M_0$}
\label{sec:MRG}

\begin{figure}[htb]
\centering
\includegraphics[width=6in]{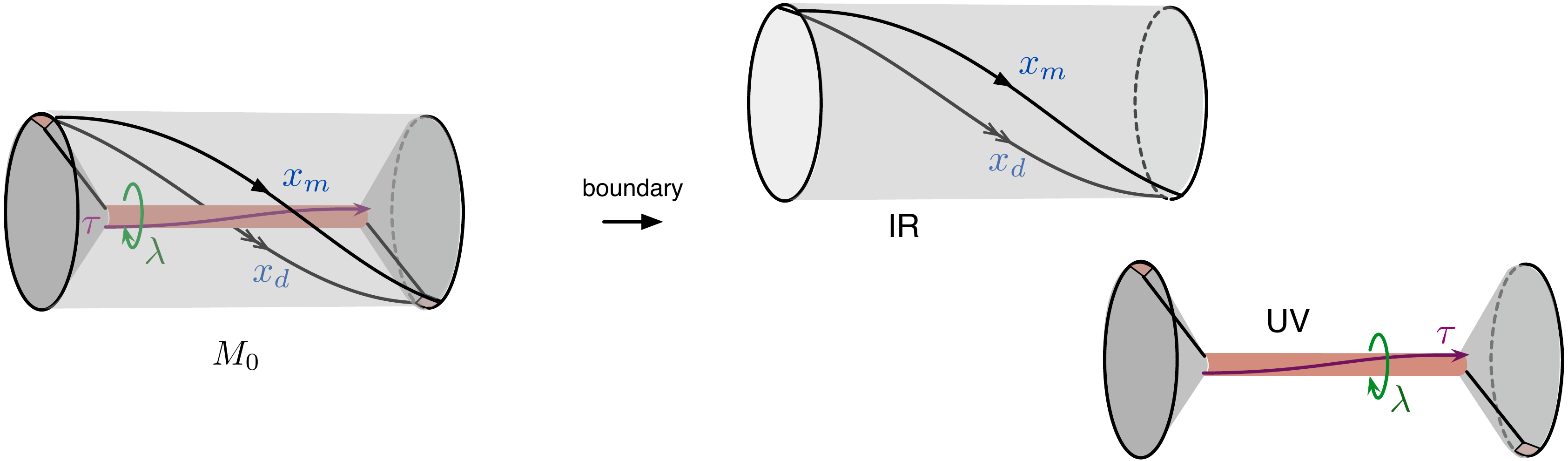}
\caption{The RG manifold for pure $SU(2)$ theory, and its IR (abelian) and UV (nonabelian) boundaries.}
\label{fig:RGbdy}
\end{figure}

The framed 3-manifold $M_0$ that representes the RG domain wall for pure $SU(2)$ theory should interpolate between an annulus $\CC$ ``in the UV'' and an annulus ``in the IR'' (Figure~\ref{fig:RGbdy}). In the IR, we represent $\CC$ as a big boundary with a WKB triangulation $\mb t$, coupling to abelianized Seiberg-Witten theory as discussed in Section \ref{sec:bdycouple}. In the UV, however, we shrink the annulus into a small boundary, which \emph{should} couple to the nonabelian $SU(2)$ theory. In the limit of infinite shrinking, the small annulus represents a defect, corresponding to Dirichlet b.c. for the UV gauge fields. The small annulus must be attached to two big cones at its ends, triangulated with degenerate triangles, which carry the irregular singularities of~$\CC$.

Topologically, the 3-manifold $M_0$ is equivalent to $\CC\times I$. Its boundary $\pd M_0$ is the union of the IR and UV boundaries, which have been ``cut'' along the two irregular singularities. Notice that $\pd M_0$ contains two small disc boundaries (on the irregular singularities) in addition to the small annulus.

To describe the space $\CP_2(\pd M_0,\mb t)$ of framed flat connections on $\pd M_0$, we choose cross-ratio coordinates $x_m,x_d$ for the magnetic and dyonic edges on the big boundary. We also use a holonomy eigenvalue $\lambda$ for the holonomy around the (oriented) A-cycle of the small annulus, and a canonically conjugate twist $\tau$. Thus the nonvanishing Poisson brackets are
\be \{\log x_d,\log x_m\} = 2\,,\qquad \{\log \tau,\log \lambda\} = 1\,.\ee
Notice that by using $\lambda$ rather than $\lambda^2$ as a coordinate, we have partially lifted from a moduli space of flat $PGL(2)$ connections to flat $SL(2)$ connections.
It is also useful to introduce an electric coordinate $x_e=-\sqrt{x_mx_d}$.\,%
\footnote{Square roots like this can be made sense of in two ways: either by using logarithmically lifted coordinates as discussed in Appendix \ref{app:coords}, or breaking 3d symmetry slightly and imposing a positive structure as in \cite{FG-Teich}.} %

The twist $\tau$ is described carefully in Appendix \ref{app:twist}. Its definition requires us to choose a path $\gamma_\tau$ from one end of the small annulus to the other. At each end of this path, the framing flags for a flat connection on the small annulus can be normalized by using the additional framing flags from the small discs of $\pd M_0$. Then the ratio of normalizations is $\tau$. Changing the path $\gamma_\tau$ by a full twist around the annulus (equivalent to twisting the entire 3-manifold, \cf\ Section \ref{sec:attractor}) simply rescales $\tau\to \lambda^2\tau$. Here, it will turn out to be more symmetric to choose two paths $\gamma_1,\gamma_2$ related by a full twist, and to set $\tau = \sqrt{\tau_1\tau_2} = \lambda\tau_1 = \lambda^{-1}\tau_2$\,.

\begin{figure}[htb]
\centering
\includegraphics[width=5.3in]{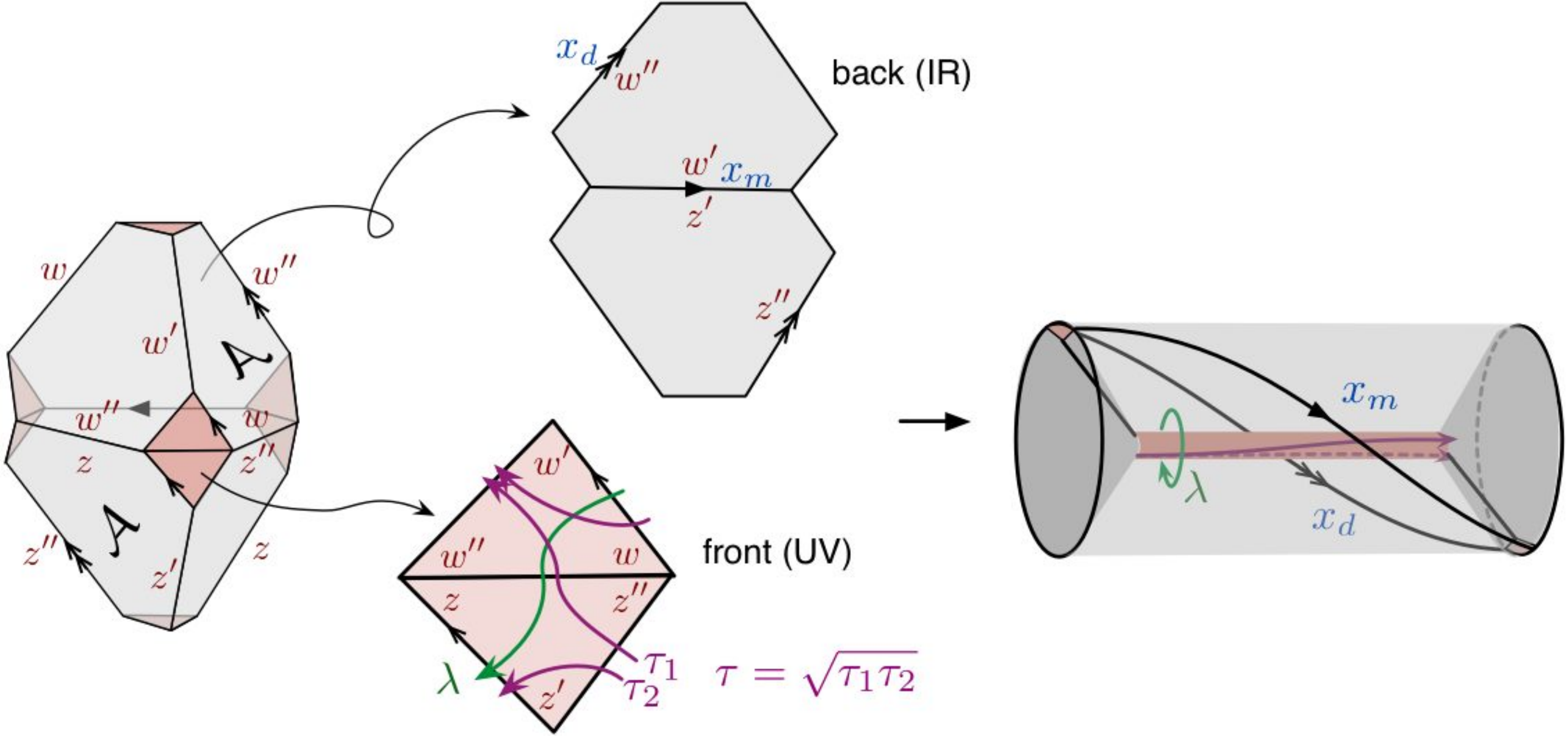}
\caption{Triangulation of the RG manifold, showing how coordinates on the big and small annuli are related to tetrahedron parameters.}
\label{fig:RG}
\end{figure}

The 3-manifold $M_0$ can be triangulated as shown in Figure \ref{fig:RG}. First two tetrahedra are glued together along a common face to form a bipyramid. Then two more faces (labelled $\CA$) are identified to form $M_0$. The small square of boundary at the front of the bipyramid gets rolled up into the small annulus of $M_0$, while the big square at the back turns into the IR boundary of $M_0$. Labeling the tetrahedra with edge parameters $z,z',z''$ and $w,w',w''$ as in the figure, and choosing $\tau$ paths as shown, we find that
\be \label{xzRG}
x_m = z'w'\,,\quad  x_d= z''w''\,,\quad x_e=\frac{1}{\sqrt{zw}}\,;\qquad
   \lambda^2 = \frac zw\,,\quad \tau = \sqrt{\frac{z''}{w''}\frac{w'}{z'}} = \lambda \frac{z''}{w''}\,.
\ee
The rules for obtaining $\lambda$ and $\tau$ are summarized in Appendix \ref{app:bdybulk}; basically one draws paths $\gamma_\lambda$, $\gamma_\tau$ on the small boundary and multiplies or divides (according to orientation) by the edge parameters on angles subtended by the paths.

In order to define a 3d theory $T_2[M_0,\mb t,\Pi]$, we must choose a polarization $\Pi$ for $\CP_2(\pd M, \mb t)$ and then compare the position coordinates in $\Pi$ to the positions/momenta of tetrahedra to figure out what symmetries get gauged. (Since there are no internal edges in the triangulation of $M_0$, there will be no superpotential terms.) For the UV boundary, we canonically choose $\lambda$ to be a position and $\tau$ its conjugate momentum. For the IR boundary, there are three natural choices of polarization:
\be \label{emd}
\Pi_e = {x_e \choose x_m}\,,\qquad \Pi_m={x_m\choose x_e^{-1}}\,,\qquad \Pi_d = {x_d \choose x_e}\,. \ee
so that the position coordinates (the top components of these vectors) are the electric, magnetic, and dyonic $x$'s, respectively.

In the polarization $\Pi_e$, the positions $\lambda$ and $x_e$ are just composed of tetrahedron positions $z,w$. So nothing gets gauged. We end up with a theory of two chirals $\phi_z,\phi_w$ with charges $(-1,-1)$ under a $U(1)_e$ flavor symmetry and charges $(1,-1)$ under a $U(1)_\lambda$ flavor symmetry, which is manifestly enhanced to $SU(2)_\lambda$. This is the nonabelian symmetry associated to the small annulus. Being more careful to keep track of background Chern-Simons levels (which depend on the precise choice of conjugate momenta, here $x_m, \tau$) and $U(1)_R$ charges (which result from a logarithmic lift of tetrahedron parameters, see Section \ref{sec:RGline}), the full theory can be described as
\be \label{TM0e}
 T_2[M_0,\mb t,\Pi_e]:\left\{\begin{array}{l}

\text{Two chirals $\phi_z,\phi_w$, with $U(2)\simeq SU(2)_\lambda\times U(1)_e$ flavor symmetry,}\\
\text{charges}\;\; \begin{array}{c|cc} &\phi_z&\phi_w \\\hline
  e& -1 & -1 \\
  \lambda & \multicolumn{2}{c}{\square} \\\hline
  R & 0 & 0 \end{array}\,,\qquad
\text{abelian CS matrix}\;\; \begin{array}{c|cc|c}
  & e & \lambda & R \\\hline
 e & 1 & 0 & 1 \\
 \lambda & 0 & 0 & 0 \\\hline
 R & 1 & 0 & * \end{array}\,. \end{array}\right.
\ee
We do not specify an R-R Chern-Simons term, since it is not uniquely determined by the geometry.

\begin{wrapfigure}{r}{2.3in}
\centering
\includegraphics[width=2.3in]{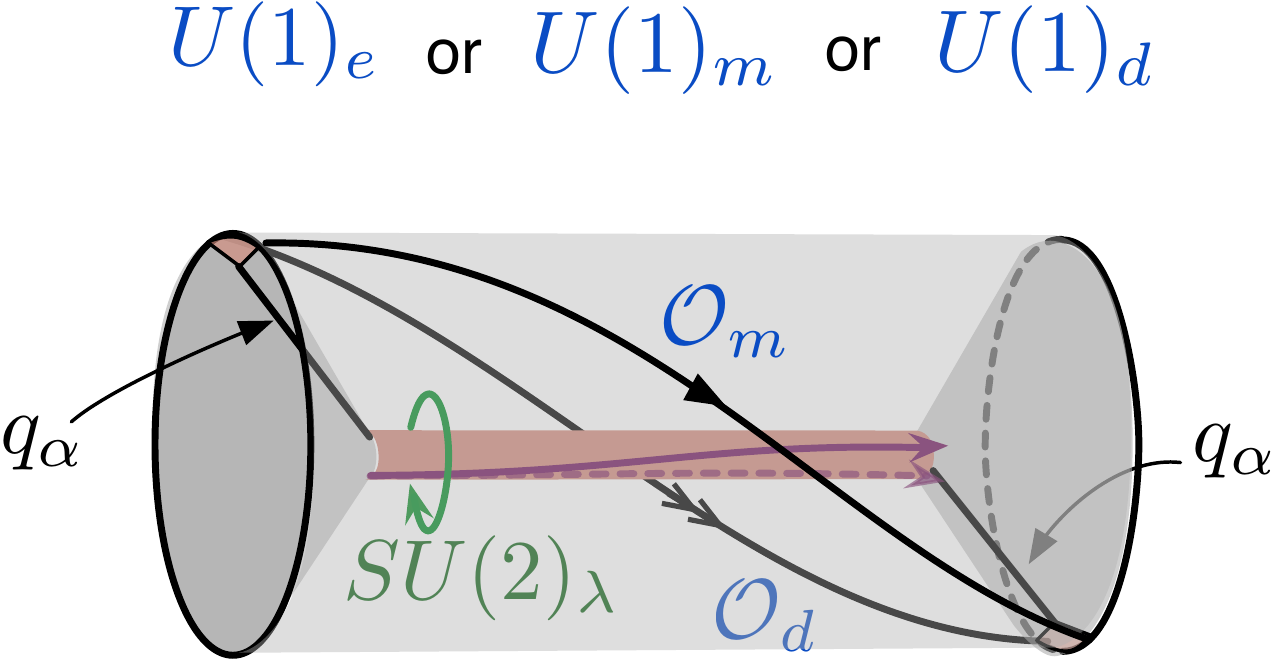}
\caption{Symmetries associated to boundaries and operators associated to edges, in $T_2[M_0,\mb t,*]$.}
\label{fig:RGops}
\end{wrapfigure}

We can couple $T_2[M_0,\mb t,\Pi_e]$ to nonabelian $SU(2)$ gauge theory on one side and an \emph{electric} $U(1)$ gauge theory on the other to obtain a full 3d-4d system with a half-BPS domain wall $\CB_2[M_0]$. The abelian coupling is just as discussed in Section \ref{sec:couplings}. In addition to gauging the $U(1)_e$ symmetry in the bulk, we note that the theory $T_2[M_0,\mb t,\Pi_e]$ has two chiral operators $\CO_m,\,\CO_d$ on which magnetic and dyonic 't Hooft lines from the bulk Seiberg-Witten theory can end. From a 3d point of view, these operators only exist in the presence of monopole flux configurations for a background $U(1)_e$ gauge field. The \emph{nonabelian} coupling similarly gauges $SU(2)_\lambda$ in the bulk. But there exists a 3d $SU(2)_\lambda$ doublet $(q_1,q_2)\equiv (\phi_z,\phi_w)$ of chiral operators as well. The doublet is associated to either of the two degenerate edges at the end of the small annulus. Had there been any hypermultiplets in the bulk $SU(2)$ theory (there aren't here), they could have coupled to $(q_1,q_2)$ at the wall.

Of course, we know from the analysis of Janus attractors that magnetic or dyonic duality frames, rather than an electric frame, are more natural for a weakly coupled $U(1)$ bulk theory. Correspondingly, we should look at 3d theories in polarizations $\Pi_m$ or $\Pi_d$. %Since $x_m$ and $x_d$ involve tetrahedron momenta ($z'',w''$),
The 3d axial $U(1)_e$ symmetry must now be gauged and replaced by topological flavor symmetries $U(1)_m$ or $U(1)_d$. The difference between the two cases is that we gauge $U(1)_e$ at bare Chern-Simons level $+1$ or $-1$.
We find
\begin{subequations} \label{RGmd}
\be T_2[M_0,\mb t,\Pi_m]:\left\{\begin{array}{l}
\text{Two chirals $\phi_z,\phi_w$, with $U(1)_e$ gauge and $SU(2)_\lambda\times U(1)_m$ flavor,}\\
\text{charges}\;\; \begin{array}{c|cc|c} &\phi_z&\phi_w & \CO_m\\\hline
  e& -1 & -1 & 0 \\\hline
  m&  0 & 0 & 1\\
  \lambda & \multicolumn{2}{c|}{\square} & - \\\hline
  R & 0 & 0 & 0\end{array}\,,\qquad
\text{abelian CS matrix}\;\; \begin{array}{c|c|cc|c}
  & e & m & \lambda & R \\\hline
 e & 1 & -1 & 0 & 1 \\\hline
 m & -1 & 0 & 0 & 0 \\
 \lambda & 0 & 0 & 0 & 0 \\\hline
 R & 1 & 0 & 0 & * \end{array}\,. \end{array}\right.
 \ee
\be T_2[M_0,\mb t,\Pi_d] :\left\{\begin{array}{l}
\text{Two chirals $\phi_z,\phi_w$, with $U(1)_e$ gauge and $SU(2)_\lambda\times U(1)_d$ flavor,}\\
\text{charges}\;\; \begin{array}{c|cc|c} &\phi_z&\phi_w & \CO_d\\\hline
  e& -1 & -1 & 0 \\\hline
  d&  0 & 0 & 1\\
  \lambda & \multicolumn{2}{c|}{\square} & - \\\hline
  R & 0 & 0 & 0\end{array}\,,\qquad
\text{abelian CS matrix}\;\; \begin{array}{c|c|cc|c}
  & e & d & \lambda & R \\\hline
 e & -1 & 1 & 0 & -1 \\\hline
 d & 1 & 0 & 0 & 0 \\
 \lambda & 0 & 0 & 0 & 0 \\\hline
 R & -1 & 0 & 0 & * \end{array}\,. \end{array}\right.
 \ee
\end{subequations}
These theories are basically the GLSM descriptions of the $\cp^1$ sigma-model that we anticipated in Section \ref{sec:flag}, aside from the bare Chern-Simons terms.
By coupling (\ref{RGmd}a-b) to respective magnetic and dyonic Seiberg-Witten theories in the IR bulk (and the standard $SU(2)$ theory in the UV bulk) we obtain alternative descriptions of the full 3d-4d system $\CB_2[M_0]$. Indeed, as discussed in Section \ref{sec:dissolve}, choices of big-boundary triangulation and polarization get dissolved after coupling to the bulk. 

We note that $T_2[M_0,\mb t,\Pi_m]$ now has a standard (dynamical) monopole operator $\CO_m$, which couples via a superpotential to the fundamental magnetic hypermultiplet of the bulk Seiberg-Witten theory. Similarly, $T_2[M_0,\mb t,\Pi_m]$ has a standard anti-monopole operator $\CO_d$ with the right charge to couple to a fundamental bulk dyon. The existence of these operators can be justified using the techniques of \cite{AHISS} --- it requires a careful analysis of Chern-Simons terms and parity anomalies.

\subsection{Line operators}
\label{sec:RGline}

As an application of the 3-manifold geometry $M_0$, we can derive the Ward identities for line operators hitting the RG wall. We add an angular momentum fugacity  to the 3d-4d system, as discussed briefly in Section \ref{sec:lineops}, so that generators of the line operator algebra $q$-commutation relations. The resulting Ward identities reproduce known relations between cross-ratio and Fenchel-Nielsen coordinates in quantum Teichm\"uller theory \cite{Kash-kernel, Teschner-TeichMod}.

To quantize, we must work in logarithmic coordinates, as reviewed in Appendix \ref{app:quant}. Logarithmic coordinates also allow an unambiguous definition of square roots. First, we express the basic relations between boundary and bulk coordinates \eqref{xzRG} as
%
%\begin{subequations}
\be \hat X_m = \hat Z'+\hat W'\,,\qquad \hat X_d = \hat Z''+\hat W''\,,\qquad \hat X_e = -\tfrac12(\hat Z+\hat W)\,; \ee
\be \hat \Lambda=\tfrac12(\hat Z-\hat W)\,,\qquad \hat \CT = \hat Z''-\hat W''+\hat \Lambda = -\hat Z'+\hat W'-\hat \Lambda\,.\quad\notag\ee
%\end{subequations}
%
The LHS of these expressions should all be viewed as operators in an algebra generated by $\hat Z,\hat Z',\hat Z'',\hat W,\hat W',\hat W''$ so that
\be [\hat Z,\hat Z']=[\hat Z',\hat Z'']=[\hat Z'',\hat Z]=\hbar\,, \ee
and similarly for the $W$'s, with $\hat Z+\hat Z'+\hat Z''=\hat W+\hat W'+\hat W''=i\pi+\tfrac\hbar2$. We can unambiguously exponentiate to find
\be \hat x_m = \hat z'\hat w'\,,\quad \hat x_d=\hat z''\hat w''\,,\quad \hat x_e = \frac{1}{\sqrt{\hat z\hat w}}\,,\qquad  \hat \lambda=\sqrt{\frac{\hat z}{\hat w}}\,,\quad \hat \tau = q^{\frac12}\hat \lambda\frac{\hat z''}{\hat w''}\,,\ee
and the inverse relations
%
%\begin{subequations}
\be \label{zxRG}
\hat w = \frac{1}{\hat \lambda\hat x_e}\,,\quad \hat z=\frac{\hat \lambda}{\hat x_e}\,,\qquad \hat z'= q^{\frac18}\frac{1}{\sqrt{\hat\lambda}}\frac1{\sqrt{\hat\tau}}\sqrt{\hat x_m}\,,\quad \hat w'=q^{\frac18}\sqrt{\hat\lambda}\sqrt{\hat \tau}\sqrt{\hat x_m}\,,\ee
\be \quad \hat z''=q^{-\frac18}\frac{1}{\sqrt{\hat \lambda}}\sqrt{\hat \tau}\sqrt{\hat x_d}\,,\quad \hat w''=q^{-\frac18}\sqrt{\hat\lambda}\frac1{\sqrt{\hat \tau}}\sqrt{\hat x_d}\,,\qquad\quad\notag\ee
%\end{subequations}%
%
where $q=e^\hbar$ and all the exponentiated operators $q$-commute: \eg\ $\hat x_m\hat x_d=q^2\,\hat x_d\hat x_m$, $\hat \lambda\hat \tau=q\,\hat\tau\hat \lambda$, etc.

Next, we use the Ward identities for tetrahedra to relate the UV and IR sides. They are $\hat z''+\hat z^{-1}-1\simeq 0$ and $\hat w''+\hat w^{-1}-1\simeq 0$, where ``$\simeq 0$'' means ``annihilates a partition function.'' By substituting \eqref{zxRG} into these Ward identities and simplifying we find
\begin{subequations} \label{WardRG}
\begin{align} \text{Wilson:}\qquad & \hat\lambda+\hat \lambda^{-1} \,\simeq\, \hat x_e+\frac1{\hat x_e}-\frac1{\hat x_e}\hat x_d \\
 &\qquad \,\simeq\, -\frac{\sqrt{\hat x_m''}\sqrt{\hat x_d''}}{q^{\frac34}} - \frac{q^{\frac34}}{\sqrt{\hat x_d''}\sqrt{\hat x_m''}}+q^{-\frac14}\sqrt{\hat x_d''}\frac{1}{\sqrt{\hat x_m''}} \notag \\
\text{'t Hooft:}\qquad& \frac{q^{-\frac18}}{\hat\lambda-\hat\lambda^{-1}}\Big(\frac{1}{\sqrt{\hat \lambda}}\sqrt{\hat\tau}-\sqrt{\hat\lambda}\frac1{\sqrt{\hat\tau}}\Big) \,\simeq\, -q^{-\frac14}\sqrt{\hat x_m}\,, \\
\text{dyonic:}\qquad& \frac{q^{\frac18}}{\hat\lambda-\hat\lambda^{-1}}\Big(\sqrt{\hat\lambda}\sqrt{\hat \tau}-\frac{1}{\sqrt{\hat \lambda}}\frac1{\sqrt{\hat\tau}}\Big) \,\simeq\, -q^{\frac14}\frac{1}{\sqrt{\hat x_d}}\,.
\end{align}
\end{subequations}
The first equation relates the fundamental (spin 1/2) $SU(2)$ Wilson line in the UV to electric and dyonic $U(1)$ line operators in the IR. The other two equations relate the spin-1/2 UV 't Hooft line and dyonic ('t Hooft-Wilson) line to IR line operators. These relations were discussed from a purely 4d point of view in \cite{GMNIII, DG-Sdual}.%
\footnote{For comparison, we should note that our edge coordinates on a big boundary are related to those used more commonly in quantum Teichm\"uller theory by (\eg) Fock and Goncharov and in \cite{GMNII,GMNIII} as $\hat X_{\rm FG} = \hat X_{\rm here}-i\pi-\hbar/2$, or in exponentiated form $\hat x_{\rm FG} = -q^{-\frac12}\hat x_{\rm here}$. Our convention is a little more natural from a 3d perspective.} %

Note that in an honest 4d $SU(2)$ gauge theory, the basic 't Hooft (and dyonic) line operators are in the spin-1 representation rather than the spin-1/2 as above; in the spin-1 Ward identities, the square roots of $\hat \lambda,\hat \tau, \hat x_m,\hat x_d$ would disappear. Alternatively, in a $PSU(2)\simeq SU(2)/\Z_2$ theory we should keep spin-1/2 magnetic operators but only use spin-1 Wilson lines.

The Ward identities \eqref{WardRG} do not require any choice of polarization: they are valid in \emph{any} duality frame describing the domain wall $\CB_2[M_0]$. On the other hand, if we wanted to write down partition functions of an isolated 3d theory on $S^3_b$ or $S^2\times_qS^1$ we would need to choose a big-boundary duality frame. In polarization $\Pi_e$, the rules of \cite{Dimofte-QRS,HHL,DGG} compute an $S^3_b$ partition function that beautifully reproduces the cross-ratio/Fenchel-Nielsen kernel of~\cite{Kash-kernel}:
\begin{align} \CZ_\hbar(\Lambda,X_e) &\,=\, e^{\frac{1}{2\hbar}(\Lambda^2+2X_e^2+2X_e(2\pi i+\hbar))}\Phi_\hbar\big(i\pi+\tfrac\hbar2+\Lambda+X_e\big)\Phi_\hbar\big(i\pi+\tfrac\hbar2-\Lambda+X_e\big) \\
&\,=\, e^{\frac{1}{2\hbar}(X_e^2+2X_e\Lambda+(X_e+\Lambda)(2\pi i+\hbar))}\frac{\Phi_\hbar\big(i\pi+\tfrac\hbar2+\Lambda+X_e\big)}{\Phi_\hbar\big(-i\pi-\tfrac\hbar2+\Lambda-X_e\big)}\,,
\notag 
\end{align}
with the quantum dilogarithm $\Phi_\hbar$ defined in Appendix \ref{app:qdl}.
This wavefunction is annihilated by \eqref{WardRG}.

%%%%%%%%%%%%%%%%%%%%%%%%%%%%%%%%%%%%%%%%%%%%%%%%%%%%%%%%%%%%%%%%%%%%%%%%%%%%%
%%%%%%%%%%%%%%%%%%%%%%%%%%%%%%%%%%%%%%%%%%%%%%%%%%%%%%%%%%%%%%%%%%%%%%%%%%%%%
%%%%%%%%%%%%%%%%%%%%%%%%%%%%%%%%%%%%%%%%%%%%%%%%%%%%%%%%%%%%%%%%%%%%%%%%%%%%%

\section{Non-abelian symmetry enhancement}
\label{sec:enhance}

We saw in Section \ref{sec:RG} that the 3d theory trapped on the RG wall for 4d $SU(2)$ Seiberg-Witten theory should have a non-abelian $SU(2)_\lambda$ flavor symmetry. We associated this theory to a framed 3-manifold $M_0$ (Figure \ref{fig:RG}), constructing it using the class-$\CR$ rules of \cite{DGG}. Despite the fact that class-$\CR$ constructions naively lead to theories with abelian flavor symmetry, we found a manifest enhancement $U(1)_\lambda\to SU(2)_\lambda$ for the symmetry associated to the small annulus of $M_0$. We would now like to argue that such an enhancement occurs for the symmetries associated to small annuli in any 3d-4d boundary condition based on a framed 3-manifold.

\begin{figure}[htb]
\centering
\includegraphics[width=5.7in]{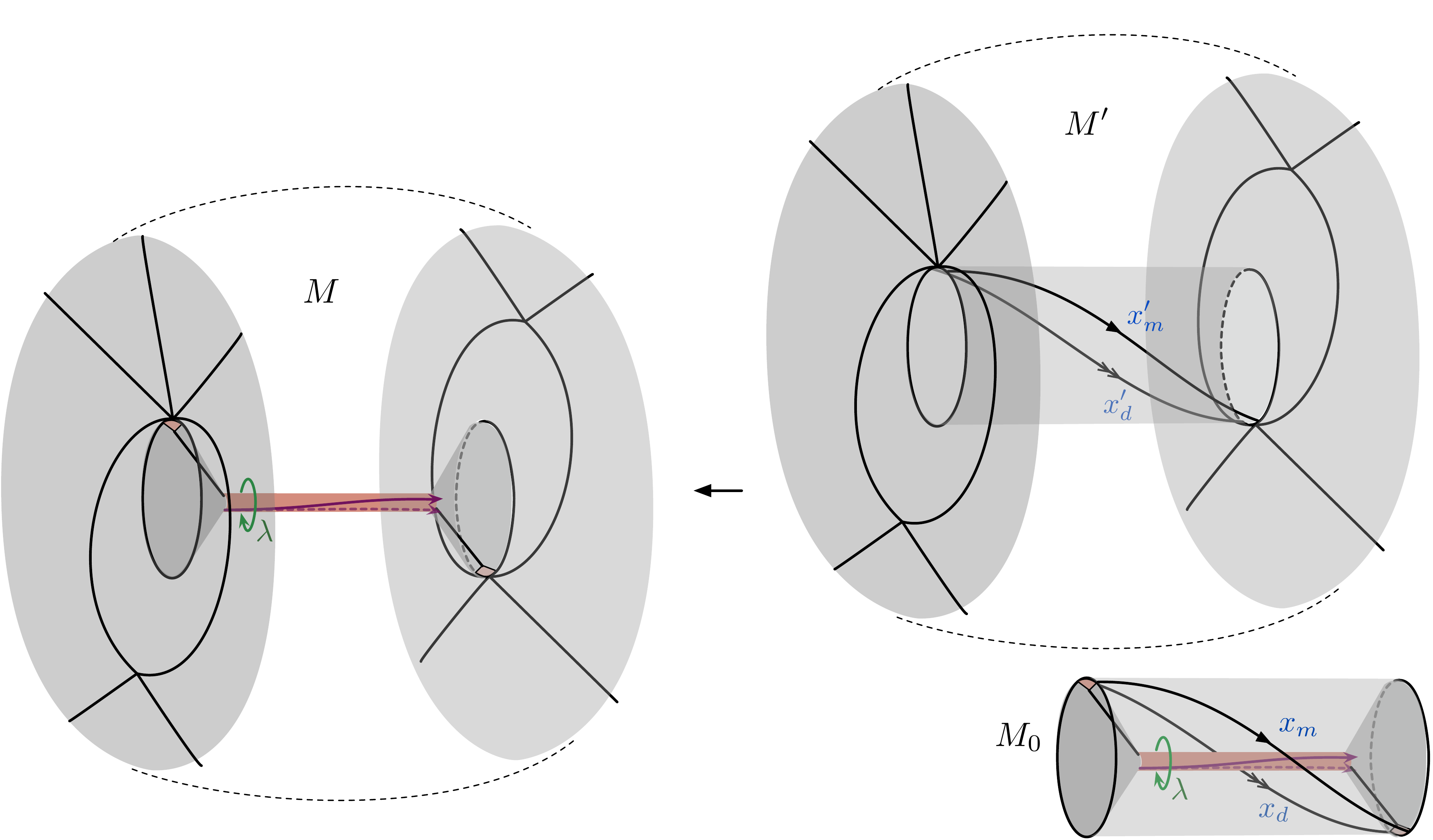}
\caption{Cutting out a basic RG manifold $M_0$ to isolate a small annulus, leaving behind a big annulus in a modified manifold $M'$.}
\label{fig:RGmodify}
\end{figure}

The basic idea is simple. First, let $M$ be a framed 3-manifold whose small boundary contains a small annulus, and assume that the annulus is attached to degenerate triangles on the big boundary (as in the UV part of Figure \ref{fig:RGbdy}). Then choose a 3d triangulation $\mb t_{3d}$ of $M$ such that the neighborhood of the small annulus looks like the basic RG manifold $M_0$. This lets us decompose $M = M'\cup M_{0}$ as in Figure \ref{fig:RGmodify}, where $M'$ contains a big annular region on its big boundary, and gluing in $M_0$ effectively shrinks this big annulus to a small one.

Now we may construct the 3d theory associated to $M$ by gluing together the theories associated to $M'$ and $M_0$, in such a way that the $SU(2)_\lambda$ symmetry of $M_0$ is inherited. Physically, we are taking the theory associated to $M'$ and colliding it with a basic RG wall (in the IR$\to$UV direction) to recover the theory associated to $M$. The RG wall provides the $SU(2)_\lambda$ flavor symmetry.

 To be more specific, we have to choose some polarizations. Let $\Pi_m$ be the magnetic polarization for $\pd M_0$, so that one of the edges on its big annulus carries a position coordinate. Then $T_2[M_0,\Pi_m]$ is basically the GLSM description of a $\cp^1$ sigma-model, as in (\ref{RGmd}a), with $SU(2)_\lambda\times U(1)_m$ flavor symmetry. For $M'$, we can choose a similar polarization $\Pi_m'$ for the annular part of its big boundary (mirroring the polarization $\Pi_m$ of $\pd M_0$), and any other polarization $\Pi$ away from this annulus. Then $T_2[M',\Pi\times \Pi_m']$ contains a $U(1)_m'$ symmetry associated to its big annulus. We form the glued theory $T_2[M,\Pi]$ by taking a product of the component theories, gauging the anti-diagonal combination $U(1)_V$ of $U(1)_m\times U(1)_m'$, and adding two superpotential terms that break the diagonal of $U(1)_m\times U(1)_m'$ and the topological symmetry $U(1)_J$ associated to $U(1)_V$. A few more details will be given in Section \ref{sec:deriveSU2}. The point, however, is that this gluing operation does not disturb the $SU(2)_\lambda$ flavor symmetry of $T_2[M_0,\Pi_m]$, which now becomes a flavor symmetry of $T_2[M,\Pi]$.

This argument shows that whenever a small annulus attaches to the centers of degenerate triangles on the big boundary of $M$, its flavor symmetry is promoted from $U(1)$ to $SU(2)$. If the big-boundary triangulation $\mb t$ is such that the annulus attaches in any other way, the flavor symmetry will generally \emph{not} be enhanced in the isolated 3d theory $T_2[M,\mb t,\Pi]$. This is easy to see by starting with a degenerate triangulation and using flips  (Figure \ref{fig:flip3d}) to change it. In the course of flipping, an $SU(2)$ doublet $q_\alpha$ of chirals associated to a degenerate triangle gets split, with only one of $q_1,q_2$ coupling to the flip operator; thus $SU(2)$ is explicitly broken. This is explained further in Section \ref{sec:doublets}.

Fortunately, we already know how to remedy this problem. If we couple $T_2[M,\mb t,\Pi]$ to an abelian 4d $\CN=2$ theory $SW_2[\pd M]$ associated to its big boundary, creating a boundary condition $\CB_{\rm SW}[M]$, the dependence on triangulation $\mb t$ and polarization $\Pi$ disappears. We can first implement the coupling with a choice of $\mb t$ so that $SU(2)$ flavor symmetry is already manifest in the 3d theory alone, and preserved by the coupling, proving that the full 3d-4d system $\CB_{\rm SW}[M]$ has $SU(2)$ symmetry. Then, by arguments of Section \ref{sec:dissolve}, the $SU(2)$ symmetry must actually be restored when coupling in a 4d duality frame specified by any other $\mb t$ and $\Pi$.

We arrive at the following picture. Given a framed 3-manifold $M$ with $a$ small annuli, we can always construct a boundary condition $\CB_{\rm SW}[M]$ whose flavor symmetry contains a subgroup $SU(2)^a$. If we want an isolated 3d theory instead, we must choose a big-boundary triangulation $\mb t$ and polarization $\Pi$. For every annulus that ends in degenerate triangles of $\mb t$, the 3d theory $T_2[M,\mb t,\Pi]$ will retain an $SU(2)$. This may not be possible for all annuli. At the ``non-enhanced'' annuli, the expected $SU(2)$ symmetry is broken by superpotential couplings involving (halves of) chiral doublets.

Note that once we have a system $\CB_{\rm SW}[M]$ with non-abelian $SU(2)^a$ symmetry, we may proceed to couple it to non-abelian (or ``UV'') 4d theories as well, just as we described in the introduction. The basic RG wall was one example of this, and more will come later.

In the remainder of this section, we fill in a few of the details from above, and also comment on symmetry enhancement and breaking at small torus boundaries.

\subsection{Gluing in the RG manifold $M_0$}
\label{sec:deriveSU2}

We begin by spelling out some of the details of the process in Figure \ref{fig:RGmodify}: reconstructing the theory associated to $M$ by colliding a basic RG wall into the theory associated to a modified manifold $M'$.

Consider first the phase spaces associated to the big boundary of $M_0$ and the annular region on the big boundary of $M'$. They are parameterized by edge coordinates $x_m,x_d$ and $x_m',x_d'$, respectively, with
\be \{\log x_m,\log x_d\} = - \{\log x_m',\log x_d'\} = 1\,. \ee
Note that the bracket has opposite sign for $M'$ due to the reversed orientation.

Two new internal edges are created in the gluing of $M_0$ to $M'$, with gluing functions $c_m=x_mx_m'$ and $c_d=x_dx_d'$. Geometrically, we want to use $c_m,c_d$ as moment maps for a symplectic reduction, enforcing the gluing constraints $c_m=c_d=1$. Correspondingly, the glued of gauge theories must involve the addition of two operators to the superpotential. In order to identify these operators, it is convenient to choose polarizations for the big-boundary phases spaces so that either $x_m,x_m'$ or $x_d,x_d'$ are positions. (It is impossible to choose all four edge coordinates as positions, since they do not commute.) We therefore take a magnetic polarization $\Pi_m$ for $\pd M_0$ and a mirror ``magnetic'' polarization $\Pi_m'$ for $\pd M$:
\be \Pi_m = {x_m\choose x_e^{-1}}\,,\qquad \Pi_m' = {x_m'\choose x_e'}\,,\ee
with $x_e=-\sqrt{x_mx_d},\,x_e'=-\sqrt{x_m'x_d'}$ as usual. We supplement $\Pi_m'$ with some other polarization $\Pi$ for the part of the boundary phase space away from the annular region. (And of course we supplement $\Pi_m$ with the usual length-twist pair ($\lambda,\tau$).)

With these polarizations, the RG theory $T_2[M_0,\Pi_m]$ has $SU(2)_\lambda\times U(1)_m$ flavor symmetry and a standard (monopole) operator $\CO_m$ charged under $U(1)_m$; while $T_2[M',\Pi\times \Pi_m']$ has $U(1)_m'\times(...)$ symmetry and must have a similar chiral operator $\CO_m'$ charged under $U(1)_m'$.
Then the operator enforcing the gluing constraint $c_m=x_mx_m'=1$ in the product theory $T_\times := T_2[M_0,\Pi_m]\times T_2[M',\Pi\times \Pi_m']$ is easy to make: it is just the product $\CO_m\CO_m'$.

The other operator enforcing $c_d = x_dx_d'=x_e^2x_e'{}^2/(x_mx_m')=1$ is trickier. We first have to change polarization for $T_\times$ so that this gluing function $c_d$ is a position. This means gauging the anti-diagonal (\ie\ vector) part $U(1)_V$ of the flavor symmetry $U(1)_m\times U(1)_m'$. Notice that $\CO_m\CO_m'$ is invariant under $U(1)_V$, so it survives the gauging. After the gauging, there is a new topological flavor symmetry $U(1)_J$, as well as the remaining diagonal (\ie\ axial) part $U(1)_A$ of $U(1)_m\times U(1)_m'$. Moreover, there must be a monopole operator $\eta_d$ charged under $U(1)_J$. To the gauged theory we add
\be W = \CO_m\CO_m' + \eta_d\,,\ee
breaking both $U(1)_A$ and $U(1)_J$, and obtaining $T_2[M,\Pi]$. The unbroken flavor symmetry is $SU(2)_\lambda\times(...)$, where $(...)$ came from the boundary of $M'$ away from the annular region.

The existence of a gauge-invariant chiral operator $\eta_d$ follows from immediately from the analysis of gluing in \cite{DGG}. One may also deduce its existence by reconstructing the theory $T_2[M,\Pi]$ from components $T_2[M',\Pi\times\Pi_d']$ and $T_2[M_0,\Pi_d]$, \ie\ starting from ``dyonic'' polarizations. Then we would find that the operator enforcing $c_d=1$ is easy to describe as $\CO_d\CO_d'$, while the operator $\eta_m$ enforcing $c_m=1$ is non-trivial:
\be W = \eta_m + \CO_d\CO_d'\,.\ee
Of course, it cannot matter which initial polarizations we start with, as long as the final one is the same.

\subsection{Matter doublets}
\label{sec:doublets}

Our next item of business is to justify a claim made both above and back in Section \ref{sec:couplings}, that whenever the big boundary of a 3-manifold contains a degenerate triangle surrounding a hole with flavor symmetry $SU(2)_\lambda$, the associated 3d theory contains a doublet of chiral operators $q_\alpha$ under $SU(2)_\lambda$. This is in contrast to all other edges of the big boundary (away from degenerate triangles), which according to \cite{DGG} get single chiral operators $\CO_E$.

\begin{figure}[htb]
\centering
\includegraphics[width=5in]{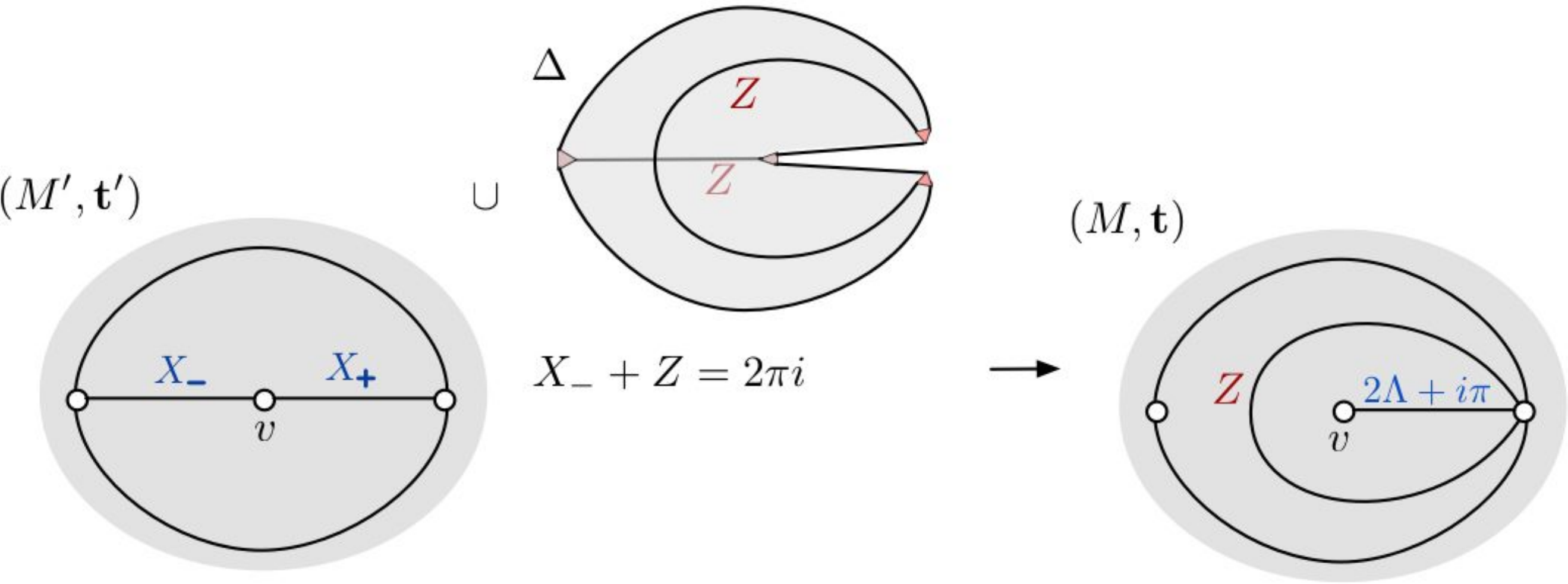}
\caption{Flipping to create a degenerate triangle on the boundary of $M$. The logarithmic edge coordinates on the LHS parametrized as $X_\pm =\Lambda\pm X+i\pi$, and obey the gluing constraint $X_-+Z=2\pi i$.}
\label{fig:doublet}
\end{figure}

We proceed by flipping. Suppose that $(M,\mb t)$ is a framed 3-manifold whose boundary triangulation has a degenerate triangle surrounding a hole $v$. The immediate neighborhood of this degenerate triangle must look like the RHS of Figure \ref{fig:doublet}. Then $(M,\mb t)$ can be obtained by flipping an edge on the boundary of another 3-manifold $(M',\mb t')$ whose boundary triangulation looks like the LHS of Figure \ref{fig:doublet} (and is equivalent to $\mb t$ away from the degenerate triangle that we are studying). Recall that, from a 3d perspective, ``flipping'' means gluing on a tetrahedron (Figure \ref{fig:flip3d}); thus $M=M'\cup \Delta$.

The two edges of $\pd M'$ that end at $v$ commute.
Looking at spaces of flat connections, we can assign to them logarithmic edge coordinates $X_\pm=\Lambda\pm X+i\pi$, obeying the constraint $(X_+-i\pi)+(X_--i\pi)=2\Lambda$ as in \eqref{sumX} or \eqref{sumhole}. We use logarithmic coordinates in order to keep track of R-charges. We choose a polarization $\Pi'$ for the boundary phase space such that $X$ and $\Lambda$ are position coordinates. Correspondingly, the theory $T_2[M',\mb t',\Pi']$ has a $U(1)_\lambda\times U(1)_x$ as part of its flavor symmetry group, and (since this is a non-degenerate boundary triangulation) it must also have two operators $\CO_+,\CO_-$ associated to our two external edges, with charges
\be  T_2[M',\mb t',\Pi']:\quad  \begin{array}{c|cc} & \CO_+ & \CO_- \\\hline
       U(1)_\lambda & 1 & 1 \\
       U(1)_x & 1 & -1 \\
       U(1)_R & 1 & 1  \end{array}
\ee
Note that our precise choice of polarization \emph{defined} what the $U(1)_\lambda$ and $U(1)_R$ currents were in this theory. It is this choice of $U(1)_\lambda$ that will get enhanced.

To perform the flip, we add a tetrahedron theory $T_\Delta$ with its chiral $\phi$ and a superpotential coupling
\be W = \phi\, \CO_-\,. \ee
The superpotential kills the operator $\CO_-$, replacing it with the operator $\phi$, which has exactly opposite flavor charges and R-charge $R(\phi)=2-R(\CO_-)=1$. Thus
\be  \label{doubletflip}
T_2[M,\mb t,\Pi]:\quad  \begin{array}{c|cc} & \CO_+ & \phi \\\hline
       U(1)_\lambda & 1 & -1 \\
       U(1)_x & 1 & 1 \\
       U(1)_R & 1 & 1  \end{array}
\ee
Now the pair of operators $(q_1,q_2):= (\CO_+,\phi)$ can become a doublet under an enhanced $U(1)_\lambda\to SU(2)_\lambda$. Notice that, by careful definition of flavor and R currents, they have exactly the same charges under all other symmetries of the theory. This is what we set out to show.

Geometrically, the flip adds a tetrahedron with edge parameter $Z$ as shown. A new internal edge is created, enforcing the constraint $Z+X_-=2\pi i$, or $Z = -\Lambda+X+i\pi$. The coordinate $Z$ labels the circular edge of the degenerate triangle on the RHS, and clearly this (standard) edge carries the operator $\phi$. The degenerate edge of the triangle on the RHS gets coordinate $X_++Z'+Z'' = 2\Lambda+i\pi$. Thus it exclusively carries the flavor symmetry of the hole, but (being degenerate) it cannot carry any operators on its own. In order to define the precise polarization $\Pi$ for the theory $T_2[M,\mb t,\Pi]$ with (potentially) enhanced flavor symmetry, we parametrized the edges near the hole in terms of $\Lambda$ and $X$, and took $\Lambda$ and $X$ as positions.

\subsection{Three-punctured spheres and UV theories of class $\CS$}
\label{sec:spheres}

The above analysis of chiral doublets and flips has a particularly nice application when big boundary are 3-punctured spheres.

\begin{figure}[htb]
\centering
\includegraphics[width=5in]{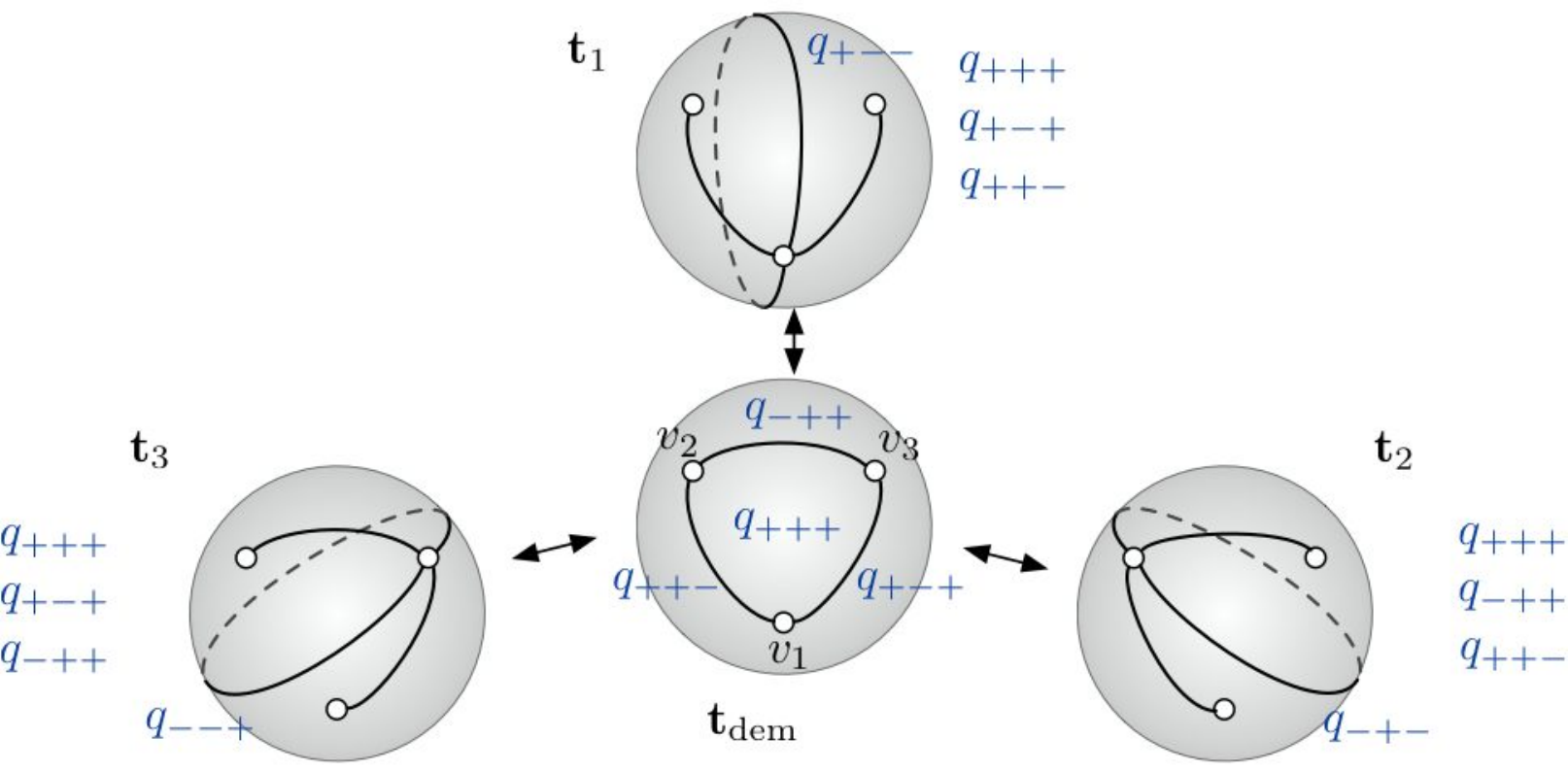}
\caption{Flipping among the four triangulations of a 3-punctured sphere.}
\label{fig:S2tq}
\end{figure}

A 3-punctured sphere $\CC$ only admits four distinct triangulations, shown in Figure \ref{fig:S2tq}, and they are precisely related by the kinds of flips we just described. Let us call the three punctures $v_i$, $i=1,2,3$, with corresponding holonomy eigenvalues $\lambda_i$ and flavor symmetries~$U(1)_i$.

One of the triangulations --- the democratic one, $\mb t_{\rm dem}$ --- has two edges ending on each of the three punctures. The edge connecting $v_1$ and $v_2$ has edge parameter $-\lambda_1\lambda_2/\lambda_3$ (logarithmically, $\Lambda_1+\Lambda_2-\Lambda_3-i\pi$), and similarly for the other edges, so that the product around any puncture $v_i$ equals $\lambda_i^2$. Note that all the edges commute; indeed, these edge coordinates are exclusively made from eigenvalues at holes, reflecting the fact that 3-punctured spheres carry no independent degrees of freedom for $SL(2)$ flat connections. If $M$ is a 3-manifold with a democratically triangulated 3-punctured sphere on its boundary, we would always choose the $\lambda_i$ as positions in a boundary polarization $\Pi$. Then we expect that $T_2[M,\mb t_{\rm dem},\Pi]$ has three chiral operators $q_{++-},\,q_{+-+},\,q_{-++}$ associated to three edges of $\mb t_{\rm dem}$, with subscripts indicating charges under $U(1)_1\times U(1)_2\times U(1)_3$. Moreover, the product
\be q_{+++} =q_{++-}q_{+-+}q_{-++} \ee
provides an important additional operator with all charges $+1$.

Now suppose that we flip the edge connecting $v_2$ and $v_3$, to move to triangulation $\mb t_1$, with two degenerate triangles. The new theory $T_2[M,\mb t_1,\Pi]$ could potentially have enhanced $U(1)_1\times SU(2)_2\times SU(2)_3$ flavor symmetry, and the matter content should reflect this. Indeed, the flip replaces the operator $q_{-++}$ with a new operator $q_{+--}$ of opposite flavor charges, leaving $q_{++-},\,q_{+-+},\,q_{+++}$ undisturbed. But these four $q$'s can succinctly be written as $q_{+\beta\gamma}$, a bifundamental of $SU(2)_2\times SU(2)_3$.

Similarly, flipping the other two edges of the democratic triangulation leads to theories with chiral operators $q_{\alpha\beta+}$ and $q_{\alpha+\gamma}$, filling out bifundamental multiplets under the other combinations of (potentially) enhanced flavor groups.

\begin{figure}[htb]
\centering
\includegraphics[width=6in]{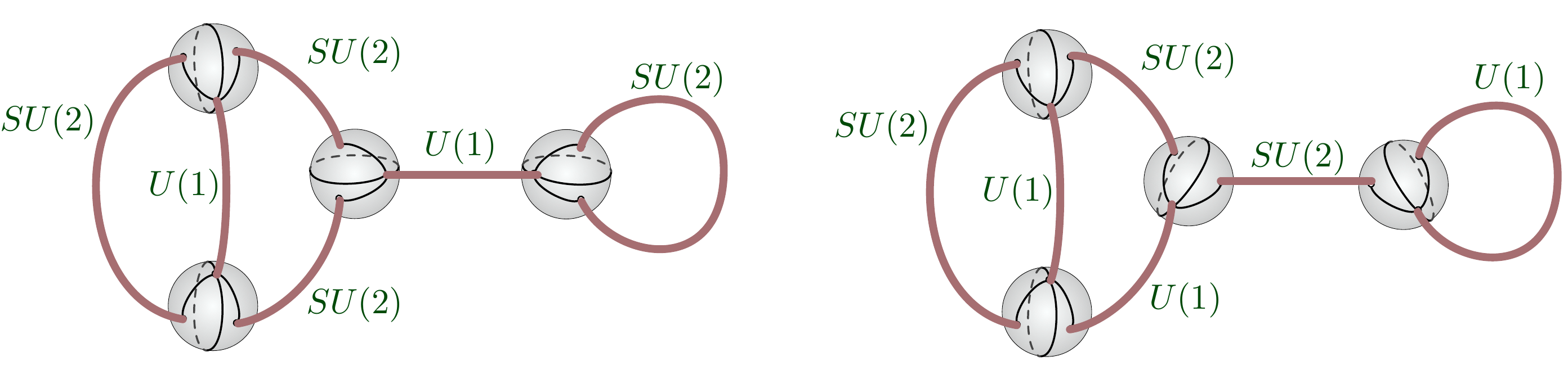}
\caption{A genus-three boundary stretched into a trivalent network of tubes connected by 3-punctured spheres. The triangulations of the 3-punctured spheres dictate whether $T_2[M,\mb t]$ will have various symmetry enhancements; two possibilities are shown.}
\label{fig:S2sym}
\end{figure}

This analysis leads to a nice combinatorial picture of when a 3-manifold whose boundary contains a trivalent network of small tubes (\emph{a.k.a.} defects) can lead to a 3d theory with enhanced flavor symmetry. At every trivalent juncture of tubes lies a 3-punctured sphere, for which we must choose one of the four triangulations of Figure \ref{fig:S2tq}. It is certainly most symmetric to choose triangulations $\mb t_{\rm dem}$ for every sphere, so that every tube ends at a bivalent hole. But then no $U(1)$ symmetries for the tubes will be enhanced! To get symmetry enhancement, we must instead choose some degenerate triangulations $\mb t_i$. Any tube that has \emph{both} ends ending at a univalent hole, in the center of a degenerate triangle, will get an $SU(2)$ enhancement (Figure \ref{fig:S2sym}). This is never possible for all tubes at once, since none of the triangulations $\mb t_i$ allow all three holes on a 3-punctured sphere to be univalent. Nevertheless, the ``obstruction'' to symmetry enhancement is minor: it comes from chiral matter that can be flipped away.

\subsubsection{4d coupling}

We expect that when the boundary of $M$ contains a trivalent network of small tubes that represents a pants decomposition of some surface $\wt \CC$, the theory $T_2[M,\mb t]$ can be coupled to the UV theory $T_2[\wt \CC]$ of class $\CS$, forming a 3d-4d boundary theory $\CB_2[M]$. We can see this explicitly here.

First, for every three-punctured sphere $\CC$ in $\pd M$ we can couple $T_2[M,\mb t]$ to a ``trinion theory'' $T_2[\CC]$. The 4d trinion theory contains eight half-hypermultiplets $Q_{\alpha\beta\gamma}$ in the trifundamental representation of $SU(2)_1\times SU(2)_2\times SU(2)_3$ \cite{Gaiotto-dualities}. We split the $Q_{\alpha\beta\gamma}$ into two sets with opposite charges according to the chosen triangulation of $\CC$. For example, for triangulation $\mb t_1$, we split $Q_{\alpha\beta\gamma} \to \big(Q^{+\beta\gamma}, Q_{+\beta\gamma}\big)$.%
\footnote{As usual, $SU(2)$ indices are raised and lowered with $\epsilon_{\alpha\beta}$.} %
Then we introduce complementary Neumann/Dirichlet b.c. for these two subsets, and a standard boundary coupling, \eg
\be \text{triangulation $\mb t_1$}:\qquad   W_{\rm bdy}= \sum_{\beta\gamma} Q^{+\beta\gamma}\big|_{\pd}\,q_{+\beta\gamma}\,,\qquad Q_{+\beta\gamma}\big|_{\pd}=q_{+\beta\gamma}\,. \qquad \ee
Let $\CB_{\rm SW}[M]$ denote the 3d theory coupled to free 4d bulk hypermultiplets like this.
By the arguments of Section \ref{sec:dissolve} the boundary condition $\CB_{\rm SW}[M]$ must have enhanced $SU(2)$ flavor symmetry for every single tube.

To finish, we just gauge all these $SU(2)$ symmetries of $\CB_{\rm SW}[M]$ in the bulk. This creates the full non-abelian theory $T_2[\wt \CC]$ of class $\CS$ in the bulk. It is manifestly presented in the duality frame corresponding to our chosen pants decomposition of $\wt \CC$. We obtain the desired boundary condition $\CB_2[M]$.

\subsubsection{Normalization of conformal blocks}

The choices made for triangulations of 3-punctured spheres map very naturally to standard choices that are often made when normalizing conformal blocks in Liouville theory. We briefly explain how this works.

Consider the Liouville conformal block corresponding to a pants decomposition of a surface $\wt \CC$, \ie\ to a Moore-Seiberg graph that is the skeleton of the pants decomposition. Specifying the conformal block requires taking a ``square root'' of the DOZZ 3-point function
at every trivalent juncture. Two options for taking this square root were explored in \cite{AGGTV, DGOT}. It is convenient to factor the 3-point function into pieces that look like quantum dilogarithm functions, as in \cite{DGOT}, since such functions obey algebraic difference equations, but there is no canonical way to do this.

By the AGT correspondence \cite{AGT}, the conformal block is a partition function of $T_2[\wt \CC]$ on half of $S^4_b$ \cite{Pestun-S4, HH-S4b}, with certain boundary conditions on the equator. The boundary conditions can pick Neumann vs. Dirichlet for the hypermultiplets of $T_2[\wt \CC]$. Algebraically, this is precisely the choice of how to split the DOZZ 3-point function into quantum dilogarithms. But we also have just argued above that the choice of boundary conditions for hypermultiplets maps, geometrically, to the different triangulations of 3-punctured spheres. The two choices are one and the same.

Of course, once we glue conformal blocks together to form a complete Liouville partition function --- or a partition function of $T_2[\wt \CC]$ on a full $S^4_b$ --- all ambiguities cancel out. In our case, it is more natural to couple the 4d theory on a half-sphere to an appropriate 3-manifold theory on the equator, and again all ambiguities must cancel out.

\begin{figure}[htb]
\centering
\includegraphics[width=5in]{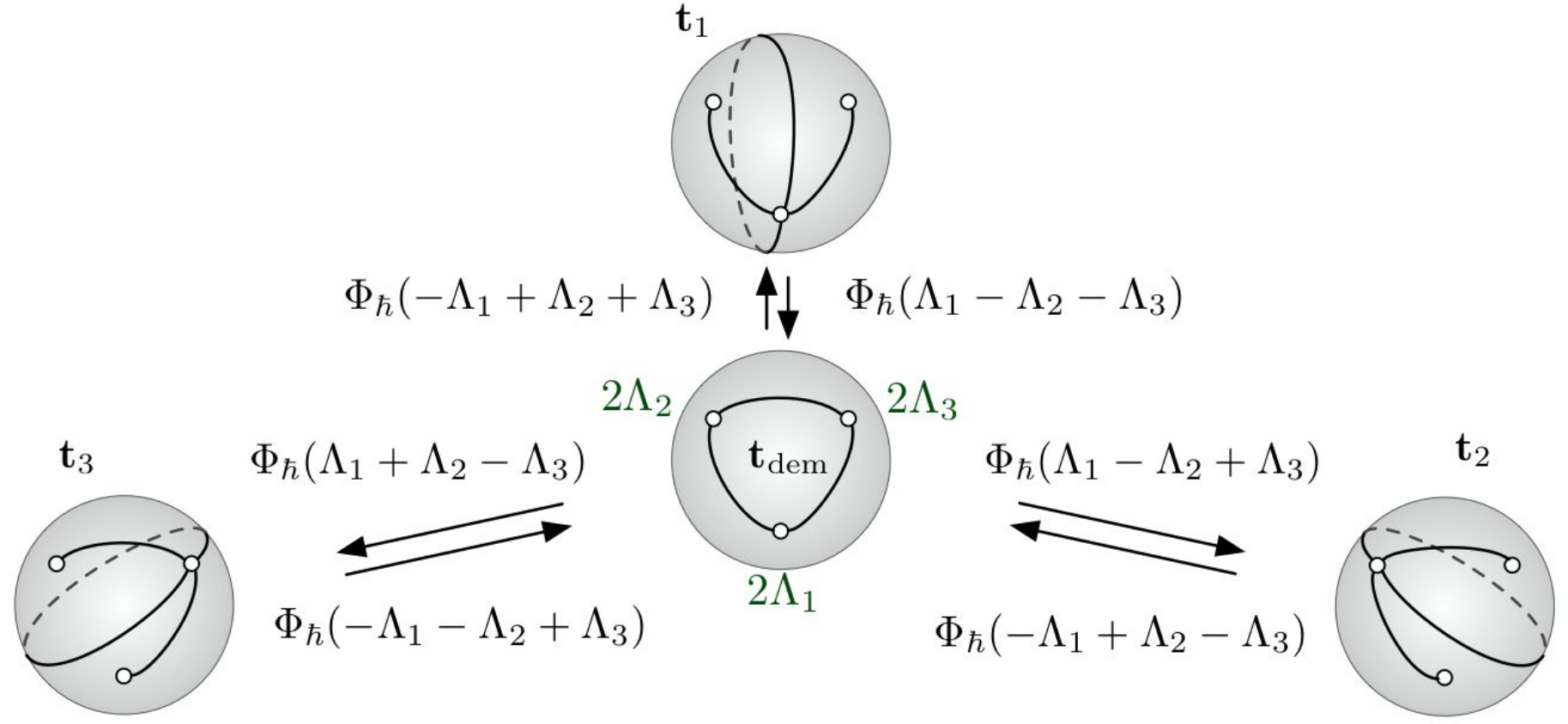}
\caption{The quantum dilogarithm prefactors that multiply an $S^3_b$ partition function when various flips are performed. These can explicitly break Weyl symmetry $\Lambda_i\to -\Lambda_i$ in the partition functions, leading to expected breaking of $SU(2)_i$ flavor symmetries. Some additional quadratic-exponential prefactors accompany these $\Phi_\hbar$ functions, and correspond to background Chern-Simons terms; they can be fixed by carefully keeping track of dual twists $\CT_i$.}
\label{fig:S2qdl}
\end{figure}

Along these lines, we can explicitly see the effect that changing triangulations has on conformal blocks by considering $S^3_b$ partition functions $\CZ_b$ of 3d theories. Each flip multiplies $\CZ_b$ by a quantum dilogarithm as shown in Figure \ref{fig:S2qdl}. These flips must simultaneously \emph{divide} conformal blocks by the same quantum dilogarithms.

\subsection{Small tori}
\label{sec:torus}

Our last special topic in symmetry enhancement concerns the small torus boundaries of a 3-manifold $M$. The situation is more subtle than with small annuli.

Small torus boundaries, just like small annuli, carry an abelian $U(1)$ flavor symmetry. From our discussion above, we might expect that it is promoted to a non-abelian $SU(2)$. After all, we expect that in six dimensions the small torus boundaries are associated with regular defects that wrap closed loops in $M$.

However, we immediately see a problem with this proposal. For $M$ with a small torus boundary, the three-dimensional theory $T_2[M,\Pi]$ depends 
depends on a choice of A and B cycles on the torus --- implicit in the polarization $\Pi$. The $U(1)\equiv U(1)_A$ symmetry is naturally associated with the A cycle. Changing the choice of cycles corresponds to performing a usual \emph{abelian} $Sp(2,\Z)$ transformation on $T_2[M,\Pi]$.
If the $U(1)$ is indeed promoted to $SU(2)$, the promotion should happen in all possible $Sp(2,\Z)$ frames. But this cannot (naively) be the case.

Suppose, for example, that we act with $T^k$ on a theory where the $U(1)_A$ is promoted to $SU(2)_A$, changing the B cycle $\gamma_B\to \gamma_B+k\gamma_A$. This adds abelian background Chern-Simons terms for $U(1)_A$, seemingly preventing enhancement in the new frame. Similarly, if we act with $ST^k$ for large $k$, corresponding to new cycles $(\gamma_A',\gamma_B')=(-k\gamma_A-\gamma_B,\gamma_A)$, we will gauge $U(1)_A$ at Chern-Simons level $k$, obtaining a new theory with a topological $U(1)_A'$ symmetry that rotates the dual photon. One might hope that quantum effects would enhance $U(1)_A'$ to $SU(2)_A'$. However, for large $k$, the gauge group $U(1)_A$ is arbitrarily weakly coupled, so such quantum effects cannot be expected to help.

A possible way to understand the lack of enhancement directly in a 3d theory $T_2[M,\Pi]$ comes from the observation that this theory has an exactly marginal superpotential coupling for every torus cusp. This can be seen geometrically. When $M$ is triangulated, every torus cusp leads to one redundant gluing constraint (Theorem \ref{thm:count}, page \pageref{thm:count}). This implies that when constructing $T_2[M,\Pi]$ one of the operators in the superpotential --- where each operator is associated to a gluing constraint --- does not break any flavor symmetries and automatically has R-charge equal to two. By the results of \cite{GKSTW}, this operator must be exactly marginal. We can imagine a scenario where $U(1)_A$ is enhanced to $SU(2)_A$ only at a special point in the exactly marginal parameter space. But $Sp(2,\Z)$ transformations need not preserve this special locus. Thus, generically, $U(1)_A$ is not enhanced.

This scenario can be matched nicely with a 6d analysis. A regular defect in six dimensions carries a triplet of moment-map operators, each in the adjoint of the $SU(2)_A$ flavor symmetry group.
After we twist the theory on $M$, with the defect wrapping a closed loop in $M$, two of these moment-map operators remain scalars and can be combined into a complex chiral operator $\mu_\C$ for the effective 3d theory $T_2[M]$. It automatically has R-charge two in the standard 3d normalization: the 3d R-charge is a $U(1)_R$ subgroup of the $SU(2)_R$ R-symmetry preserved by a defect in flat space, and the triplet of moment maps has spin one under $SU(2)_R$, while the supercharges have spin $1/2$.%
\footnote{These properties of the moment-map operators can be derived by embedding the entire setup in M-theory. While dynamical branes wrap $M\times \R^3$ in $T^*M\times \R^3\times\R^2$, the defect branes wrap the conormal bundle $N^*\CK\times\R^3$ of a curve $\CK\subset M$. The complex scalar moment-map operator parametrizes motion of the defect branes in $\R^2$, and must have R-charge two since $U(1)_R$ rotates this $\R^2$.} %
We can now choose a Cartan subgroup of $SU(2)_A$, and add to the 3d superpotential the $\mu^3_\C$ component of the complex moment-map operator in the direction of this Cartan. We get an exactly marginal 
deformation along which we break $SU(2)_A$ flavor symmetry down to $U(1)_A$. This is compatible with the constraints of \cite{GKSTW}: the two remaining components $\mu^\pm_\C$ 
of the complex chiral moment map combine with the broken $SU(2)_A$ flavor currents to give non-protected multiplets.

We now have a consistent picture of how non-abelian symmetry enhancement/breaking could happen. It is also interesting to observe that protected quantities that are insensitive to marginal deformations --- such as $S^3_b$ partition functions or sphere indices --- always have manifest Weyl invariance $\lambda_A\to \lambda_A^{-1}$ in the parameter corresponding to a $U(1)_A$ symmetry. This supports the idea that there exists a point of enhancement in the marginal parameter space (much as in the examples of \cite{DG-E7}), though it is certainly not a proof.

The analysis of moment maps and marginal deformations here can be extended to a theory $T_K[M]$ of $K$ M5 branes in a straightforward fashion. This was done in \cite{DGG-Kdec}.

%%%%%%%%%%%%%%%%%%%%%%%%%%%%%%%%%%%%%%%%%%%%%%%%%%%%%%%%%%%%%%%%%%%%%%%%%%%%
%%%%%%%%%%%%%%%%%%%%%%%%%%%%%%%%%%%%%%%%%%%%%%%%%%%%%%%%%%%%%%%%%%%%%%%%%%%%
%%%%%%%%%%%%%%%%%%%%%%%%%%%%%%%%%%%%%%%%%%%%%%%%%%%%%%%%%%%%%%%%%%%%%%%%%%%%

\section{Example 1: $\CN=2^*$ interfaces}
\label{sec:N=2*}

We would now like to use the technology of the previous sections to construct RG walls and UV duality walls for some 4d theories of class $\CS$ with matter. Our approach will be to first build framed 3-manifolds representing the desired interfaces, and then to read off effective 3d interface theories (and their 4d couplings) from triangulations of the manifolds.

We begin here by looking at $\CN=2^*$ theory, \ie\ four-dimensional $\CN=2$ theory with gauge group $SU(2)$ and an adjoint hypermultiplet $\Phi_{\alpha\dot\alpha}$. Let's review some of its geometric features. The complex mass of the hypermultiplet is associated with an $SU(2)_\mu$ flavor symmetry. To see this flavor symmetry, we must split the hypermultiplet into chirals $\Phi_{\alpha\dot\alpha}=(X_{\alpha\dot\alpha},Y_{\alpha\dot\alpha})$; then $SU(2)_\mu$ acts on $(X,Y)$ as a doublet. If the mass is tuned to zero, the theory gains maximal $\CN=4$ supersymmetry.

As a theory of class $\CS$, $\CN=2^*$ theory is well known to be realized by compactification on a punctured 2-torus, $\CC=T^2_*$ \cite{Gaiotto-dualities, GMN}. The $SU(2)_\mu$ flavor symmetry is associated to the puncture, while the $SU(2)_\lambda$ gauge symmetry (in given weakly coupled description) is associated to an A-cycle for the torus. If we stretch $T^2_*$ into a trivalent network of tubes as on the RHS of Figure \ref{fig:N2RGbdy} below, the network contains a single three-punctured sphere. Correspondingly, $T_2[T^2_*]$ should contain eight half-hypermultiplets $Q_{\alpha\dot\alpha\gamma}$. They combine into the traceless adjoint hyper $(X_{\alpha\dot\alpha},Y_{\alpha\dot\alpha})=\big((Q_{\alpha\dot\alpha+},Q_{\alpha\dot\alpha-})-\text{trace}\big)$ and an additional decoupled gauge-invariant hyper $\hat\Phi = (\hat X,\hat Y) = (Q_\alpha{}^\alpha{}_+,Q_\alpha{}^\alpha{}_-)$. Thus $T_2[T^2_*]$ is a small extension of ``standard'' $\CN=2^*$ theory.

The usual modular group $PSL(2,\Z)$ acts on the A and B cycles of the punctured torus, and provides a UV duality group for $T_2[T^2_*]$. In particular, the S element exchanges the A and B cycles, acting as electric-magnetic duality. The 3d degrees of freedom living on an S-duality wall --- where dual $SU(2)$ gauge couplings become weak on the two sides --- were extracted in a theory called $T[SU(2)]$ in \cite{GW-Sduality}. We will reconstruct this theory from a framed 3-manifold $M_S$ that implements the S cobordism. The $T[SU(2)]$ theory of \cite{GW-Sduality} had a manifest $SU(2)_\lambda\times U(1)_{\lambda'}$ flavor symmetry that was enhanced to $SU(2)_\lambda\times SU(2)_{\lambda'}$ by quantum effects, which could then couple to two copies of 4d $\CN=2^*$ theory in the bulk. By choosing a convenient triangulation of $M_S$ in Section \ref{sec:TSU22}, we will discover a mirror-symmetric description of $T[SU(2)]$ that has manifest $SU(2)_\lambda\times SU(2)_{\lambda'}$ already in the Lagrangian.

First, however, we will describe the more elementary $\CN=2^*$ RG wall.

\subsection{The RG wall}
\label{sec:N2*RG}

\begin{figure}[htb]
\centering
\includegraphics[width=6in]{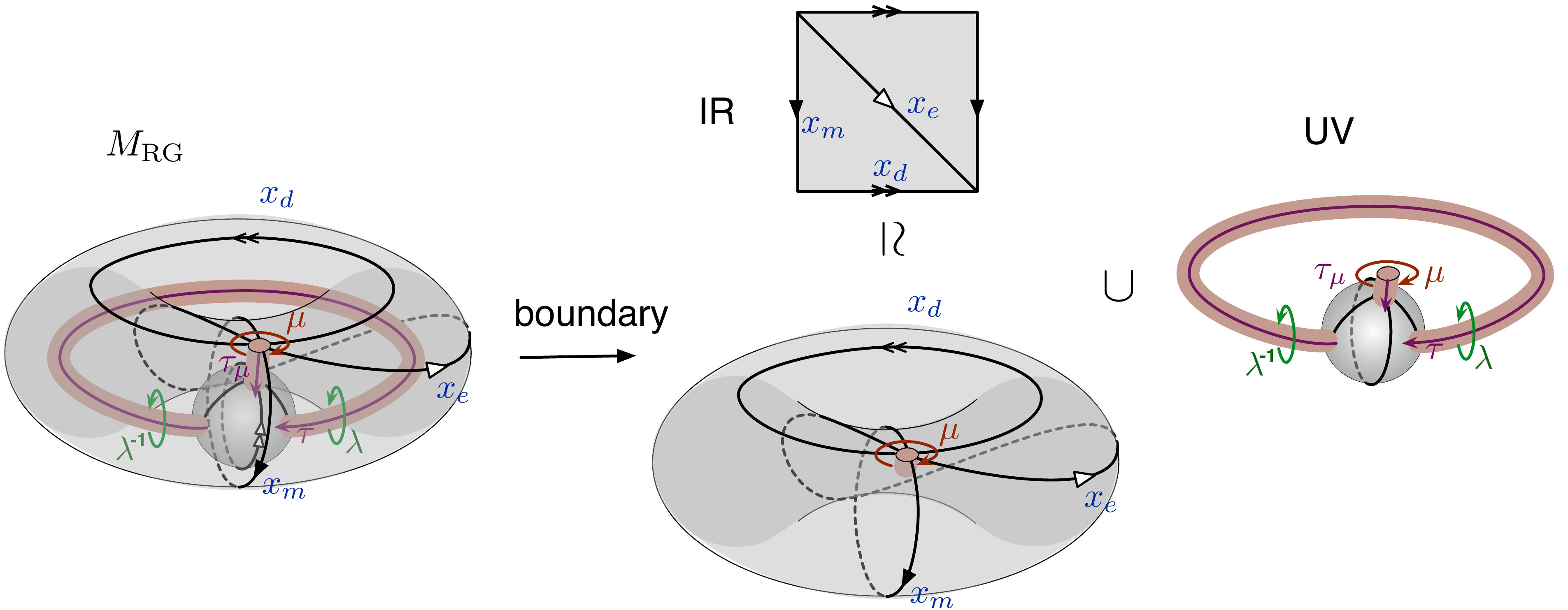}
\caption{The RG manifold for $\CN=2^*$ theory (left), interpolating between IR and UV boundaries.}
\label{fig:N2RGbdy}
\end{figure}

The RG manifold $M_{\rm RG}$ for $T_2[T^2_*]$ should interpolate between a trivalent network of tubes (representing the UV) and a WKB triangulation of $T^2_*$ (representing the IR). These boundaries are shown in Figure \ref{fig:N2RGbdy}. Modulo the action of the modular group, there is a unique triangulation of the one-punctured torus. Its three edges can be labelled by electric, magnetic, and dyonic charges of the $U(1)$ Seiberg-Witten theory in the infrared, satisfying
\be \langle \gamma_m,\gamma_e\rangle = \langle \gamma_e,\gamma_d\rangle=\langle \gamma_d,\gamma_m\rangle = 2\,,\qquad \gamma_m+\gamma_d+\gamma_e = -\gamma_\mu\,,\ee
where $\gamma_\mu$ is the flavor charge of states under a Cartan subgroup $U(1)_f\subset SU(2)_\mu$ (normalized such that a doublet has charges $\pm \gamma_\mu$). Correspondingly, we associate parameters $x_m,x_d,x_e$ to the edges of the triangulation, satisfying
\be \label{xmu}
 \{\log x_m,\log x_e\}=\{\log x_e,\log x_d\}=\{\log x_d,\log x_m\}=2\,,\qquad x_mx_dx_e = -\mu^{-1}\,.
 \ee
These parametrize (a partial lift) framed flat $PSL(2)$ connections on the triangulated $T^2_*$, with fixed holonomy eigenvalue $\mu^{-1}$ at the puncture.%
\footnote{Logarithmically, the central constraint in \eqref{xmu} is $2(X_m+X_d+X_e-3\pi i) = -2M$. As usual, this tells us how to ``lift'' from specifying the squared-eigenvalue at the puncture to specifying the eigenvalue itself. Dividing by two and exponentiating, we find the minus sign in \eqref{xmu}.}

On the UV side, we find a big 3-holed sphere, two of whose holes are connected to each other by a small annulus. We define a pair of eigenvalue-twist coordinates $(\lambda,\tau)$ for this annulus, as in Section \ref{sec:MRG} or Appendix \ref{app:twist}. We will want the 3d theory $T_2[M_{\rm RG}]$ by itself to have manifest $SU(2)_\lambda$ symmetry. Therefore (according to Section \ref{sec:spheres}) we choose the triangulation for the 3-punctured sphere with two degenerate triangles surrounding the holes with eigenvalues $\lambda^{\pm 1}$. The third hole of the 3-punctured sphere must be connected by another small annulus to the hole on the IR boundary. This annulus has holonomy eigenvalue $\mu$ (from the perspective of the 3-punctured sphere, the eigenvalue is inverted), and we may also give it a dual twist coordinate $\tau_\mu$.

Altogether, the RG manifold of Figure \ref{fig:N2RGbdy} has a 6-dimensional boundary phase space, with $\C^*$ coordinates $\{x_m,x_d,x_e;\lambda,\tau;\mu,\tau_\mu\}$ satisfying the Poisson brackets \eqref{xmu} as well as
\be \{\log \tau,\log\lambda\} = \{\log \tau_\mu,\log\mu\}=1\,. \ee
The UV coordinates $(\lambda,\tau)$ commute with the flavor coordinates $(\mu,\tau_\mu)$ and with all the external edge coordinates. The flavor twist-parameter $\tau_\mu$, however, cannot commute with all three IR edge coordinates simultaneously. We will choose%
\footnote{We emphasize that there is no canonical choice for $\tau_\mu$, because the IR boundary does not have a triangulation with a degenerate triangle surrounding its hole (on which the flavor annulus ends). However, a canonical choice is \emph{not necessary}! The basic reason for this is that in coupling $T_2[M_{\rm RG},\mb t,\Pi]$ to 4d UV and IR theories, we will never gauge the $U(1)_\mu$ symmetry associated to the flavor annulus --- we will simply identify it with flavor symmetries in the bulk. The ambiguity in choosing $\tau_\mu$ translates to ambiguous background Chern-Simons couplings involving $U(1)_\mu$ gauge fields, which we do not really care about if this symmetry is not gauged.} %
it to commute with $x_e$ and~$x_m$. Then one natural polarization $\Pi_e$ for $\CP_2(\pd M_{\rm RG},\mb t)$ is
\be \label{N2Pi} \Pi_e:\quad \text{positions $(x_e,\lambda,\mu)$}\,,\qquad \text{conjugate momenta $(\sqrt{x_m},\tau,\tau_\mu)$}\,. \ee

\begin{figure}[htb]
%\centering
\hspace{-.1in}\includegraphics[width=6.2in]{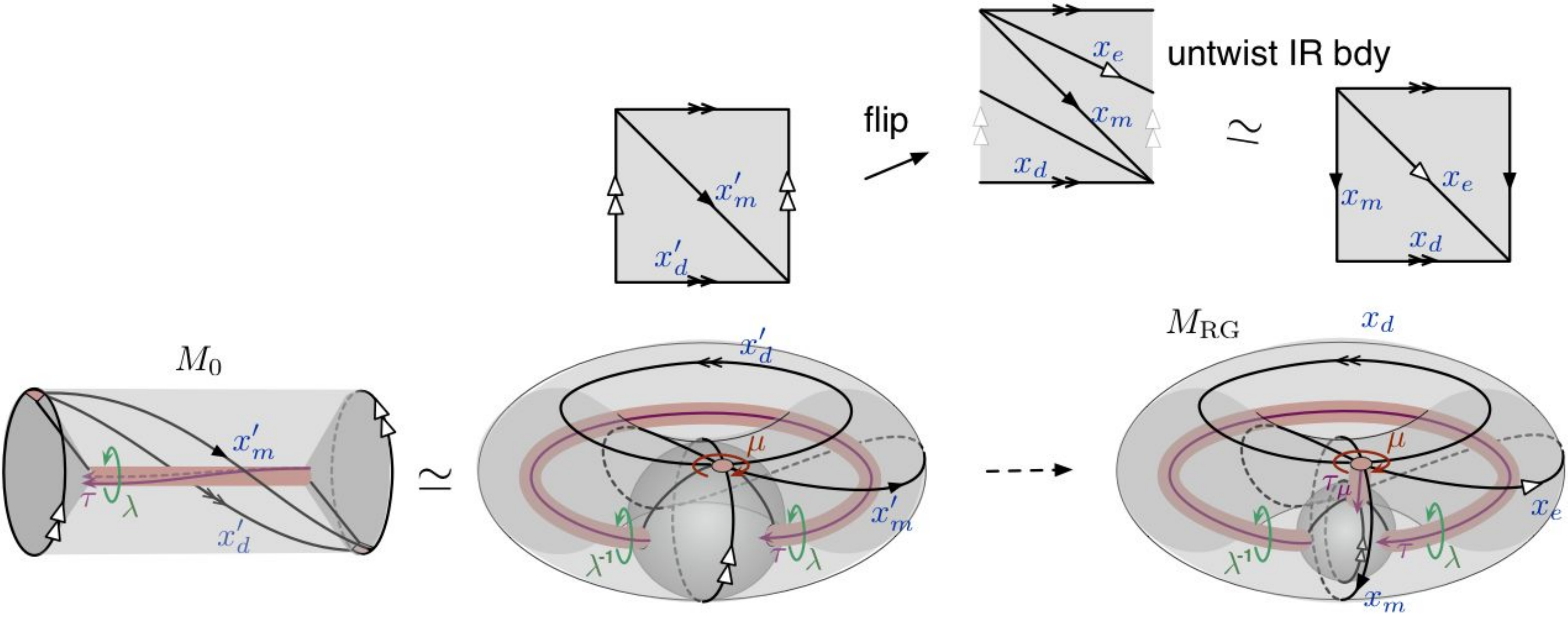}
\caption{Identifying the ends of a basic RG manifold $M_0$ (for pure $SU(2)$ theory) to obtain the RG manifold for $\CN=2^*$. An extra flip thickens out the manifold, separating the UV and IR boundaries.}
\label{fig:N2RGflip}
\end{figure}

The simplest way to triangulate $M_{\rm RG}$ is to take the basic RG manifold $M_0$ for pure $SU(2)$ theory and identify the two circular edges at its ends. This turns the big annulus on the IR boundary into a big one-punctured torus. However, the resulting manifold is a little too ``thin'' at the newly identified edge --- its UV and IR boundaries are pinched together there, so that they touch. We can fix this by flipping the thin edge on the outside, as in Figure \ref{fig:N2RGflip}. The edge created by the flip gets the electric coordinate $x_e$.

\begin{figure}[htb]
%\centering
\hspace{-.1in}\includegraphics[width=6.2in]{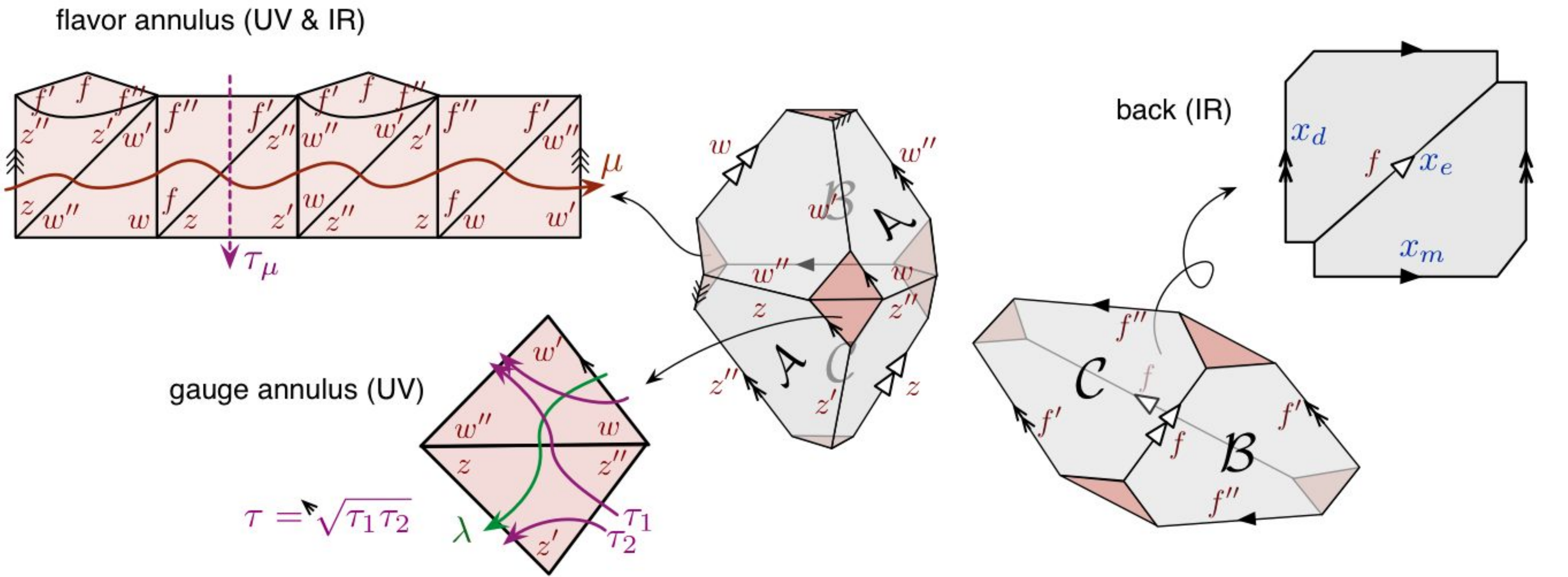}
\caption{Triangulation of the RG manifold for $\CN=2^*$ theory.}
\label{fig:N2RG}
\end{figure}

By combining the triangulation of the basic RG manifold $M_0$ (two tetrahedra) with the extra flip (one more tetrahedron), we arrive at the triangulation shown in Figure \ref{fig:N2RG}. We label the edges of the third tetrahedron with parameters $f,f',f''$. The piece of small boundary that forms the small annulus of $M_0$ is unchanged, and we read off relations
\begin{subequations} \label{N2zx}
\be \lambda^2 = \frac zw\,,\qquad \tau = \lambda\frac{z''}{w''} \ee
just as for pure $SU(2)$ theory. The three edges on the IR boundary are changed a little bit by the flip; they get relations
\be x_e = f\,,\qquad x_m = x_m'f''{}^2=z'w'f''{}^2\,,\qquad x_d = x_d'f'{}^2 = z''w''f'{}^2\,,\ee
where $x_m',x_d'$ denote the edge coordinates of $M_0$ from \eqref{xzRG}.

The remaining ten small triangles of the three tetrahedra come together to tile the small flavor annulus. Then we may read off the eigenvalue $\mu$ and also define a dual twist $\tau_\mu$%
\footnote{Logarithmically, $M = Z+W+F-i\pi$ and $\CT_\mu := -\frac12(Z'+W')$\,.}
\be \mu^2 = \frac{zwf^2}{z'z''w'w''}\quad\Rightarrow\quad \mu = -zwf\,;\qquad \tau_\mu := \frac{f''}{\sqrt{x_m}}=\frac1{\sqrt{z'w'}}\,. \ee
\end{subequations}
Notice that the expected relation $\mu^{-1} = -x_ex_mx_d$ holds, and could actually have been used to bypass the computation of $\mu$ from a path on the small annulus.
Here the twist $\tau_\mu$ is simply chosen as an appropriate combination of tetrahedron parameters that satisfies the desired commutation relations from above (in particular, $\tau_\mu$ commutes with $x_e$ and $x_m$); it could also be expressed as a certain average of paths on the flavor annulus.

\subsubsection{Gauge theory and couplings}

We can now build the theory $T_2[M_{\rm RG},\mb t,\Pi_e]$ in a fairly intuitive manner, mirroring the construction of $M_{\rm RG}$. We simply take the basic RG theory $T_2[M_0,\mb t,\Pi_e']$ from \eqref{TM0e} of Section \ref{sec:MRG} --- a theory of two chirals $q_\alpha = (\phi_z,\phi_w)$ with $SU(2)_\lambda\times U(1)_e'$ flavor symmetry --- and add one more chiral $\phi_f$ from the third tetrahedron. There is no superpotential, since no internal edges are created in the gluing. Moreover, since the positions $x_e,\lambda,\mu$ in our polarization $\Pi_e$ only involve tetrahedron positions $z,w,f$, there is no gauging. Thus, $T_2[M_{\rm RG},\mb t,\Pi_e]$ is basically a theory of three free chirals, whose flavor symmetry we write as $SU(2)_\lambda\times U(1)_e\times U(1)_\mu \subset U(3)$.

The flavor charges of $q_\alpha$ and $\phi_f$ can be read off by inverting the relations $\lambda^2 = z/w,\,\mu = -zwf$, and $x_e=f$. We find
\be \begin{array}{c|cc} & q_\alpha & \phi_f \\\hline
  SU(2)_\lambda & \square & - \\
  U(1)_\mu & \frac12 & 0 \\
  U(1)_e & -\frac12 & 1  \end{array}
\ee
In addition, there is a $U(1)_R$ R-symmetry, whose charge assignment is dictated by the logarithmic form of the coordinate relations. A more careful analysis reveals that $R(q_\alpha)=1/2$ and $R(\phi_f)=0$.

Of course this theory of free chirals is accompanied by various background Chern-Simons terms. As usual, these can be determined from our choice of momenta in $\Pi_e$. All we really need to know, however, is that there is no abelian Chern-Simons term for $U(1)_\lambda$ that would prevent its enhancement to $SU(2)_\lambda$; such a term does not exist because it is absent in the basic RG theory $T_2[M_0,\mb t,\Pi_e']$, and we have made no changes to the polarization that involve $\lambda$ or $\tau$.

The theory $T_2[M_{\rm RG},\mb t,\Pi_e]$ can couple both to non-abelian $\CN=2^*$ theory $T_2[T^2_*]$ and to the abelianized Seiberg-Witten theory $SW_2[T^2_*]$, by gauging the $SU(2)_\lambda$ and $U(1)_e$ flavor groups, respectively. On the abelian side, our 3d theory can couple to an electric BPS hypermultiplet $\Phi_e = (X_e,Y_e)$ via the superpotential
\be  W_{\text{bdy, IR}} = X_e\big|_\pd\,\CO_e \,,\qquad \CO_e \equiv \phi_f\,, \ee
where $\CO_e$ is the standard operator associated to the electric edge on the big boundary.
More interestingly, on the non-abelian side the 3d theory can couple to the adjoint hypermultiplet $\Phi_{\alpha\dot\alpha} = (X_{\alpha\dot\alpha},Y_{\alpha\dot\alpha})$ via
\be W_{\text{bdy, UV}} = X^{\alpha\dot\alpha}\big|_\pd\, q_\alpha q_{\dot \alpha}\phi_f\,.\ee
In this case, the three operators $q_\alpha q_{\dot\alpha}\phi_f$ are associated to the three-punctured sphere of $\pd M_{\rm RG}$, a special case of the general structure of Section \ref{sec:spheres}. Notice in particular that the operators $q_\alpha q_{\dot\alpha}\phi_f$ are neutral under the IR symmetry $U(1)_e$, and have the right $SU(2)_\lambda\times U(1)_\mu$ charges to couple to $X^{\alpha\dot\alpha}$.

Quite beautifully, the bulk-boundary couplings give us a full set of relations between elementary operators in the UV and the IR. Namely, after integrating out $\phi_f$, the full bulk-boundary superpotential $W_{\text{bdy, IR}}+ W_{\text{bdy, UV}}$ imposes the relation $X^{\alpha\dot\alpha}\big|_\pd\, q_\alpha q_{\dot \alpha} = X_e\big|_\pd$ between half-hypermultiplets on the two sides. In addition, the above couplings must be accompanied by modified Dirichlet b.c. for $Y_{\alpha\dot\alpha}$ and $Y_e$, which set $Y_{\alpha\dot\alpha}\big|_\pd = q_\alpha q_{\dot \alpha} \phi_f$ and $Y_e\big|_\pd = \phi_f$, in other words $Y_{\alpha\dot\alpha}\big|_\pd= q_\alpha q_{\dot \alpha}Y_e\big|_\pd$. These relations supersymmetrize nicely the boundary conditions for the vectormultiplet scalar fields.

Finally, we should note that even though $U(1)_\mu$ is not enhanced to $SU(2)_\mu$ in the isolated 3d theory $T_2[M_{\rm RG},\mb t,\Pi_e]$ (we would not expect an enhancement, since the flavor annulus does not end on degenerate triangles), it must be enhanced in the full coupled 3d-4d system. The mechanism for enhancement was discussed in Section \ref{sec:enhance}.

\subsubsection{Line operators}
\label{sec:N2RG-ops}

We can use the 3d geometry of $M_{\rm RG}$ to relate coordinates for flat $SL(2)$ connections on the IR boundary to those on the UV boundary. This translates as usual to formulas relating UV and IR line operators. These formulas for $\CN=2^*$ theory are already well known from gauge/Liouville theory \cite{DGOT, AGGTV, GMNIII} and from Teichm\"uller theory (\eg) \cite{Fock-Teich, FG-Teich, FG-qdl-cluster}; and were interpreted three-dimensionally in \cite{DG-Sdual}. It is nice to see now that the formulas arise naturally from our triangulations as well.%
\footnote{Interestingly, an ideal triangulation of the geometry $M_{\rm RG}$ (or rather the thin $M_{\rm RG}'$) was also used in \cite{Kabaya-pants} to find the classical (un-quantized) version of relations \eqref{N2WardRG}. There, however, it was done without reference to moduli spaces of framed flat connections, or to the symplectic structures needed for quantization.}

As in Section \ref{sec:RGline},%
\footnote{The full quantization procedure is summarized in Appendix \ref{app:quant}.} %
the first step is to invert the relations between tetrahedron parameters and boundary coordinates. Both quantization and roots can be handled unambiguously by first using logarithmic relations and then exponentiating. From \eqref{N2zx} we find
\be \label{N2xz}
\begin{array}{c@{\qquad}c@{\qquad}c}
 \hat z = iq^{\frac14}\hat\lambda\sqrt{\frac{\hat\mu}{\hat x_e}}\,,& \hat w=iq^{\frac14}\frac{1}{\hat\lambda}\sqrt{\frac{\hat\mu}{\hat x_e}}\,,& \hat f=\hat x_e\,,\\[.2cm]
 \hat z''=iq^{-\frac18}\sqrt{\frac{\hat x_e}{\hat\mu\hat\lambda}}\sqrt{\hat\tau}\hat\tau_\mu\,,&
 \hat w''=iq^{-\frac18}\sqrt{\frac{\hat\lambda}{\hat x_e\hat\mu}}\frac{\hat\tau_\mu}{\sqrt{\hat\tau}}\,,&
 \hat f''=\sqrt{\hat x_m}\hat\tau_\mu\,.
\end{array}
\ee
Then we can substitute these into the (quantum) tetrahedron Lagrangians $\hat z''+\hat z^{-1}\simeq 0$, etc., and simplify, to obtain the desired relations between different sets of boundary coordinates.

The structure of the resulting equations turns out to be a little simpler if we do another flip of the edge labelled `$x_e$' on the IR boundary of the RG manifold. Basically this is because the flip we used in Figure \ref{fig:N2RGflip} to thicken the manifold twisted the triangulation of the IR boundary with respect to the UV $\lambda$-annulus; we want to flip again to undo this twist. The new flip, a standard one, adds a new tetrahedron (with parameters $g,g',g''$) and creates an internal edge (gluing constraint $c:=fg=1$, or $C=F+G-2\pi i=0$). Call the new IR boundary edge-coordinates $x_e'',x_m'',x_d''$, with $\{\log x_e'',\log x_m''\}=\{\log x_m'',\log x_d''\}=\{\log x_d'',\log x_e''\}=2$ and (upon quantization) $\hat x_d''\hat x_m''\hat x_e''=-q^{\frac12}\hat\mu^{-1}$.
The bulk-boundary relations become
\be \label{N2xz-flip}
\begin{array}{c@{\qquad}c@{\qquad}c@{\qquad}c}
\ds \hat z = -iq^{-\frac14}\lambda\sqrt{\frac{\hat x_e''\hat\mu}{\hat c}}\,,&
\ds \hat w = -\frac{i}{q^{\frac14}\hat\lambda}\sqrt{\frac{\hat x_e''\hat \mu}{\hat c}}\,,& 
\ds \hat f = q\frac{\hat c}{\hat x_e''}\,,&
\ds \hat g = \hat x_e''\,,\\[.2cm]
\ds \hat z''=-iq^{\frac38}\sqrt{\frac{\hat c}{\hat x_e''\hat\lambda\hat\mu}}\sqrt{\hat\tau}\hat\tau_\mu\,,&
\ds \hat w''=iq^{\frac38}\sqrt{\frac{\hat c\hat\lambda}{\hat x_e''\hat\mu}}\frac{\hat\tau_\mu}{\sqrt{\hat\tau}}\,,&
\ds \hat f''=\hat \gamma\hat \tau_\mu\,,&
\ds \hat g''=-\frac{q\hat \gamma}{\hat x_e''}\frac{1}{\sqrt{\hat x_m''}}\,,
\end{array}
\ee
where $\hat \gamma=\hat f''\sqrt{\hat z'\hat w'}$ is an operator canonically conjugate to the gluing operator $\hat c$.

Upon substituting \eqref{N2xz-flip} into the four tetrahedron equations $\hat z''+\hat z^{-1}-1\simeq 0$, etc.; eliminating $\hat \gamma$ and $\hat \tau_\mu$; and setting $\hat c=1$; we obtain the desired UV--IR relations. (We eliminate the conjugate $\hat \tau_\mu$ to the flavor symmetry parameter $\mu$ because we want to keep this as a non-dynamical flavor symmetry --- we will not consider flavor 't Hooft loops that could shift $\mu$.) After some rearrangement, we find
\begin{subequations} \label{N2WardRG}
\be \text{(Wilson)}\quad\; \hat\lambda+\hat \lambda^{-1} \,\simeq\, 
\frac{iq^{\frac14}}{\sqrt{\mu\hat x_e''}}+\frac{\sqrt{\mu\hat x_e''}}{iq^{\frac14}}-\frac{i q^{\frac14}}{\sqrt{\mu\hat x_e''}}\frac{1}{\hat x_m''}
 \,\simeq\,  -\frac{\sqrt{\hat x_m''}\sqrt{\hat x_d''}}{q^{\frac34}} - \frac{q^{\frac34}}{\sqrt{\hat x_d''}\sqrt{\hat x_m''}}+q^{-\frac14}\sqrt{\hat x_d''}\frac{1}{\sqrt{\hat x_m''}}\,,
\ee \be
\frac{q^{-\frac18}}{\hat\lambda-\hat\lambda^{-1}}\Big(\frac{1}{\sqrt{\hat \lambda}}\sqrt{\hat\tau}-\sqrt{\hat\lambda}\frac1{\sqrt{\hat\tau}}\Big) \,\simeq\, -q^{-\frac14}\sqrt{\hat x_m''}\,,
\ee \be
 \frac{q^{\frac18}}{\hat\lambda-\hat\lambda^{-1}}\Big(\sqrt{\hat\lambda}\sqrt{\hat \tau}-\frac{1}{\sqrt{\hat \lambda}}\frac1{\sqrt{\hat\tau}}\Big) \,\simeq\, -q^{\frac14}\frac{1}{\sqrt{\hat x_d''}}\,.
\ee
\end{subequations}
In the IR variables $\hat x_m$ and $\hat x_d$, these are identical to the Ward identities \eqref{WardRG} for pure $SU(2)$ theory --- the mass $\mu$ drops out. Thus is no surprise, since the RG manifold for $\CN=2^*$ theory is built from the basic RG manifold $M_0$. The first equation in \eqref{N2WardRG} relates the UV Wilson line to IR line operators. The remaining equations, however, do not immediately correspond to UV 't Hooft (or dyonic) line operators; some further manipulation is required to put them in that form.

\subsection{The S-wall: $T[SU(2)]$ with manifest $SU(2)\times SU(2)$ flavor symmetry}
\label{sec:TSU2}

The 3-manifold $M_{\rm S}$ that implements S-duality for the non-abelian $\CN=2^*$ theory was described back in Figure \ref{fig:toruspp} of the Introduction. It interpolates between two trivalent networks representing two different pants decompositions (related by $S\in Sp(2,\Z)$) of the one-punctured torus $T^2_*$. Geometrically, $M_{\rm S}$ is the complement of the so-called ``Hopf network'' in $S^3$.

The boundary of the framed manifold $M_{\rm S}$ consists of two big 3-punctured spheres and three small annuli. Two of the small annuli connect a 3-punctured sphere back to itself. We will choose triangulations for the 3-punctured sphere such that both of these small annuli end on degenerate triangles, and thus carry enhanced flavor symmetries $SU(2)_\lambda,\,SU(2)_{\lambda'}$. These are the symmetries that get gauged when coupling $T_2[M_{\rm S},\mb t]$ to 4d $\CN=2^*$ theory in its two S-dual frames. The third small annulus forms a ``bridge'' between the three-punctured spheres; it will carry a $U(1)_\mu$ flavor symmetry, identified with the Cartan of the flavor group of $\CN=2^*$ theory in the bulk.

The phase space $\CP_2(\pd M_{\rm S},\mb t)$ is six-dimensional and can be parametrized by three pairs of eigenvalue-twist coordinates $(\lambda,\tau)$, $(\lambda',\tau')$, and $(\mu,\tau_\mu)$, one for each small annulus. The twists $\tau$ and $\tau'$ are canonically defined (modulo multiplication by $\lambda$ and $\lambda'$, respectively), and because the bridge connects two three-punctured spheres it turns out that the twist $\tau_\mu$ also has a canonical definition (modulo multiplication by $\mu^{\frac12}$). The coordinates obey
\be \{\log\tau,\log \lambda\}=\{\log\tau',\log \lambda'\}=\{\log\tau_\mu,\log \mu\}=1\,,\ee
with all other commutators vanishing, and there is a canonical polarization $\Pi_{\rm S}$ with $\lambda,\lambda',\mu$ as positions and $\tau,\tau',\tau_\mu$ as conjugate momenta. So we want to find $T_2[M_{\rm S},\mb t,\Pi_{\rm S}]$ and its bulk-boundary couplings.

\begin{figure}[htb]
\centering
\includegraphics[width=5.8in]{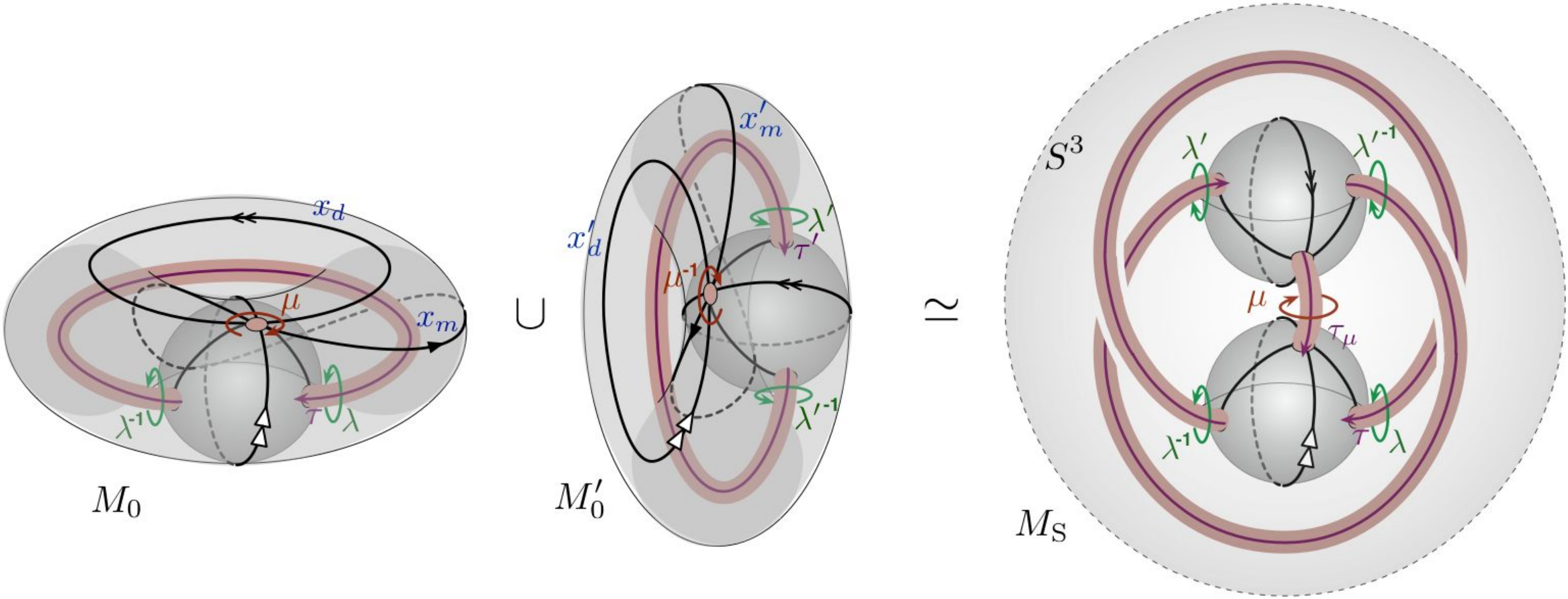}
\caption{Gluing together two solid tori, each representing an RG manifold for $\CN=2^*$ theory (with no extra thickening), to form the S-duality manifold.}
\label{fig:MSglue}
\end{figure}

The manifold $M_{\rm S}$ can be glued together from two basic RG manifolds $M_0$ and $M_0'$. Geometrically, $M_0$ and $M_0'$ first have their ends identified, forming two solid tori (as in the first part of Figure \ref{fig:N2RGflip}), each with a three-punctured sphere and a small annulus running through its core. These are two copies of the RG manifold for $\CN=2^*$ theory, without the extra flip that was added in Section \ref{sec:N2*RG}; the thickening from the extra flip is not needed.  The boundaries of the two solid tori are then identified with an S transformation, forming a 3-sphere $S^3$ with the Hopf network in its interior (Figure \ref{fig:MSglue}).

\begin{figure}[htb]
%\centering
\hspace{-.2in}\includegraphics[width=6.5in]{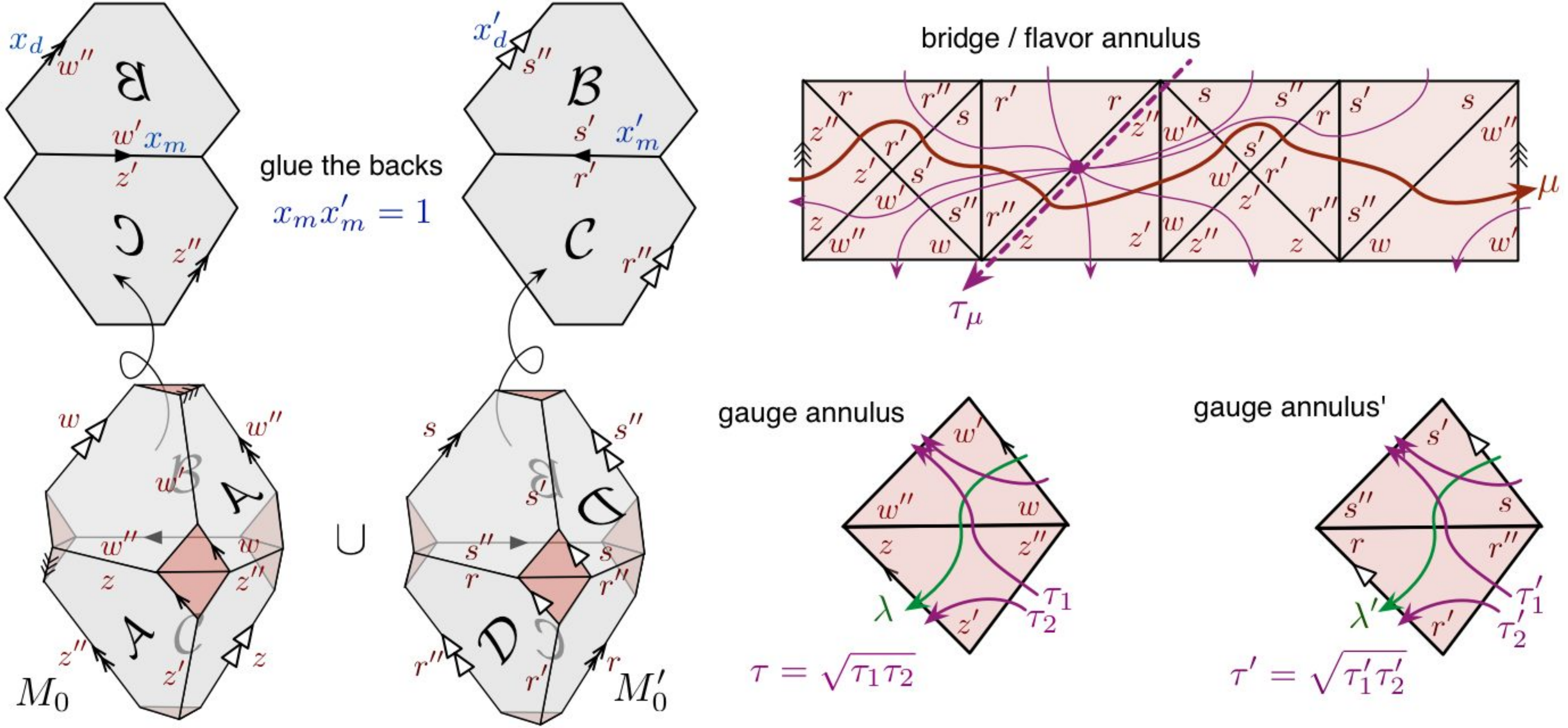}
\caption{A triangulation of $M_{\rm S}$ obtained by triangulating $M_0$ and $M_0'$.}
\label{fig:MS}
\end{figure}

By triangulating $M_0$ and $M_0'$ the standard way and then gluing them together, we obtain the triangulation of $M_{\rm S}$ shown in Figure \ref{fig:MS}, involving four tetrahedra. We can actually guess immediately what one Lagrangian description of $T_2[M_{\rm S},\mb t,\Pi_{\rm S}]$ should be. One internal edge is formed in the gluing, enforcing the constraint that the product of magnetic IR coordinates for $M_0$ and $M_0'$ is one:
\be c := x_mx_m' = 1\,.\ee
(Here $x_m,x_d,x_e$ and $x_m',x_d',x_e'$ will denote IR coordinates for the two basic RG manifolds, defined just as in Section \ref{sec:MRG}.) Recall that in its magnetic polarization, the theory $T_2[M_0,\mb t,\Pi_m]$ is $U(1)_e$ gauge theory at level $+1$, coupled to a doublet of chirals $q_\alpha$ (both with $U(1)_e$ charge $+1$); and this theory has a monopole operator $\CO_m$ transforming under the topological $U(1)_m$ flavor symmetry. Therefore, to form $T_2[M_{\rm S},\mb t,\Pi_{\rm S}]$ we simply take a product $T_2[M_0,\mb t,\Pi_m]\times T_2[M_0',\mb t',\Pi_m']$ and add a superpotential
\be W = \CO_m\CO_m' \ee
to break the diagonal of the flavor symmetry $U(1)_m\times U(1)_m'$ and enforce the gluing. Of course $SU(2)_\lambda\times SU(2)_\lambda'$ flavor symmetry, acting on doublets $q_\alpha$ and $q_\alpha'$, is inherited.
It turns out that to make the third flavor symmetry $U(1)_\mu$ manifest, we should also gauge the anti-diagonal of $U(1)_m\times U(1)_m'$, under which $\CO_m\CO_m'$ is invariant; then $U(1)_\mu$ is its associated topological symmetry.

To justify the claim about gauging the anti-diagonal of $U(1)_m\times U(1)_m'$, we should be a bit more careful about the relation between coordinates in $\CP_2(\pd M_{\rm S},\mb t)$ and the coordinates of $M_0,M_0'$, as well as individual tetrahedra. For the two independent ``gauge'' annuli of $M_0$ and $M_0'$ we have the usual relations, \cf\ \eqref{xzRG},
\be \label{ltrs}
\lambda^2 =\frac zw\,,\qquad \tau = \lambda\frac{z''}{w''}\,;\qquad
 \lambda'{}^2 = \frac rs\,,\qquad \tau' = \lambda'\frac{r''}{s''}\,.\ee 
The remaining twelve small triangles of our four tetrahedra come together to tile the ``bridge'' annulus of $M_{\rm S}$, as shown in the figure. We can find $\mu$ from a path around the A-cycle of this annulus,
\be \mu = -zws''r'' = -\frac{x_d'}{x_e^2} \simeq -\frac{x_e'{}^2}{x_d}\,, \label{murs} \ee
where the two expressions on the RHS are equivalent modulo the gluing constraint $c=x_mx_m' = z'w'r's'=1$. 
We could also easily obtain $\mu$ by noting that $\mu^{\pm1}$ must be the coordinates of the circular edges on the three-punctured spheres, labelled by $\blacktriangleright\!\blacktriangleright$ and $\rhd\!\rhd$ in the figures.
Finally, a canonical $\tau_\mu$ coordinate can be defined by first drawing a straight line%
\footnote{There is some freedom in choosing this line, but different choices simply rescales $\tau_\mu\to \mu^{\pm 1}\tau_\mu$, provided that we use the averaged $\tau_\mu$ as described below!} %
connecting the circular edge on one three-punctured sphere to the circular edge on the other (dashed in figure \ref{fig:MS}), and then averaging the path-coordinates for \emph{sixteen} paths evenly distributed to the left and right of this line. Using the geometric dictionary of Appendix \ref{app:twist}, this translates to an averaging of normalizations of flags along the bridge annulus. A short calculation produces
\be \tau_\mu = \sqrt{\frac{w''z''}{r''s''}}=\sqrt{\frac{x_d}{x_d'}}\,. \label{tmurs} \ee
It is the relation \eqref{murs}, involving both position $(x_m=-x_e^2/x_d)$ and momentum $(x_e$) coordinates of the $M_0$ manifolds, that tells us to gauge the anti-diagonal of $U(1)_m\times U(1)_m'$ in the product of $M_0$ theories and to identify $U(1)_\mu$ with the new topological symmetry.

\subsubsection{Bulk-boundary couplings}

Now that we have the full map between tetrahedron coordinates and the $M_{\rm S}$ boundary phase space, we can actually give a slightly simpler description of $T_2[M_{\rm S},\mb t,\Pi_{\rm S}]$. It corresponds roughly to gluing together two \emph{electric} theories $T_2[M_0,\mb t,\Pi_e]$, which had no dynamical gauge symmetry to begin with. Ultimately, it does not matter whether we glue together the magnetic theories or the electric theories, or individual tetrahedra: the \emph{only} difference in these procedures is how the $ISp(8,\Z)$ transformation to the final polarization $\Pi_{\rm S}$ is implemented --- how it is broken up into elementary generators. The very fact that the group $ISp(8,\Z)$ acts on 3d theories guarantees that after a flow to the IR the results will be the same.

The slightly simpler theory involves two doublets of chirals $q_\alpha$ and $q_\alpha'$ where both the $U(1)_e$ rotating $q_\alpha$ and the $U(1)_e'$ rotating $q_\alpha'$ are gauged, with a \emph{mixed} Chern-Simons coupling. In detail:
\be \label{defSelec}
T_2[M_{\rm S},\mb t,\Pi_{\rm S}]: \left\{\begin{array}{l}
\text{$U(1)_e\times U(1)_e'$ gauge theory coupled to 4 chirals $q_\alpha,q_\alpha'$\,,} \\
\text{with $SU(2)_\lambda\times SU(2)_\lambda'\times U(1)_\mu$ flavor symmetry and $U(1)_R$,} \\
\text{and a superpotential $W = \CO_c$}\\
\text{charges:}\quad \begin{array}{c|cc|ccc}
 & q_\alpha & q_\alpha' & \eta & \eta' & \CO_c\\\hline
e & 1 & 0 & 0 & -2 & 0 \\
e' & 0 & 1 & -2 & 0 & 0 \\\hline
\lambda & \square & - & - & - & - \\
\lambda' & - & \square  & - & - & - \\
\mu & 0 & 0  & -1 & 1 & 0 \\\hline
R & 0 & 0  & 1 & 1 & 2\end{array} \qquad
\text{CS levels:}\quad \begin{array}{c|cc|c|c}
 & e & e' & \mu & R \\\hline
 e & -1 & 2  & 1 & -1 \\
 e' & 2 & -1  & -1 & 1 \\\hline
 \mu & 1 & -1  & -1 & 0 \\\hline
 R & -1 & 1  & 0 & * \end{array}
\end{array}\right.
\ee
There are several useful things to notice about this theory. It has a gauge-invariant monopole operator $\CO_c$ associated to the $U(1)_A$ gauge group that is a diagonal of $U(1)_e\times U(1)_e'$, and this operator breaks a topological flavor symmetry that enforces the gluing of $M_{\rm S}$. The other topological symmetry, associated to the anti-diagonal of $U(1)_e\times U(1)_e'$ is our flavor symmetry $U(1)_\mu$.

The theory also has two monopole operators $\eta$ and $\eta'$ for $U(1)_e$ and $U(1)_e'$, respectively, which are not fully gauge invariant: $\eta$ is charged under $U(1)_e'$, and vice versa. This is actually ideal, because it allows us to create good bulk-boundary couplings! Let $\Phi_{\alpha\dot\alpha}=(X_{\alpha\dot\alpha},Y_{\alpha\dot \alpha})$ and $\Phi_{\alpha\dot\alpha}'=(X_{\alpha\dot\alpha}',Y_{\alpha\dot \alpha}')$ denote the adjoint hypermultiplets of two S-dual copies of $\CN=2^*$ theory. Then we may introduce $SU(2)_\lambda\times SU(2)_\lambda'\times U(1)_\mu$--invariant couplings
\be \label{Sbb} \begin{array}{c} W_{\rm bdy} = X^{\alpha\dot\alpha}\big|_\pd\, q_\alpha q_{\dot\alpha}\eta + X^{\alpha\dot\alpha}{}'\big|_\pd\, q_\alpha' q_{\dot\alpha}'\eta'\,;\\[.2cm]
Y_{\alpha\dot\alpha}\big|_\pd=q_\alpha q_{\dot\alpha}\eta\,,\qquad Y_{\alpha\dot\alpha}'\big|_\pd= q_\alpha' q_{\dot\alpha}'\eta'\,.\end{array} \ee
The ``dressed'' monopole operators $q_\alpha q_{\dot\alpha}\eta$ and $q_\alpha' q_{\dot\alpha}'\eta'$ are fully invariant under the 3d gauge group $U(1)_e\times U(1)_e'$, and are associated geometrically to the two three-punctured spheres in $\pd M_{\rm S}$, as per Section \ref{sec:spheres}. With boundary conditions \eqref{Sbb}, the 3d $U(1)_\mu$ flavor symmetry, identified with the Cartans of the bulk flavor symmetry in two $\CN=2^*$ theories, is promoted to $SU(2)_\mu$.
We may complete our construction of an S-duality domain wall by gauging $SU(2)_\lambda\times SU(2)_\lambda'$ in the bulk.

\subsection{Reproducing standard $T[SU(2)]$}
\label{sec:TSU22}

The theory $T_2[M_{\rm S},\mb t,\Pi_{\rm S}]$ of Section \ref{sec:TSU2} is a 3d mirror of (\ie\ is infrared dual to) the standard 3d S-wall theory known as $T[SU(2)]$. It has a beautiful feature: its $SU(2)_\lambda\times SU(2)_\lambda'$ flavor symmetry is classically manifest. In contrast, in the standard description of $T[SU(2)]$, only $SU(2)_\lambda\times U(1)_\lambda'$ is manifest; the $U(1)_\lambda'$, being a topological symmetry, gets promoted to $SU(2)_\lambda'$ by quantum corrections. Of course, the classical $SU(2)_\lambda\times SU(2)_\lambda'$ symmetry comes at a price: the superpotential of our theory in \eqref{defSelec} must contain a monopole operator.

The equivalence between $T_2[M_{\rm S},\mb t,\Pi_{\rm S}]$ above and standard $T[SU(2)]$ can be seen by applying a sequence of elementary dualities. These dualities correspond geometrically to tetrahedron rotations (replacing a $U(1)$ gauge theory with a charged chiral with a free chiral, or vice verse) and 2--3 moves (replacing an XYZ model with $N_f=1$ SQED). We could discuss the sequence of dualities at the level of Lagrangians, but it is much nicer to understand the duality geometrically. To this end, we will exhibit a different 3d triangulation of $M_{\rm S}$ that produces standard $T[SU(2)]$ theory (almost) directly. Since two labelled triangulations of a framed 3-manifold must be related by a sequence of 2--3 moves and tetrahedron rotations,%
\footnote{This is not quite a mathematical theorem since, since we require our triangulations to satisfy a few non-degeneracy constraints. In the presence of these constraints, a full result about relating 3d triangulations is not yet known. Nevertheless, in the present case it is very easy to find the sequence of non-degenerate 2--3 moves (and tetrahedron rotations) relating the different triangulations of $M_{\rm S}$; it boils down to 2--3 moves on the octahedron that were discussed in \cite[Sec 4.5]{DGG}.} %
this amounts to a proof that $T_2[M_{\rm S},\mb t,\Pi_{\rm S}]\simeq T[SU(2)]$.

\begin{figure}[htb]
\centering
\includegraphics[width=6in]{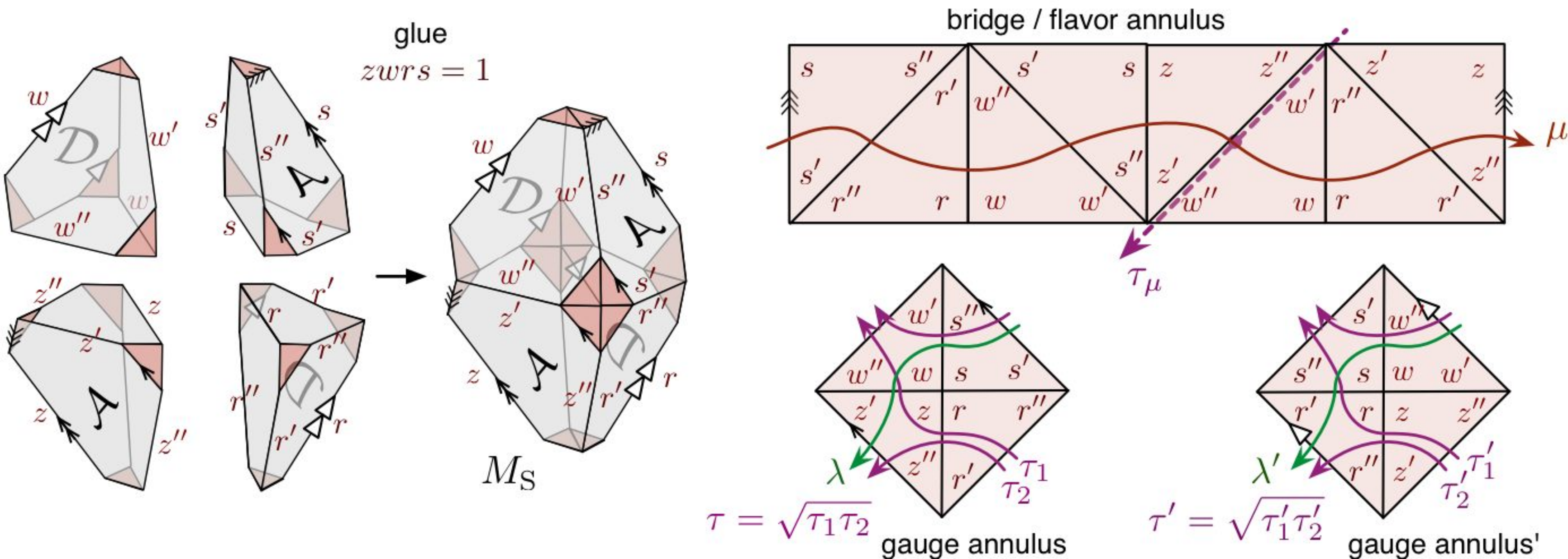}
\caption{Gluing four tetrahedra a different way to form $M_{\rm S}$.}
\label{fig:MS2}
\end{figure}

Consider the triangulation in Figure \ref{fig:MS2}. First an octahedron is constructed by gluing together four tetrahedra around a central edge. This is the same octahedron as the one glued together from $M_0$ and $M_0'$ in Figure \ref{fig:MS}, but with a different triangulation. Then two front faces (marked $\CA$) and two back faces (marked $\CD$) are glued to form the manifold $M_{\rm S}$.

We expect to find a theory of four chirals $\phi_z,\phi_w,\phi_r,\phi_s$, with a superpotential
\be W = \phi_z\phi_w\phi_r\phi_s \label{WTS0} \ee
enforcing the gluing. Moreover, some of the flavor symmetries that leave \eqref{WTS0} invariant should get gauged in the process of changing polarization to match the canonical $\Pi_{\rm S}$. To determine what gets gauged, we express the positions $\lambda,\lambda',\mu$ in terms of tetrahedron parameters using the standard rules, and find
\be \label{TSpos} \lambda^2 = \frac{s''z'}{w}\,,\qquad \lambda'{}^2=\frac{w''r'}{s}\,,\qquad \mu^2 = \frac{rw}{sz}\,. \ee
Similarly, we may compute the conjugate twist coordinates
\be \tau = \frac{\lambda}{\lambda'{}^2}\frac zs\,,\qquad \tau' = \frac{\lambda'}{\lambda^2}\frac{r}{w}\,,\qquad \tau_\mu^2 = \frac{r''w''}{s''z''}\,.\ee

The relation for $\mu^2$ to tetrahedron positions indicates that $U(1)_\mu$ is the symmetry that rotates the chirals with charges $(-\tfrac12,\tfrac12,\tfrac12,-\tfrac12)$. The other two $U(1)$ flavor symmetries get gauged, so that $U(1)_\lambda$, and $U(1)_\lambda'$ become  topological symmetries. Unraveling everything, we find a $U(1)_e\times U(1)_e'$ dynamical gauge theory with
\be \text{charges:}\;\begin{array}{c|cccc}
  & \phi_z & \phi_s & \phi_r & \phi_w \\\hline
e & 1 & -1 & 0 & 0 \\
e' & 0 & 0 & 1 & -1 \\
\mu & -\frac12 & -\frac12 & \frac12 & \frac12 \\\hline
R & \frac12 & \frac12 & \frac12 & \frac12\end{array}
 \qquad \text{CS level matrix:}\; \begin{array}{c|cc|ccc}
  & e & e' & \mu & \lambda & \lambda' \\\hline
 e & 0 & -1 & 0 & 2 & 0 \\
 e'& -1 & 0 & 0 & 0 & 2 \\\hline
 \mu &0 & 0 & -\tfrac{1}{2} & 0 & 0 \\
 \lambda &2 & 0 & 0 & 1 & -2 \\
 \lambda'&0 & 2 & 0 & -2 & 1 \end{array}
\ee
This is basically two copies of $N_f=1$ SQED, with a mixed Chern-Simons term coupling the $U(1)_e$ and $U(1)_e'$ photons. We also see from the Chern-Simons (\ie\ FI) couplings that the topological symmetries $U(1)_\lambda$ and $U(1)_\lambda'$ are directly associated to $U(1)_e$ and $U(1)_e'$. Evidently, the nonabelian enhancement to $SU(2)_\lambda\times SU(2)_\lambda'$ must be a quantum phenomenon.

To get from this theory --- which must be mirror to $T_2[M_{\rm S},\mb t,\Pi_{\rm S}]$ --- to standard $T[SU(2)]$ we perform one \emph{non-geometric} 2--3 move. We isolate the first copy of SQED, with gauge group $U(1)_e$, and dualize it to an XYZ model. This duality is non-geometric because it does not correspond to an actual 2--3 move performed on the tetrahedra of $M_{\rm S}$; nevertheless, it makes perfect sense in gauge theory.

In SQED$\,\to$XYZ duality \cite{AHISS}, the gauge-invariant meson $\phi_z\phi_s$ is replaced by a single fundamental chiral $M$. Moreover, the monopole and anti-monopole operators $\eta,\wt \eta$ of $U(1)_e$ become fundamental chirals. They have charges $\pm2$ under $U(1)_\lambda$ and $\mp 1$ under the remaining gauge group $U(1)_e'$ due to the mixed Chern-Simons terms. We find, then, a new superpotential
\be W \to \phi_rM\phi_w + \eta M\tilde \eta\,,\ee
where the second term is the standard XYZ contribution.
After shifting the $U(1)_\lambda$ flavor current by a multiple of the gauge current $U(1)_e'$, we kill the background Chern-Simons couplings for $U(1)_\lambda$ and obtain a gauge theory with charges
\be \begin{array}{c|ccccc}
 & \phi_r & \phi_w & \wt \eta & \eta & M \\\hline
 e' & 1 & -1 & 1 & -1 & 0 \\\hline
 \lambda & 1 & -1 & -1 & 1 & 0 \\
 \mu & \tfrac12 & \tfrac12 & \tfrac12 & \tfrac12 & -1\end{array}\;.
\ee 
This is the $\CN=2$ deformation of the $T[SU(2)]$ theory of \cite{GW-Sduality}. Now there is a classical enhancement $U(1)_\lambda\to SU(2)_\lambda$, while $U(1)_\lambda'$ remains a topological symmetry, with quantum enhancement.

\subsection{S-duality of line operators}

We can use the triangulation of the S-wall geometry $M_{\rm S}$ to reproduce the map between Wilson and 't Hooft line operators in $\CN=2^*$ theory and its S-dual. From the current perspective, this is actually the easiest way to identify how the 't Hooft operator should look in terms of $(\lambda,\tau)$.
Mathematically, we are finding the action of the mapping class group of the one-punctured torus on Fenchel-Nielsen coordinates.

The S transformation on line operators could be obtained directly by ``gluing together'' two copies of the RG-wall relations \eqref{N2WardRG}. This was done in \cite{DG-Sdual}. But it is actually easier to work directly with tetrahedra.

Following Appendix \ref{app:quant}, we first quantize and invert relations \eqref{ltrs}--\eqref{tmurs}, along with the gluing relation $c=z'w'r's'$ and its canonical conjugate $\gamma=1/\sqrt{r''s''}$:
\begin{align}
\hat z=iq^{\frac14}\sqrt{\hat\mu}\hat\lambda\hat\gamma\,,\qquad
\hat w = iq^{\frac14}\frac{\sqrt{\hat\mu}\hat\gamma}{\hat\lambda}\,,\qquad
\hat r = iq^{\frac14}\frac{\hat \lambda'}{\sqrt{\hat c\hat \mu}}\frac{\hat\gamma}{\hat \tau_\mu}\,,\qquad
\hat s = \frac{iq^{\frac14}}{\hat\lambda'\sqrt{\hat c\hat \mu}}\frac{\hat\gamma}{\hat\tau_\mu}\,,\quad \\
\hat z'' =\frac{1}{q^{\frac18}\sqrt{\hat\lambda}}\frac{\hat\tau_\mu\sqrt{\hat \tau}}{\hat\gamma}\,,\qquad
\hat w''= \frac{\sqrt{\hat\lambda}}{q^{\frac18}}\frac{\hat\tau_\mu}{\hat\gamma\sqrt{\hat\tau}}\,,\qquad
\hat r''=\frac{1}{q^{\frac18}\sqrt{\hat\lambda'}}\frac{\sqrt{\hat\tau'}}{\hat\gamma}\,,\qquad
\hat s''=\frac{\sqrt{\hat\lambda'}}{q^{\frac18}}\frac{1}{\hat\gamma\sqrt{\hat\tau'}}\,.\notag
\end{align}
We then take the left ideal generated by $\hat z''+\hat z^{-1}-1\simeq0$ (etc.) for each tetrahedron, rewrite it in terms of $\hat\lambda,\hat\tau,\hat\lambda',\hat \tau',\hat\mu,\hat\tau_\mu',\hat c,\hat \gamma$, eliminate $\hat\gamma$, and set $\hat c\to 1$. To simplify further, we eliminate two out of three of the twist operators $\hat\tau,\hat \tau',\hat \tau_\mu$, obtaining
\begin{subequations} \label{SWT}
\begin{align}
\hat \lambda'+\hat\lambda'{}^{-1} &\,\simeq\, \frac{\mathfrak s\big(\hat \lambda\sqrt{\hat\mu}/(iq^{\frac14})\big)}{\mathfrak s(\hat \lambda)}q^{-\frac18}\sqrt{\hat\lambda}\frac{1}{\sqrt{\hat\tau}}
  +\frac{\mathfrak s\big(iq^{\frac14}\hat \lambda/\sqrt{\hat\mu}\big)}{\mathfrak s(\hat \lambda)}q^{-\frac18}\frac{1}{\sqrt{\hat\lambda}}\sqrt{\hat\tau} \\
\hat \lambda+\hat\lambda{}^{-1} &\,\simeq\, \frac{\mathfrak s\big(\hat \lambda'/(iq^{\frac14}\sqrt{\hat\mu})\big)}{\mathfrak s(\hat \lambda')}q^{-\frac18}\sqrt{\hat\lambda'}\frac{1}{\sqrt{\hat\tau'}}
  +\frac{\mathfrak s\big(iq^{\frac14}\hat\lambda'\sqrt{\hat\mu}\big)}{\mathfrak s(\hat \lambda')}q^{-\frac18}\frac{1}{\sqrt{\hat\lambda'}}\sqrt{\hat\tau'} 
\end{align}
\end{subequations}
where
\be  \mathfrak s(x) := x-x^{-1}\,. \ee
The first relation sets the spin-1/2 Wilson loop on one side of the S-wall equal to the spin-1/2 't Hooft loop on the other, and the second does the opposite. They appear in \cite{DGOT} and (with slightly different normalization) in \cite{AGGTV}.

Note that the relations are invariant separately under $(\hat\lambda,\hat\tau)\mapsto (\hat\lambda^{-1},\hat\tau^{-1})$. This Weyl invariance reflects the enhanced symmetry $SU(2)_\lambda\times SU(2)_{\lambda'}$. On the other hand, Weyl invariance in $\hat\mu$ is (predictably) absent. To get from one equation to the other we send $(\hat \lambda,\hat\tau)\leftrightarrow(\hat\lambda',\hat\tau)$ along with $\hat\mu\leftrightarrow\hat\mu^{-1}$.

If we eliminate $\hat \tau,\hat \tau'$ from our left ideal but not $\hat \tau_\mu$, we get a third independent equation associated to ``bridge'' between the two loops of the Hopf network $M_{\rm S}$:
\begin{align} \label{Wmu}
 &\fs(q^{\frac14}\sqrt{\hat\mu})\bigg(\hat\lambda^2+\frac{1}{\hat\lambda^2}+\frac{\hat\mu}{q^{\frac12}}+\frac{q^{\frac12}}{\hat\mu} \bigg) \frac{q^{\frac14}}{\sqrt{\hat\mu}}\frac{1}{\hat\tau_\mu}
-  \fs(\hat\mu)\bigg(\hat\lambda+\frac{1}{\hat \lambda}\bigg)\bigg(\hat\lambda'+\frac1{\hat\lambda'}\bigg) \\
&\hspace{1.7in}
+ \fs(q^{-\frac14}\sqrt{\hat\mu})\bigg(\hat\lambda'{}^2+\frac{1}{\hat\lambda'{}^2}+q^{\frac12}\hat\mu+\frac{1}{q^{\frac12}\hat\mu} \bigg) q^{\frac14}\sqrt{\hat\mu}\hat\tau_\mu\;\simeq 0\,. \notag
\end{align}
This equation has the following physical interpretation. So far we have considered the 3d theory $T_2[M_{\rm S},\mb t,\Pi_{\rm S}]$. to be a duality wall between two copies of 4d $\CN=2^*$ theory. However, if we additionally gauge the $SU(2)_\mu$ flavor symmetry in the bulk (which is enhanced from $U(1)_\mu$ after the usual coupling to bulk hypers), we obtain a 3d boundary condition for the 4d theory of class $\CS$ associated to a closed genus-2 surface. This surface is the complete topological boundary $\pd M_{\rm S}$ of the S-manifold, so the 4d theory is $T_2[\pd M_{\rm S}]$, and the full boundary condition is the complete/canonical $\CB_2[M_{\rm S}]$ (\cf\ Section \ref{sec:spheres}). Equation \eqref{Wmu} is a Ward identity involving the dynamical (spin-1) 't Hooft operator of $SU(2)_\mu$, brought to the boundary.

The three equations \eqref{SWT}, \eqref{Wmu} annihilate the index and ellipsoid partition functions of $T_2[M_{\rm S},\mb t,\Pi_{\rm S}]$, \ie\ the index and ellipsoid partition functions of $T[SU(2)]$. These are well known expressions, having been obtained both via supersymmetric localization \cite{HLP-wall, HHL} and relations to Liouville theory and quantum groups (\cf\ \cite{Teschner-TeichMod, DGOT}). Now we can obtain the obtain partition functions as $SL(2)$ Chern-Simons wavefunctions of $M_{\rm S}$, using methods of \cite{Dimofte-QRS, DGG, DGG-index}.
For example, using the triangulation from Section \ref{sec:TSU2}, the ellipsoid partition function is written as
\begin{align} \CZ_\hbar[M_{\rm S},\mb t,\Pi_{\rm S}] &= \int dR dS\,e^{\frac{1}{\hbar}\big[\frac12(\Lambda^2+\Lambda'{}^2-M^2)+2RS+M(R-S)+(i\pi+\frac\hbar2)(R+S)\big]} \notag \\ & \hspace{1in}\times  \Phi_\hbar( - R\pm\Lambda)\Phi_\hbar(-S\pm \Lambda')\,. %\\
%&= 
\end{align}
It is equivalent to the more standard partition function of $T[SU(2)]$ from \cite{HLP-wall, HHL}.
Here $(\lambda,\lambda',\mu')=\exp(\Lambda,\Lambda',M)$, and the respective operators $\hat\tau,\hat\tau',\hat\tau_\mu$ act by shifting these three variables: $\hat\tau f(\Lambda)=f(\Lambda+\hbar)$, etc.

\section{Example 2: interfaces for $SU(2)$ $N_f=4$}
\label{sec:N4}

Our last set of examples includes the RG and duality walls for 4d $\CN=2$ $SU(2)$ theory with $N_f=4$ flavors of hypermultiplet matter.

The $SU(2)$ $N_f=4$ theory has a  rank-$1$ six-dimensional realization based on a four-punctured sphere $\CC$. 
This construction only manifests an $SU(2)_a \times SU(2)_b \times SU(2)_c \times SU(2)_d$ subgroup of the full $SO(8)$ flavor group of the four-dimensional theory. The full flavor symmetry is only recovered in the infrared. Correspondingly, generic boundary conditions and domain walls for $SU(2)$ $N_f=4$ built from three-manifolds will only preserve the $SU(2)_a \times SU(2)_b \times SU(2)_c \times SU(2)_d$ subgroup of the flavor symmetry. Special examples such as S-duality or RG walls, though, should preserve the full $SO(8)$ flavor symmetry once coupled to the bulk. This expected symmetry enhancement will provide be beautiful consistency checks throughout this section. 
Momentarily, we will discuss carefully the possible flavor symmetry enhancements of the various theories we will 
encounter in this section.

\begin{figure}[htb]
\centering
\includegraphics[width=5in]{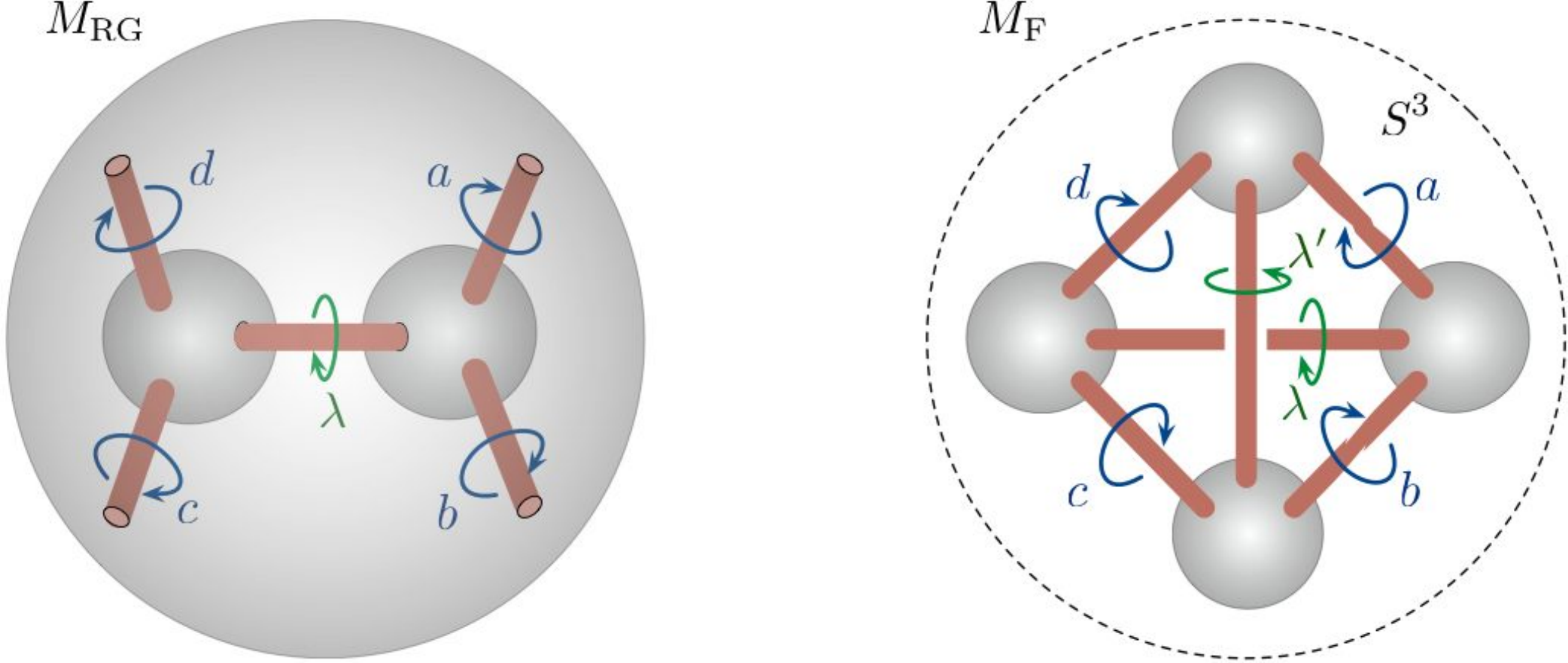}
\caption{The RG and S-duality manifolds for $SU(2)$ $N_f=4$ theory.}
\label{fig:RGF}
\end{figure}

In order to couple with a UV copy of $SU(2)$ $N_f=4$, a framed 3-manifold must include two three-punctured spheres, joined by a small annulus which produces the flavor symmetry $SU(2)_\lambda$ to be gauged in the bulk. The remaining punctures of the three-punctured spheres must be endpoints of annuli associated to appropriate $SU(2)$ flavor subgroups. 
We can readily describe the 3-manifolds $M_{\rm RG}$ and $M_{\rm F}$ which correspond to an RG domain wall and to an S-duality wall. 
The former must have a big boundary composed of a four-punctured sphere and two three-punctured spheres. We will have an $SU(2)_\lambda$ small annulus joining the three-punctured spheres and the four $SU(2)_{a,b,c,d}$ small annuli joining the three-punctured spheres and the four-punctured sphere. We can envision the four-punctured sphere as the boundary of a solid ball, with the three-punctured spheres and the annuli carved out of it (Figure \ref{fig:RGF}). The boundary phase space $\CP_2(\pd M_{\rm RG})$ is twelve-dimensional and it includes the holonomy and twist parameters for the five annuli and an extra pair of edge coordinates from the four-punctured sphere. 

In order to obtain the 3-manifold $M_{\rm F}$ for an S-duality wall, we can glue together two RG domain wall geometries, 
permuting the four punctures accordingly. The two balls glue into a solid three-sphere, with a network of annuli and three-punctured sphere carved out of it. The network has the topology of a tetrahedral 
graph, with a pair of opposite edges labeled by the to-be-gauged $SU(2)_\lambda$ and $SU(2)_{\lambda'}$ symmetries an other opposite pairs labelled by $SU(2)_a$ and $SU(2)_c$ or $SU(2)_b$ and $SU(2)_d$ respectively. The boundary phase space $\CP_2(\pd M_{\rm F})$ is again twelve-dimensional
and is parameterized by holonomy and twist coordinates for all six annuli in the network. We use the subscript `F' for this manifold because its $SL(2)$ Chern-Simons wavefunction coincides with the fusion kernel in Liouville theory.

Of course, the actual flavor symmetries of the corresponding three-dimensional theories will depend on the choices of triangulations of the big boundaries. In order to identify useful choices of triangulations, we should consider the full flavor symmetry of the problem.

\subsection{Enhanced flavor symmetry and SO(12) surprises}

\subsubsection{UV considerations}

The four $SU(2)$'s that are manifest in six dimensions combine pairwise into a block-diagonal $SO(4) \times SO(4)$ subgroup of the infrared $SO(8)$. Actually, all three possible pairwise combinations are OK: they correspond to $SO(4) \times SO(4)$ subgroups related by triality. 
The three basic representations of $SO(8)$ decompose in terms of doublet representations of the $SU(2)_a \times SU(2)_b \times SU(2)_c \times SU(2)_d$ flavor subgroup as
\begin{equation}
8_v = 2_a \otimes 2_b + 2_c \otimes 2_d \qquad 8_s = 2_a \otimes 2_d + 2_b \otimes 2_c \qquad 8_c = 2_a \otimes 2_c + 2_b \otimes 2_d 
\end{equation}
Correspondingly, the fundamental hypermultiplets in the standard S-duality frame decompose into two trinion blocks 
\begin{equation}
2_\lambda \otimes 8_v  =  2_\lambda \otimes 2_a \otimes 2_b + 2_\lambda \otimes 2_c \otimes 2_d
\end{equation}
The theory has two other S-duality frames where the fundamental hypers transform in $8_s$ and $8_c$, 
which correspond to the other two pair of pants decompositions of the four-punctured sphere.  

We can ask how to insure that a 3d theory associated to some three-manifold will give rise to a boundary condition or domain wall that preserves the full $SO(8)$ flavor symmetry.
If we consider a coupling to the UV description of the theory, the triangulations of three-punctured spheres (partially) determine which symmetries are present. 
At the very least, we can try to pick triangulations to preserve all flavor symmetry: 
$U(1)_\lambda \times SU(2)_a \times SU(2)_b \times SU(2)_c \times SU(2)_d$. This choice has the price that the $SU(2)_\lambda$
symmetry to be gauged only appears in the infrared, after coupling to the bulk hypers. One may then hope to find a description of the isolated 3d theory where the $SU(2)_a \times SU(2)_b \times SU(2)_c \times SU(2)_d$
is explicitly extended to $SO(8)$, in such a way that the operators used in the coupling to 4d hypers form a vector $8_v$ of $SO(8)$,
which is coupled to an $SO(8)$-invariant half of the bulk hypermultiplets. 

In our concrete examples, however, we will use a different mechanism to show the full $SO(8)$ invariance of the interface.
One can pick the three-punctured sphere triangulations to preserve an 
$SU(2)_\lambda \times SU(2)_a \times U(1)_b \times SU(2)_c \times U(1)_d$, \ie\ the $U(2) \times U(2)$ subgroup of $SO(4) \times SO(4)$. 
The splitting of the bulk hypermultiplets $Q_{\alpha \beta \ell}$ and $Q_{\ell \gamma \delta}$ into halves actually preserves more than 
$U(2) \times U(2)$: we can combine the halves into a fundamental of $U(4) \subset SO(8)$ and an anti-fundamental (both doublets of the gauge group).
It is then natural to look for 3d theories whose $U(2) \times U(2)$ symmetry group is correspondingly enhanced to a full $U(4)$,
in such a way that the operators we couple to the hypermultiplets form a fundamental of $SU(4)$ and the bulk-boundary superpotential coupling 
manifestly preserves $SU(4)$ as well. If that is the case, the standard promotion of $U(1)_b$ and $U(1)_d$ to $SU(2)_b$ and $SU(2)_d$ (upon coupling to the bulk) will insure that $U(4)$ is promoted to a full $SO(8)$

\subsubsection{IR considerations}

If we consider the coupling to an IR Seiberg-Witten description of the theory, we can make flavor symmetry enhancements manifest only if the 
spectrum of BPS particles is organized in representations of the desired flavor symmetry group.  
 In order to understand why this is a subtle requirement, observe that the BPS spectrum for generic values of the mass parameters has no reason to form full representations of $SO(8)$, as the mass parameters themselves break the $SO(8)$ flavor symmetry.
Making all mass parameters small (compared to the Coulomb branch parameter) makes the BPS spectrum very intricate and the association to WKB triangulations unpredictable.

A simple compromise is to give all four (UV) hypers 
the same large mass, which preserves $U(4)$. Then the BPS spectrum organizes itself in $U(4)$ representations. For example, the Coulomb branch has three singular loci. At one locus we find 4 light BPS particles in a fundamental representation of $SU(4)$, which are identified with 
fundamental hypers in the UV theory. The other two loci are basically the usual monopole and dyon points for the pure $SU(2)$ theory,
each with a light BPS particle that is a $SU(4)$ singlet. The BPS spectrum is then robust against adding some small $U(4)$ breaking mass, and we can easily find WKB triangulations 
whose edges carry appropriate $U(4)$ quantum numbers.

\begin{figure}[htb]
\centering
\includegraphics[width=5in]{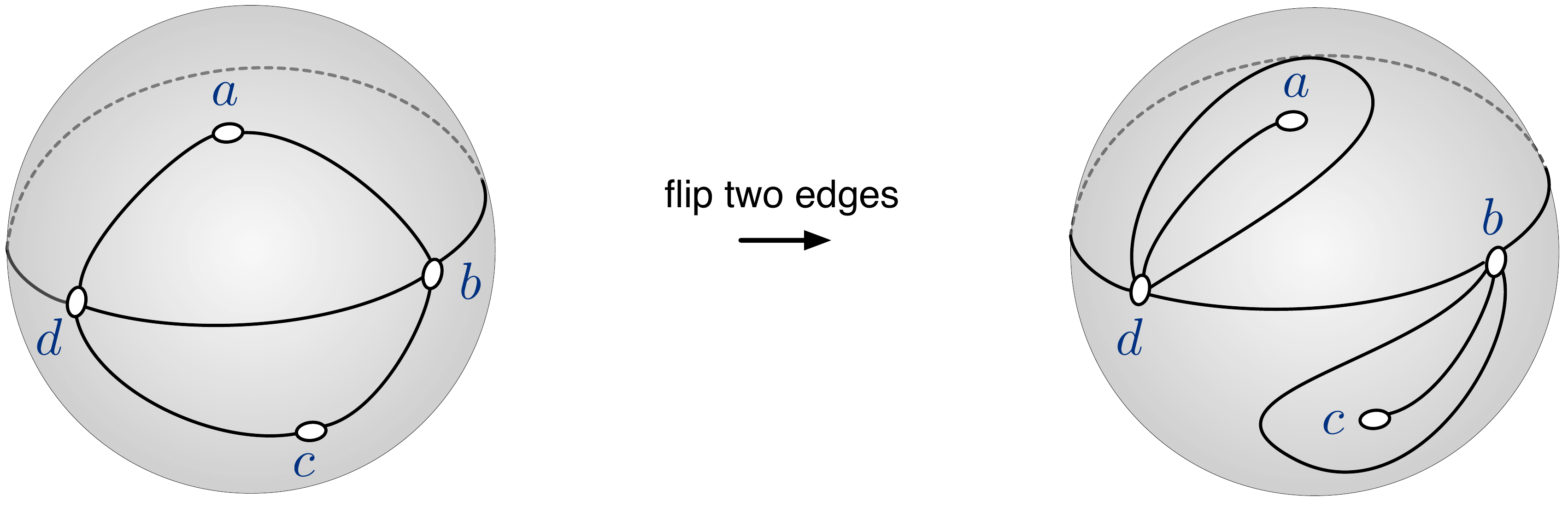}
\caption{A triangulation of the IR boundary of $M_{\rm RG}$ that can preserve $SU(2)_a\times SU(2)_c$.}
\label{fig:N4RG-big}
\end{figure}

It is useful to start from a ``square'' triangulation, where the four punctures are the vertices of a square. We add two diagonal edges between the $b$ and $d$ vertices on opposite sides of the sphere to complete the triangulation. Then we flip opposite edges of the square to get degenerate triangles around $a$ and $c$, so that $SU(2)_a$ and $SU(2)_b$ can (potentially) be preserved in a 3d theory (Figure \ref{fig:N4RG-big}).

The degenerate triangles will be associated to two doublets of 3d operators that have the correct quantum numbers
to couple to the 4d BPS particles in a fundamental of $SU(4)$. The operators associated to the diagonal edges do not carry $SU(4)$ charge. 
This is the triangulation of the four sphere that we will use to realize the RG domain wall. If we can present the three-dimensional theory 
and the couplings to bulk fields in a way that preserves $U(4)$ rather than just $SU(2)_a \times U(1)_b \times SU(2)_c \times U(1)_d$, we will recover $SO(8)$ in the IR. We will do so in the next section.

\subsubsection{Surprises}

In the case of the S-duality wall, say between the ``$8_v$'' and ``$8_s$'' duality frames, we can simply triangulate the three-spheres 
to preserve  $SU(2)_\lambda \times SU(2)_{\lambda'} \times SU(2)_a \times U(1)_b \times SU(2)_c \times U(1)_d$ (see Figure \ref{fig:N4F} below).
The corresponding theory has a chance to preserve $U(4)$ and thus give a domain wall with $SO(8)$ flavor symmetry. 
The possibility of $SO(8)$ enhancement upon coupling to the bulk hypers has a very entertaining consequence. 
Suppose that we ignore the fact that at some point we will want to gauge the $SU(2)_\lambda \times SU(2)_{\lambda'}$ flavor symmetries, 
and we simply use the three-dimensional theory to define a boundary condition for four trinion theories.
The tetrahedral graph has a high degree of symmetry, and the final boundary condition should treat democratically all vertices and all pairs of edges of the tetrahedron. 

Even before we couple the three-dimensional theory to the bulk hypers, the tetrahedral symmetry combines with symmetry enhancements 
in a neat way. Both the subgroups $ SU(2)_a \times U(1)_b \times SU(2)_c \times U(1)_d$
and $SU(2)_\lambda \times SU(2)_{\lambda'}  \times U(1)_b \times U(1)_d$ want to be promoted to $U(4)$s.
This is possible only if the overall symmetry of the 3d theory is enhanced to $SU(4) \times SU(4)$.

\begin{figure}[htb]
\centering
\includegraphics[width=5in]{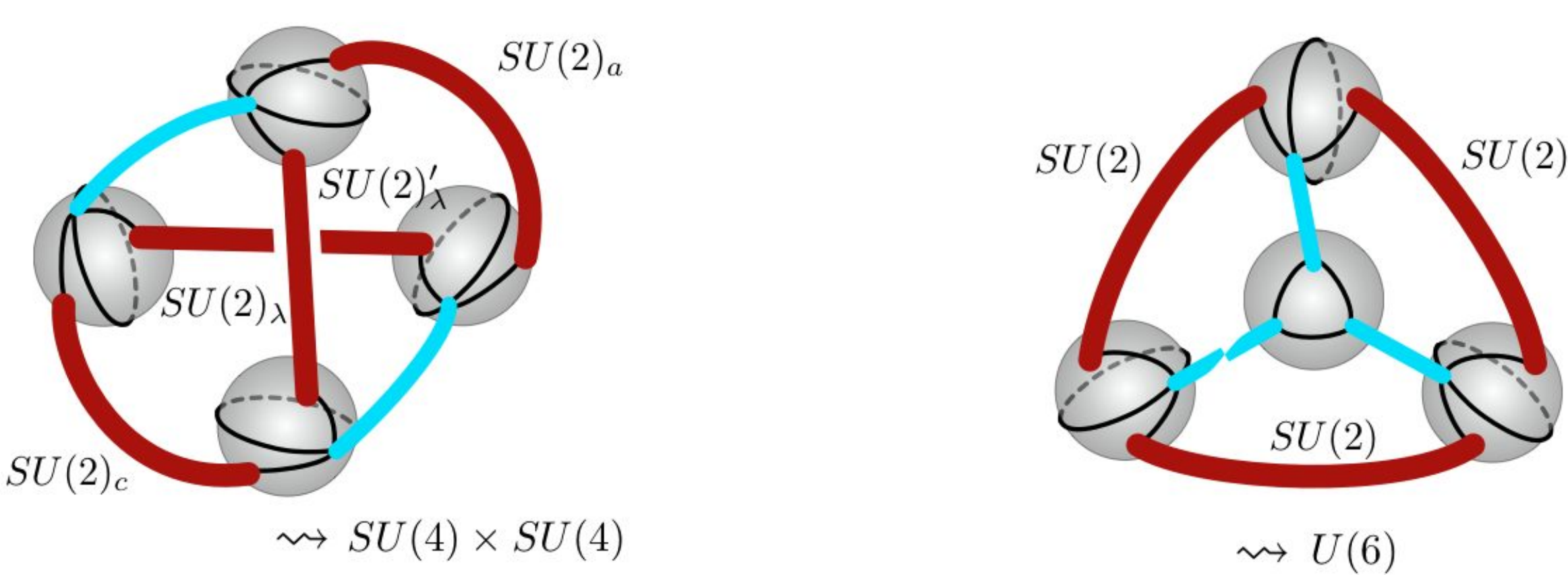}
\caption{Triangulations of 3-punctured spheres in the tetrahedral graph that should lead to 3d theories $T_2[M_{\rm F}]$ with $SU(4)\times SU(4)$ and $U(6)$ enhanced symmetry.}
\label{fig:FSU4}
\end{figure}

Once we couple to hypers, the flavor enhancement is striking. 
The $SU(2)$ flavor symmetries associated to any two pairs of opposite edges 
should belong to some $SO(8)$. We have three $SO(4)\simeq SU(2)\times SU(2)$ groups, and each pair of $SO(4)$'s must sit block-diagonal in an $SO(8)$ flavor group. 
This is possible, provided that the flavor group of the boundary condition is actually enhanced to $SO(12)$! Notice that the bulk theory of four trinions indeed has a $SO(12)$ flavor symmetry (subgroup of the full $USp(32)$): we can assemble them all into 
a (pseudo real, $32$ dimensional) chiral spinor of $SO(12)$. Indeed, such a spinor decomposes under $SO(4) \times SO(4) \times SO(4)$  
as 
\begin{equation}
(32) = 2_L \otimes 2_L \otimes 2_L + 2_L \otimes 2_R \otimes 2_R +  2_R \otimes 2_L \otimes 2_R +2_R \otimes 2_R \otimes 2_L\,.
\end{equation}
Notice that each $SU(2)$ group links two trinions, and the whole structure reproduces the tetrahedral network. 
The 3d geometry conjecturally represents a boundary condition for the theory of $32$ half-hypermultiplets which preserves this $SO(12)$ subgroup of their full $USp(32)$ flavor symmetry. This is a beautiful object, whose existence was proven in \cite{DG-E7}. 

The $SO(12)$ symmetry enhancement suggests a second interesting way to pick the triangulations of the three-punctured spheres. 
The first choice we mentioned should give a 3d theory with $SU(4) \times SU(4) \sim SO(6) \times SO(6)$ symmetry, 
block-diagonal inside $SO(12)$. A second choice is to pick the triangulations so to preserve the $SU(2)$'s at three edges surrounding a single face of the tetrahedron. These are the three right-chiral $SU(2)$'s in $SO(4)^3$. All these $SU(2)$'s can fit inside an $SU(6)$ subgroup of $SO(12)$. Furthermore, the spinor of $SO(12)$ decomposes as $(1) + (15) + (\ol{15}) + (1)$ under $SU(6)$. 
All the 4d half-hypers that receive a modified Dirichlet b.c. happen to sit inside one $(15)$ and one singlet of $SU(6)$.
Correspondingly, we expect the 3d SCFT to have an $SU(6)$ flavor symmetry in the IR.

It is interesting to observe that the authors of \cite{TV-6j} propose a candidate theory that should be related to 
this second option. The proposal is based on the identification between the ellipsoid partition function 
of the 3d theory and the fusion kernel in Liouville theory. 
Up to a prefactor%
\footnote{Modulo a typo in \cite{TV-6j}, this prefactor corresponds to flipping the triangulations of 3-punctured spheres that should preserve $SU(6)$ back to democratic ones, using the rules of Figure \ref{fig:S2qdl}. The actual fusion kernel in its most standard normalization corresponds to all democratic triangulations.}%
, the partition function coincides with the expression 
one would get from a 3d $SU(2)$ gauge theory with six quark flavors, which has explicit $U(6)$ flavor symmetry. 
Looking at the fugacities in the partition function, denoted there as $\alpha_i$, with $i=1,2,3,4,s,t$, 
such that the four vertices of the tetrahedron are attached to the defects with label 
$(1,2,s)$, $(3,4,s)$, $(2,3,t)$, $(1,4,t)$ and $(s,t)$, $(1,3)$, $(2,4)$ are opposite pairs,
one can see explicitly the $SU(2)_s \times SU(2)_2 \times SU(2)_3$ block-diagonal subgroups of $U(6)$,
as expected. The other fugacities give charge $1/2$ to four quarks, $-1/2$ to two of them, cyclically. 

The 3d $SU(2)$ gauge theory with six quark flavors has meson operators in the $(15)$ of the $SU(6)$ flavor symmetry. These 
chiral operators may well be coupled to the $(15)$ half-hypermultiplets in the bulk. A monopole operator of the 3d $SU(2)$ gauge theory
will only carry a charge under the $U(1)$ diagonal subgroup of $U(6)$. This means that it will have charge $1$ under all three 
$U(1)_t \times U(1)_1 \times U(1)_4$. This may couple to the remaining half-hypermultiplet in a singlet of $SU(6)$. 
The existence of the correct set of chiral operators is a strong suggestion that the theory proposed by the authors of \cite{TV-6j}
is truly a mirror description of the (abelian!) class ${\cal R}$ theory associated with our 3d geometry, 
in the $\Z_3$--symmetric boundary triangulation.

There is another well known formula for the fusion kernel in Liouville theory that inspires a 3d field theory interpretation. This is given, \eg\ in \cite[Eqn 2.17]{TV-6j}.
Translated directly, the formula suggests a 3d $N_f=4$ SQED, possibly with an extra superpotential coupling 
which breaks both the topological and axial flavor symmetries. This can be done, say, by adding the 
basic monopole and anti-monopole operators. Looking up the charge assignments of monopoles in 
\cite{AHISS} we verify that the axial fugacity has been fixed so that the monopoles are marginal. 

Notice that the residual flavor symmetry of the $N_f=4$ SQED is $SU(4)_L \times SU(4)_R$, i.e. $SO(6) \times SO(6)$, 
just as we expect from the first triangulation of 3-punctured spheres discussed above. 
Indeed, the ellipsoid partition function of the theory suggests that $SU(2)_s \times SU(2)_t$ is block diagonal inside $SU(4)_R$, 
i.e. an $SO(4)$ block in $SO(6)$, and the same for $SU(2)_1 \times SU(2)_3$.

Next, we provide the gory details of the triangulated manifolds and the corresponding abelian CSM descriptions of the 
RG and S-duality kernels.

\subsection{The RG wall theory}
\label{sec:N4RG}

\begin{figure}[htb]
%\centering
\hspace{-.2in}\includegraphics[width=6.4in]{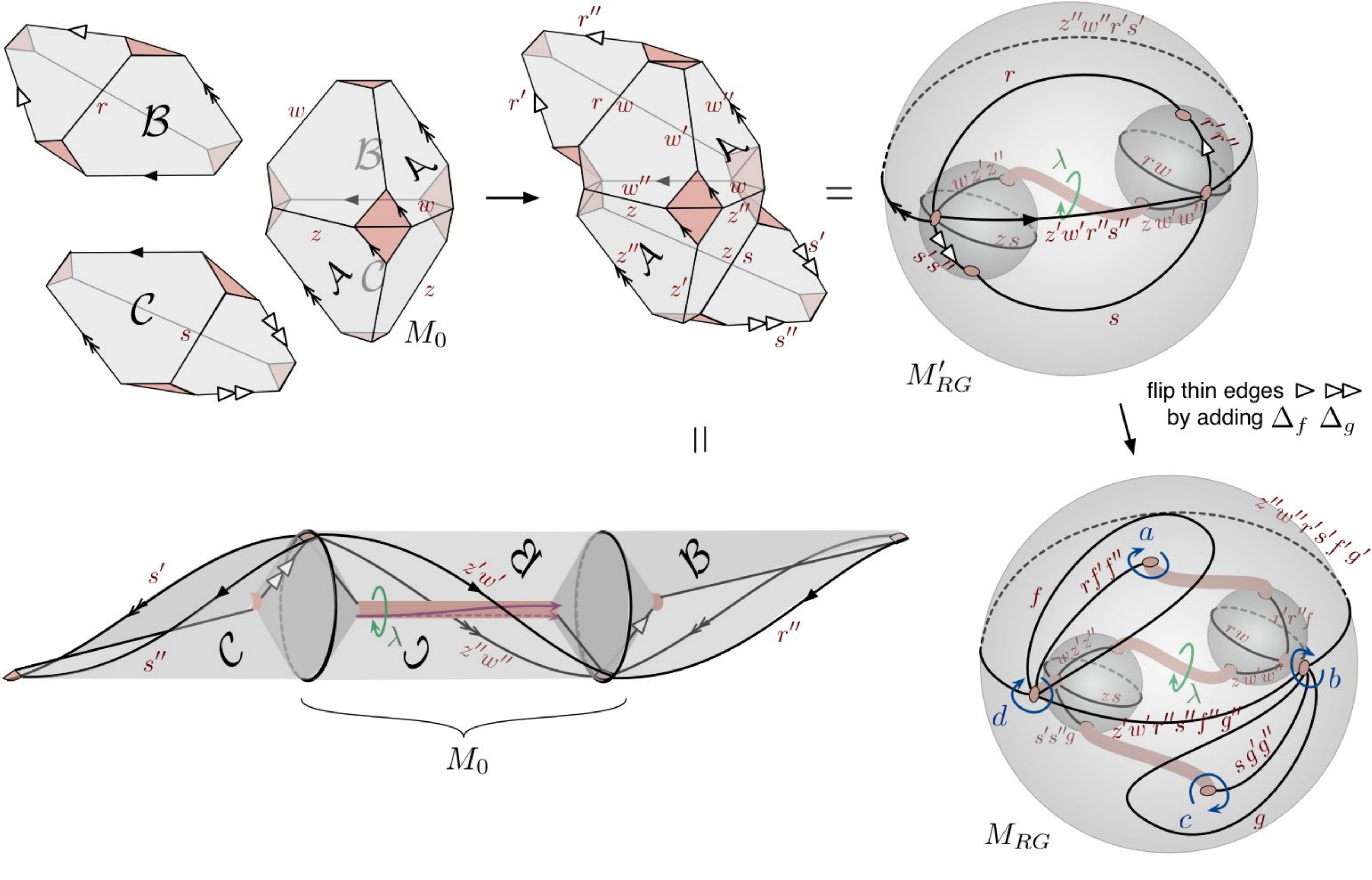}
\caption{Construction of the $N_f=4$ RG manifold $M_{\rm RG}$ from six tetrahedra, starting with a basic $M_0$ building block. The big-boundary edges of $M_{\rm RG}'$ and $M_{\rm RG}$ are labelled by the products of tetrahedron edge-parameters incident there.}
\label{fig:N4RG}
\end{figure}

We can triangulate the RG manifold $M_{\rm RG}$ with six tetrahedra in such a way that the induced big-boundary triangulation $\mb t$ is compatible with 
$SU(2)_\lambda \times SU(2)_a \times U(1)_b \times SU(2)_c \times U(1)_d$. We show the triangulation in Figure \ref{fig:N4RG}.
Although the triangulation looks complicated, and the careful determination of the various couplings will take a little work, the result 
will be amazingly simple, and can be described right away. Indeed, the 3d theory basically coincides with the RG 
theory for the pure $SU(2)$ gauge theory, together with four auxiliary chiral fields whose only purpose is to glue together 
at the boundary the four electrically charged BPS particles in the $SU(4)$ fundamental and the corresponding UV hypermultiplets,
much as we saw happen for the $N=2^*$ theory.

The triangulation is simply built upon the basic RG manifold $M_0$ for pure $SU(2)$ theory. Its two tetrahedra are assigned parameters $z$ and $w$. Two more tetrahedra (parameters $r,s$) are attached to the ends of $M_0$, initially identifying only edges and not faces, in such a way that two three-punctured spheres get trapped in the gluing. 
These extra tetrahedra can be seen as halves of the basic $M_0$ building block.  Subsequently, the half-$M_0$ tetrahedra are folded back onto the full $M_0$ core, making the face identifications labeled $\CB$ and $\CC$. This leads to a geometry $M_{\rm RG}'$ that is almost correct, but has two thin edges $\rhd$ and $\rhd\!\rhd$, just like the thin edge in the $\CN=2^*$ RG manifold (Figure \ref{fig:N2RGflip}). The thin edges are thickened out by adding two more tetrahedra (parameters $f$ and $g$), thus flipping the IR boundary triangulation precisely in the manner of Figure \ref{fig:N4RG-big}.

The triangulation has no internal edges and a brief inspection shows a natural polarization of the IR boundary such that no dynamical gauge fields are needed:
the theory consists of six free chiral fields!
The eigenvalues and dual twists are related to tetrahedron parameters as
\be \label{N4RGxz}
\lambda^2 = \frac zw\,,\qquad a^2 = \frac fr\,,\qquad (-b)^2 =zwrf\,,\qquad c^2 = \frac gs\,,\qquad (-d)^2=zwsg\,,\qquad x_e=\frac{1}{\sqrt{zw}}\,, \notag\ee
\be \hspace{-.15in}\tau = \lambda\frac{z''}{w''}\,,\quad\;\; \tau_a = a\frac{f''}{r''}\,,\quad\;\; \tau_b = bx_er''f''\,,\quad\;\; \tau_c=c\frac{g''}{s''}\,,\quad\;\; \tau_d=dx_es''g''\,,\quad\;\; x_m=bdz'w'r''s''f''g''\,,
\ee
where we emphasize that only the twists $\tau,\tau_a,\tau_c$ for the annuli that end in degenerate triangles (and have enhanced $SU(2)$ symmetry) are canonically defined. We have also chosen an electric coordinate $x_e$ to associate to the IR $U(1)$ gauge charge (well defined up to flavor charges), and a conjugate magnetic coordinate $x_m$. We let $\Pi_e$ be the polarization with positions $\lambda,a,b,c,d,x_e$ and the $\tau$'s and $x_m$ as momenta. The theory $T_2[M_{\rm RG},\mb t,\Pi_e]$ requires no gauging because all the position coordinates are made from tetrahedron positions.

The charge assignment for the six chiral fields $(\phi_z,\phi_w,\phi_r,\phi_s,\phi_f,\phi_g)$ is obtained by inverting the position relations in \eqref{N4RGxz}. It is most easily expressed by grouping the chirals into three doublets $q_\alpha=(\phi_z,\phi_w),\, q_\beta'=(\phi_f,\phi_r),\,q_\gamma''=(\phi_g,\phi_s)$. Then charges are
\be \label{N4RG-charges}
\begin{array}{c|ccc}
 & q_\alpha & q_\beta' & q_\gamma'' \\\hline
SU(2)_\lambda & \square & - & - \\
SU(2)_a & - & \square & - \\
U(1)_b & 0 & 1 & 0 \\
SU(2)_c & - & - & \square \\
U(1)_d & 0 & 0 & 1 \\ 
U(1)_x & -1 & 1 & 1 \\\hline
U(1)_R & 0 & 1 & 1 \end{array}
\ee
With momenta as above, there is also a level -1 background Chern-Simons term for the combination $A_x-A_R$ of abelian gauge fields; but there are no Chern-Simons terms for any of the other symmetries.
 
%%
%\be \begin{array}{c|cccccc}
% & \phi_z & \phi_w & \phi_f & \phi_r & \phi_g & \phi_s \\\hline
%\lambda & 1 & -1 & 0 & 0 & 0 & 0 \\
%a & 0 & 0 & 1 & -1 & 0 & 0 \\
%b & 0 & 0 & 1 & 1 & 0 & 0 \\
%c & 0 & 0 & 0 & 0 & 1 & -1 \\
%d & 0 & 0 & 0 & 0 & 1 & 1 \\
%x_e & -1 & -1 & 1 & 1 & 1 & 1 \\\hline
%R & 0&0&1&1&1&1\,. \end{array} \ee
%
We see manifestly the pure $SU(2)$ domain wall theory built from the first pair of chirals $q_\alpha$, together with a fundamental representation $u_A=(q_\beta',q_\gamma'')$ of a full $U(4)$ flavor group, of opposite $U(1)_x$ abelian gauge charge. Then the eight bilinear operators $q_\alpha u_A$ are associated to the three-punctured spheres
and couple neatly to the UV hypermultiplets. Moreover, for the four-punctured sphere on the IR boundary, the four operators $u_A$ are associated to the degenerate triangles, while the two long diagonal edges are associated to appropriate operators in a monopole background. 

If we focus on the electric superpotential couplings, we will find a situation analogous to the $\CN=2^*$ case:
\be \label{Snf} W_{\rm bdy} = X^{\alpha A}\big|_\pd\, q_\alpha u_A + X_e^{A}\big|_{\pd}\, u_A\, \ee
which enforces the notion that the quadruplet of light abelian particles $X_e^A$ coincide with the projection of the 
UV hypers along the direction $q_\alpha$ of the IR abelian gauge field inside the UV $SU(2)$ gauge field.

\subsection{The S-duality wall}
\label{sec:N4F}

\begin{figure}[htb]
\centering
\includegraphics[width=6in]{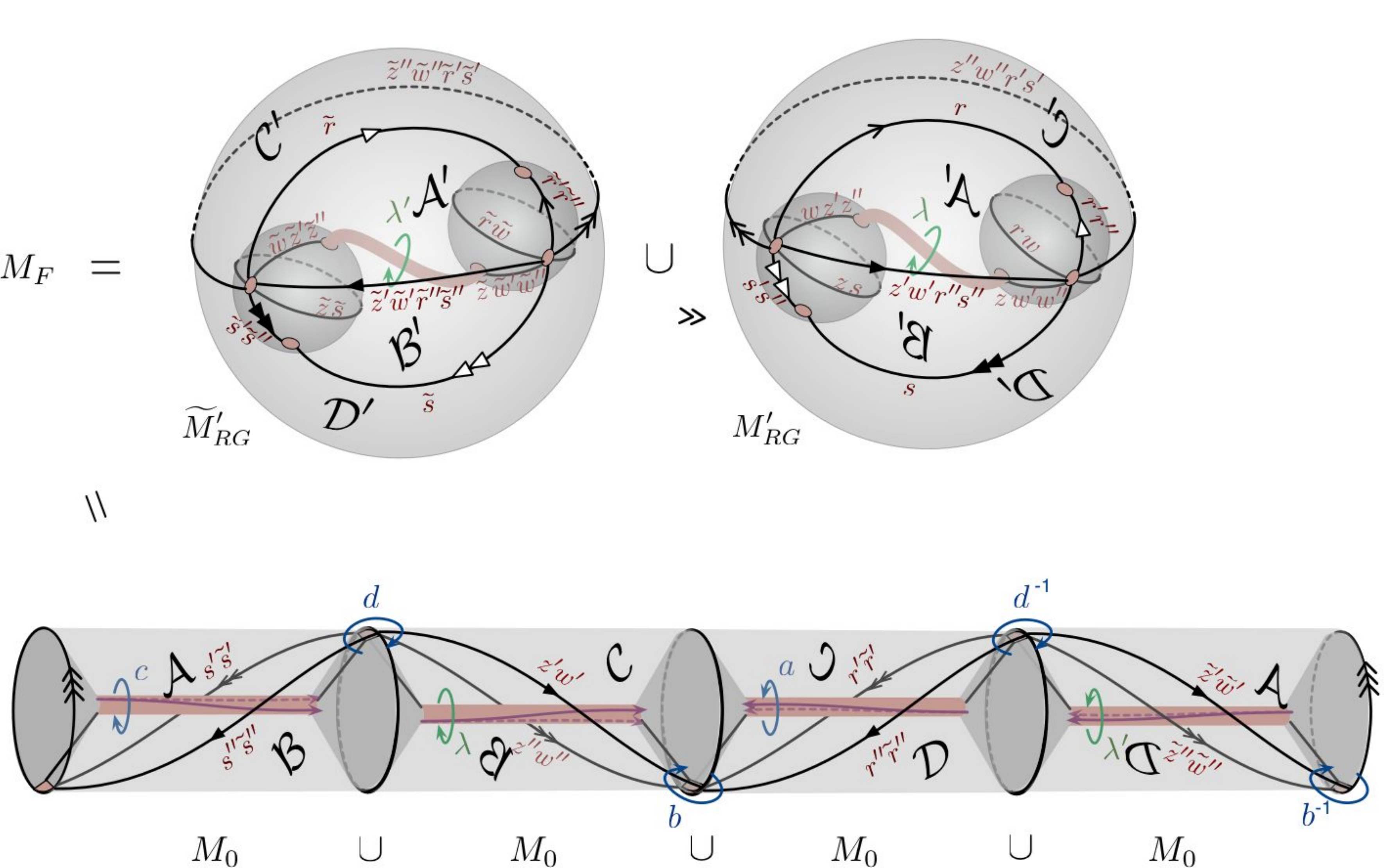}
\caption{}
\label{fig:N4F}
\end{figure}

We can triangulate the geometry $M_{\rm F}$ in order to reproduce 3d $N_f=4$ SQED on the nose, with the monopole and anti-monopole superpotentials 
arising from two internal edges of the triangulation. Notice that as all chirals appear in doublets of a flavor symmetry, 
the geometry should somehow be built out of four copies of the basic RG manifold $M_0$. It takes a certain amount of geometric imagination to 
guess the construction. Luckily, the physics does the job for us: the gluing constraints must take the form 
$\prod_{i=1}^4 z'_{i,+}z''_{i,-}=1$ and $\prod_{i=1}^4 z''_{i,+}z'_{i,-}=1$ in order to correspond to the monopoles of SQED. 
This allows us to get the correct geometry, depicted in Figure \ref{fig:N4F}.

To build $M_{\rm F}$, we place four copies of $M_0$ end-to-end in a loop, trapping four three-punctured spheres at their interfaces. The four annuli in these $M_0$'s carry the $SU(2)_\lambda\times SU(2)_a\times SU(2)_c\times SU(2)_{\lambda'}$ symmetries that we want to enhance. Then we fold this loop over on itself (making face identifications $\CA,\CB,\CC,\CD$) to obtain the complement of the tetrahedral graph in $S^3$. Alternatively, the gluing of $M_{\rm F}$ can be obtained by taking two copies of the RG manifold $M_{\rm RG}'$, before its edges were thickened, and identifying their IR boundaries, as at the top of Figure \ref{fig:N4F}. We'll give the four tetrahedra in one copy of $M_{\rm RG}$ parameters $z,w,r,s$ and in the other copy $\tilde z,\tilde w,\tilde r,\tilde s$.

With a little work, we find the relations between eigenvalue/twist coordinates and tetrahedron parameters
\be \label{N4F-xz}
 \lambda^2 = \frac{z}{w}\,,\qquad a^2 = \frac{\tilde r}{r}\,,\qquad (-b)^2 = zwr\tilde r\,,\qquad c^2=\frac{\tilde s}{s}\,,\qquad (-d)^2=zws\tilde s\,,\qquad \lambda'{}^2=\frac{\tilde z}{\tilde w}\,; \notag\ee
\be \tau=\lambda\frac{z''}{w''}\,,\quad\; \tau_a=a\frac{\tilde r''}{r''}\,,\quad\;
\tau_b^2 = \frac{z''w''s'\tilde s'}{z'w's''\tilde s''}\,,\quad\; \tau_c=c\frac{\tilde s''}{s''}\,,\quad\; \tau_d^2=\frac{z''w''r'\tilde r'}{z'w'r''\tilde r''}\,,\quad\; \tau'=\lambda'\frac{\tilde z''}{\tilde w''}\,.\ee
Now all of the twists are canonically defined, because all annuli begin and end at three-punctured spheres. Moreover, the gluing constraints $c_1$ and $c_2$, coming from internal edges labelled $\blacktriangleright$ and $>\!\!>$\hspace{-.17in}--- in the figure, take the promised form
\be \label{N4F-c}
 c_1 = z'\tilde z'w'\tilde w'r''\tilde r''s''\tilde s''\,,\qquad c_2= z''\tilde z''w''\tilde w''r'\tilde r's'\tilde s'\,.\ee

With a polarization $\Pi_{\rm F}$ in which all the eigenvalues are position coordinates, we use the relations \eqref{N4F-xz}-\eqref{N4F-c} to find that $T_2[M_{\rm F},\mb t,\mb \Pi]$ is a copy of $\CN_f=4$ SQED with eight chirals $u_\alpha=(\phi_z,\phi_w),\,u_\beta'=(\phi_{\tilde z},\phi_{\tilde w}),\,v_\gamma=(\phi_{\tilde r},\phi_r),\, v_\delta'=(\phi_{\tilde s},\phi_s)$ of charges
\be \begin{array}{c|cccc}
 & u_\alpha & u_\beta' & v_\gamma & v_\delta' \\\hline
SU(2)_\lambda & \square & - & - & - \\
SU(2)_{\lambda'} & - & \square & - & - \\
SU(2)_a & - & - & \square & - \\
SU(2)_c & - & - & - & \square\\
U(1)_b & \frac12 & -\frac12 & \frac12 & -\frac12 \\
U(1)_d & \frac12 & -\frac12 & -\frac12 & \frac12 \\\hline
U(1)_{\rm gauge} & 1 & 1 & -1 & -1 \\\hline
U(1)_R & 1&1&1&1 \end{array}
\ee
The monopole and anti-monopole in the superpotential break the axial and topological symmetries associated to $c_1$ and $c_2$, and there are no background Chern-Simons terms of any kind.

The couplings of the bulk hypermultiplets to 3d $N_f=4$ SQED are readily described. The natural gauge-invariant chiral bilinears, 
or mesons, carry an index of both $SU(4)$ fundamental representations, \ie\ a spinor index of both $SO(6)$'s inside the 
$SO(12)$ symmetry that is expected to emerge after the coupling. This is half of the $SO(12)$ spinor representation of the bulk operators, 
and thus we have a natural coupling, which can be reproduced from the geometry. Notice that in the tetrahedral graph each of the quartets of bilinears $u_\alpha v_\gamma,\, u_\alpha v_\delta',\, u_\beta'v_\gamma,\,u_\beta'v_\delta'$ is associated to one of the three-punctured spheres.

\subsubsection{Line operators}

We may compute the Ward identities for line operators at the S-duality domain wall using a logarithmic version of relations \eqref{N4F-xz}-\eqref{N4F-c} and the standard elimination procedure. The classical result, at $q=1$, sets the Wilson loop $\lambda+\lambda^{-1}$ on one side equal to the 't Hooft loop on the other,
\begin{align} \label{opsF} \lambda+\lambda^{-1}\,=\,& \frac{1}{\fs(\lambda')^2}\Big[-\big(\fc(c)+\fc(\lambda'/b)\big)\big(\fc(a)+\fc(\lambda'/d)\big)\frac{1}{\tau'} \notag\\
 & \hspace{.7in}+ \fc(\lambda')\big(\fc(a)\fc(c)+\fc(b)\fc(d)\big)+2\big(\fc(a)\fc(b)+\fc(c)\fc(d)\big) \\
 & \hspace{.7in} -\big(\fc(c)+\fc(b\lambda')\big)\big(\fc(a)+\fc(d\lambda')\big)\tau'\Big]\,,\notag
\end{align}
where $\fs(x)=x-x^{-1}$ and $\fc(x)=x+x^{-1}$. In Teichm\"uller theory, this encodes the action of the nontrivial generator of the mapping class group of a 4-punctured sphere in Fenchel-Nielsen-like coordinates.

The fully quantized operator identity appeared in \cite[Sec 5.4]{DGOT}, and is equivalent to \eqref{opsF} in the classical limit after flipping the three-punctured spheres to democratic triangulations. Using the dictionary from Figure \ref{fig:S2qdl} and being more careful about Chern-Simons prefactors, one sees that the flips have the effect of rescaling
\be \tau \mapsto \frac{\fc\big(\sqrt{d\lambda/c}\big)}{\fc\big(\sqrt{b/(a\lambda)}\big)}\tau\,,\qquad \tau' \mapsto \frac{\fc\big(\sqrt{c\lambda'/b}\big)}{\fc\big(\sqrt{a/(d\lambda')}\big)}\tau' \ee
inside classical operator relations.

\subsection{The 6j symbol and volumes of non-ideal tetrahedra}
\label{sec:6j}

The ellipsoid partition function of the S-wall theory $T_2[M_{\rm F}, \mb t,\Pi]$ can readily be computed as%
\footnote{The quantum dilogarithms $\Phi_\hbar$ used here are equivalent to the ``double-sine'' functions that one encounters in work on supersymmetric localization, Liouville theory, etc., as reviewed in Appendix \ref{app:qdl}.}
\begin{align} \label{ZF}
 \CZ_\hbar(\Lambda,A,B,C,D,\Lambda') \;&=\; \int_\R du\, e^{\frac1{2\hbar}(u^2+\Lambda^2+\Lambda'{}^2+A^2+B^2+C^2+D^2+(i\pi+\frac\hbar2)^2)} \notag \\
  &\hspace{-1.2in}\times \Phi_\hbar(-u\pm \Lambda+\tfrac12(B+D))\,\Phi_\hbar(-u\pm \Lambda'-\tfrac12(B+D))\,\Phi_\hbar(u\pm A+\tfrac12(B-D))\,\Phi_\hbar(u\pm C-\tfrac12(B-D))\,,
\end{align}
where as usual the upper-case variables are logarithms of the eigenvalues, $\lambda=\exp(\Lambda)$, etc.
Alternatively, this is an $SL(2)$ Chern-Simons wavefunction for the manifold $M_F$, obtained through quantization and symplectic gluing of tetrahedra.
In terms of Chern-Simons theory, two observations from \cite{TV-6j} about the function \eqref{ZF} acquire a beautiful geometric interpretation.

The first observation is that \eqref{ZF} coincides with a formula for the  6j symbols of the quantum group $U_q(sl_2)$, first calculated in \cite{PonsotTeschner}.
Specifically, the 6j symbols in (2.17) of \cite{TV-6j} are obtained from \eqref{ZF} by relating $(\Lambda,A,B,C,D,\Lambda')\leftrightarrow (\alpha_s,-\alpha_1,-\alpha_2,-\alpha_3,-\alpha_4,\alpha_t)$ and flipping the triangulation of 3-punctured spheres to the democratic one. Following Figure \ref{fig:S2qdl}, the flips multiply the partition function by
\be \Phi_\hbar(B-\Lambda-A)\,\Phi_\hbar(C+\Lambda'-B)\,\Phi_\hbar(D+\Lambda-C)\,\Phi_\hbar(A-D-\Lambda')\,,\ee
modulo some quadratic exponentials.

One may compare the relation to 6j symbols with a familiar situation in the compact case. The 6j symbols for $U_q(su(2))$ can be obtained as an $SU(2)$ Chern-Simons partition function of a tetrahedral network of Wilson lines ``colored'' by representations of $SU(2)$ that extend to representations of $U_q(su(2))$ \cite{Witten-6j, Witten-vertex, Turaev-1994}. This is sometimes called a quantum spin network.
The tetrahedral network precisely encodes how the tensor product of representations of $U_q(su(2))$ can be decomposed in two different ways. Now we are encountering a noncompact analogue of this relation. The $SL(2)$ Chern-Simons wavefunction of a tetrahedral network is equivalent to the partition function of our manifold $M_{\rm F}$. We are simply replacing the ``colors'' on edges of the tetrahedron with boundary conditions for Chern-Simons theory at the tubes of $M_{\rm F}$, in the form of fixed holonomy eigenvalues $\lambda,a,b,c,d,\lambda'$. Correspondingly, we fully expect to recover the 6j symbols for $U_q(sl(2))$, where the holonomy eigenvalues are weights of principal-series representations.

We may also understand the Chern-Simons partition function of a tetrahedral network in terms of conformal field theory. In the compact case, the $SU(2)$ partition function would reproduce the Moore-Seiberg fusion kernel for the $SU(2)$ WZW model. Now, in the $SL(2)$ case, with our particular Chern-Simons boundary conditions, we would expect to get the fusion kernel for Liouville theory. But, due to work of \cite{Tesch-Liouv}, this is equivalent to the 6j symbol of $U_q(sl_2)$.

The second observation of \cite{TV-6j} is that in the classical $\hbar\to0$ limit, the leading asymptotics of \eqref{ZF} reproduce the hyperbolic volume of a \emph{non-ideal} tetrahedron.
This can easily be understood from Chern-Simons theory by noting that the semi-classical limit of a partition function on $M_{\rm F}$ must reproduce the volume of a flat $SL(2)$ connection, \ie\ the (complexified) hyperbolic volume of $M_{\rm F}$. The boundary conditions at the six small annuli of $M_{\rm F}$ are such that a hyperbolic metric would have deformed cusps, with deficit angles equal to the argument of the eigenvalues $\lambda,a,b,c,d,\lambda'$. We can then cut $M_{\rm F}$ symmetrically into two halves along geodesic surfaces lying on the \emph{faces} of the tetrahedral network, obtaining two non-ideal hyperbolic tetrahedra $\Delta,\bar\Delta$ of opposite orientation. The dihedral angles of each tetrahedron are half the arguments of $\lambda,a,b,c,d,\lambda'$. Then for $\hbar\in i\R_{>0}$ we expect
\be \big|\CZ_\hbar(\Lambda,A,B,C,D,\Lambda')\big|\overset{\hbar\to 0}{\sim} e^{-\frac1{|\hbar|}{\rm Vol}(M_{\rm F})}\,,\qquad \text{Vol}(M_{\rm F}) = \text{Vol}(\Delta)+ \text{Vol}(\ol\Delta)=2\text{Vol}(\Delta)\,,\ee
where the absolute value isolates the real hyperbolic volume of the geometries $M_{\rm F}$ and $\Delta$.

%%%%%%%%%%%%%%%%%%%%%%%%%%%%%%%%%%%%%%%%%%%%%%%%%%%%%%%%%%%%%%%%%%%%%%%%%%
%%%%%%%%%%%%%%%%%%%%%%%%%%%%%%%%%%%%%%%%%%%%%%%%%%%%%%%%%%%%%%%%%%%%%%%%%%
%%%%%%%%%%%%%%%%%%%%%%%%%%%%%%%%%%%%%%%%%%%%%%%%%%%%%%%%%%%%%%%%%%%%%%%%%%

\section*{Acknowledgements}

We would like to thank Christopher Beem, Clay Cordova, Alexander Goncharov, Sergei Gukov, Lotte Hollands, Gregory Moore, Andrew Neitzke, and Edward Witten for enlightening discussions during the course of this project.
TD would like to thank the U.C. Berkeley mathematics and physics departments and the Perimeter Institute for Theoretical Physics for generously hosting his visits there in 2012--2013 to collaborate on this project.
RV would similarly like to thank the Institute for Advanced study for hosting him in March 2012.
The research of TD was supported in part by a William D. Loughlin membership at the Institute for Advanced Study and in part by the Friends of the Institute for Advanced Study, with additional support from DOE grant DE-FG02-90ER-40542.
The research of DG was supported in part by the Roger Dashen membership at the Institute for Advanced Study, and by the Perimeter Institute for Theoretical Physics. Research at Perimeter Institute is supported by the Government of Canada through Industry Canada and by the Province of Ontario through the Ministry of Economic Development and Innovation.
The research of RV was supported by the Netherlands Organisation for Scientific Research.

%%%%%%%%%%%%%%%%%%%%%%%%%%%%%%%%%%%%%%%%%%%%%%%%%%%%%%%%%%%%%%%%%%%%%%%%%%
%%%%%%%%%%%%%%%%%%%%%%%%%%%%%%%%%%%%%%%%%%%%%%%%%%%%%%%%%%%%%%%%%%%%%%%%%%
%%%%%%%%%%%%%%%%%%%%%%%%%%%%%%%%%%%%%%%%%%%%%%%%%%%%%%%%%%%%%%%%%%%%%%%%%%

\appendix

\section{Coordinates from hybrid triangulations}
\label{app:coords}

In this appendix, we discuss the construction of coordinates for spaces of framed flat connections on framed 3-manifolds and their boundaries. We follow the framework of \cite{FG-Teich} in two dimensions and \cite{DGG-Kdec} in three dimensions.

Recall that a framed, oriented 3-manifold $M$ (as defined in Section \ref{sec:6ddict} or \cite{DGG-Kdec}) has two types of boundary components, ``small'' and ``big.'' The big boundary consists of surfaces of any genus, at least one hole/puncture, and with negative Euler characteristic, and can be given a 2d ideal triangulation $\mb t$. The small boundary consists of tori, annuli, or discs, which connect to the holes on the big boundary.

Also recall that a framed flat connection on $M$ is a standard flat connection together with a choice of flat section of an associated flag bundle at every small boundary. This means a \emph{global} flat section at each small boundary --- \ie\ a flag that's invariant under the boundary holonomy. Similarly, a framed flat connection on $\pd M$ is (by convention) a standard flat connection on $\pd M^* :=(\pd M$ with small discs removed) with unipotent holonomy around every excised disc, together with a choice of flat section of the flag bundle along every remaining small boundary component and every $S^1$ boundary of an excised disc.

Our first goal will be to consider the algebraic moduli space $\CP_2(\pd M)$ of framed flat $PGL(2,\C)$ connections on the boundary $\pd M$, and to show that Zariski-open patches $\CP_2(\pd M,\mb t)$ (which depend on a big-boundary triangulation $\mb t$) can be given cluster $\C^*$ coordinates. In fact, there is a factorization
\be \label{Pfactors} \CP_2(\pd M,\mb t) \simeq (\C^*)^{2d}\times(\C^*)^{2a}\times(\C^*)^{2t}\,.  \ee
Here the $(\C^*)^{2d}$ factor is the standard $\CX$-coordinate space of Fock and Goncharov \cite{FG-Teich} for a big boundary with an ideal triangulation $\mb t$, where
\be d=3g-3+2h\,,\qquad g=\text{genus(big bdy)}\,,\quad h=\text{\# holes(big bdy)}\,;\ee
the middle factor consists of pairs of coordinates for each of the $a$ small annuli of $\pd M$, which are a novel generalization of Fenchel-Nielsen coordinates \cite{Wolpert-deformation, Wolpert-symplectic}; and the final factor contains pairs of coordinates for each of the $t$ small tori.

We also describe a natural holomorphic symplectic form $\Omega$ on $\CP_2(\pd M,\mb t)$, which on every patch $\CP_2(\pd M,\mb t)$ splits into a sum of three canonical pieces $\Omega = \Omega_{\rm big} + \Omega_a+\Omega_t$ according to the decomposition \eqref{Pfactors}. This form admits K-theoretic and motivic ``avatars'' as in \cite{FG-Teich}. The form $\Omega$ projects to the standard -Atiyah-Bott-Goldman form $\Omega = \int_{\pd M} \Tr \delta \CA\wedge \delta \CA$ \cite{AtiyahBott-YM, Goldman-symplectic} on \emph{smooth} parts of the standard moduli space of unframed flat connections $\CA$.

Our second goal is to define a holomorphic Lagrangian submanifold $\CL_2(M)\subset \CP_2(\pd M)$, via its intersections with the patches
\be \CL_2(M,\mb t) \subset \CP_2(\pd M,\mb t)\,. \ee
The Lagrangian $\CL_2(M)$ describes the subset of framed flat connections on $\pd M$ that extend to framed flat connections in the bulk $M$ and obey certain regularity conditions --- in particular that the holonomy of 3d connections is irreducible, and that choices of framing flags are generic.

Given a 3d ideal triangulation $\mb t_{\rm 3d}$ for $M$ (compatible with the boundary triangulation $\mb t$), we will relate $\C^*$ coordinates on $\CP_2(\pd M,\mb t)$ to edge coordinates of tetrahedra, and thus show that
\be \CP_2(\pd M,\mb t) = \big(\prod_i\CP_{\pd \Delta_i}\big) \big/\!\!\big/(\C^*)^{N_I} \ee
is a symplectic reduction of a product of tetrahedron phase spaces $\CP_{\pd \Delta}\simeq \C^*\times \C^*$ by $\C^*$ actions. (This statement is independent of $\mb t_{\rm 3d}$.) We will also see that pulling the product of tetrahedron Lagrangians $\prod_i \CL_{\Delta_i}$ through the symplectic quotient produces an open subvariety $\CL_2(M,\mb t,\mb t_{\rm 3d})\subseteq \CL_2(M,\mb t)$. The closure of $\CL_2(M,\mb t,\mb t_{\rm 3d})$ equals $\CL_2(M,\mb t)$ (and thus is independent of $\mb t_{\rm 3d}$) when the 3d triangulation is sufficiently refined. Note that expressing $\CL_2(M,\mb t)$ via symplectic reduction makes manifest not only that it is a Lagrangian submanifold but that the $K_2$ avatar of the symplectic form $\Omega$ vanishes on it --- \ie\ it is a $K_2$ Lagrangian.

\subsection{Boundary coordinates}
\label{app:bdy}

\begin{wrapfigure}{r}{2.5in}
\centering
\includegraphics[width=2.5in]{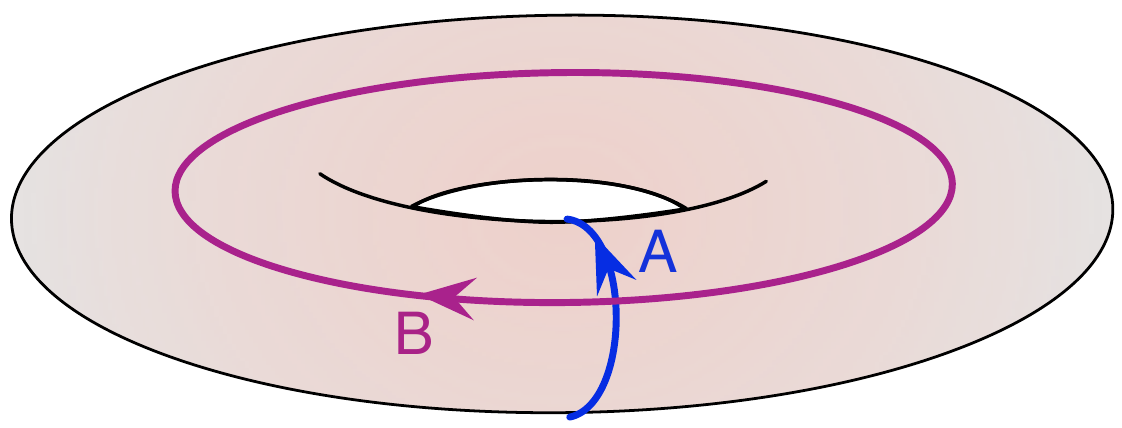}
\caption{A- and B-cycles with $\langle \gamma_B,\gamma_A\rangle=1$, if viewing the small torus boundary from \emph{inside} $M$.}
\label{fig:torusAB}
\end{wrapfigure}

Let us begin to describe the components of \eqref{Pfactors}.

\subsubsection{Small tori}
\label{app:tori}

Since small tori are disjoint from the rest of the topological boundary of $M$, they can be considered separately. They contribute canonical factors to $\CP_2(\pd M,\mb t)$, independent of~$\mb t$.

Let us choose A- and B-cycles for a small torus $T^2$, oriented so that the intersection product $\langle \gamma_B,\gamma_A\rangle =1$ (see Figure \ref{fig:torusAB}).
Since $\pi_1(T^2)\simeq \Z^2$ is abelian, the holonomies of any flat $PGL(2)$ connection on $T^2$ can be simultaneously conjugated into Jordan normal form. The flat connection is then uniquely determined by the two A- and B-cycle eigenvalues $\lambda_A,\lambda_B$, modulo any residual action of the Weyl group, which would send $\lambda\to \lambda^{-1}$. A choice of framing flag on $T^2$ is simply a choice of simultaneous eigenline for the two holonomies, and thus \emph{orders} the eigenvalues, breaking the Weyl symmetry.

Suppose that upon parallel transport around $\gamma_A$ and $\gamma_B$ the 1-dimensional subspace in the framing flag is multiplied by $\pm\lambda_A^{-1}$ and $\pm\lambda_B^{-1}$, respectively. Another way to say this would be that any projective basis $(f_1,f_2)$ of the two-dimensional space $V_2=\C^2$ that's compatible with the flag $F$ (\ie\ $f_1\in F$) is multiplied by the holonomy matrix
\be \label{eline}
 \begin{pmatrix}f_1\\ f_2 \end{pmatrix} \mapsto \begin{pmatrix} 1 & 0  \\ * & \lambda^2  \end{pmatrix} \begin{pmatrix}f_1\\ f_2 \end{pmatrix} \ee
after being parallel-transported around a respective cycle $\gamma$. The matrix here is unnormalized, in $PGL(2)$.

The $\C^*$ coordinates for framed flat $PGL(2)$ on the small torus are simply $\lambda_A^2,\lambda_B^2$. The symplectic form and Poisson bracket are
\be \Omega = \frac 12\, \frac{d \lambda_B^2}{\lambda_B^2} \wedge\frac{d \lambda_A^2}{\lambda_A^2} \,,\qquad \{\log \lambda_B^2,\log \lambda_A^2\} = 2\,.\ee
The factors of $2$ and $1/2$ here ultimately come from the Cartan matrix of $sl_2$ and its inverse, \cf\ \cite{DGG-Kdec}. Note that only the squares of eigenvalues are well-defined in $PGL(2)$. We can lift ``partway'' to $SL(2)$ by taking a square root of one eigenvalue or the other, as discussed physically in Section \ref{sec:PGL}. Then the Poisson bracket becomes more canonical, \eg, $\{\log \lambda_B,\log \lambda_A^2\} = 1$.

\subsubsection{Big boundary}
\label{app:bigbdy}

\begin{wrapfigure}{r}{1.5in}
\centering
\includegraphics[width=1.5in]{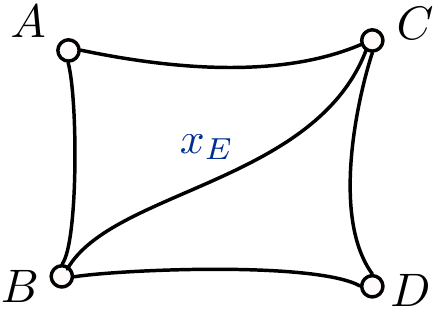}
\caption{Taking a cross-ratio to define $x_E$.}
\label{fig:xrat2}
\end{wrapfigure}

Suppose that a component of the big boundary of $M$ has genus $g$ and $h$ holes. Then every 2d ideal triangulation (like the WKB triangulations of Section \ref{sec:couplings}) has exactly $2d+h = 3(g-1+h)$ edges. Each edge $E$ is assigned a coordinate $x_E$.
Namely, the sections of the flag bundle at the four vertices of the quadrilateral containing $E$ are parallel-transported to a common point inside the quadrilateral and evaluated there to obtain four flags $A,B,C,D$ in the two-dimensional space $V_2\simeq \C^2$. (These are just lines in $\C^2$.) The orientation convention is as in Figure \ref{fig:xrat2} Then one takes a cross-ratio
\be \label{defxEa}
x_E = \frac{\langle a\wedge b\rangle \langle c\wedge d\rangle}{\langle a\wedge c\rangle\langle b\wedge d\rangle}\,,
\ee
where $\langle *\wedge *\rangle$ denotes an $SL(2)$-invariant volume form on $V_2$, and $a,b,c,d$ are any vectors in the lines $A,B,C,D$. Note that the boundary in Figure \ref{fig:xrat} is viewed from \emph{outside} the manifold $M$, and that the orientation of $M$ determines the sign in \eqref{defxEa}. This sign is opposite the ``positive'' convention of Fock and Goncharov \cite{FG-Teich} but is more natural for relations to 3d tetrahedra.%
\footnote{Asking for full 3d symmetry of the coordinates on tetrahedra --- for example, that coordinates are invariant with respect to $\Z_3$ cyclic rotations --- seems to necessarily break positivity. It can be restored by choosing a global ``time'' direction in a 3-manifold $M$, so that a pair of opposite edged of each tetrahedron is distinguished. This choice is automatically present when just studying framed flat connections a big boundary and building 3-manifolds from sequences of flips. It would be interesting to study the positive structure of the coordinates we discuss here more generally.}

\begin{figure}[htb]
\centering
\includegraphics[width=4.6in]{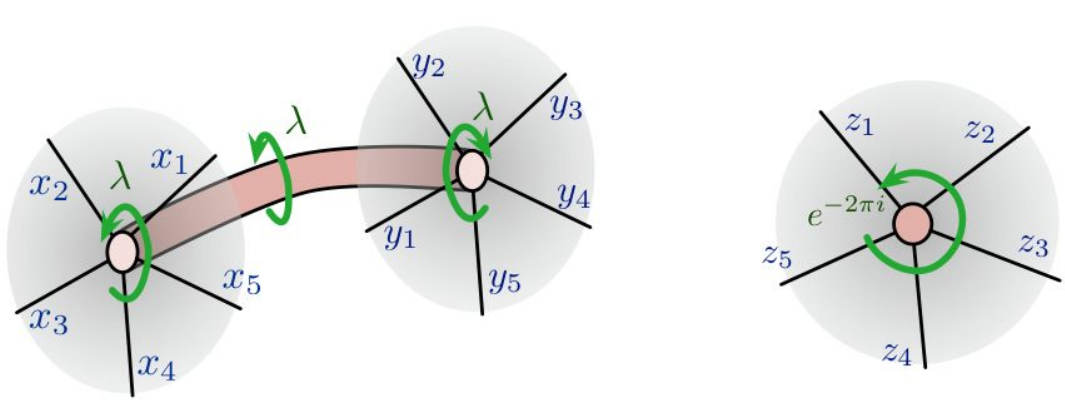}
\caption{Central elements on the big boundary, coming from holonomies around holes. Logarithmically, $\sum_i (X_i-i\pi) = 2\Lambda$ while $\sum_i(Y_i-i\pi) = -2\Lambda$. Around a small disc, with unipotent holonomy, the relation is $\prod_i (-z_i) = 1$, which lifts to $\sum_i (Z_i-i\pi)=-2\pi i$.}
\label{fig:central}
\end{figure}

Now consider a hole on the big boundary and a flat section $F$ of the flag bundle there. 
Suppose that after a \emph{counterclockwise} monodromy around the hole, as viewed from outside of $M$, the 1-dimensional subspace in the flag $F$ is rescaled by a factor $\pm\lambda^{-1}$, as in \eqref{eline}. Then it is not hard to see that the product of edge-coordinates around the hole equals $\lambda^2$,
\be \prod_{\text{$E$ around hole}} (-x_{E}) = \lambda^2\,. \label{prodhole} \ee
as in Figure \ref{fig:central}.
For future reference, it is convenient to lift this to a logarithmic relation
\be \sum_{\text{$E$ around hole}} (X_E-i\pi) = 2\Lambda\,. \label{sumhole} \ee

Aside from the $h$ relations \eqref{prodhole}, the edge-coordinates $x_E$ are completely independent. This leaves $2d$ coordinates to parameterize the space of framed flat connections on the big boundary. The symplectic and Poisson structures are given by
\be \Omega = \sum_{E,E'}\epsilon_{E,E'}^{-1} \frac{dx_E}{x_E}\wedge \frac{dx_{E'}}{x_{E'}}\,,\qquad \{\log x_E,\log x_{E'}\} = \epsilon_{E,E'}\,,\ee
where $\epsilon_{E,E'} \in \{0,\pm 1,\pm 2\}$ counts the number of oriented triangles shared by edges $E$ and $E'$. The contribution is $+1$ if $E'$ occurs to the left of $E$ in a triangle, as viewed from outside of~$M$. Note that all the hole constraints \eqref{prodhole} are central with respect to the Poisson brackets.

\begin{wrapfigure}{l}{2.4in}
\centering
\includegraphics[width=2.4in]{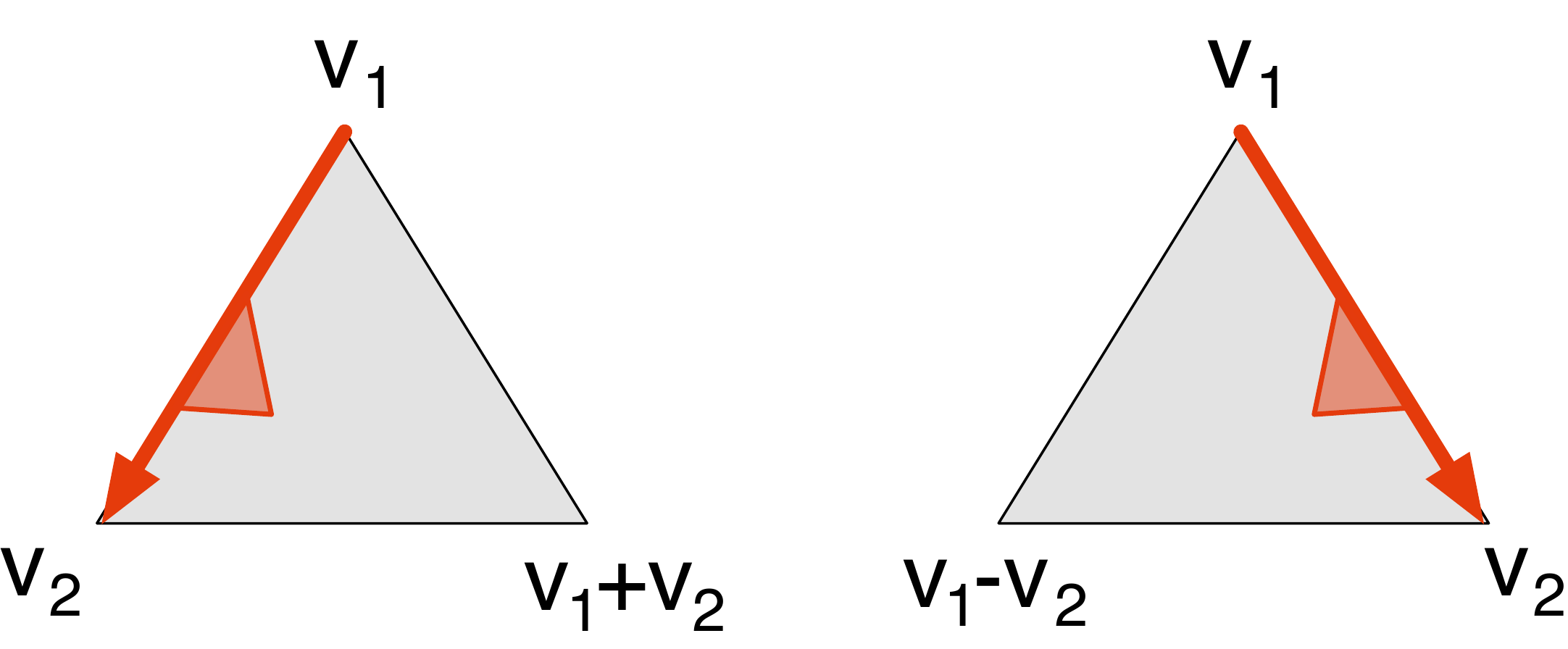}
\caption{Projective bases associated to snakes.}
\label{fig:snakedef}
\end{wrapfigure}

\begin{figure}[htb]
\centering
\includegraphics[width=5in]{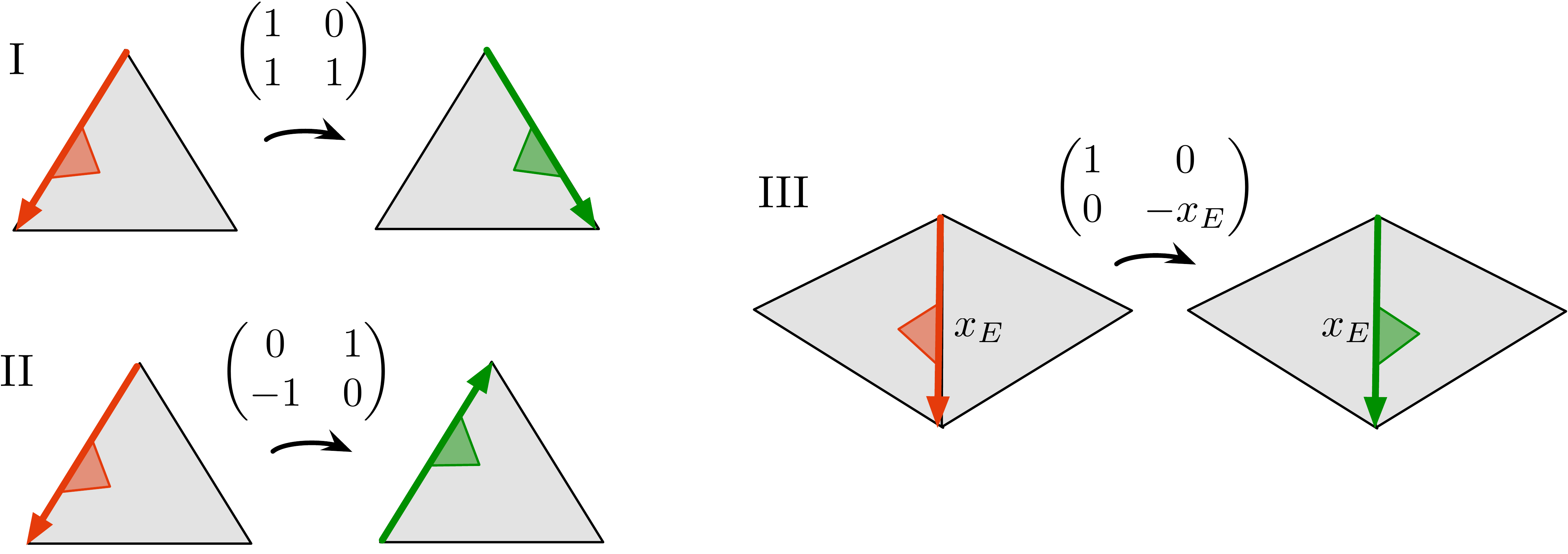}
\caption{Transformation matrices associated to fundamental snake moves: I and II within a triangle and III between neighboring triangles. The convention is such that an (unnormalized) PGL(2) transformation matrix $M$ acts on a projective basis $\ds {v_1 \choose v_2}$ by left multiplication.}
\label{fig:snakemoves2d}
\end{figure}

In order to calculate the $PGL(2)$ holonomy around \emph{any} path on the big boundary, one can use 2d snakes.%
\footnote{2d snakes were introduced by \cite{FG-Teich} to study $PGL(K)$ connections. For $PGL(2)$, however, such holonomy rules go way back to work of Thurston, Fock, and others on 2d hyperbolic geometry.} %
The idea is that in each triangle one can parallel-transport the three framing flags from the vertices to a common point in the interior, evaluate them there, and use them to define six projective bases for the space $V_2$. Graphically, each basis is associated to a \emph{snake}: an arrow going from one vertex of the triangle to another, with a ``fin'' pointing towards the interior of the triangle. (There are six ways to draw a snake in a given triangle.) The projective basis $(v_1,v_2)$ associated to a snake is given by any vector $v_1$ from the line in the flag at the tail of the snake and the unique vector $v_2$ from the line at the head of the snake such that either $v_2+v_1$ or $v_2-v_1$ (according to orientation, as in Figure \ref{fig:snakedef}) lies in the line at the third vertex of the triangle.
Then, from this definition, one discovers rules I and II of figure \ref{fig:snakemoves2d} for transforming the projective basis from one snake to another.
In addition, if a snake is moved from one triangle to an adjacent one (rule III), the vector $v_2$ at the head of the snake gets rescaled by (minus) the cross-ratio coordinate $x_E$ at the edge that the snake sits on.

By moving snakes around a closed loop on the big boundary, and multiplying the transformation matrices that relate consecutive projective bases along the way, any holonomy can be calculated. Indeed, by simply starting with a set of abstract edge-coordinates $x_E$ and using the snake rules, one can completely reconstruct both the holonomy representation of a certain flat $PGL(2)$ connection and its choice of framing flags. This constitutes the proof that edge-coordinates really do parametrize an open subset of framed flat connections on the big boundary.

\subsubsection{Small annuli}
\label{app:annuli}

Finally, we come to the small annular parts of $\pd M$. We already know how to define one coordinate for framed flat $PGL(2)$ connections on a small annulus. We choose an orientation for the non-contractible A-cycle $\gamma_A$ (as in Figure \ref{fig:central} above) and take the eigenvalue-squared $\lambda^2$ for the holonomy along that cycle (as in \eqref{eline}). In $SL(2)$, we would use $\lambda$ itself. We have just explained that this coordinate Poisson-commutes with all big-boundary coordinates $x_E$. 

What we need is a second coordinate $\tau$ that is canonically conjugate to $\lambda$ and also commutes with (almost) all big-boundary coordinates. It will be defined next. The pair $(\lambda,\tau)$ generalize classical Fenchel-Nielsen length and twist coordinates, respectively.

Notice that the total number of independent big-boundary coordinates $x_E$, small torus coordinates $(\lambda_A,\lambda_B)$, and small-annulus coordinates $(\lambda,\tau)$ exactly equals the expected dimension of $\CP_2(\pd M)$. For example, suppose that the topological boundary $\pd M$ has a single connected component, of genus $\wt g$, with no holes. If this component is split into two big boundaries $\CC_1$, $\CC_2$ of genus $g_1,g_2$ and $h_1,h_2$ holes, such that $a$ small annuli connect all the holes pairwise, then the total genus is
\be \wt g = g_1+g_2 + a -1 = g_1+g_2 +\tfrac12 (h_1+h_2) -1\,, \ee
which lets us rewrite the expected complex dimension of $\CP_2(\pd M)$ as
\be 6\wt g-6 = (6g_1-6+2h_1) + (6g_2-6+2h_2) + 2a\,. \ee
On the RHS, we see the sum of big-boundary dimensions $d_1+d_2$, and the small annulus contribution $2a$.

\subsection{Twists for small annuli}
\label{app:twist}

\begin{wrapfigure}{r}{2.4in}
\centering
\includegraphics[width=2.4in]{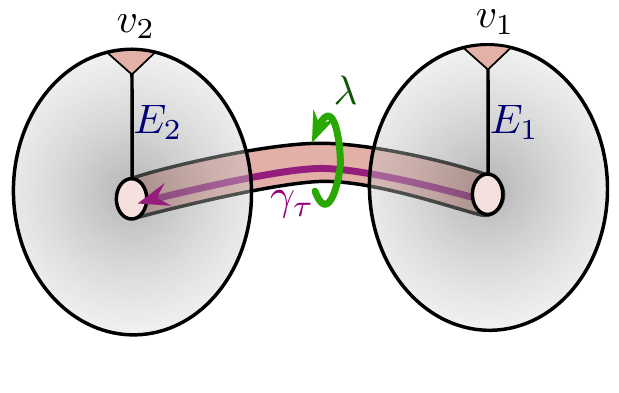}
\caption{The setup for defining a twist coordinate.}
\label{fig:twistsetup}
\end{wrapfigure}

In order to define the twist coordinate $\tau$ for a small annulus, let us first assume that the big-boundary triangulation $\mb t$ is such that the annulus ends in two degenerate triangles $t_1$ and $t_2$, as shown in Figure \ref{fig:twistsetup}. We call the degenerate edges of these triangles $E_1$, $E_2$; and we call the second vertex of each triangle (the one disjoint from the small annulus) $v_1$ and $v_2$. We will use the framing flags at $v_1$ and $v_2$ to normalize the framing flag on the annulus in two different ways, and take a ratio of these normalizations to define $\tau$.%
\footnote{The complex generalization of Fenchel-Nielsen coordinates that we describe here are closely related to coordinates studied by Kabaya \cite{Kabaya-pants}, and previously by \cite{Tan-cxFN, Kourouniotis-cxFN, Goldman-cxFN} and others. The novel feature here is use of framed moduli spaces and their eventual quantization. A discussion of length/twist coordinates in the context of framed moduli spaces also appeared in the first arXiv version of \cite{FG-Teich}.}

We must choose an oriented open path $\gamma_\tau$ along the annulus that has intersection number one with the A-cycle $\gamma_A$. We require that the endpoints of $\gamma_\tau$ lie anywhere along the $S^1$ boundaries of the annulus \emph{except} at the points where the degenerate edges $E_1,E_2$ attach. We take two paths $\gamma_\tau$, $\gamma_\tau$' to be equivalent if there exists a homotopy relating them \emph{without} crossing the endpoints of $E_1$ and $E_2$. Thus, any two inequivalent paths are related by extra twists around the A-cycle of the annulus.

\begin{figure}[htb]
\centering
\includegraphics[width=6in]{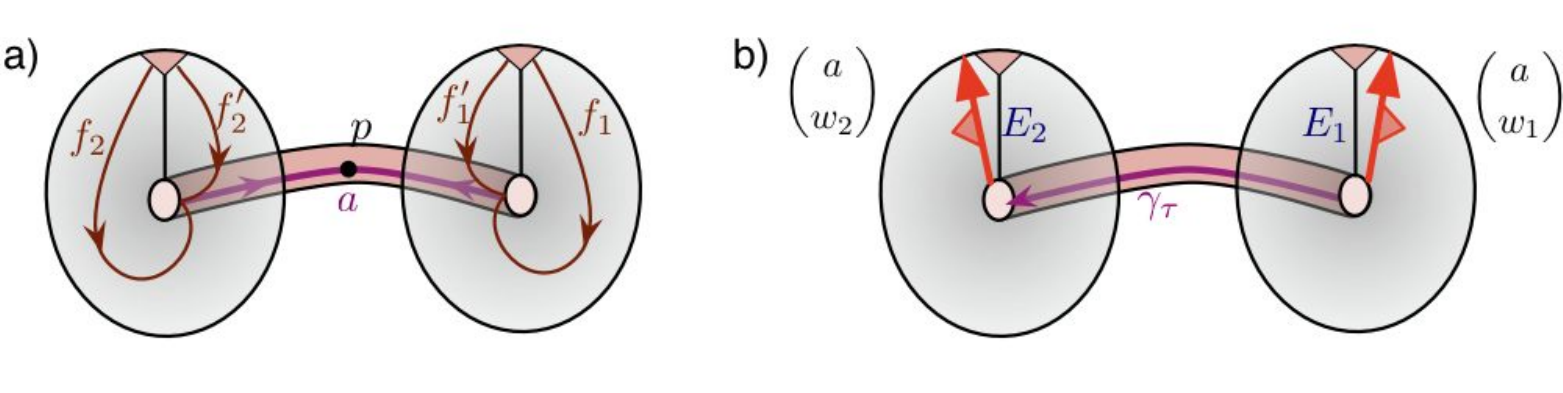}
\caption{The parallel transport of vectors in the framing flags to a common point $p$ on the annulus. The twist can be defined by either (a) comparing normalizations given by these framing flags; or b) comparing projective snake-bases at the ends of the annulus.}
\label{fig:twistdef}
\end{figure}

Now, let us use the flat connection on $\pd M$ to parallel-transport the framing flags $F_1$ and $F_2$ from vertices $v_1$ and $v_2$ to the endpoints of the curve $\gamma_\tau$ and then to some common point $p$ around the middle of $\gamma_\tau$. In fact, inside each triangle $t_i$ there are two inequivalent ways to transport the flags from vertex $v_i$ to the endpoint of $\gamma_\tau$ on the central hole, as shown in Figure \ref{fig:twistdef}a. Therefore, we obtain four different lines in the fiber $V_2\simeq \C^2$ at $p$, denoted $F_1|_p,\, F_1'|_p,\, F_2|_p,\, F_2'|_p$. Notice that if we let $M_\lambda$ denote the $PGL(2)$ holonomy matrix around the A-cycle starting and ending at $p$, then
\be F_1'|_p = M_\lambda\, F_1|_p\,,\qquad F_2'|_p = M_\lambda\, F_2|_p\,. \ee

Of course, in addition to the flags we have parallel-transported to the annulus, we also have the framing flag $A$ on the annulus itself. Let $a$ be any vector in the one-dimensional subspace of $A|_p$. Let $f_1,f_1',f_2,f_2'$ be any vectors in the lines $F_1|_p$, etc. Then
\be a_1 = \sqrt{\frac{\langle f_1\wedge f_1'\rangle}{\langle f_1\wedge a\rangle\langle f_1'\wedge a\rangle}}\,a\qquad\text{and}\qquad a_2 = \sqrt{\frac{\langle f_2\wedge f_2'\rangle}{\langle f_2\wedge a\rangle\langle f_2'\wedge a\rangle}}\,a\ee
provide two different invariant normalizations of the vector $a$. The square of the ratio of normalizations,
\be \label{deftau}
\tau = - \frac{\langle f_1\wedge f_1'\rangle}{\langle f_1\wedge a\rangle\langle f_1'\wedge a\rangle}\cdot \frac{\langle f_2\wedge a\rangle\langle f_2'\wedge a\rangle}{\langle f_2\wedge f_2'\rangle} 
 = -\frac{\langle f_1\wedge M_\lambda f_1\rangle \langle f_2\wedge a\rangle^2}{\langle f_1\wedge a\rangle^2\langle f_2\wedge M_\lambda f_2\rangle} \ee
is invariant under rescaling $a,f_1,f_1'$, etc., the $SL(2)$ action on $V_2$, and the chosen volume form $\langle*\wedge *\rangle$; it defines the twist coordinate $\tau$. (To obtain the expression on the RHS of \eqref{deftau} we used $\langle M f\wedge a\rangle = \langle f\wedge M^{-1} a\rangle$ for any $M\in SL(2)$, and $M_\lambda^{\pm 1}a = \lambda^{\mp 1}a$.)

Alternatively, we may define $\tau$ by comparing two projective bases at the ends of the annulus. Let us form two bases associated to the snakes in Figure \ref{fig:twistdef}b. We still evaluate all flags at the common point $p$ in the middle of the annulus. We can use the same vector $a$ as the first vector in both bases, which then uniquely selects the second components $w_1$ and $w_2$ obeying
\be w_i \in F_i|_p\,,\qquad a+w_1 \in F_1'|_p\,,\quad a+w_2 \in F_2'|_p\,. \ee
Then we have
\be w_2 = -\tau w_1 + \alpha\, a \ee
for some $\alpha$; in other words,
\be \label{deftau2} \tau = -\frac{\langle a\wedge w_2\rangle}{\langle a\wedge w_1\rangle}\,. \ee
Note again that this definition of $\tau$ is independent of the choice of vector $a_1$, or the precise choice of point $p$.
To see that the definitions \eqref{deftau} and \eqref{deftau2} are equivalent, it suffices to take $f_i=w_i,\, f_i'=a+w_i$ in \eqref{deftau} and simplify the RHS.

The definition of $\tau$ does depend on the equivalence class of the path $\gamma_\tau$. Indeed, if $\gamma_\tau$ is given an extra turn around the A-cycle of the annulus, the twist coordinate transforms as
\be \tau \mapsto \lambda^2 \tau\,. \ee

From the techniques for calculating Poisson brackets developed in \cite{FG-Teich} and \cite{GMNII, GMNIII}, it is straightforward to show that
\be \{\log \tau,\log \lambda^2\} = 2\,. \label{LT} \ee
This is basically a consequence of the fact that both the A-cycle eigenvalue and the twist can be expressed by comparing projective snake-bases at the beginning and end of two paths $\gamma_A$ and $\gamma_\tau$ on the annulus, and these paths have intersection number one.
To get a canonical bracket, we should lift to $SL(2)$ and use $\lambda$ rather than $\lambda^2$, or take a square root of $\tau$. Since the bracket \eqref{LT} is nontrivial, whereas $\lambda$ commutes with all edge coordinates $x_E$ on the big boundary, it must be that $\tau$ is a new coordinate, completely independent of the $x_E$. Thus it fills out the required dimension of $\CP_2(\pd M)$ as desired. The twist $\tau$ commutes with all big-boundary coordinates \emph{except} those on the central edges of the degenerate triangles $x_{E_1}=x_{E_2}^{-1}=-\lambda^2$ and the circular outside edges of the degenerate triangles. The latter implies that the Poisson bracket on the phase space \eqref{Pfactors} is not completely independent between the big-boundary and small-annulus factors. However, the twist $\tau$ can be adjusted by a Laurent monomial of $x_E$ coordinates (on a case-by-case basis) to obtain complete factorization, if desired.

Having defined all the coordinates on $\CP_2(\pd M,\mb t)$, we see now exactly which patch of $\CP_2(\pd M)$ it describes. It is the open subset of $\CP_2(\pd M)$ for which all cross-ratios in the definitions of edge coordinates $x_E$ and twists $\tau$ are well-defined in $\C^*$. This means that the configurations of framing flags entering the cross-ratios are generic. It implies in particular that in the presence of small annuli the holonomy representation of the flat $PGL(2)$ connection must not be reducible to a proper reductive subgroup (\ie\ to $GL(1)$).

\subsubsection{General boundary triangulation and canonical twists}

In general, a chosen big-boundary triangulation $\mb t$ may not have degenerate triangles at the ends of all small annuli. If so, it is still possible to define \emph{some} twist coordinates $\tau$ the same way as above. Namely, we arbitrarily choose two triangles $t_1$, $t_2$ in the triangulation at the ends of each annulus; choose paths $\wt \gamma_\tau$ that begin and end inside these triangles; and then compare normalizations of flags along the path. The resulting coordinates $\wt \tau$ will always satisfy the Poisson bracket \eqref{LT}, but may in general have very complicated Poisson brackets with the various $x_E$ on the big boundary. If desired, one can then proceed to modify $\tau$ by Laurent monomials in big-boundary coordinates (or even their roots) in order to simplify or factorize the Poisson bracket.

\begin{wrapfigure}{r}{2.3in}
\centering
\includegraphics[width=2.3in]{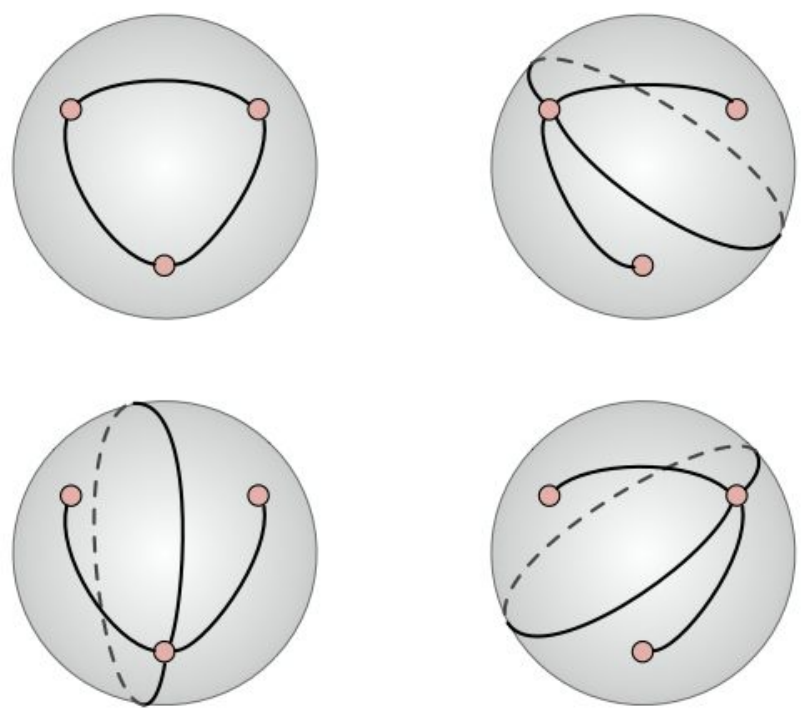}
\caption{The four triangulations of a 3-punctured sphere.}
\label{fig:S2t}
\end{wrapfigure}

When all big boundaries are three-punctured spheres, connected by small annuli --- such as in a ``pants degeneration'' of some surface $\CC$ --- much more canonical twist coordinates can be defined. Indeed, three-punctured spheres only admit four distinct triangulations (Figure \ref{fig:S2t}). Correspondingly, an annulus may attach to a hole incident to one, two, or four edges of the big-boundary triangulation.

If only one (degenerate) edge touches the end of an annulus, the choice of endpoint for $\gamma_\tau$ there is canonical. If two edges touch the end of an annulus, one can take an average over two different paths $\gamma_\tau, \gamma_\tau'$ ending on the triangles separated by these edges, and related (locally) by a twist around the A-cycle. If four edges touch, then one can take an average over four paths, differing from each other by quarter-twists, as illustrated in Figure \ref{fig:MS} on page \pageref{fig:MS}. Globally, one must also choose an overall (possibly fractional) twisting in the bulk of each annulus.

The resulting coordinates $\{\lambda_i,\tau_i\}_{i=1}^a$ have the pleasant property that they diagonalize the Poisson bracket:
\be \{\log \tau_i,\log \lambda_j\} = \delta_{ij}\,.\ee
Also notice that since three-punctured spheres carry no independent degrees of freedom, the entire phase space is parametrized by these pairs: $\CP_2(\pd M,\mb t)\simeq (\C^*)^{2a}$. These are the most direct generalizations of Fenchel-Nielsen lengths and twists.

\subsection{Tetrahedra and coordinates in the bulk}

Having defined coordinates for framed flat connections on the boundary of $M$, we proceed to extend the flat connections into the bulk. We explain how to relate boundary coordinates to edge-parameters of tetrahedra, and how to obtain $\CL_2(M,\mb t)$ via symplectic gluing. This is largely review (covering aspects of \cite{NZ, Neumann-combinatorics, Dimofte-QRS, DGG-Kdec}), aside from the computation of the twist $\tau$.

\subsubsection{A single tetrahedron}

For a single tetrahedron, the phase space $\CP_2(\pd \Delta)$ is the space of framed flat connections with unipotent holonomy on a 4-punctured sphere; while the subspace $\CL_2(\Delta)$ is simply the set of configurations of four flags at the punctures (since flat connections become trivial in the bulk).

\begin{wrapfigure}{r}{1.5in}
\centering
\includegraphics[width=1.3in]{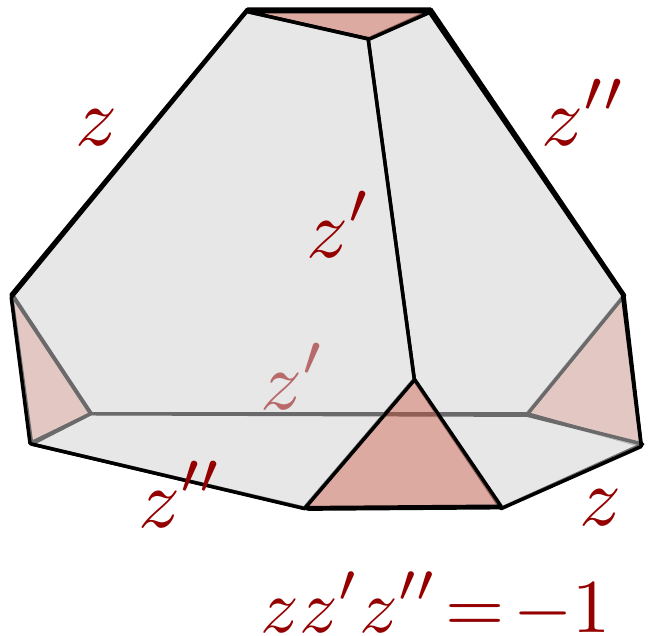}
\end{wrapfigure}

A tetrahedron comes with a conical boundary triangulation $\mb t$. Then $\CP_{\pd \Delta}:=\CP_2(\pd \Delta,\mb t)$ describes the framed flat connections whose framing flags are generic --- giving well-defined cross-ratios at the six edges. From Section \ref{app:bigbdy} we know that the product of edge-coordinates around any vertex must equal $-1$ (since the holonomy is unipotent). It's then standard to call the six edge-coordinates $z,z',z''$, equal on opposite edges, occurring in the same counter-clockwise order around any vertex, and satisfying $zz'z''=-1$. We find
\be \CP_{\pd \Delta} = \{z,z',z''\in \C^*\,|\,zz'z''=-1\}\simeq (\C^*)^2\,,\ee
with $\{\log z,\log z'\}=\{\log z',\log z''\}=\{\log z'',\log z\}=1$, or $\Omega_{\pd \Delta} = \frac{dz}{z}\wedge \frac{dz'}{z'}$. For future reference, we also define a logarithmic lift of phase space coordinates that satisfy
\be Z+Z'+Z'' = i\pi\,,\qquad \{Z,Z'\}=\{Z',Z''\}=\{Z'',Z\}=1\,. \label{Plog} \ee

One can compute any holonomy on the boundary of the tetrahedron by moving snakes around closed loops and using the rules of Figure \ref{fig:snakemoves2d}. For example, the holonomy around any vertex is conjugate to
\be  M_{\rm vertex} = \begin{pmatrix} 1 & 0 \\ -zz'(z''+z^{-1}-1)\; & -zz'z'' \end{pmatrix}\,. \ee
We see explicitly that $zz'z''=-1$ is the unipotent constraint. If we want to extend a flat connection into the bulk of the tetrahedron, all holonomies must be trivial, so we also find that the Lagrangian must be given by
\be \label{LDapp} \CL_\Delta:= \CL_2(\Delta,\mb t) :=\{z''+z^{-1}-1=0\}\;\subset\, \CP_{\pd \Delta}\,.\ee
The Lagrangian condition is trivially satisfied, and moreover this is the canonical $K_2$-Lagrangian for the form $z\wedge z'$.

Alternatively, we can obtain the Lagrangian constraint \eqref{LDapp} by computing all three cross-ratio coordinates $z,z',z''$ from flags that have been parallel-transported to a single common point in the bulk of the tetrahedron. The three cross-ratios obey a Pl\"ucker relation, which is precisely $z''+z^{-1}-1=0$. Thus $\CL_\Delta$ describes the configuration space of four generic flags in $V_2\simeq \C^2$.

\subsubsection{Triangulated 3-manifold}
\label{app:t3d}

Now suppose that $M = \cup_{i=1}^N \Delta_i$ has a 3d triangulation $\mb t_{\rm 3d}$ compatible with the boundary triangulation $\mb t$, and suppose that we fix a framed flat connection on $M$.
We can define edge-parameters $z_i,z_i',z_i''$ for every tetrahedron $\Delta_i$ by parallel-transporting the framing flags at the vertices of $\Delta_i$ to a common point in its interior, and taking cross-ratios. (For each tetrahedron, we must choose which edges to label $z_i,z_i',z_i''$, consistent with the cyclic order induced by the orientation of $M$; but since all constructions are invariant under cyclic permutations, the choice is immaterial.) The parameters will automatically satisfy
\be \label{Lbulk} z_iz_i'z_i''=-1\,,\qquad z_i''+z_i^{-1}-1=0\qquad \forall\; i\,.\ee
In addition, for every internal edge $E_I$ of the triangulation, the product of edge-parameters incident to $E_I$ automatically equals $1$:
\be \label{defc} c_I := \prod \text{($z_i,z_i',z_i''$ around $E_I$)} = 1\,.\ee
The edge constraints $c_I=1$ hold algebraically due to a perfect cancellation in the numerators and denominators of cross-ratios computed around the edge $E_I$. They reflect the fact that the holonomy on a contractible loop in $M$ that winds around $E_I$ is trivial.
We will lift \eqref{defc} to the logarithmic constraints
\be C_I := \sum \text{($Z_i,Z_i',Z_i''$ around $E_I$)}-2\pi i = 0\,. \label{defC} \ee

\begin{wrapfigure}{r}{2.3in}
\centering
\includegraphics[width=2.3in]{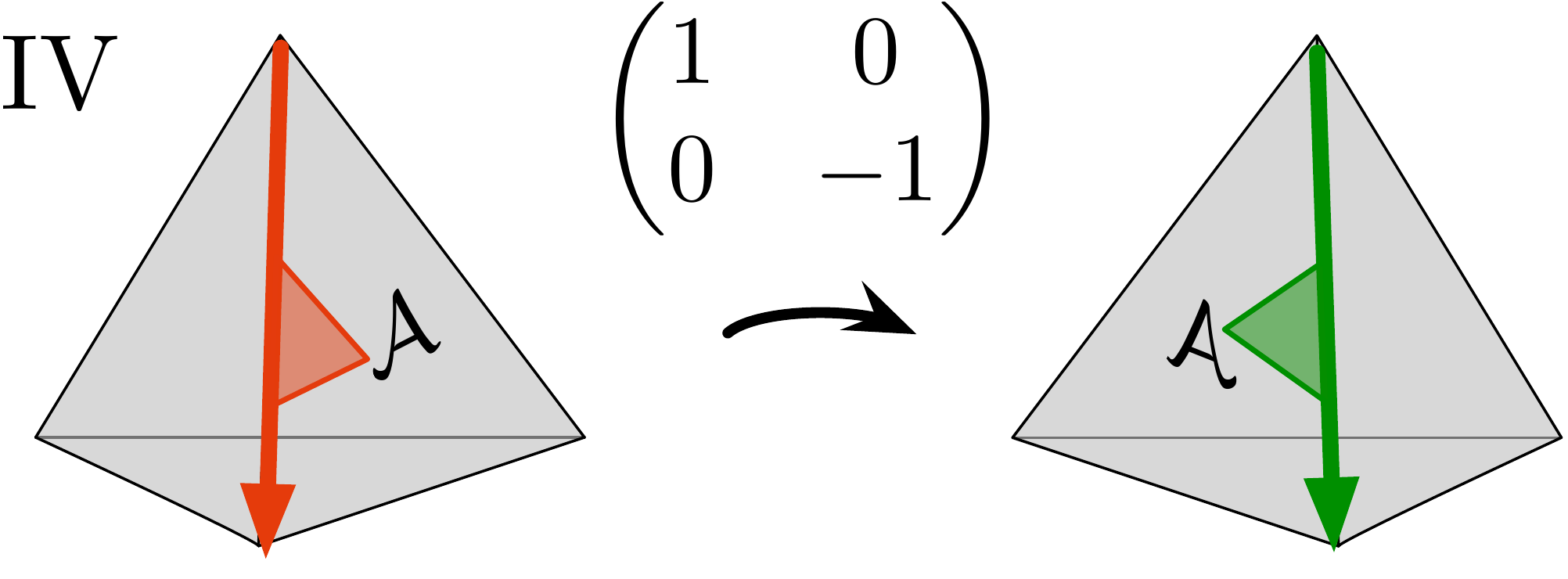}
\caption{Moving a snake between two tetrahedra glued along face $\CA$.}
\label{fig:snakemove3d}
\end{wrapfigure}

The holonomy around any closed loop in $M$ can be calculated with the help of 3d snakes. We already explained how to use snakes to define local projective bases for $V_2\simeq \C^2$, and how to move them around the boundary of a single tetrahedron. To move them between adjacent tetrahedra inside $M$, we simply supplement the rules of Figure \ref{fig:snakemoves2d} with one more rule in Figure \ref{fig:snakemove3d}.

Conversely, if we start with abstract edge-parameters $z_i,z_i',z_i''$ for every tetrahedron that satisfy \eqref{Lbulk} and \eqref{defc}, we can use the four snake rules to fully reconstruct the holonomy representation of a flat $PGL(2)$ connection and its framing flags. Conditions \eqref{Lbulk}--\eqref{defc} are precisely the ``cocycle constraints'' that ensure flatness locally.

Thus, the set of tetrahedron parameters satisfying \eqref{Lbulk} and \eqref{defc} parameterizes an open algebraic subset $\CL_2(\Delta,\mb t,\mb t_{\rm 3d})$ in the space of framed flat connections on $M$. This subset contains those flat connections whose framing data is generic with respect to the 3d triangulation $\mb t_{\rm 3d}$: the parallel-transported flags inside each tetrahedron define non-degenerate cross-ratios $z_i,z_i',z_i''\in \C^*\bs\{1\}$. The dependence of $\CL_2(\Delta,\mb t,\mb t_{\rm 3d})$ on the actual 3d triangulation is very mild; for sufficiently refined triangulations, the closures of all $\CL_2(\Delta,\mb t,\mb t_{\rm 3d})$'s agree. Therefore, we will usually just write $\CL_2(\Delta,\mb t)$.%
\footnote{A more careful way to define $\CL_2(\Delta,\mb t)$ would be to take the union of all $\CL_2(\Delta,\mb t,\mb t_{\rm 3d})$ over all 3d triangulations $\mb t_{\rm 3d}$ that agree with $\mb t$.} %
In order to show that $\CL_2(\Delta,\mb t)$ is a $K_2$-Lagrangian, we will show in Section \ref{app:red} that it comes from symplectic reduction.

\subsubsection{Boundary and bulk coordinates}
\label{app:bdybulk}

The boundary coordinates $x_E$, $(\lambda_A,\lambda_B)$, $(\lambda,\tau)$ for big and small boundaries of a framed 3-manifold can \emph{all} be written as Laurent monomials in the tetrahedron edge-parameters. These monomials provide an explicit embedding of $\CL_2(\Delta,\mb t)$ in $\CP_2(\pd \Delta,\mb t)$, and are also the key to symplectic gluing.

First, for an edge $E$ on the big boundary of $M$, it is not hard to see from the definition of various cross-ratios that
\be x_E = \prod \text{($z_i,z_i',z_i''$ incident to $E$)}\qquad\text{or}\qquad X_E = \sum \text{($Z_i,Z_i',Z_i''$ incident to $E$)}\,. \label{xz}\ee

\begin{wrapfigure}{r}{2in}
\centering
\includegraphics[width=2in]{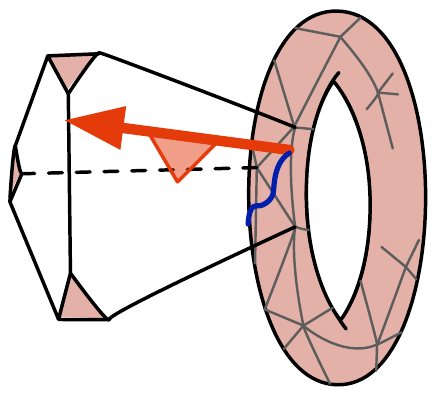}
\caption{Snake extending from a path into $M$.}
\label{fig:snakepath}
\end{wrapfigure}

We can then use 3d snakes to calculate holonomy eigenvalues and twists on the small boundary. In particular, given \emph{any} closed or open path $\gamma$ on the small boundary, the projective basis defined by a snake at the end of the path (meaning a snake whose tail lies at the endpoint of $\gamma$ and whose head extends into $M$ along the face of a tetrahedron, \cf\ Figure \ref{fig:snakepath}) will be related to the projective basis defined by a snake at the beginning of the path by a transformation matrix of the form
\be M_\gamma = \begin{pmatrix} 1 & 0 \\ * & \lambda_\gamma^2 \end{pmatrix}\,. \ee
The matrix must be triangular, because the holonomy along a small boundary must preserve the framing flag there. The matrix element $\lambda_\gamma^2$ is \emph{either} the holonomy eigenvalue or the twist coordinate associated to the path, and turns out to be a product of $z_i^{\pm 1},z_i'{}^{\pm 1},z_i''{}^{\pm 1}$.

It is very convenient to rewrite the snake rules for such a coordinate $\lambda_\gamma$ more directly in terms of the edge-parameters encountered by the path. Recall that the small boundary is tiled by small triangles from  truncated tetrahedra vertices in a 3d triangulation $\mb t_{\rm 3d}$. The angles of these small triangles are labelled by edge-parameters $z_i$, $z_i'$, $z_i''$.
Let $\gamma$ be an oriented path drawn on the small boundary so that inside every triangle $\gamma$ looks like
\be     \includegraphics[width=.7in]{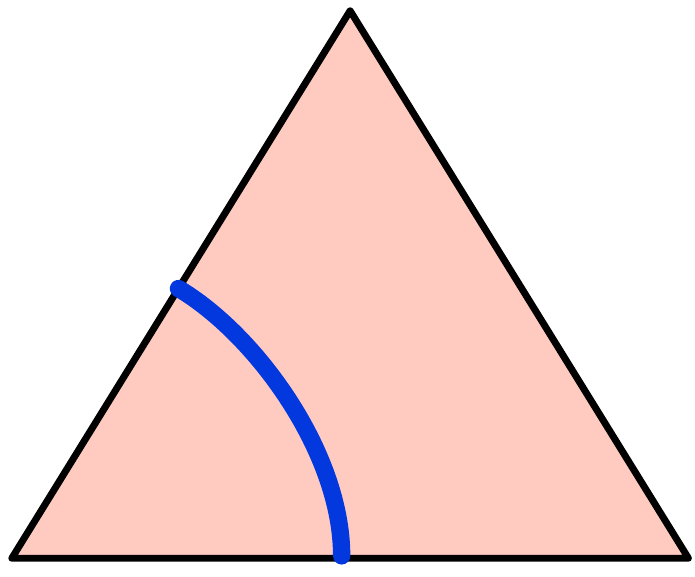}\qquad\raisebox{.2in}{\text{or}}\qquad\includegraphics[width=.7in]{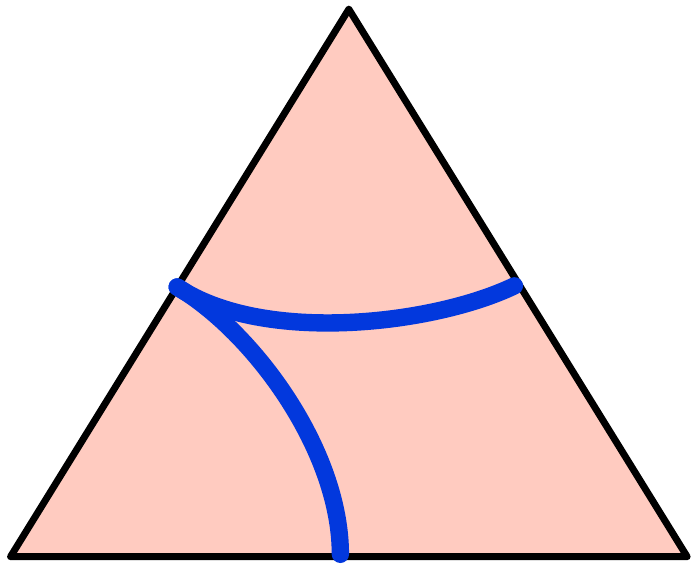}\quad\raisebox{.2in}{,} \notag\ee
with either orientation. We assume that $\gamma$ has no self-intersections and that it is either closed (as relevant for A- and B-cycle coordinates) or begins and ends at $S^1$ boundaries of the small boundary (as relevant for twists).

\begin{figure}[htb]
\centering
\includegraphics[width=4.4in]{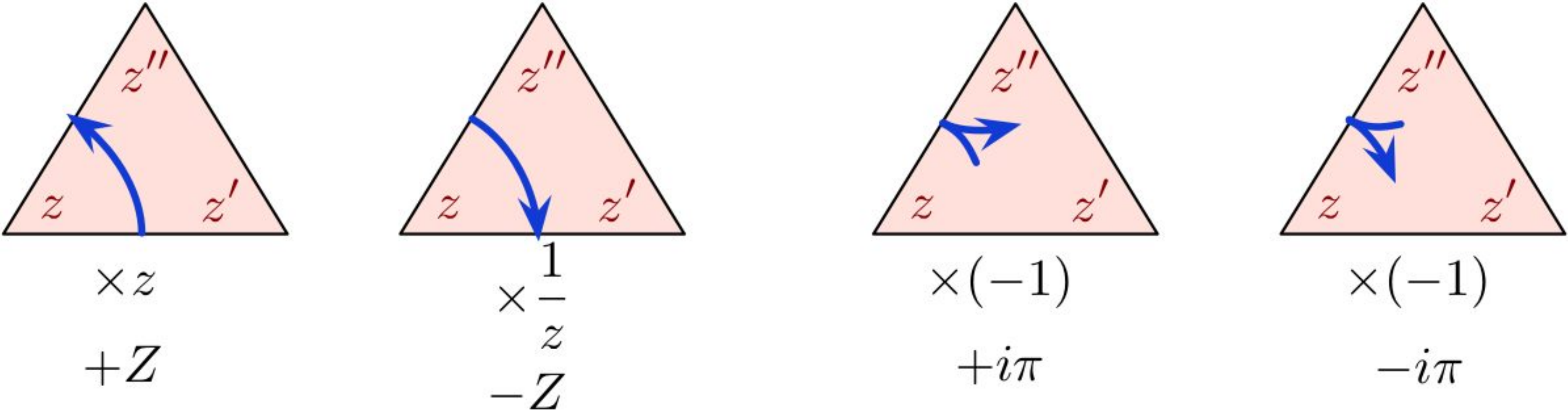}
\caption{Standard factors associated to segments of an oriented path, and their logarithmic lifts.}
\label{fig:pathrules}
\end{figure}

\begin{figure}[htb]
\centering
\includegraphics[width=3.3in]{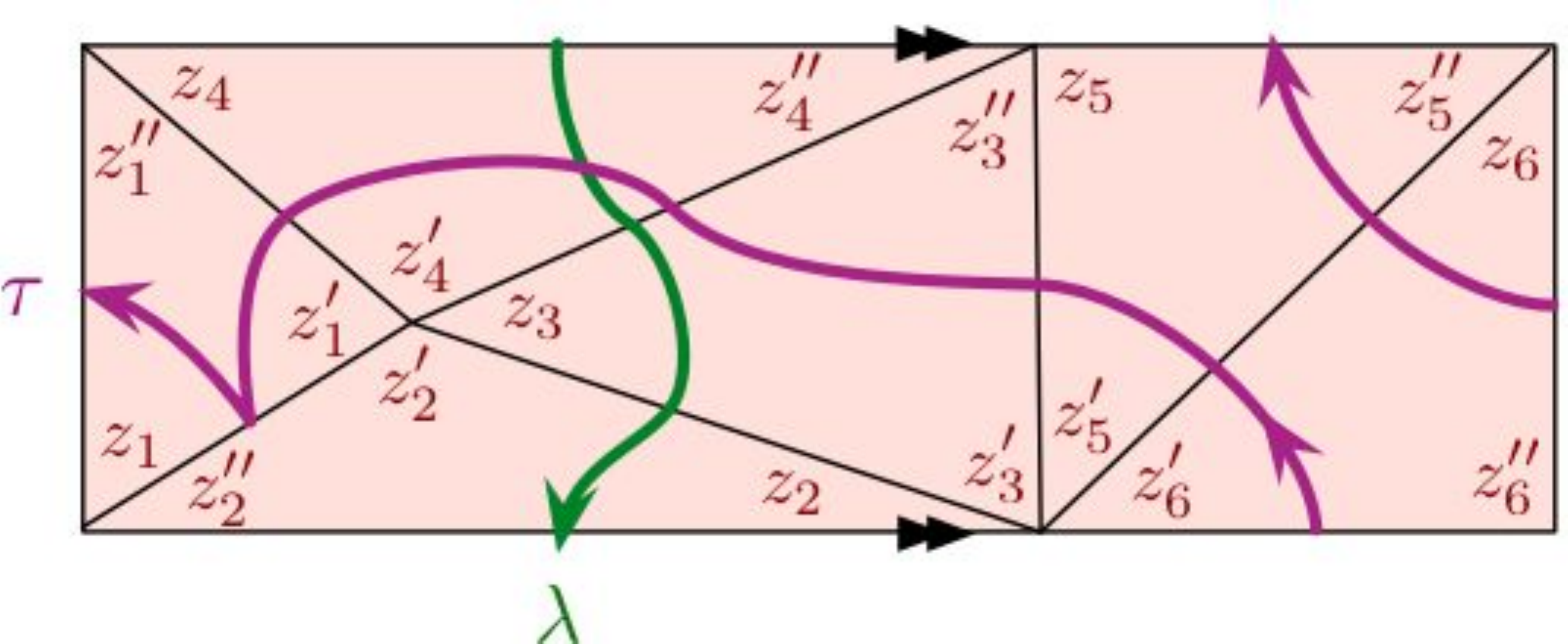}
\caption{Length and twist for a small annulus as combinations of tetrahedron edge-parameters. Here $\lambda^2=z_3/(z_2z_4'')$ and $\tau=-z_6z_5''(z_6'z_5')^{-1}z_3''(z_4'z_1'z_1)^{-1}$. Or, Logarithmically, $2\Lambda=-Z_4''+Z_3-Z_2$, while $\CT = Z_6+Z_5''-Z_6'-Z_5'+Z_3''-Z_4'-Z_1'+i\pi-Z_1$. Here the small boundary is viewed from \emph{outside} of $M$.}
\label{fig:pathLT}
\end{figure}

Then the parameter $\lambda_\gamma^2$ can be computed by multiplying and dividing by edge-parameters on angles subtended by $\gamma$, using the rules of Figure \ref{fig:pathrules}. These rules can consistently be lifted to calculate logarithmic parameters $2\Lambda_\gamma$ (with $\lambda_\gamma^2=\exp(2\Lambda_\gamma)$), as shown. An example appears in Figure \ref{fig:pathLT}.

Notice that the parameter $2\Lambda_\gamma$ is independent of path homotopies, generated by moves of the type
\be  \raisebox{-.5in}{\includegraphics[width=5in]{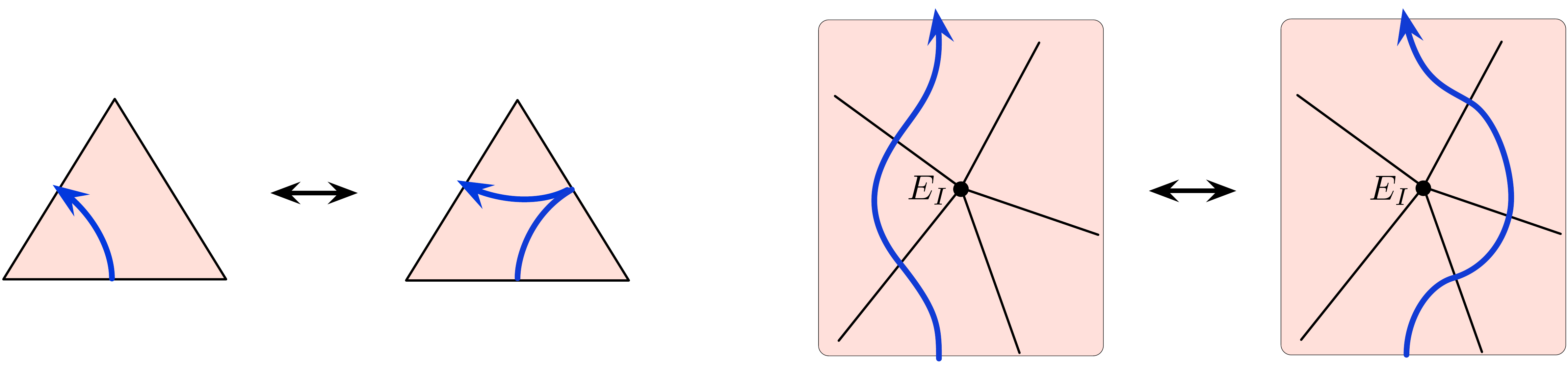}} \ee
The first move holds due to the linear logarithmic relation \eqref{Plog} among $Z_i,Z_i',Z_i''$. The second move adds a sum of edge-parameters $C_I$ around an internal edge of the triangulation, as in \eqref{defC}, to $2\Lambda_\gamma$; but $C_I$ should be set to zero according to the gluing constraint. One can understand the logarithmic gluing constraint as a \emph{consistency condition} that allows unambiguous parameters $2\Lambda_\gamma$, independent of path homotopy, to be defined.
Also note, in general, that any contractible clockwise (resp., counterclockwise) loop on the small boundary must be assigned logarithmic parameter $2\pi i$ ($-2\pi i$), modulo gluing functions $C_I$. 

\begin{figure}[htb]
\centering
\includegraphics[width=3in]{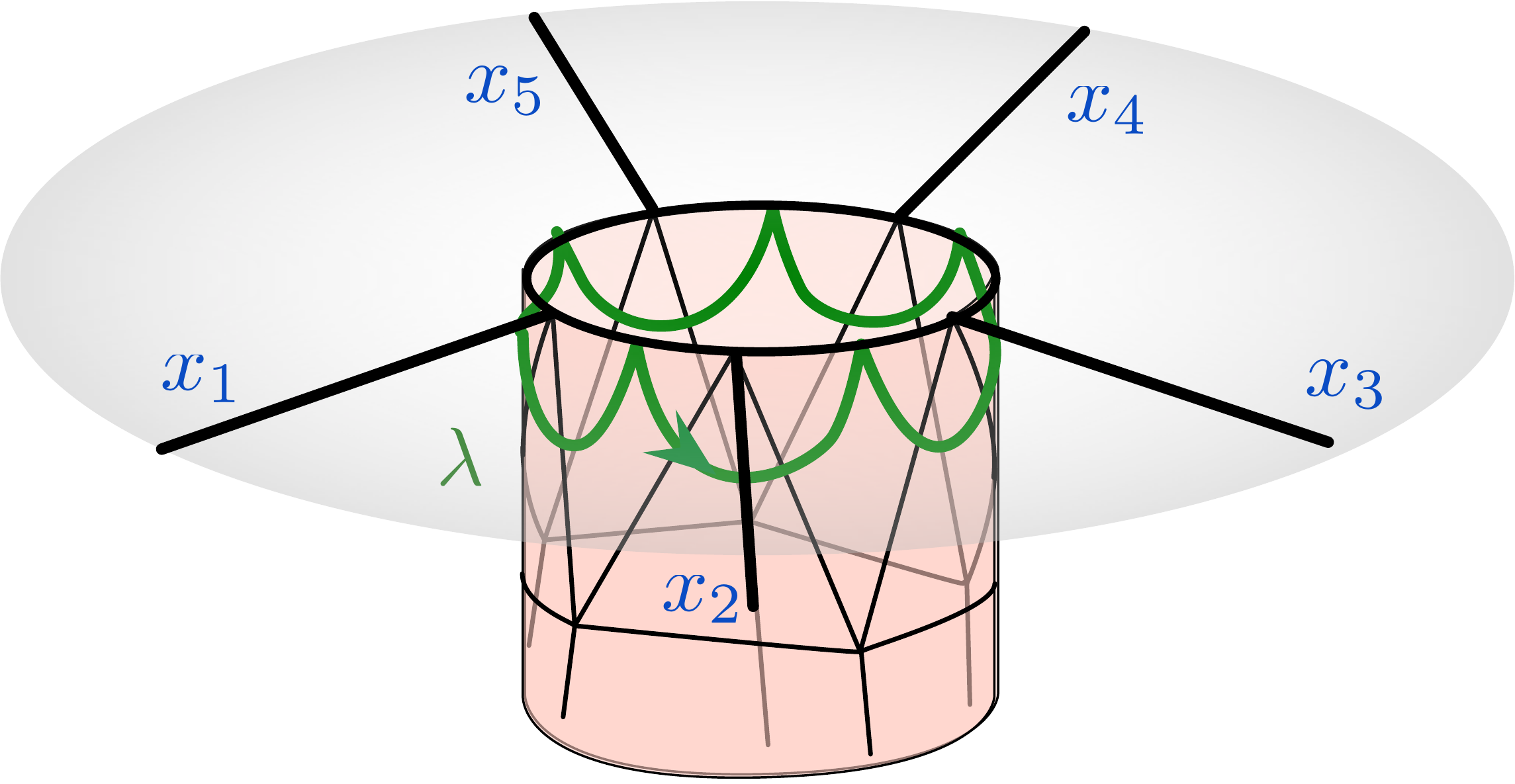}
\caption{The immediate neighborhood of a hole on the big boundary, connected to a small annulus. Calculating the logarithmic path parameter $2\Lambda$ shows that $2\Lambda=\sum_i (X_i-i\pi)$.}
\label{fig:pathsumX}
\end{figure}

Invariance under path homotopy can be used to motivate the logarithmic relation \eqref{sumhole} between the sum of big-boundary coordinates around a hole and the eigenvalue there. Consider the setup in Figure \ref{fig:pathsumX}. One one hand, the path $\gamma$ is in the homotopy class of the A-cycle of the annulus, so its total parameter is $2\Lambda$. On the other hand, $\gamma$ is composed of semi-circular segments that sum up edge-parameters around external edges --- contributing external coordinates $X_i$ --- and ``bounces'' that subtract $i\pi$. Therefore, $2\Lambda = \sum (X_E - i\pi)$.

\subsubsection{Symplectic reduction}
\label{app:red}

One would like to show that the phase space $\CP_2(\pd M,\mb t)$ defined in Sections \ref{app:bdy}--\ref{app:twist} can be obtained as the symplectic reduction of a product of tetrahedron phases spaces in \emph{any} 3d triangulation of $M$. This requires two basic results.

The first result is that the symplectic structure on boundaries of tetrahedra is compatible with the symplectic structure on $\pd M$. It can be stated as follows. Fix a triangulation $\mb t_{\rm 3d}$ with $N$ tetrahedra and let $\CP_\times =\prod_{i=1}^N \CP_{\pd \Delta_i} \simeq \{z_i,z_i',z_i''\in \C^*|z_iz_i'z_i''=-1\} \simeq (\C^*)^{2N}$ be the product of tetrahedron phase spaces, with the product symplectic and Poisson structures, \ie\ $\{\log z_i,\log z_j'\}=\{\log z_i',\log z_j''\}=\{\log z_i'',\log z_j\}=\delta_{ij}$. For every internal edge $E_I$ in the triangulation, we define a function $c_I$ on $\CP_\times$ via \eqref{defc}. We also introduce a set of ``boundary'' functions on $\CP_\times$, consisting of 1) functions $x_E$ for every external edge, defined by \eqref{xz}; 2) functions $\lambda_A^2$, $\lambda_B^2$ associated to A- and B-cycles on every small torus boundary, defined by choosing representative paths $\gamma_A,\gamma_B$ and using the path rules of Figure \ref{fig:pathrules}; and similarly 3) length and twist functions $\lambda^2,\tau$ for every small annulus. These are all Laurent monomials in the coordinates $z_i,z_i',z_i''$. (Note that the definitions of $\gamma_A,\gamma_B;\gamma,\tau$ are unique up to multiplication by $c_I$'s.) Then

\begin{theorem} \label{thm:Poisson} The functions $x_E,\lambda_A,\lambda_B,\lambda,\tau$ on the symplectic space $\CP_\times$ satisfy the standard Poisson brackets described in Sections \ref{app:bdy}--\ref{app:twist}. Namely,
\begin{subequations} \label{brackets}
\be \{\log x_E,\log x_E'\} = \epsilon_{EE'}\,,\qquad \{x_E,\lambda_A\}=\{x_E,\lambda_B\}=\{x_E,\lambda\}=0\,; \ee
\be \{\log \lambda_B^2,\log\lambda_A^2\}=\{\log \lambda^2,\log \tau\}= 2\,,\ee
\end{subequations}
with functions $\lambda_A,\lambda_B$ or $\lambda,\tau$ on different tori or annuli commuting with each other. Moreover, the internal-edge functions $c_I$ commute among themselves and with all of the boundary functions. 
\end{theorem}

The proof follows from basic combinatorics, and is a direct consequence of arguments in the Appendix of \cite{DGG-Kdec}. One way of phrasing this theorem in terms of path parameters (Section \ref{app:bdybulk}) is that the commutation relation between any two path parameters $\{\log \lambda_\gamma^2,\log \lambda_{\gamma'}^2\}$ equals the intersection number $2\langle \gamma,\gamma'\rangle$ of paths on the small boundary --- as long at least one of the paths is closed. For example, the $c_I$ are path parameters for closed contractible loops on the small boundary, so their intersection number with any other path is zero, which implies $\{c_I,*\}=0$.

Now suppose that the big boundary has a phase space of expected dimension $2d=6g-6+2h$; and that there are $a$ small annuli and $t$ small tori. 
The nontrivial brackets \eqref{brackets} show immediately that the $2t$ functions $(\lambda_A,\lambda_B)$, the $2a$ functions $(\lambda,\tau)$ and $2d$ of the functions $x_E$ must be linearly independent, and independent of all the $c_I$'s.%
\footnote{We mean linearly independent with respect to multiplication, since everything is $\C^*$-valued.} %
Moreover, it follows from calculating Euler characters that the number of internal edges is
\be \text{\# internal edges $(E_I)$} \;= N-d-a\,. \ee
The corresponding $N-d-a$ functions $c_I$ \emph{cannot} all be independent because there are too many of them. Indeed, an argument given in \cite[Sec 5]{DGG-Kdec} (based on homotopy-invariance of path parameters) shows that there is a linear relation among the $c_I$'s for every small torus boundary of $M$. The second result we need is that there are no other relations:

\begin{theorem} \label{thm:count} The number of linearly independent gluing functions $c_I$ is exactly $N-d-a-t$.
\end{theorem}

It follows immediately from the two Theorems that
\begin{align} \label{redapp}
\CP_2(\pd M,\mb t) &= \Big(\prod_{i=1}^N\CP_{\pd \Delta_i}\Big)\big/\!\!\big/(\C^*)^{N-d-a-t} = \Big(\prod_{i=1}^N\CP_{\pd \Delta_i}\Big)\big/\raisebox{-.1cm}{$(\C^*)^{N-d-a-t}$}\Big|\raisebox{-.1cm}{$(c_I=1)$} \\
&\simeq (\C^*)^{2d+2a+2t}\,. \notag
\end{align}
The $(\C^*)^{N-d-a-t}$ action in the symplectic quotient is generated by any independent subset of the commuting moment maps $c_I$. There is also an obvious logarithmic version of the symplectic reduction, using logarithmic functions $X_E,\Lambda,\CT, Z_i$, etc., and an additive $(\C)^{N-d-a-t}$ action generated by the $C_I$.

Having \eqref{redapp}, we may attempt to pull the canonical product Lagrangian $\prod_i\CL_{\Delta_i} \subset \prod_i \CP_{\pd\Delta_i}$ through the reduction. This makes sense as long as the moment maps and their flows are transverse to the Lagrangian equations, and will automatically define a $K_2$ Lagrangian submanifold $\CL_2(M,\mb t,\mb t_{\rm 3d})$. It must (tautologically) be the same space we described in Section \ref{app:t3d}, but now the Lagrangian property is manifest.

A better way to phrase this second Theorem above (and incorporate it with the first) is by generalizing a beautiful combinatorial result of Neumann \cite[Thm 4.2]{Neumann-combinatorics}, which in turn generalized a classic result of Neumann and Zagier \cite{NZ}. This will appear elsewhere, along with a full proof.

\subsubsection{Logarithms and quantization}
\label{app:quant}

The description of the symplectic pair $\CL_2(M,\mb t)\subset \CP_2(\pd M,\mb t)$ as a symplectic quotient \eqref{redapp} enables it to be quantized using methods of \cite{Dimofte-QRS}.%
\footnote{It is strongly suspected that such quantization ``via symplectic reduction'' is completely equivalent to other new quantization methods, such as topological recursion \cite{EO, GS-quant, BorotEynard}, as well quantization via recursion relations for knot polynomials \cite{gukov-2003, garoufalidis-2004}. Showing equivalence of these different approaches should be a fascinating endeavor.} %
By quantization here we mean promoting $\CP_2(\pd M,\mb t)$ to a $q$-commutative ring (an algebra of operators) $\hat \CP_2(\pd M,\mb t)$ generated by $\hat x_E,\hat \lambda_A,\hat\lambda_B,\hat \lambda,\hat \tau$ with commutation relations
\be \hat x \hat y  = q^{\{\log x,\log y\}}\hat y\hat x \ee
(for instance $\hat \tau\hat \lambda^2 = q\,\hat \lambda^2\hat \tau$); and promoting $\CL_2(M,\mb t)$ to a left ideal $\hat \CL_2(M,\mb t)$ in this ring. One can also consider various wavefunctions annihilated by the left ideal $\hat \CL_2(M,\mb t)$, as discussed in \cite{Dimofte-QRS, KashAnd, DGG-index, Gar-index, BDP-blocks} and earlier in \cite{hikami-2006, DGLZ}. We note that the quantization of boundary phase spaces alone, along with some simple Lagrangians, intertwines beautifully with quantization in $PGL(2)$ and higher Teichm\"uller theory \cite{Kash-Teich, FockChekhov, FG-qdl-cluster}.

We emphasize that no matter what method of quantization is used, an effective lift to logarithmic coordinates such as we have described above must arise. It was explained combinatorially in \cite{Dimofte-QRS} that the seemingly artificial counting of $2\pi i$'s in logarithms translates to keeping track of nontrivial powers of $q$ in quantization. From the perspective of wavefunctions, the imaginary parts of coordinates $\Im\, Z_i$, etc. are treated as dihedral angles in the geometry of $M$, and finiteness of wavefunctions depends crucially on the existence of ``positive angle structures'' \cite{KashAnd, Gar-index}. Indeed, it was noticed much earlier (\cf\ \cite{Neumann-combinatorics}) that logarithms are necessary for a consistent definition of ``complex volumes'' for flat connections on $M$, which appear as the semi-classical piece of any quantum wavefunction.

Physically, the importance of the $2\pi i$'s is also clear. By the 3d-3d correspondence of \cite{DGG}, operators and wavefunctions become quantized when a 3d gauge theory is placed in a twisted and/or compactified 3d geometry (\cf\ Section \ref{sec:lineops} here). This requires a well-defined $U(1)_R$ symmetry. Moreover, in the 3d-3d correspondence, it is precisely the imaginary parts of logarithms $Z_i$, etc. that are mapped to a physical R-charge assignment. Thus the logarithms tell us how to compactify a 3d gauge theory, and how to quantize.

The combinatorial quantization of \cite{Dimofte-QRS} basically says to quantize tetrahedron Lagrangians and then to pull them through a quantum symplectic reduction. The basic rules are as follows.

First, given a 3d triangulation of $M$, one forms a product algebra $\hat \CP_\times$ generated by $\{\hat z_i,\hat z_i',\hat z_i''\}_{i=1}^N$ with
\be \hat z_i\hat z_i'\hat z_i'' = -q\,\,,\qquad \hat z_i\hat z_i'=q\hat z_i'\hat z_i\,,\quad\text{etc.} \ee
The central constraint here is most naturally interpreted as an exponentiation of the logarithmic $\hat Z_i+\hat Z_i'+\hat Z_i''=i\pi+\hbar/2$, where $\hat z_i = \exp\hat Z_i$ (etc.), $q = e^\hbar$, and $[\hat Z_i,\hat Z_i']=[\hat Z_i',\hat Z_i'']=[\hat Z_i'',\hat Z_i]=\hbar$. (In general, every logarithmic $i\pi$ gets quantum-corrected to $i\pi+\hbar/2$.) Inside the $q$-commutative ring $\hat \CP_\times$, one identifies a canonical left ideal
\be \hat \CL_\times  = \big(\hat z_i''+\hat z_i-1\big)_{i=1}^N\,.\ee

Next, one defines logarithmic gluing operators $\hat C_I = \sum(\text{$\hat Z_i,\hat Z_i',\hat Z_i''$ surrounding $E_I$})-2\pi i-\hbar$, and exponentiates them to obtain $\hat c_I = \exp \hat C_I$. The $\hat c_I$'s are  Laurent monomials in the $\hat z_i,\hat z_i',\hat z_i''$, along with a well-defined $q$-correction. Similarly, one unambiguously quantizes the logarithmic formulas for $X_E,\Lambda_A,\Lambda_B,\Lambda,\CT$ (correcting $i\pi\to i\pi+\hbar/2$) and exponentiates to obtain expressions for $\hat x_E, \hat \lambda_A,\hat\lambda_B,\hat\lambda,\hat \tau$.

Finally comes the reduction. One takes the centralizer of the product ideal $\hat \CL_\times$ with respect to the $\hat c_I$. This means eliminating elements of $\hat\CL_\times$ that do not commute with the $\hat c_I$. In the centralizer, one may unambiguously set $\hat c_I$, obtaining a left ideal $\hat \CL_2(M,\mb t,\mb t_{\rm 3d})$. This ideal can be written entirely in terms of the boundary coordinates $\hat x_E, \hat \lambda_A,\hat\lambda_B,\hat\lambda,\hat \tau$ (since these generate the centralizer of $\hat c_I$ in the ring $\hat \CP_\times$). Indeed, $\hat \CL_2(M,\mb t,\mb t_{\rm 3d}) \subset \hat\CP_2(M,\mb t)$ is the sought-after quantization of the Lagrangian. By construction, taking the classical limit $q\to 1$ reduces $\hat \CL_2(M,\mb t,\mb t_{\rm 3d})$ to a classical ideal in $\CP_2(M,\mb t)$ (viewed as a ring), which must contain the classical Lagrangian $\CL_2(M,\mb t,\mb t_{\rm 3d})$.

%%%%%%%%%%%%%%%%%%%%%%%%%%%%%%%%%%%%%%%%%%%%%%%%%%%%%%%%%%%%%%%%%%%%%%%%%%%%%%
%%%%%%%%%%%%%%%%%%%%%%%%%%%%%%%%%%%%%%%%%%%%%%%%%%%%%%%%%%%%%%%%%%%%%%%%%%%%%%
%%%%%%%%%%%%%%%%%%%%%%%%%%%%%%%%%%%%%%%%%%%%%%%%%%%%%%%%%%%%%%%%%%%%%%%%%%%%%%

\section{Triangulation of general RG and duality manifolds}
\label{app:triang}

In this final section, we describe explicitly the triangulation of a framed 3-manifold describing an RG wall or a duality wall for (almost) any 4d $\CN=2$ theory of class $\CS[A_1]$ --- obtained from compactifying the $A_1$ (2,0) theory on a punctured Riemann surface.
This is actually not too hard to do; there are plenty of 3d triangulations of these manifolds. However, it is useful to find triangulations that make certain properties of the corresponding gauge theories $T_2[M]$ particularly simple or manifest. Here, following the philosophy from previous sections, we will build the triangulations out of basic RG manifolds $M_0$ (Section \ref{sec:MRG}), which describe free chiral theories with automatic $SU(2)$ flavor enhancements.

\begin{wrapfigure}{r}{2.1in}
\centering
\includegraphics[width=2in]{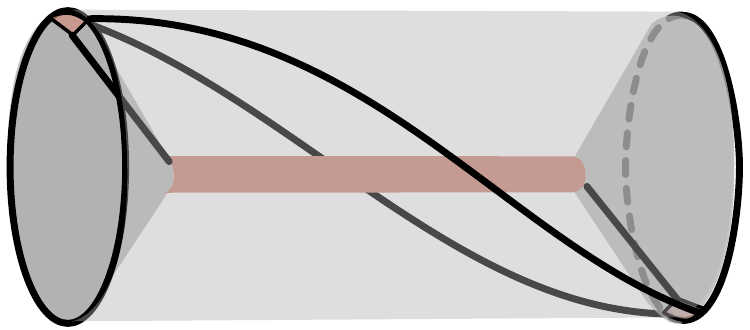}
\caption{$M_0$, repeated from Figure \ref{fig:RG}}
\label{fig:M0app}
\end{wrapfigure}

Mathematically, the 3-manifolds provide 1) the relation between cross-ratio coordinates and Fenchel-Nielsen-type coordinates, and their quantization, in Teichm\"uller theory; and 2) the mapping class group action and its quantization in (especially) Fenchel-Nielsen-type coordinates.
Out constructions can be viewed as a 3d analog of the work of Teschner \cite{Teschner-TeichMod}, in which he described 2d ideal triangulations adapted to pants decompositions of surfaces for this same purpose.

It is actually sufficient to consider RG manifolds only. Suppose, for example, that we want to create a duality manifold $M_{\mb p\mb p'}$ that interpolates between two different pants decompositions $\mb p$, $\mb p'$ of a punctured surface $\CC$. We can first create the RG manifolds $M_{\mb p}$ and $M_{\mb p'}$ that interpolate between big triangulated boundaries $\CC$ and pants decompositions (a usual network of small annuli). Then we flip the big-boundary triangulations of $M_{\mb p}$ and $M_{\mb p'}$ so that they match, according to any desired identification (any element of the mapping class group), and we glue $M_{\mb p}$ and $M_{\mb p'}$ together to get $M_{\mb p\mb p'}$.

\subsection{The construction}

Let $\CC$ be a surface of genus $g$ with $h$ regular or irregular punctures, and let $\mb p$ be a pants decomposition of $\CC$. We require as usual that $2g-2+h+\sum(\text{ranks of irregular punctures})>0$, and in addition that $h> 0$. This is our only constraint. It is necessary, since framed 3-manifolds cannot represent RG walls if $h=0$.%
\footnote{Framed 3-manifolds \emph{can} represent duality walls for surfaces with no holes ($h=0$), \ie\ for theories of class $\CS$ with no continuous flavor symmetry in the UV. But these duality manifolds cannot be obtained by gluing RG manifolds, and we will not consider them here.} %
We want to create a triangulated RG 3-manifold $M_{\mb p}$.

We assume that if $\CC$ is a sphere it has at least four punctures. The case of three regular punctures is uninteresting, and $M_{\mb p}$ for three or fewer irregular punctures can be obtained by small modifications of the basic $M_0$ geometry.

\subsubsection*{Step 1: Irregular punctures}

First we observe that, given a triangulated RG manifold $M'$ for a surface $\CC'$ with regular punctures, it is easy to modify $M'$ to make any of the punctures irregular.

\begin{figure}[htb]
\centering
\includegraphics[width=5in]{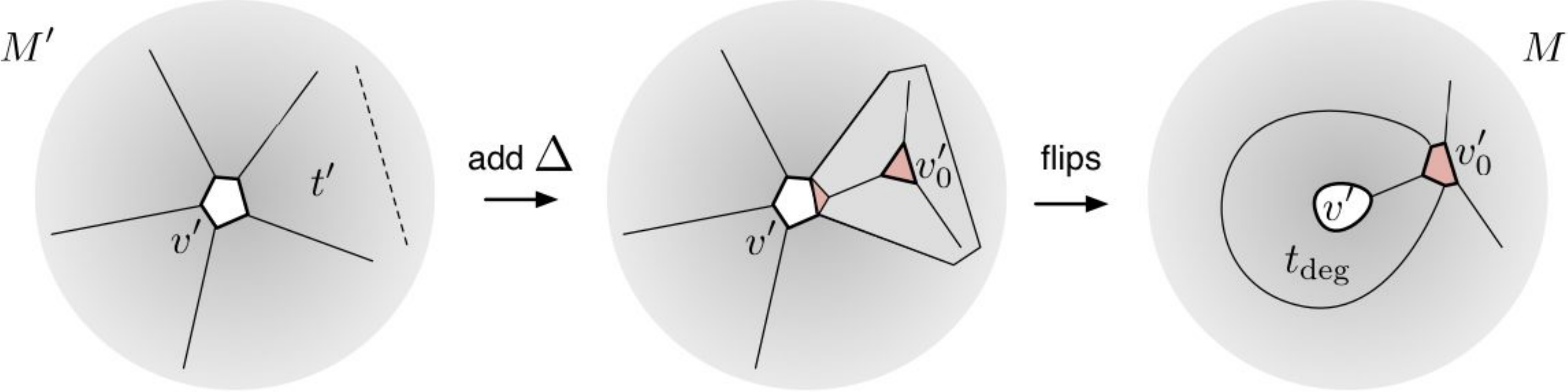}
\caption{A degenerate modification of a hole $v'$ on a big boundary. First glue a single tetrahedron onto a triangle adjacent to $v'$, and then flip to surround $v'$ by a degenerate triangle.}
\label{fig:modify}
\end{figure}

Let us focus on a particular regular puncture that we want to make irregular of minimal rank 1/2. The big IR boundary of $M'$ has a hole $v'$ on which a small annulus (carrying the regular mass parameter) ends. We choose a triangle $t'$ in the big-boundary triangulation $\mb t'$ that is adjacent to $v'$, and we glue a single tetrahedron onto this triangle, as in Figure \ref{fig:modify}. This introduces a new hole $v'_0$ on the big boundary, now filled in by a small disc (one of the truncated vertices of the new tetrahedron). We flip the big-boundary triangulation by gluing on additional tetrahedra (two faces at a time) to obtain a triangulation that has a degenerate triangle $t_{\rm deg}$ surrounding $v'$, with additional vertex $v'_0$. Let $M$ be the resulting triangulated 3-manifold.

This is enough. We now can ``cut'' along the circular edge of the degenerate triangle $t_{\rm deg}$, identifying the big boundary outside this triangle as the IR boundary of $M$, and $t_{\rm deg}$ itself together with the old UV boundary as the new UV boundary. The circular edge of $t_{\rm deg}$ (connecting the IR and UV boundaries) represents a rank--1/2 irregular puncture.

It is clear that we can promote any regular punctures of an RG manifold to rank--1/2 irregular ones in this way. To increase the rank further, we simply continue to glue more tetrahedra around the puncture on the IR boundary, adding new holes filled by small discs, and then redefine the UV/IR splitting so $\pd M$ is ``cut'' by polygons that pass through the new small discs.

Note that irregular punctures often arise naturally in practice (as in the basic RG manifold $M_0$ itself), or can be obtained by \emph{un}gluing some tetrahedron faces, rather than adding new tetrahedra. Otherwise, the above method can be applied systematically.

For future reference, we will refer to the procedure of Figure \ref{fig:modify} --- adding a tetrahedron next to a puncture $v'$ and flipping to surround $v'$ by a degenerate triangle --- as a \emph{degenerate modification of $v'$}.

\subsubsection*{Step 2: Choosing $SU(2)$ enhancements}

We may now assume that all punctures are regular.

\begin{figure}[htb]
\centering
\includegraphics[width=5.5in]{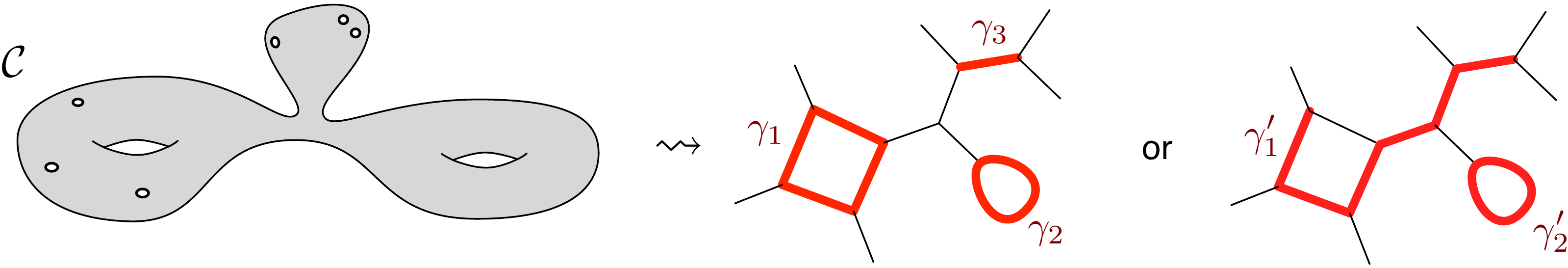}
\caption{Drawing the skeleton of $\CC$ according to a fixed pants decomposition, then choosing a locally maximal set of internal edges to enhance (highlighted in red). The enhanced edges form open or closed paths $\gamma_i$.}
\label{fig:skeleton}
\end{figure}

Let us draw a trivalent skeleton of $\CC$ according to the pants decomposition $\mb p$ (Figure \ref{fig:skeleton}), and choose a maximal set of internal edges to promote to non-abelian flavor symmetry $SU(2)$. Recall from Section \ref{sec:spheres} what this means. Each trivalent vertex of this graph will become a 3-punctured sphere in the UV boundary of $M_{\mb p}$. The edges of the graph become small annuli connecting the spheres.
We want to choose un-democratic triangulations for each 3-punctured sphere ($\mb t_1,\mb t_2,$ or $\mb t_3$ of Figure \ref{fig:S2tq}) so that as many small annuli as possible begin and end at holes surrounded by degenerate triangles.
In the 3d theory $T_2[M,...]$ these annuli will correspond to enhanced symmetries $SU(2)_{\lambda_i}$ --- which in turn get gauged in coupling to a 4d theory of class $\CS$.

Combinatorially, we choose a subset of internal edges of the skeleton of $\CC$ to ``enhance'' so that at most two enhanced edges end at any vertex. Some examples appear in Figure \ref{fig:skeleton}. We want to this subset to be (locally) maximal, meaning that no additional internal edges can be added to it.

We will not worry about external edges of the skeleton --- corresponding to small annuli in $M$ that connect the UV and IR boundaries, and whose flavor symmetries in $T_2[M]$ are identified with bulk flavor symmetries of $T_2[\CC]$. If these small annuli happen to end in degenerate triangles of 3-punctured spheres on the UV boundary, the triangulation of the IR boundary can always be flipped to create degenerate triangles (and hence $SU(2)$ enhancement) there as well.

\begin{figure}[htb]
\centering
\includegraphics[width=5.8in]{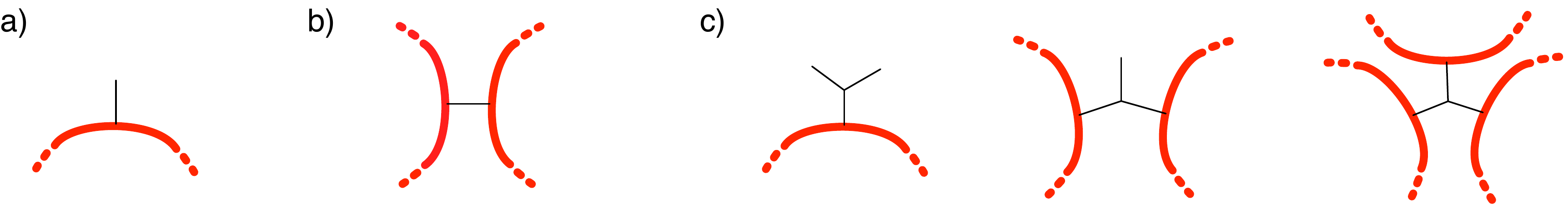}
\caption{The three options for unenhanced edges: extrenal, isolated, or trinion.}
\label{fig:unenhanced}
\end{figure}

Notice that the enhanced edges of the skeleton form either 1) closed loops; or 2) straight open paths. In particular, enhanced edges cannot branch. Moreover, the complement of the enhanced edge paths $\gamma_i$ consists of: \\
\indent a) external edges of the skeleton (connecting to punctures of $\CC$);\\
\indent b) isolated internal edges that connect two enhanced paths; or\\
\indent c) ``trinions'' that have one, two, or three legs connecting to enhanced paths. \\
\noindent See Figure \ref{fig:unenhanced}. 
We may (and will) require that there are no un-enhanced trinions (c) with just one leg connecting to an enhanced path (and two legs connecting to punctures). This can be arranged by slightly modifying the set of enhanced edges.

\subsubsection*{Step 3: Surround enhanced edges with $M_0$}

\begin{figure}[htb]
\centering
\includegraphics[width=5.7in]{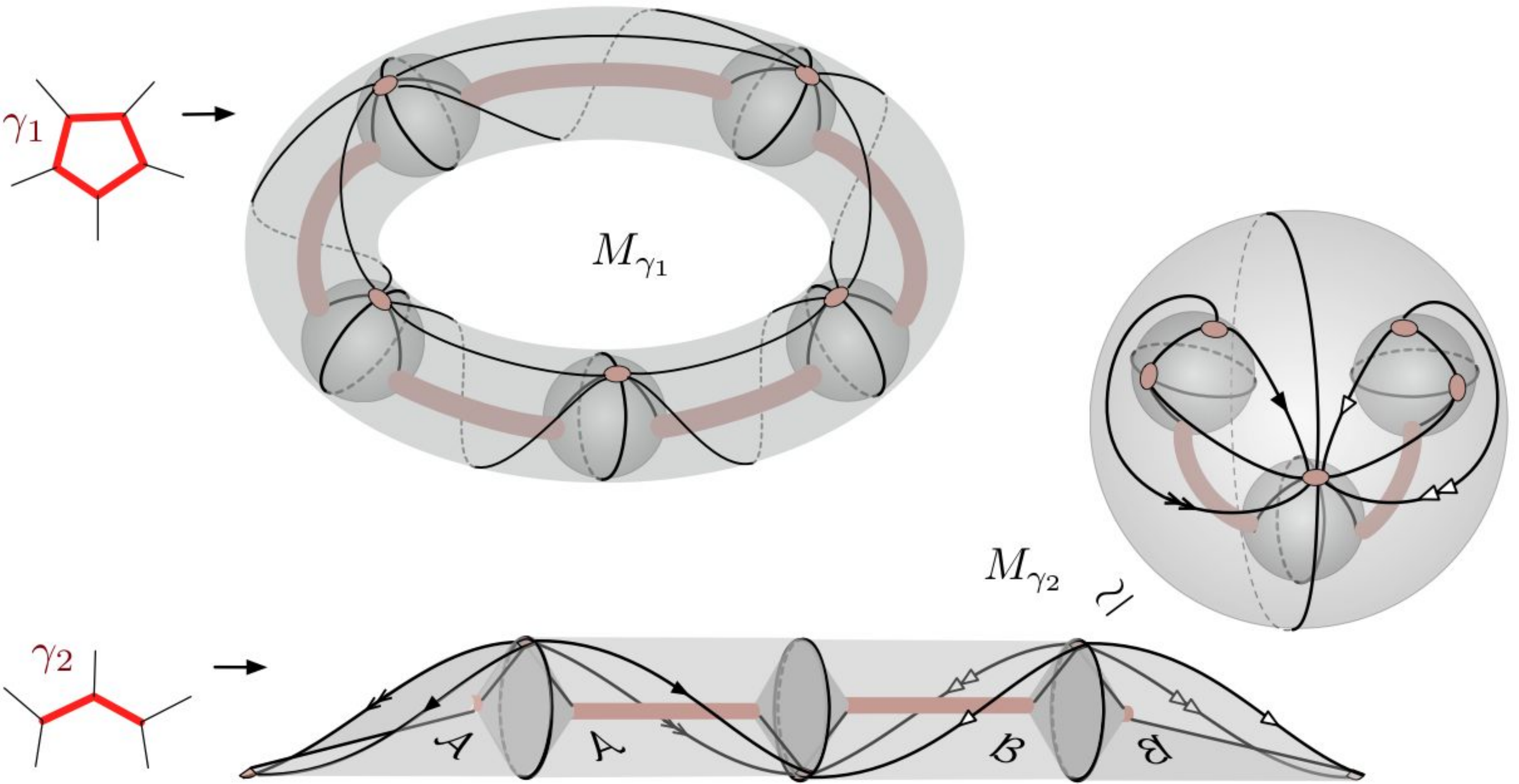}
\caption{Constructing RG manifolds for the skeletons of a five-punctured torus (top) and a 5-punctured sphere (bottom), with $SU(2)$ enhancement along the internal paths indicated.}
\label{fig:Mg}
\end{figure}

For every closed loop $\gamma$ containing $n$ enhanced edges in the skeleton of $\CC$, build a 3-manifold $M_\gamma$ by gluing $n$ copies of $M_0$ end-to-end. An example is shown in Figure \ref{fig:Mg}. Recall that $M_0$ is the basic RG manifold from Section \ref{sec:MRG}, built from two tetrahedra. Each time two copies of $M_0$ are connected, the circular edges on their ends are identified, trapping a 3-punctured sphere on the inside. These newly identified edges are ``thin'' in the sense of Section \ref{sec:N2*RG}: they don't quite fully separate the UV and IR boundaries of $M_\gamma$; but they can be thickened out with extra flips at the end.

For every open path $\gamma$ of $n$ enhanced edges in the skeleton, we can first build a manifold $M_\gamma'$ in a similar way, by gluing $n$ copies of $M_0$ end-to-end (but not closing the loop). Now we need to decorate the ends of this manifold the same way was done for the $N_f=4$ RG manifold in Section \ref{sec:N4RG}. We add a self-glued tetrahedron (half of $M_0$) to each end of $M_\gamma'$, trapping 3-punctured spheres; then we fold these tetrahedra back over the big IR boundary of $M_\gamma'$, identifying a face of each with a face on the big IR boundary. See Figure \ref{fig:Mg}. Call the resulting manifold $M_\gamma$.

Note that the manifolds $M_\gamma$ associated to open or closed paths $\gamma$ are themselves RG manifolds corresponding to some surfaces $C_\gamma$. Here $C_\gamma$ is an $n$-punctured torus if $\gamma$ is closed, and an $(n+3)$-punctured sphere if $\gamma$ is open.

\subsubsection*{Step 4: Connect the enhanced RG manifolds}

Finally we glue together our manifolds $M_\gamma$ into an RG manifold $M_{\mb p}$.

Due to the remarks at the end of Step 2, any two paths $\gamma,\gamma'$ in the skeleton that are to be connected must be connected by either (b) a single un-enhanced edge, or (c) an internal un-enhanced trinion. We will deal with the trinions first. Note that $\gamma$ and $\gamma'$ need not be distinct.

\begin{figure}[htb]
\centering
\includegraphics[width=5.5in]{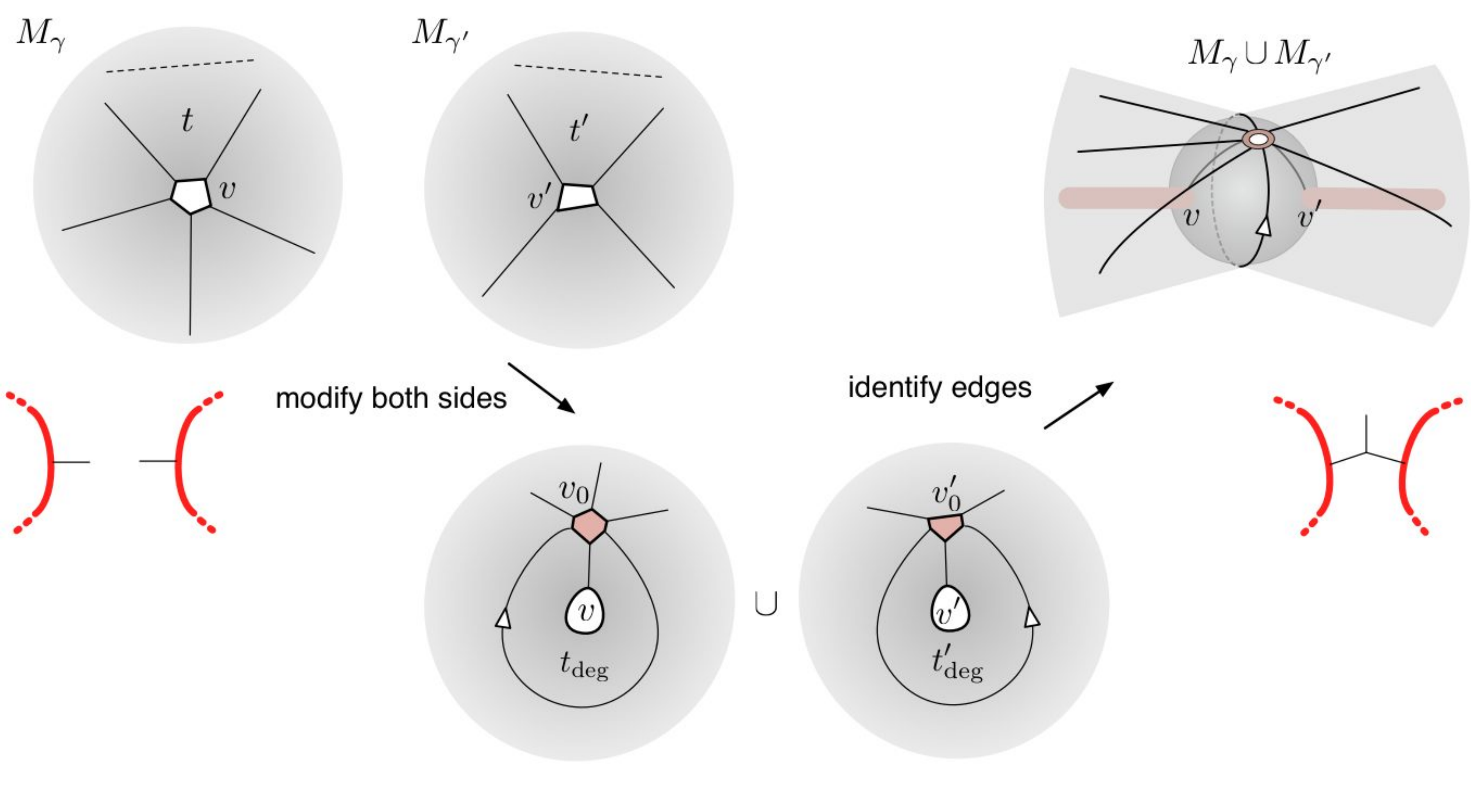}\vspace{-.3in}
\caption{Gluing pieces of RG manifolds together to produce a trinion in the skeleton.}
\label{fig:connect-t}
\end{figure}

Select a trinion, and two of its legs that connect to enhanced paths. (If a third leg also connects to an enhanced path, we leave it alone for now.) Let $v$, $v'$ be the corresponding holes on the IR boundaries of $M_\gamma$ and $M_{\gamma'}$ that are to be connected. $M_\gamma$ may coincide with $M_{\gamma'}$. Perform a degenerate modification of each hole $v,v'$ as in Figure \ref{fig:connect-t}, introducing two new degenerate triangles $t_{\rm deg}$, $t_{\rm deg}'$ on the IR boundaries, along with two new holes $v_0,v_0''$ filled by small discs. Then identify the circular edge of $t_{\rm deg}$ with the circular edge of $t_{\rm deg}'$. This traps a 3-punctured sphere inside a new manifold $M_\gamma\cup M_{\gamma'}$; two of its holes are identified with $v, v'$, and the third hole, an amalgamation of $v_0$ and $v_0''$, connects to the new IR boundary. We have reproduced the trinion.

We repeat this procedure for all trinions, attaching two of their three legs. If in the skeleton of $\CC$ there were no single un-enhanced edges (b) or trinions (c) with all three legs on enhanced paths, we are done. Otherwise, we arrive a new collection of manifolds $\wt M_i$, which, in terms of the skeleton, must be connected to each other by single un-enhanced internal edges. (These might be either isolated un-enhanced edges, or the third legs of trinions that we neglected to connect above.) Since $\CC$ has $h>0$ punctures, at least one of the $\wt M_i$ has a hole $v_0$ that doesn't get connected to anything. Call this manifold $\wt M^{(0)}$.

\begin{figure}[htb]
\centering
\includegraphics[width=5.5in]{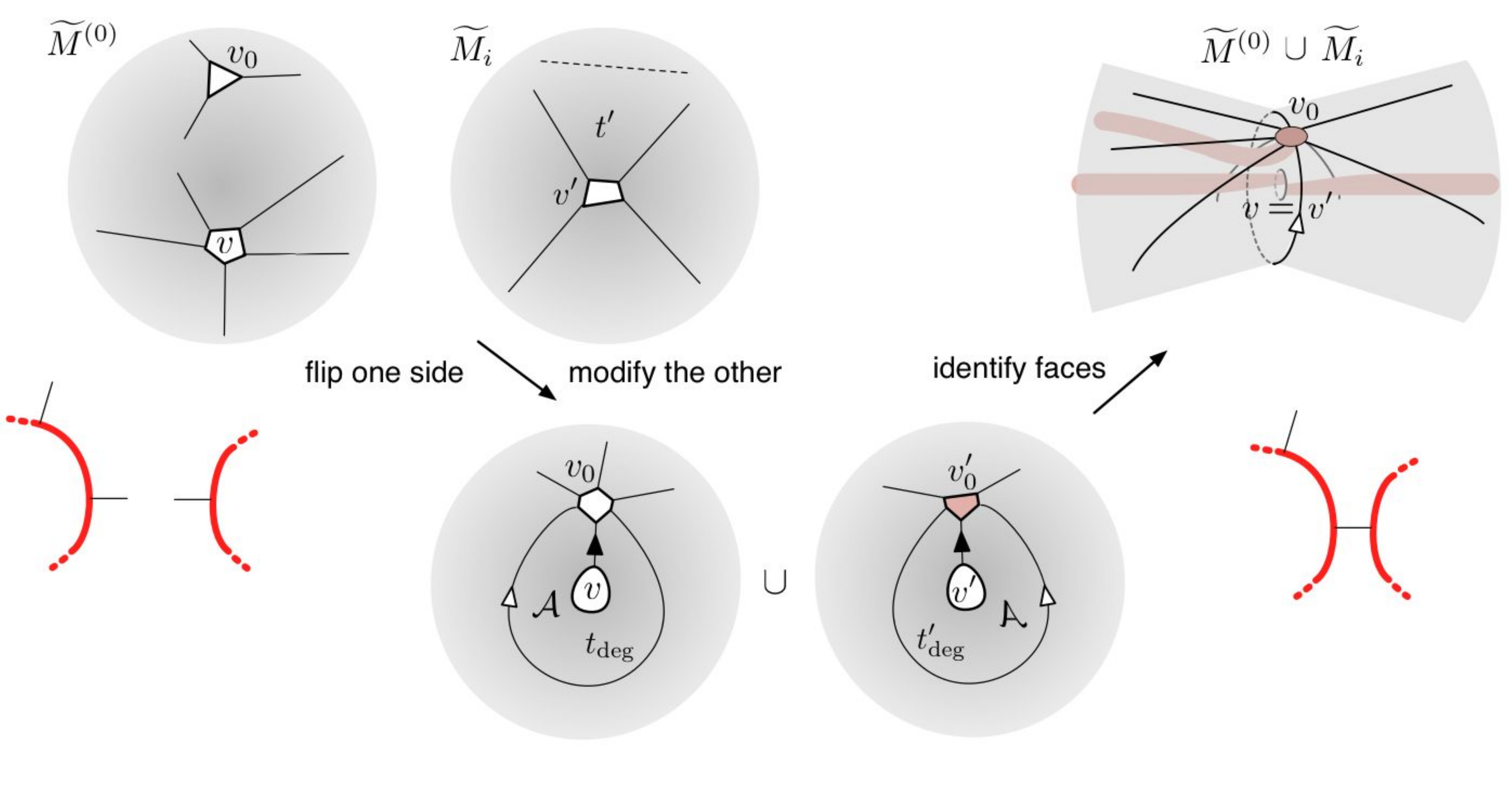}\vspace{-.3in}
\caption{Gluing pieces of RG manifolds together to produce a direct connection in the skeleton. This must be done in the presence of an external puncture.}
\label{fig:connect-d}
\end{figure}

We now choose pairs of holes $v,v'$ on the $\wt M_i$ and glue them together. Sometimes this is obvious and easy to do. Otherwise, we proceed systematically as follows.
We require that at least one of $v,v'$ (say $v$) lies on $\wt M^{(0)}$, or, after subsequent gluing, on the connected 3-manifold that contains $\wt M^{(0)}$. To glue, we flip the big-boundary triangulation of $\wt M^{(0)}$ to create a degenerate triangle $t_{\rm deg}$ surrounding $v$ (with additional vertex at $v_0$). We also perform a degenerate modification of $v'$ (Figure \ref{fig:modify}), creating a degenerate triangle $t_{\rm deg}'$ surrounding $v'$. Then we identify the faces of $t_{\rm deg}$ and $t_{\rm deg}'$, as in Figure \ref{fig:connect-d}. The connection is made, and we repeat.

\subsubsection*{Step 5: Touching up}

We arrive at a triangulated framed 3-manifold $M_{\mb p}$ representing the RG wall. It may have two technical problems.

The separation between the UV and IR boundaries may be ``thin'' in places, where only the edges of tetrahedra were identified to create $M_{\rm p}$. We fix this as usual either by flipping the IR boundary or by flipping 3-punctured spheres on the UV boundary.

The gluing of holes corresponding to single un-enhanced edges of the skeleton of $\CC$ may also create ``hard'' internal edges in the 3d triangulation of $M_{\mb p}$, in the sense of \cite[Sec. 4.1]{DGG}. These are not a problem for mathematical applications or computations of partition functions, line operators, etc. However, in order to define the full 3d gauge theory $T_2[M_{\mb p}]$, the ``hard'' edges must be removed by changing the 3d triangulation, \ie\ by performing a sequence of 2--3 moves.

%%%%%%%%%%%%%%%%%%%%%%%%%%%%%%%%%%%%%%%%%%%%%%%%%%%%%%%%%%%%%%%%%%%%%%%%%%%%%%%

\section{Quantum dilogarithms}
\label{app:qdl}

We briefly recall the definition of the quantum dilogarithm function $\Phi_\hbar(Z)$. It is used to build $SL(K,\R)$--like Chern-Simons wavefunctions for framed 3-manifolds $M$, and ellipsoid partition functions for theories $T_K[M]$. (See \cite{DGG-index} for the related quantum dilogarithm used to build 3d indices.)

For $Z\in \C$ and $\Re\,\hbar<0$ we set
\be \label{qdl-def}
\Phi_\hbar(Z) = \prod_{r\geq 0} \frac{1-q^{r+\frac12}e^Z}
{1-\tilde q^{-r-\frac12}e^{\tilde Z}}\,,\qquad q=e^{\hbar}\,,\quad \tilde q=e^{-\frac{4\pi^2}{\hbar}}\,,\quad \tilde Z=\frac{2\pi i}{\hbar}Z\,.
\ee
This is basically the double-gamma function of Barnes \cite{Barnes-qdl}, rediscovered by Faddeev when studying nonperturbative completions of Weyl algebras \cite{Fad-modular}. It has many remarkable properties, most of which we will not discuss here (see \eg\ \cite{DGLZ, FG-qdl-cluster} for reviews). We observe, though, that $\Phi_\hbar(Z)$ can be extended to a meromorphic function of $Z\in \C$ for all $\hbar$ in the cut plane $\C\bs \{i\R_{\leq 0}\}$, and in particular for $\hbar\in i\R_{>0}$. The poles and zeroes of \eqref{qdl-def} can easily be found from the product \eqref{qdl-def}.

The function $\Phi_\hbar(i\pi+\frac\hbar2-Z)$ is the $SL(2)$ Chern-Simons wavefunction of a tetrahedron, in a certain natural polarization $\Pi_z$ \cite{Dimofte-QRS}. It is also the ellipsoid ($S^3_b$) partition function of the tetrahedron theory $T_2[\Delta,\mb t,\Pi_Z]$, the theory of a free 3d chiral multiplet $\phi$ with $U(1)_R$ charge zero, charge $+1$ under a $U(1)_F$ flavor symmetry, and a (supersymmetrized) background Chern-Simons coupling at level -1/2 for the combination $A_F-A_R$ of $U(1)_F$ and $U(1)_R$ gauge fields. To interpret $\Phi_\hbar(i\pi+\frac\hbar2-Z)$ as an $S^3_b$ partition function, one identifies $\hbar=2\pi ib^2$ ($b\in \R$) and sets the effective complexified mass of $\phi$ on $S^3_b$ to be $m_Z=2\pi b Z$.

One finds several other forms of $\Phi_\hbar(Z)$ in the math and physics literature. In particular, one encounters a ``double sine'' function $s_b$ in the study of Liouville theory and representation theory of $U_q(sl_2)$, as well as $S^3_b$ partition functions. It is related to $\Phi_\hbar(Z)$ as
\be \label{defsb}
 s_b(m) = \exp\big[ -\tfrac{i\pi}{24}(b^2+b^{-2})-\tfrac{i\pi}{2}m^2\big]\, \Phi_{2\pi ib^2}(2\pi b m)\,.\ee
The function $s_b(m)$ has a balanced reflection property: $s_b(-m)=s_b(m)^{-1}$. In \cite{TV-6j} (which is used for some comparisons in Section \ref{sec:N4}) the authors use a non-standard convention, replacing $s_b(m)$ with $s_b(m)^{-1}$. The easiest way to distinguish definitions of the quantum dilogarithm is to compare lattices of poles and zeroes.

\bibliographystyle{JHEP_TD}
\bibliography{toolbox}

\end{document}